\documentclass[12pt]{article}

\newlength{\abstractwidth}
\usepackage{amsmath, nccmath}
\usepackage{amsfonts}
\usepackage{amssymb}
\usepackage{latexsym}

\usepackage{color}
\usepackage{graphicx}
\usepackage{tikz}
\usepackage{dsfont}

\usetikzlibrary{arrows,shapes,positioning}
\usetikzlibrary{decorations.markings}
\usepackage[rightcaption]{sidecap}
\tikzstyle arrowstyle=[scale=1]
\tikzstyle directed=[postaction={decorate,decoration={markings,
    mark=at position .65 with {\arrow[arrowstyle]{stealth}}}}]
\tikzstyle reverse directed=[postaction={decorate,decoration={markings,
    mark=at position .65 with {\arrowreversed[arrowstyle]{stealth};}}}]
\usetikzlibrary{positioning}

\usepackage{hyperref}
\definecolor{darkred}{rgb}{0.8,0.1,0.1}
\hypersetup{colorlinks=true, linkcolor=darkred, citecolor=blue, linktoc=page}

\renewcommand{\thefootnote}{\fnsymbol{footnote}}
\renewcommand{\thanks}[1]{\footnote{#1}}
\newcommand{\starttext}{
\setcounter{footnote}{0}
\setcounter{section}{0}
\renewcommand{\thefootnote}{\arabic{footnote}}}

\newcommand{\bea}{\begin{eqnarray}}
\newcommand{\eea}{\end{eqnarray}}
\newcommand{\be}{\begin{eqnarray}}
\newcommand{\ee}{\end{eqnarray}}
\newcommand{\bma}{\begin{matrix}}
\newcommand{\ema}{\cr\end{matrix}}
\newcommand{\<}{\langle}
\renewcommand{\>}{\rangle}

\newtheorem{thm}{Theorem}[section]
\newtheorem{lem}[thm]{Lemma}
\newtheorem{prop}[thm]{Proposition}
\newtheorem{cor}[thm]{Corollary}

\def\cH{{\cal H}}
\def\cI{{\cal I}}

\def\cO{{\cal O}}

\def\cS{{\cal S}}
\def\cT{{\cal T}}

\def\bA{{\bf A}}
\def\bB{{\bf B}}

\def\bH{{\bf H}}

\def\bK{{\bf K}}
\def\bL{{\bf L}}
\def\bM{{\bf M}}
\def\bN{{\bf N}}

\def\bS{{\bf S}}

\def\bX{{\bf X}}
\def\bY{{\bf Y}}

\def\bd{{\bf d}}

\def\bu{{\bf u}}

\def\bx{{\bf x}}

\def\mA{\mathfrak{A}}
\def\mB{\mathfrak{B}}
\def\mC{\mathfrak{C}}

\def\mJ{\mathfrak{J}}

\def\mM{\mathfrak{M}}
\def\mN{\mathfrak{N}}

\def\mS{\mathfrak{S}}

\def\ZZ{{\mathbb Z}}
\def\RR{{\mathbb R}}
\def\NN{{\mathbb N}}
\def\CC{{\mathbb C}}

\def\det{{\rm det \,}}

\def\half{{1\over 2}}
\def\thalf{{\tfrac{1}{2}}}
\def\p{\partial}

\def\a{\alpha}
\def\b{\beta}

\def\eps{\epsilon}

\def\tet{\vartheta}
\def\ep{\varepsilon}
\def\om{\omega}

\def\ndots{\cdots}

\def\see{{\boldsymbol \ell}}
\def\CS{{\bf M}_1}
\def\CM{{\bf M}}
\def\SZ{C_\delta^{\rm sym}}

\def\BL{{\cal T}}
\def\mux{\kappa^{(n)}}
\def\LL{\boldsymbol{\Lambda}}
\def\FF{F}
\def\bee{L}
\def\eecyc{{\cal L}}

\def\pbz{\p _{\bar z}}

\def\no{\nonumber}
\def\sm{\smallskip}

\begin{document}
\starttext
\setcounter{footnote}{0}

\begin{flushright}
2022 November 16  \\
revised 2023 Jan 16\\
UUITP-51/22
\end{flushright}

\bigskip

\begin{center}

{\Large \bf Cyclic products of Szeg\"o kernels}

\vskip 0.1in

{\Large \bf and spin structure sums}

\vskip 0.1in

{ \bf Part I: hyper-elliptic formulation }

\vskip 0.4in

{\large Eric D'Hoker$^{(a)}$, Martijn Hidding$^{(b)}$, and Oliver Schlotterer$^{(b)}$}

\vskip 0.1in

 ${}^{(a)}$ {\sl Mani L. Bhaumik Institute for Theoretical Physics}\\
 { \sl Department of Physics and Astronomy }\\
{\sl University of California, Los Angeles, CA 90095, USA}\\

\vskip 0.1in

 ${}^{(b)}$ { \sl Department of Physics and Astronomy,} \\ {\sl Uppsala University, 75108 Uppsala, Sweden}
 
 \vskip 0.1in
 
{\tt dhoker@physics.ucla.edu},  {\tt martijn.hidding@physics.uu.se}, {\tt oliver.schlotterer@physics.uu.se}

\hskip 0.5in

\begin{abstract}
The summation over spin structures, which is required to implement the GSO projection in the RNS formulation of superstring theories, often presents a significant impediment to the explicit evaluation of superstring amplitudes.   In this paper we discover that, for Riemann surfaces of genus two and even spin structures, a collection of novel identities leads to a dramatic simplification of the spin structure sum.  Explicit formulas for an arbitrary number of vertex points are obtained in two steps. First, we show that the spin structure dependence of a cyclic product of Szeg\"o kernels (i.e.\ Dirac propagators for worldsheet fermions)  may be reduced to the spin structure dependence of the four-point function. Of particular importance are certain \textit{trilinear relations} that we shall define and prove.  In a second step, the known expressions for the genus-two even spin structure measure are used to perform the remaining spin structure sums. The dependence of the spin summand on the vertex points is reduced to simple building blocks that can already be identified from the two-point function. The  hyper-elliptic formulation of genus-two Riemann surfaces is used to derive these results, and its $SL(2,\mathbb C)$ covariance is employed to organize the calculations and the structure of the final formulas. The translation of these results into the language of Riemann $\vartheta$-functions, and applications to the evaluation of higher-point string amplitudes,  are relegated to  subsequent companion papers.
\end{abstract}

\end{center}

\newpage

\setcounter{tocdepth}{2} 
\tableofcontents

\newpage

\baselineskip=16pt
\setcounter{equation}{0}
\setcounter{footnote}{0}

\newpage

\section{Introduction}
\label{sec:intro}
\setcounter{equation}{0}

In the Ramond-Neveu-Schwarz (RNS) formulation of superstring theory, supersymmetry of the spectrum and of  amplitudes is achieved by implementing the Gliozzi-Scherk-Olive (GSO) projection.  The gauge sector of the Heterotic string in terms of 32 Majorana-Weyl fermions similarly requires an extension of the GSO projection 
to produce the $E_8 \times E_8$ or the ${\rm Spin}(32)/\ZZ_2$ anomaly free gauge groups.
In both cases, the projections are implemented by summing over the spin structures of the corresponding worldsheet fermions.  On a Riemann surface of genus $g$, the number of different spin structures is $2^{2g}$ and grows rapidly with increasing genus.\footnote{We recall that, for arbitrary genus, a spin structure is even (resp.\ odd) if and only if the number of Dirac zero modes of a single worldsheet spinor is even (resp.\ odd). The number of even spin structures is $2^{g-1} (2^{g}+1) $ while the number of odd spin structures is $2^{g-1}(2^g-1)$, amounting to 10 even and 6 odd in genus two. For all points in the genus-two moduli space, and at generic points for higher genus, the number of zero modes is actually~0 for even spin structures, and~1 for odd spin structures.} For genus one, a plethora of readily available Jacobi $\tet$-function identities (see for example \cite{Mumford}) greatly facilitates carrying out the summation explicitly, as was shown in \cite{Tsuchiya:1988va, Stieberger:2002wk, Bianchi:2006nf, Tsuchiya:2012nf, Broedel:2014vla, Berg:2016wux} for computations with external NS-sector states and \cite{Atick:1986rs, Lin:1988xb, Lee:2017ujn} for external R-sector states. As soon as the genus exceeds one, however, the corresponding identities between Riemann $\tet$-functions are considerably more involved and not necessarily available. As a result, the need to carry out the summation over spin structures is often regarded as a drawback of string amplitude computations in the RNS formulation. 

\sm

In this paper, we shall introduce a collection of novel identities for genus-two Riemann surfaces that dramatically simplify  the sum over even spin structures in multi-particle amplitudes. Our techniques apply to the spin structure sums for an arbitrary number of massless  external NS states such as gravitons and, with minor modifications, also to massive external NS states. The even spin structure contribution to the chiral amplitude of these states precisely corresponds to the even parity contribution to the chiral amplitude, while the odd spin structure part corresponds to the odd parity part. Thus, our results will apply to the  even parity part of Type I, Type II, and Heterotic string amplitudes with external NS-sector states.

\sm

In the modern approach to genus-two string amplitudes in the RNS formulation based on chiral splitting and the super-period matrix \cite{RMP,DHoker:1989cxq} (reviewed in the lecture notes \cite{DHoker:2002hof}), the even spin structure dependence enters via three different ingredients. Firstly, spin structure dependence  enters through the measure factor, which is universal and independent of the number of external states; it was evaluated in \cite{DP1,DP2,DP4} and re-derived using purely algebraic geometry methods in \cite{Witten:2013tpa}. Secondly, spin structure dependence enters through the Szeg\"o kernel (namely the Dirac propagator of the worldsheet fermions), which is used to evaluate the correlators of the NS vertex operators and worldsheet supercurrent operators \cite{DP5,DP6}. Thirdly, spin structure dependence enters through the gauge-choice made for the worldsheet gravitino field. Unitary gauge  in which the gravitino slice is supported at two spin-structure independent points $q_1, q_2$,  was used successfully in the explicit calculation of the 4-point functions in \cite{DP5,DP6} and again in the calculation of the 5-point functions in \cite{DHoker:2021kks}.\footnote{See  also \cite{DHoker:2020prr} for an earlier construction of genus-two 5-point amplitudes based on pure-spinor methods and chiral splitting as well as \cite{DHoker:2020tcq} for their low-energy expansion and S-duality properties.} The spin structure dependence of the Szeg\"o kernels enters either via cyclic products of Szeg\"o kernels, or via a concatenated product of Szeg\"o kernels along a linear chain stretching from $q_1,$ to $q_2$. It is the spin structure dependence of the cyclic product of Szeg\"o kernels that will be analyzed  in this paper, while that of the linear chain will be relegated to a future paper.

\sm

The protagonist of this paper is the cyclic product of $n$ Szeg\"o kernels $S_\delta(z_i,z_j)$, 
anchored at generic insertion points $z_i, z_j\in \Sigma$ of worldsheet fermions ($i,j=1,2,\cdots, n$)
on an arbitrary  genus-two Riemann surface $\Sigma$ with an arbitrary even spin structure~$\delta$, denoted as follows,
\bea
\label{Cdelta}
C_\delta (z_1, z_2,\cdots, z_n) = S_\delta(z_1,z_2) S_\delta(z_2,z_3) \cdots S_\delta(z_{n-1},z_n) S_\delta(z_n,z_1)
\eea
We recall from \cite{RMP,fay,Alvarez-Gaume:1986rcs} that the Szeg\"o kernel $S_\delta(z,w)$ is a $(\thalf, 0)$ form in $z$ and in $w$ which, for even spin structure $\delta$, is defined as the inverse of the chiral Dirac operator on $\Sigma$,
\bea
\pbz S_\delta(z,w) = 2 \pi \delta (z,w)
\eea
For genus two and even spin structure $\delta$ the inverse is well-defined throughout moduli space since the Dirac operator has no zero modes.  The explicit expression for $S_\delta(z,w)$  will be given in the hyper-elliptic formulation in (\ref{Szego}) below.

\sm

The first key result of this paper will be to prove that the spin structure dependence of the cyclic products $C_\delta$ may be reduced to a universal form which is independent of the number $n$ of insertion points. Specifically, we shall produce an explicit algorithm to reduce all dependence of $C_\delta$ on the spin structure $\delta$ to \textit{quadratic polynomials} in $z_i$-independent universal building blocks $\see_\delta^{ab}$ symmetric in its $SL(2,\mathbb C)$ indices $a,b=1,2$ to be defined below. The remaining dependence on the points $z_i$ is through spin structure independent functions and differential forms that we shall construct. We shall present explicit formulas for this reduction of $C_\delta$ in the cases $n=2,3,4,5,6,7,8$.

\sm

The second key result of this paper will be to produce explicit formulas for the spin structure sums of the remaining spin structure dependence, namely through {\it quadratic polynomials} in $ \see_\delta^{ab}$. The spin structure sums will be carried out against the genus-two superstring measure for the supersymmetric chiral part of the Type II or Heterotic strings and against the measure for the fermions representing the gauge algebras for the Heterotic strings. Actually, these sums reduce to those of the $n\leq 4$ case, which are well-known. Prior to the present work, the spin structure sums for $n \leq 5$ had been carried out at the cost of laborious calculations using the Riemann identities, Fay identities, and other specialized relations \cite{DP6,DHoker:2021kks}, while for $n>5$ they were beyond the reach of available methods.

\sm

The constructions leading to the above results will be carried out in the hyper-elliptic formulation of genus-two Riemann surfaces, which we summarize in section \ref{sec:hyper} below. A fundamental role will be played by the $SL(2,\CC)$ group-theoretic structure of the hyper-elliptic formulation. The results will subsequently be translated into the language of Riemann $\tet$-functions, the prime form, and Thomae-type formulas in a companion paper \cite{DHS}.

\sm

The corresponding reduction of the cyclic products of Szeg\"o kernels at genus one was obtained by Tsuchiya in \cite{Tsuchiya:2012nf, Tsuchiya:2017joo}, where the role of the $\see_\delta^{ab}$ is played by the three branch points $e_\delta$ of the genus-one curve and the role of the trilinear relations is played by the cubic equation $4 e_\delta ^3 - g_2 e_\delta - g_3=0$ they satisfy. Along with \cite{Tsuchiya:2022lqv} these references also present partial results for  genus two, and propose a reduction of the spin structure dependence of $C_\delta$ to  polynomials of degree $\lfloor n/2 \rfloor$ in a quantity $P_{IJ}(\Omega_\delta)$ that is similar to our $\see_\delta^{ab}$. 
However, neither the further reduction of the spin structure dependence via trilinear relations, nor a general algorithm for obtaining the coefficients of the polynomials in $P_{IJ}(\Omega_\delta)$, nor the explicit form of the spin structure independent terms, nor the summation over spin structures beyond five points, was obtained there.

\subsection*{Organization}

The remainder of this paper is organized as follows. In section \ref{sec:hyper}, we present a brief review of the hyper-elliptic formulation of genus-two Riemann surfaces, including the Szeg\"o kernel for even spin structures, and the role of modular and $SL(2,\CC)$ transformations. In section~\ref{sec:cyclic}, we formulate the problem of reducing the spin structure dependence of the cyclic product $C_\delta$ to the universal building blocks $\see_\delta ^{ab}$. In section \ref{sec:trilinear}, we prove the \textit{trilinear relations}.  In section~\ref{sec:low-n}, we carry out the reduction of the spin structure dependence of $C_\delta$ to the universal form, explicitly for the cases $n=3,4,5,6$ (see appendix \ref{apponCdelta} for $n=7,8$), and  in the form of a conjecture for the case of arbitrary $n$. Finally, in section \ref{sec:sum}, we use the results obtained in the preceding sections for arbitrary $n$ to sum $C_\delta$  over spin structures against the measure of the chiral supersymmetric sector and the Heterotic chiral gauge sector. A crucial role will be played by finite-dimensional tensors and representations of the group  $SL(2,\CC)$ of conformal automorphisms of the Riemann sphere underlying the hyper-elliptic construction; its representation theory will be reviewed in appendix \ref{sec:A}. Various derivations, proofs and examples are relegated to further appendices.

\subsection*{Main theorems}

In order to complement the organization section above, we gather here a summary of the main theorems where we obtain the reduction of the spin structure dependence of cyclic products $C_\delta$ of Szeg\"o kernels and the spin structure sums over these cyclic products. 
\begin{itemize}
\itemsep = 0.0in
\item In Theorem \ref{thm:3.1} all spin structure dependence of the cyclic product $C_\delta(1, \cdots, n)$ of an arbitrary number $n$ of Szeg\"o kernels is reduced to certain polynomials  $Q_\delta$. 
\item In Theorem \ref{thm:1} all spin structure dependence for the two-point function $C_\delta(i,j)$ is reduced to a symmetric bi-holomorphic form $L_\delta(i,j)$, which will be the fundamental building block for all spin structure dependence of  cyclic products of Szeg\"o kernels  for arbitrary~$n$. The  bi-holomorphic form $L_\delta(i,j)$ uniquely corresponds to  a spin structure dependent  rank-two symmetric $SL(2,\CC)$ tensor $\see_\delta$.  A key ingredient in the spin structure independent contribution for arbitrary $n$ is a polynomial $Z(i,j)$ which is determined already by the two-point function. An explicit reduction of the polynomials $Q_\delta$ to simple combinations of $L_\delta(i,j)$ and $Z(i,j)$ can be found in the all-multiplicity conjecture (\ref{Qdel.17}).
\item The trilinear relations in components of $\see_\delta$, obtained in Theorem \ref{4.tril} and Corollary~\ref{sping3.55}, are simplified in the form of $SL(2,\CC)$ covariant tensorial relations in Theorem \ref{thm:6}. This theorem guarantees that all spin structure dependence of an arbitrary cyclic product of Szeg\"o kernels $C_\delta(1,\cdots, n)$ may be reduced to degree-two polynomials in components of $\see_\delta$ with $\delta$-independent coefficients.
\item The spin structure sums of monomials in $\see_\delta$ of degree $\leq 2$ against the measures relevant to Type I, Type II and Heterotic strings are obtained in  Theorems \ref{thm:6.1}, \ref{thm:6.11}, and \ref{thm:6.10}. The analogous sums for higher powers of $\see_\delta$ may then be deduced using the trilinear relations, and are presented for the supersymmetric sector in Corollaries \ref{cor:6.2}---\ref{cor:6.4}.
\item Using the spin structure sums over multi-linears in $\see_\delta$ we obtain the spin structure sums of cyclic products of Szeg\"o kernels for the supersymmetric sector in Theorems \ref{thm:6.3} for up to five points; \ref{thm:6.7} for six points;  \ref{thm:6.8} for seven points; and \ref{thm:6.9} for eight points. Spin structure sums for products of cyclic products are evaluated for the supersymmetric measure in Corollaries \ref{cor:6.15} and \ref{cor:6.16} for up to six points. The resulting expressions contain Parke-Taylor type poles and the key polynomial $Z(i,j)$.
\end{itemize}

\newpage

\subsection*{Acknowledgments}

The research of ED is supported in part by NSF grants PHY-19-14412 and  PHY-22-09700.
The research of MH and OS is supported by the European Research Council under ERC-STG-804286 UNISCAMP. 
MH and OS are grateful to UCLA and the  Mani Bhaumik Institute for kind hospitality and creating a stimulating 
atmosphere during initiation of this work. We gratefully acknowledge the hospitality of the KITP during early stages of this work,  and support from National Science Foundation  grant PHY-17-48958.

\newpage

\section{The hyper-elliptic representation for genus two}
\label{sec:hyper}
\setcounter{equation}{0}

Every genus-two Riemann surface $\Sigma$ is hyper-elliptic:  it  may be represented by a double cover of the Riemann sphere $\hat \CC= \CC \cup \{ \infty\}$ ramified over six branch points $u_1 ,  \cdots, u_6 \in \hat \CC$. In the hyper-elliptic  representation, every point $z \in \Sigma$  may be parametrized by a pair $z=(x,s)$ where $x \in \hat \CC$ and $s^2$ is given in terms of $x$ and $u_r$ by,
\bea
\label{shyper}
s^2 = \prod _{r=1}^6 (x-u_r)
\eea
Away from the branch points $(u_r,0)$ for $r=1,\cdots, 6$, every point $x \in \hat \CC$ maps to two distinct points $(x, \pm s)$ in $\Sigma$ corresponding to the two possible signs of $s$ given that $s^2$ is fixed in terms of $x$ and $u_r$ by (\ref{shyper}). As a result,  $\hat \CC$ maps to the two sheets of $\Sigma$ which intersect at the six branch points $(u_r,0)$, as shown in Figure \ref{fig:1}.  In particular, each sheet has its own point at infinity $P_{\pm \infty}$.   The  surface $\Sigma$ is invariant under the holomorphic involution $\cI$ that interchanges the two sheets of $\Sigma$ and acts by $\cI (x, s) = (x, -s)$ where $s$ obeys (\ref{shyper}). 

\sm 

\begin{figure}[h]
\begin{center}
\tikzpicture[scale=1]
\scope[xshift=10cm,yshift=0cm]
\draw[very thick] (0,0) -- (12,0);
\draw[very thick] (0,0) -- (3,4);
\draw[very thick] (12,0) -- (15,4);
\draw[very thick] (3,4) -- (15,4);
\draw[very thick] (0,-0.5) -- (12,-0.5);
\draw[very thick] (0,-0.5) -- (0.4,0);
\draw[thick, dashed] (0.4,0) -- (3,3.5);
\draw[very thick] (12,-0.5) -- (15,3.5);
\draw[thick, dashed] (3,3.5) -- (14.6,3.5);
\draw[very thick] (14.6,3.5) -- (15,3.5);
\draw (11.6,0.3) node{\small $P_{+\infty}$};
\draw (11.5,-0.27) node{\small $P_{-\infty}$};
\draw[very thick] (3,1) -- (3.5,2);
\draw[very thick] (6,2.5) -- (7,1.5);
\draw[very thick] (9,1) -- (10.5,2.5);
\draw[thick, color=blue,  rotate=-25] (2.2,2.7)  ellipse (19pt and 33pt);
\draw[thick, color=blue,  rotate=40] (6.4,-2.6)  ellipse (23pt and 36pt);

\draw[thick, color=red,  rotate=00] (3.5,2)  arc (140:75:2.5 and 1.5);
\draw[thick, color=red, dotted,  rotate=00] (3.5,2)  arc (-90:-35:3 and 1);
\draw[thick, color=red,  rotate=00] (7,1.5)  arc (83:43:3.2 and 1.5);
\draw[thick, color=red, dotted,  rotate=00] (7,1.5)  arc (-160:-55:1.4 and 1);
\draw [blue] (3.8,0.6) node{\small $\hat \mA_1$};
\draw [blue] (6.1,0.95) node{\small $\hat \mA_2$};
\draw[red]  (4.4,2.8) node{\small $\mB_1$};
\draw[red]  (8.4,1.65) node{\small $\mB_2$};
\draw (3,1) node{$\bullet$};
\draw (3.5,2) node{$\bullet$};
\draw (6,2.5) node{$\bullet$};
\draw (7,1.5) node{$\bullet$};
\draw (9,1) node{$\bullet$};
\draw (10.5,2.5) node{$\bullet$};
\draw (2.7,1.1) node{\small $u_1$};
\draw (3.1,2) node{\small $u_2$};
\draw (7.3,1.8) node{\small $u_4$};
\draw (6.4,2.6) node{\small $u_3$};
\draw (9.6,0.9) node{\small $u_5$};
\draw (11,2.4) node{\small $u_6$};
\draw (14.3,1.3) node{\large $\Sigma$};
\draw[thick, blue,->] (3.824,1.5) -- (3.825,1.51);
\draw[thick, blue,->] (7.31,2.5) -- (7.3,2.51);
\draw[thick, red,->] (4.51,2.45) -- (4.5,2.45);
\draw[thick, red,->] (8.01,1.35) -- (8,1.352);
\endscope
\endtikzpicture
\end{center}
\caption{The hyper-elliptic curve $\Sigma$ is represented in terms of a double cover of the Riemann sphere $\hat \CC$, with distinct points at infinity $P_{\pm \infty}$. A choice of branch cuts and canonical homology cycles $\mA_1, \mA_2, \mB_1, \mB_2$ is indicated in terms of the cycles $\mA_1=\hat \mA_1$ and $\mA_2 = \hat \mA_2-\hat \mA_1$. \label{fig:1}}
\end{figure}

\subsection{$SL(2,\CC)$ transformations and tensors}

In this paper, a fundamental role will be played by the group $SL(2,\CC)$ of conformal automorphisms of the Riemann sphere $\hat \CC$ which is isomorphic to the complex projective space ${\rm CP}^1$. The points in ${\rm CP}^1$ may be parametrized by a doublet $\bx$ of complex homogeneous coordinates subject to the equivalence relation $\sim$ of rescaling by a non-zero complex number $\lambda$,
\bea
\hat \CC = {\rm CP}^1= \left \{ \bx 
= \left ( \bma \bx^1 \cr  \bx^2 \ema \right )  \in \CC^2 \setminus  \{ 0   \} , \, \hbox{ with } \lambda \bx \sim \bx, \, \lambda \in \CC \setminus \{ 0 \} \right \}
\eea
The group $SL(2,\CC)$ of conformal automorphisms of $\hat \CC$ acts linearly on doublets,
\bea
\label{2.gam}
\bx \to \gamma \bx \hskip 1in \gamma = \left ( \bma a & b \cr c & d \ema \right ) \in SL(2,\CC)
\eea
The tensor $\ep_{ab}=-\ep_{ba}$, normalized by $\ep_{12}=1$, defines the anti-symmetric pairing $\bx_1^a \, \ep _{ab} \, \bx_2^b $, and may be used to lower and raise doublet indices,\footnote{Throughout, 
we shall use the Einstein convention for the summation over repeated upper and lower doublet indices $a,b=1,2$ or $I,J=1,2$. The relation $\ep_{ab} = \ep_{ac} \, \ep_{bd} \, \ep^{cd}$ along with $\ep_{12}=1$ implies  $\ep^{12}=1$.}
\bea
\bx_a = \ep _{ab} \, \bx^b \hskip 1in \bx ^a = - \ep^{ab} \, \bx_b = \bx_b \, \ep ^{ba}
\label{defbx}
\eea 
In the standard manner, the raising and lowering operations on doublet indices may be generalized by tensor products to tensors of arbitrary rank.

\sm

In the coordinate patch on $\hat \CC$ where $\bx^2 \not=0$, we may parametrize a doublet $\bx$ in terms of the ratio $x=\bx^1/\bx^2$ by choosing $\lambda = 1/\bx^2$ in the definition of ${\rm CP}^1$. The doublets for a generic point $\bx$ and for the branch points $\bu_r$ for $r=1,\cdots, 6$ are then given as follows, 
\bea
\bx = \left ( \bma \bx^1 \cr \bx^2 \ema \right )  = \left ( \bma x \cr 1 \ema \right ) 
\hskip 1in 
\bu_r = \left ( \bma \bu_r^1 \cr \bu_r^2 \ema \right ) = \left ( \bma u_r \cr 1 \ema \right )
\label{doubleteq}
\eea
In this coordinate patch all of $\hat \CC$ is covered except for the point at infinity; the anti-symmetric pairing reduces to the difference of their top entry, $\bx_1^a \, \ep _{ab} \, \bx_2^b = x_1 - x_2$; the doublets transform as follows under $SL(2,\CC)$, 
\bea
\bx \to  (cx+d)^{-1} \, \gamma \bx
\hskip 1in
\bu_r \to (cu_r +d)^{-1} \, \gamma \bu_r
\eea
which reduce to the familiar M\"obius transformation rules for the top entries of the doublets, 
\bea
\label{xuSL}
x \to  { ax + b \over cx + d} 
\hskip 0.9in 
u_r \to  { a u_r + b \over c u_r +d}
\eea
The difference and the differential transform by, 
\bea
\label{dSL}
x_1 - x_2 \to { x_1 - x_2 \over (cx_1+d) (cx_2+d)}
\hskip 1in
dx \to { dx \over (cx+d)^2} 
\eea
while the function $s(x)$, used to define the genus-two curve in (\ref{shyper}), transforms as follows,
\bea
\label{sSL}
s(x) \to { s(x) \over (cx+d)^3 J_\gamma}
\hskip 0.9in
J_\gamma ^2 = \prod _{r=1}^6 (cu_r+d)
\eea
$SL(2,\CC)$ transformations are generated by infinitesimal translations $\BL$, dilatations, and special conformal transformations. The latter may be obtained from translations $\BL$ by applying the inversion $\cS: x \to -1/x$ and thus are generated by $\cS \BL \cS$, while dilations may be obtained from the commutator of $\BL$ and $\cS \BL \cS$.  Thus,  all infinitesimal $SL(2,\CC)$ transformations are generated  by combining $\BL $ and $\cS$, whose action on $x$ and $u_r$ is given by,
\begin{align}
\label{SLTS}
\BL x  & =  \epsilon & \BL u_r  & =  \epsilon  & 
 \cS x & = -1/x &
 \cS u_r & = - 1/u_r
\end{align}
Throughout, it will be convenient to verify $SL(2,\CC)$ invariance or covariance properties by analyzing the behavior under the transformations  $\BL$ and $\cS$. We may set $\epsilon =1$ and  recall that $\BL$ acts as a vector field via Leibniz's rule of differentiation. Since there is
no infinitesimal version of the inversion ${\cal S}$, its action on products departs from
the Leibniz property of ${\cal T}$,
\bea
{\cal T}(a\cdot b) = {\cal T}(a ) \cdot b + a \cdot {\cal T}(b)
\, , \ \ \ \ \ \ 
\cS(a\cdot b) = \cS(a )\cdot \cS(b)
\eea

\subsection{Holomorphic Abelian differentials}
\label{sec:B2}

A standard choice for the canonical homology basis of $\mA_I$ and $\mB_I$ cycles for the intersection pairing $\mJ(\mA_I, \mA_J)=\mJ(\mB_I, \mB_J)=0$ and $\mJ(\mA_I, \mB_J) = \delta_{IJ}$ on a genus-two surface $\Sigma$ with $I,J=1,2$ is depicted in Figure \ref{fig:1}. The space of holomorphic $(1,0)$-forms is two-dimensional and  a standard basis $\om_I$ may be normalized on $\mA$-cycles as follows, 
\bea
\label{holo-I}
\oint _{\mA_I} \om _J = \delta_{IJ} 
\hskip 1in 
\oint _{\mB_I} \om _J = \Omega_{IJ} 
\eea
where $\Omega_{IJ}$ are the components of the period matrix. By the Riemann relations the matrix~$\Omega$ is symmetric and has a positive definite imaginary part. A modular transformation ${\mM \in Sp(4,\ZZ)}$ maps the cycles $\mA_I$ and $\mB_I$ into linear combinations with integer coefficients that leave the intersection pairing $\mJ$ invariant, namely we have $\mM^t \, \mJ \, \mM = \mJ$ with,
\bea
\left ( \bma \tilde \mB \cr \tilde \mA \ema \right ) = \mM \left ( \bma  \mB \cr  \mA \ema \right )
\hskip 0.6in 
\mJ = \left ( \bma 0 & - I  \cr I & 0 \ema \right )
\hskip 0.6in 
\mM = \left ( \bma A & B  \cr C & D \ema \right )
\eea
The row-matrix of $(1,0)$-forms $\om$ and period matrix $\Omega$ transform as follows under $\mM$, 
\bea
\om & \to & \tilde \om = \om (C \Omega+D)^{-1} 
\no \\
\Omega & \to & \tilde \Omega = (A \Omega +B) (C \Omega +D)^{-1}
\eea
The action of the modular group $Sp(4,\ZZ)$ reduces to the action of the permutation group  $\mS_6$ on the six branch points $u_r$, as is shown in appendix \ref{sec:F.1}.\footnote{The action of the modular group $Sp(4,\ZZ)$ on the branch points should not be confused with the action of the automorphic group $SL(2,\CC)$. Although $SL(2,\CC)$ allows one to fix three of the six branch points at arbitrary points in $\hat \CC$ (leaving the remaining three branch points to parametrize the three complex moduli of a genus-two surface),  we shall refrain from making this choice, or any other choice, here and instead maintain manifest $SL(2,\CC)$ covariance by leaving all six branch points free. Doing so, the action of $Sp(4,\ZZ)$ indeed reduces to the group $\mS_6$ of permutations of the branch points, as stated in the body of the text.}
 
\sm

In the hyper-elliptic formulation, a  natural basis of holomorphic $(1,0)$-forms is provided instead  in terms of  the  forms $\varpi^a(z)$ and $\varpi_a(z) =  \varepsilon_{ab} \varpi^b(z)$ given  by,
\bea
\varpi ^1 = { x dx \over s} 
\hskip 0.4in
\varpi^2 = { dx \over s}
\hskip 1in
\varpi _1 =  {  dx \over s} 
\hskip 0.4in
\varpi_2 = - { x \, dx \over s}
\label{defvarpis}
\eea
where $z=(x,s)$ and $s$ is related to $x$ by (\ref{shyper}). The forms $\varpi^a$ and $\varpi_a $ are modular invariant, since 
$x$ and $s$ are invariant under arbitrary permutations of the branch points. They transform as a doublet under  $\gamma \in SL(2,\CC)$, 
\bea
\left ( \bma \varpi ^1 \cr \varpi^2 \ema \right ) \to 
J_\gamma \, \gamma \left ( \bma \varpi ^1 \cr \varpi^2 \ema \right )
\eea
where $\gamma$ was given in (\ref{2.gam}) and the multiplicative factor $J_\gamma$ was defined in (\ref{sSL}). 

\sm

The basis forms $\om_I$ and $\varpi_a$ are linearly related to one another with $z$-independent, but moduli dependent, coefficients  $\sigma^a{}_I$ given by,\footnote{Throughout this paper, in order to clearly distinguish modular transformations from M\"obius transformations, we shall use uppercase letters $I,J,K, \ndots=1,2$ for indices in the $Sp(4,\ZZ)$ frame (such as $\mathfrak{A}_I,\mathfrak{B}_I,\omega_I$) and reserve  lowercase letters $a,b,c,\ndots=1,2$ for $SL(2,\CC)$ doublet indices.}
\bea
\label{2.sigma}
\om_I (z) =   \varpi_a(z) \, \sigma ^a {}_I
\hskip 1in 
\delta _{IJ} =   \oint _{\mA_J} \varpi_a \, \sigma ^a{}_I
\eea
One may think of $\sigma ^a{}_I$ as analogous to a \textit{zweibein} relating $Sp(4,\ZZ)$ and $SL(2,\CC)$ frames.
The transformations of $\sigma$ and $\det \sigma$ under $\mM \in Sp(4,\ZZ)$ are given by,
\bea
\sigma^a{}_I & \to &  \sigma^a{}_J  \big [ (C \Omega +D)^{-1} \big ]^J{}_I
\no \\
\det \sigma & \to &  (\det \sigma) \,  \det (C \Omega +D)^{-1} 
\eea
which promotes $\det \sigma$ (to be encountered in numerous equations of section \ref{sec:sum}) into an $Sp(4,\ZZ)$ Siegel modular form of weight $-1$. The transformation law of $\sigma$ under $SL(2,\CC)$ is given by, 
\bea
\left ( \bma \sigma ^1{}_I \cr \sigma^2{}_I \ema \right ) \to 
J_\gamma \, \gamma \left ( \bma \sigma ^1{}_I \cr \sigma^2{}_I \ema \right )
\hskip 1in 
\det \sigma \to J_\gamma ^2 \, (\det \sigma)
\eea
The ubiquitous anti-symmetric bi-holomorphic form $\Delta$, which is defined in terms of $\om_I$ by,
\bea
\Delta (z,w) = \om_1(z) \om_2(w) - \om_2(z) \om _1(w)
\label{defdelta}
\eea  
may be expressed in terms of the hyper-elliptic basis for $z_i=(x_i, s_i)$, as follows,
\bea
\label{DelHyper}
\Delta (z_1,z_2) = (\det \sigma) \, { (x_1-x_2)  \, dx_1 \, dx_2 \over s_1 \, s_2}
\eea
Therefore, as a function of $z_1=(x_1, s_1)$, the two zeros of the holomorphic $(1,0)$-form $\Delta (z_1,z_2)$ are at $z_2=(x_2,s_2) $  and its image under involution $\cI z_2=(x_2,-s_2)$. The form $\Delta (z_1,z_2)$ is a Siegel modular form of weight $-1$ under $Sp(4,\ZZ)$ in view of the factor $\det \sigma$.

\subsection{The Szeg\"o kernel for even spin structures}

The expression for the Szeg\"o kernel for even spin structure  in terms of $\tet$-functions and the prime form may be found, for example, in \cite{DHoker:2002hof}. Our  analysis of the cyclic products of Szeg\"o kernels $C_\delta$ in (\ref{Cdelta}) will start  from the hyper-elliptic representation of the Szeg\"o kernel, which we now review in detail. 

\sm

For genus two there exists a one-to-one map between the 10 even spin structures, generically denoted by $\delta$, and the ten inequivalent partitions of the six branch points $u_r$ for $r=1,\cdots, 6$ into two disjoint sets $A,B$ of three branch points,\footnote{We note that  for genus two all even spin structure are \textit{regular spin structures} \cite{fay}. The choice of even and odd labels for the subscript is a matter of convenient convention, following \cite{DP4}.}
\bea
\label{deltapart}
\delta  \equiv  A\cup B 
& \qquad &
A = \{ r_1 , r_3, r_5 \} 
\qquad
A \cap B = \emptyset 
\no \\ && 
B = \{ r_2, r_4 , r_6 \}
\qquad
A \cup B = \{ 1, \cdots, 6\}
\eea
We take the opportunity here to express also the odd spin structures, generically denoted by $\nu$,  in terms of the hyper-elliptic representation. For genus two there exists a one-to-one map between the six odd spin structures and the six branch points,
\bea
\nu_r \equiv u_r \hskip 1in r=1,\cdots, 6
\eea
which may be viewed as all inequivalent partitions of the six branch points into a set of one branch point and its complement of five branch points. The functional extension of the space of hyper-elliptic functions, needed to give the Szeg\"o kernel,  is by the square roots $s_A(x)$ and $s_B(x)$ of the following cubic polynomials,
\bea
s_A(x)^2 = \prod _{r \in A} (x-u_r) 
\hskip 0.7in
s_B(x)^2 = \prod _{r \in B} (x-u_r)
\label{sabpoly} 
\eea
The relative sign in the square roots is fixed by requiring their product to be $s(x)$, where $s(x)^2$ was given earlier in (\ref{shyper}), 
\bea
\label{sAsB}
s_A(x) s_B(x) = s(x)
\eea
The Szeg\"o kernel for even spin structure $\delta$ and two arbitrary points $z_i\!=\!(x_i,s_i) \! \in\! \Sigma$ is given~by,
\bea
\label{Szego}
S_\delta(z_1,z_2) = { s_A(x_1) s_B(x_2) + s_B(x_1) s_A(x_2) \over 2 \, x_{12} } 
\left ( { dx_1 \over s_1 } \right )^{1 \over 2}
\left ( { dx_2 \over s_2 } \right )^{1 \over 2}
\eea
where we employ the familiar notation $x_{12}=x_{1,2}= x_1-x_2$. The analytic structure of this expression may be verified by noting that the denominator $x_{12}$ produces simple poles at $(x_1,s_1) = (x_2, \pm s_2)$ but the pole at $s_1=-s_2$ cancels since the numerator vanishes in view of the relation (\ref{sAsB}). The remaining normalization is such that the pole $(dx_1 \, dx_2)^{1/2}/x_{12}$ has unit residue. 

\sm

As mentioned earlier, $Sp(4,\ZZ)$ modular transformations act on the hyper-elliptic representation by the permutation group $\mS_6$ acting on the branch points. Since even and odd spin structures are uniquely  labeled by partitions of the branch points, the action of modular transformations on spin structures is induced from their action  on the branch points. As is well-known, and may be easily verified in terms of the representation via partitions of the branch points, modular transformations map even spin structures to even spin structures and odd to odd. The precise expressions for these actions, along with the translation in terms of half-integer characteristics, is provided in detail in appendix \ref{sec:F.1} and \cite{DP4}. The action of an $Sp(4,\ZZ)$ modular transformation maps the Szeg\"o kernel $S_\delta(z_1,z_2)$  into the Szeg\"o kernel $S_{\tilde \delta} (z_1,z_2)$ where $\tilde \delta$ is the image of $\delta$ under  the modular transformation. 

\sm 

Finally the Szeg\"o kernel is invariant under $SL(2,\CC)$ transformations, 
\bea
\gamma S_\delta(z_1,z_2) = S_\delta(z_1,z_2)
\eea
provided the points $z_1, z_2$, their differentials, the branch points $u_r$, and the spin structure in terms of the branch points are all transformed according to (\ref{xuSL}), (\ref{dSL}), and (\ref{sSL}).

\newpage

\section{Reduction of cyclic products of Szeg\"o kernels}
\label{sec:cyclic}
\setcounter{equation}{0}

In this section, we shall initiate the process of reducing  the spin structure dependence of the cyclic product $C_\delta(z_1, \cdots, z_n)$ of Szeg\"o kernels, defined in (\ref{Cdelta}), to a universal structure that is independent of the number of points $n$. This process will be carried out in a number of steps, and with the help of 
several intermediary functions that we shall introduce and discuss below. The proposed reduction of $C_\delta$ will be carried out completely in this section only for $n=2$. But the structures uncovered for $n=2$ will constitute the fundamental building blocks for the reduction of the higher point functions in section \ref{sec:low-n}, to be carried out with the help of the results of section \ref{sec:trilinear} on the \textit{trilinear relations}.

\sm 

To begin, we express the Szeg\"o kernels in $C_\delta$ in terms of  the hyper-elliptic representation using (\ref{Szego}), and organize the cyclic product as follows,\footnote{When no confusion is expected to arise, we shall often use the shorthand $S_\delta(i,j) = S_\delta(z_i, z_j)$ in the Szeg\"o kernels and elsewhere, so that, 
$ C_\delta (1,2, \cdots, n) = S_\delta(1,2) S_\delta(2,3) \cdots S_\delta(n{-}1,n) S_\delta(n,1)$.}
\bea
C_\delta(1, \cdots, n)  & = & N_\delta (1,\cdots, n)  \prod _{i=1}^n   { dx_i \over 2 \, x_{i, i+1}  \, s_i}
\no \\
N_\delta (1,\cdots, n) & = &  \prod _{i=1}^n \Big ( s_A(i) s_B(i{+}1) + s_B(i) s_A(i{+}1) \Big )
\label{againdefC}
\eea 
where we identify the labels $n{+}1 \equiv 1$. The simple poles at $x_i=x_{i+1}$ arise through the Parke-Taylor factor $(x_{12}x_{23}\ndots x_{n-1,n}x_{n1})^{-1}$ well-known from tree-level correlators. All the dependence on the spin structure $\delta  \equiv  A\cup B $ resides in the cyclically invariant numerator $N_\delta$. An equivalent but more useful expression for $N_\delta$ is given as follows,
\bea
N_\delta(1,\cdots, n) = \sum_{{\sigma _i =A,B \atop i=1,\cdots, n}} s_{\sigma_1}(1) s_{\bar \sigma_1}(2)
s_{\sigma_2}(2) s_{\bar \sigma_2}(3) \cdots s_{\sigma_n}(n) s_{\bar \sigma_n}(1)
\label{numexpansion}
\eea
where $\sigma _i$ takes the values $A$ or $B$ and $\bar \sigma _i$ takes the value opposite to $\sigma _i$.

\subsection{Isolating the spin structure dependence of the numerators $N_\delta$}
\label{sec:qdel}

Each term in the expansion (\ref{numexpansion}) of $N_\delta$ has precisely $n$ factors  $s_A$ and $n$ factors  $s_B$. Products $s_A(i) s_B(i)$ are equated with $s_i$, which is independent of the spin structure.  The number of $s_i$ factors thus obtained must always be even when $n$ is even or odd when $n$ is odd. The term with the largest number of $s_i$ factors is always twice the product of the $n$ different $s_i$ factors.

\sm

Using the above observations we decompose $N_\delta$ into a  linear combination of products involving spin structure independent factors $s_{i_1} \cdots s_{i_p}$ times spin structure dependent factors. To capture the latter, we define the following combinations,
\bea
Q_\delta(i_1, \cdots, i_m|j_1, \cdots, j_m) = s_A (i_1)^2 \cdots s_A(i_m)^2 s_B(j_1)^2  \cdots s_B(j_m)^2+ (A \leftrightarrow B)
\label{defQdelta}
\eea
where the labels $i_1, \cdots, i_m, j_1, \cdots, j_m$ are all distinct from one another.
The properties of $Q_\delta$ follow from those of the polynomials $s_A(x)^2$ and $s_B(x)^2$ and may be conveniently summarized as follows. The functions $Q_\delta(i_1, \cdots, i_m|j_1, \cdots, j_m)$ are,
\begin{itemize}
\itemsep=-0.07in
\item polynomial in $x_{i_1}, \cdots ,x_{i_m}, x_{j_1}, \cdots, x_{j_m}$ of degree 3 in each variable;
\item invariant under all permutations of $i_1, \cdots, i_m$. and all 
permutations of $j_1, \cdots, j_m$;
\item invariant under swapping the set $\{i_1, \cdots, i_m\}$ with the set $\{j_1, \cdots, j_m\}$.
\end{itemize}
For $n\leq 6$, the decomposition may be easily worked out and  we have,
\bea
\label{NQ}
N_\delta(1,2) & = &  2 s_1 s_2  + Q_\delta(1|2)
 \\
N_\delta(1,2,3) & = & 2 s_1 s_2 s_3 
+ \big \{ s_1 Q_\delta(2|3) + \hbox{cycl} (1,2,3) \big \}
\no \\
N_\delta(1,2,3,4) & = & 2 s_1 s_2 s_3 s_4  + Q_\delta(1,3|2,4) 
\no \\ && \quad
+ \big \{ s_1s_2 Q_\delta(3|4) + \thalf s_1s_3 Q_\delta(2|4)  + \hbox{cycl} (1,2,3,4) \big \} 
\no \\
N_\delta(1,2,3,4,5) & = & 2 s_1 s_2 s_3 s_4 s_5 + \big \{ s_1 Q_\delta(2,4|3,5) +  s_1s_2s_3 Q_\delta(4|5)
\no \\ && \quad
 + s_1s_2s_4 Q_\delta(3|5) +\hbox{cycl} (1,2,3,4,5) \big \} 
\no \\
N_\delta(1,2,3,4,5,6) & = & 
2 s_1 s_2 s_3 s_4 s_5 s_6 + Q_\delta(1,3,5|2,4,6) + \big \{ s_1s_2 s_3s_4 Q_\delta(5|6) 
\no \\ && \quad
+ s_1s_2 s_3s_5 Q_\delta(4|6) +\thalf s_1s_2 s_4s_5 Q_\delta(3|6)   +s_1s_2 Q_\delta(3,5|4,6) 
\no \\ && \quad
+ s_1s_3 
Q_\delta(2,5|4,6)+\thalf s_1s_4   Q_\delta(2,5|3,6)+\hbox{cycl} (1,2,3,4,5,6) \big \}
\no \eea
The instruction to add cyclic permutations applies to every term enclosed by the same curly brackets. The factors of $\thalf$ account for the symmetry of the configuration of points under consideration and avoid overcounting 
when the cyclic images are added.

{\thm
\label{thm:3.1}
For arbitrary $n$, the  decomposition of $N_\delta$ into a linear combination of polynomials $Q_\delta$ with coefficients that are polynomials in the square roots $s_i$ are obtained as follows, 
\bea
\label{NtoQdelta}
N_\delta (1,\cdots, n) & = &   2 s_1  \cdots s_n + \sum_{\ell=1}^{ \lfloor n/2\rfloor }  
\sum_{1\leq i_1<j_1<\ndots <i_\ell < j_\ell \leq n}
\! \! \! \! Q_\delta(i_1,\ndots, i_\ell | j_1,\ndots,j_\ell) 
\! \! \! \prod_{k\neq i_1,\ndots,i_\ell \atop{k \neq j_1,\ndots,j_\ell}} \! \! \! s_k
\qquad
\eea }
The proof of this theorem is elementary and proceeds from the representation (\ref{numexpansion}) of $N_\delta$ and by rearranging the summands in terms of $Q_\delta$ of various orders.

\sm

The sum in (\ref{NtoQdelta}) terminates after $\lfloor n/2\rfloor{+}1$ terms,  the last   term being given as follows, 
\begin{align}
n & \hbox{~even} & & Q_\delta  (1,3, \cdots, n{-}1| 2, 4, \cdots, n)
\label{lastterms} \\
n & \hbox{~odd} & &
s_1 Q_\delta  (2,4,, \cdots, n{-}1| 3,5, \cdots, n)+\hbox{cycl} (1,2,\cdots,n)
\no
\end{align}

\subsection{The two-point cyclic product $C_\delta(i,j)$ and its $\delta$ dependence}

The decomposition carried out in the preceding subsection isolates all dependence of the cyclic products $C_\delta$ in the polynomial functions $Q_\delta$. In the present subsection, we shall show that all spin structure dependence in the two-point cyclic product,
\bea
C_\delta(i,j) = { Q_\delta(i|j) + 2 s_i s_j \over 4 x_{ij} x_{ji}} \times { dx_i \, dx_j \over s_i \, s_j}
\label{bitisolated}
\eea
 may be further isolated to reside in a symmetric bi-holomorphic form in $z_i$ and $z_j$, that we shall identify explicitly, and that will play a central role in the sequel.

\sm

Given an even spin structure $\delta$ corresponding to the partition $(u_{r_1},u_{r_3},u_{r_5} | u_{r_2},u_{r_4},u_{r_6})$  (\ref{deltapart}) of branch points, the polynomials $s_A^2,s_B^2$ and $s^2$ may be expressed as follows,
 \bea
 s_A(x)^2 & = & (x-u_{r_1}) (x-u_{r_3}) (x-u_{r_5}) = x^3 - \a_1 x^2 + \a_2 x - \a_3
  \\
 s_B(x)^2 & = & (x-u_{r_2}) (x-u_{r_4}) (x-u_{r_6}) = x^3 -\b_1 x^2 + \b_2 x - \b_3
 \no \\
 s(x)^2 & = & (x-u_1) \cdots (x-u_6) = \mu_0 x^6 - \mu_1 x^5 + \mu_2 x^4 - \mu_3 x^3 +\mu_4 x^2 -\mu_5 x +\mu_6
 \no
 \eea
 where $\a_1,\a_2,\a_3,\b_1,\b_2,\b_3$ are \textit{partially symmetric polynomials} in  triplets of branch points, 
\begin{align}
\label{albet}
\a_1&= u_{r_1}+u_{r_3}+u_{r_5} & 
\b_1&= u_{r_2}+u_{r_4}+u_{r_6} 
\no \\
\a_2 & = u_{r_1} u_{r_3}+u_{r_3}  u_{r_5}+u_{r_5} u_{r_1} & 
\b_2 & = u_{r_2} u_{r_4}+u_{r_4}  u_{r_6}+u_{r_6} u_{r_2} 
\no \\
\a_3 & = u_{r_1} u_{r_3} u_{r_5} &
\b_3 & = u_{r_2} u_{r_4} u_{r_6} 
\end{align}
and the following \textit{symmetric polynomials} of the 6 branch points are defined by $\mu_0=1$ and,\footnote{More precisely, $\mu_m$ denote the unique symmetric degree-$m$ polynomials
in the $u_{i}$ with $i=1,\ndots,6$ that are at most linear in each branch point.} 
\bea
\mu_m = \sum_{1 \leq r_1 < \cdots < r_m \leq 6 } u_{r_1} \cdots u_{r_m} \hskip 1in 1\leq m \leq 6
\eea
The polynomials are related by the following identities,
\begin{align}
\label{mualpha}
\mu_1 & =   \a_1+\b_1 & 
\mu_4 & =  \a_1 \b_3 + \a_3 \b_1 + \a_2 \b_2
\no \\
\mu_2 & =  \a_2 +\b_2 +\a_1 \b_1 &
\mu_5 & =  \a_2 \b_3 + \a_3 \b_2
\no \\
\mu_3 & =  \a_3 + \b_3 + \a_1 \b_2 + \a_2 \b_1 &
\mu_6 & =  \a_3 \b_3
\end{align}
While the symmetric polynomials $\mu_m$ are invariant under the full group $\mS_6$ of permutations of the six branch points, the partially symmetric polynomials $\a_1,\a_2,\a_3,\b_1,\b_2,\b_3$ are invariant only under the subgroup $\mS_3 \times \mS_3$ of $\mS_6$. This subgroup leaves the spin structure $\delta$ invariant, as does the swap $(\a_1,\a_2,\a_3) \leftrightarrow (\b_1,\b_2,\b_3)$. The dependence of $\alpha_m$ and $\beta_m$ on the spin structure $\delta$  will not be exhibited in order to avoid cluttering.

\subsubsection{The auxiliary polynomial $P_\delta(i,j)$}

To isolate the spin structure dependence of $C_\delta(i,j)$ further than we have already done so far
in (\ref{bitisolated}), we shall use a convenient auxiliary polynomial function $P_\delta(i,j)$ defined by,
\bea
P_\delta (i,j) = \big ( s_A(i)^2 - s_A(j)^2 \big ) \big ( s_B(j)^2 - s_B(i)^2 \big )
\label{defPtwo}
\eea
Expanding the product, we see that this function is closely related to $Q_\delta(i|j)$, as follows,
\bea
\label{PQ2}
P_\delta (i,j) = Q_\delta(i|j) - s_i^2 -s_j^2
\eea
The advantage of $P_\delta$ is that it is automatically divisible by $x_{ij}^2$ since we have,
 \begin{align}
 \label{2.ratios}
 s_A(i)^2-s_{A}(j)^2  & =  x_{ij} \big (X_{i,j} + \a_1 Y_{i,j} +\a_2 \big ) & X_{i,j} & = x_i^2 + x_i x_j + x_j^2
\no \\
s_{B}(j)^2-s_{B}(i)^2  & =  x_{ji} \big ( X_{i,j} + \b_1 Y_{i,j} +\b_2 \big ) & Y_{i,j} & =  - x_i -x_j
\end{align}
The spin structure dependence of these factors is simplified further in view of the fact that they no longer depend on either  $\a_3$ or $\b_3$.  The disadvantage of using $P_\delta$ as an auxiliary combination is  that it does not enjoy good transformation properties under $SL(2,\CC)$, in contrast with $C_\delta$ and $Q_\delta$ which transform homogeneously. Good manifest transformation properties will be easily restored in  the final results, as we shall see below. 

\sm

Carrying out the product of the factors in (\ref{2.ratios}) to obtain $P_\delta$, we obtain,
\bea
\label{Pdelta2}
P_\delta(i,j) & = &  x_{ij} x_{ji} \Big (X_{i,j} + \a_1 Y_{i,j} +\a_2 \Big ) \Big ( X_{i,j} + \b_1 Y_{i,j} +\b_2 \Big )
\no \\ & = &
x_{ij} x_{ji} \Big (X_{i,j}^2 + \mu_1 X_{i,j} Y_{i,j} + \mu_2 X_{i,j} + x_i x_j \phi_2 -(x_i+x_j) \phi_3 + \phi_4 \Big )
\eea
The second line has been derived from the first line using the relations of (\ref{mualpha}) to combine all spin structure dependence in terms of the following three \textit{partially symmetric polynomials} in the branch points, 
\bea
\label{phi234}
\phi_2 & = & \a_1 \b_1
\no \\
\phi_3 & = & \a_1 \b_2 + \a_2 \b_1
\no \\
\phi_4 & = & \a_2 \b_2
\eea
where the subscripts of $\phi_m,\alpha_m,\beta_m$ indicate their polynomial degree in the branch points, and we again keep the dependence on the spin structure implicit to avoid cluttering.
The key property of the polynomials $\phi_2, \phi_3, \phi_4$ is that they share the symmetry under the group of permutations $\mS_3 \times \mS_3 \times \ZZ_2$ with the even spin structure $\delta$. Recall that the two $\mS_3$  factors generate the permutations of the branch points that leave $\alpha_m$ and $\beta_m$ invariant, respectively, while the $\ZZ_2$ factor swaps $\alpha_m$ and $\beta_m$ variables. Therefore the variables $\phi_2, \phi_3, \phi_4$ provide a \textit{minimal parametrization} of the spin structure dependence, as will be further clarified below. Substituting $P_\delta$ into $Q_\delta$ and then into $C_\delta$, we obtain, 
\bea
C_\delta(i,j) = { P_\delta(i,j) 
+(s_i{+}s_j)^2 \over 4 x_{ij} x_{ji}} \times { dx_i \, dx_j \over s_i \, s_j}
\label{CwithP}
\eea
Using the fact that the spin structure dependence of $P_\delta$ in (\ref{Pdelta2}) is localized in the last three terms on the second line, namely those that involve the $\phi_m$, we obtain upon substitution into the above formula for $C_\delta$, 
\bea
\label{Cdel1}
C_\delta (i,j) = \Big (  x_i x_j \phi_2 -(x_i+x_j) \phi_3 + \phi_4 \Big )  { dx_i \, dx_j \over 4 \, s_i \, s_j}
+ \hbox{ spin structure independent}
\eea
We readily observe that the spin structure dependence is entirely localized in a symmetric bi-holomorphic form in $z_i, z_j$, as was announced at the beginning of this section, while the double pole of $C_\delta (i,j)$ at $x_i=x_j$ resides in its spin structure independent part.

\subsubsection{Restoring manifest $SL(2,\CC)$ invariance} 

A central guiding principle of this work is to organize the cyclic products $C_\delta$ and their spin structure sums into tensors of $SL(2,\CC)$. A key advantage of tensorial equations in irreducible representations of $SL(2,\CC)$ is that they can be derived or verified from a single component. Moreover, $SL(2,\CC)$ tensors will facilitate the translation of our results to $\tet$-functions and modular tensors in the $Sp(4,\mathbb Z)$-frame in follow-up work \cite{DHS}.

\sm

While $C_\delta(i,j)$ is manifestly $SL(2,\CC)$ invariant, the use of the auxiliary function $P_\delta(i,j)$ in (\ref{CwithP}) has obscured this invariance at intermediate stages. In particular, the spin structure dependent bi-holomorphic form, identified in (\ref{Cdel1}), is not $SL(2,\CC)$ invariant. 
To restore manifest $SL(2,\CC)$ invariance, we begin by recording the transformations of $\phi_2, \phi_3,\phi_4$ and $\mu_m$ for $0\leq m \leq 6$, under infinitesimal translations $\BL$  and inversions $\cS$,\footnote{Recall from section \ref{sec:hyper} that invariance under both $\BL$ and $\cS$ implies invariance under the full $SL(2,\CC)$.}
\begin{align}
\label{Tmuphi}
\BL \mu_m & = (7-m) \mu_{m-1} & \cS \mu_m & = (-)^m \mu_{6-m} / \mu_6
\no \\
\BL \phi_2 & = 3 \mu_1 & \cS \phi_2 & = \phi_4/\mu_6
\no \\
\BL \phi_3 & = \phi_2 + 3 \mu_2 & \cS \phi_3 & = - \phi_3/\mu_6
\no \\
\BL \phi_4 & = 2 \phi_3 & \cS \phi_4 & = \phi_2/\mu_6
\end{align}
where it is understood that $\mu_0=1$ and $\mu_{-1}=0$. Using these transformation laws, along with $\BL x_i =1$ and the Leibniz property of ${\cal T}$, we readily evaluate the translation of the holomorphic part in (\ref{Cdel1}), 
\bea
\label{T2}
\BL \big (  x_i x_j \phi_2 -(x_i+x_j) \phi_3 + \phi_4 \big )  = 3 \mu_1 x_i x_j - 3 \mu_2 (x_i+x_j)
\eea
The right side of the above equation is manifestly independent of $\phi_2, \phi_3, \phi_4$ and thus independent of the spin structure $\delta$. It may be expressed as the translation of a spin-structure independent combination that is uniquely  determined as follows,
\bea
\label{T3}
3 \mu_1 x_i x_j - 3 \mu_2 (x_i+x_j) = \BL \left ( \tfrac{ 3}{5} \mu_2 x_i x_j - \tfrac{ 9}{10 } \mu_3 (x_i+x_j) + \tfrac{3}{5} \mu_4 \right )
\eea
Combining (\ref{T2}) with (\ref{T3}) we obtain a translation-invariant combination, that we shall normalize and express as follows, 
\bea
\label{Ldel2}
L_\delta(i,j) = \bL_\delta (i,j) \cdot { dx_i \, dx_j \over s_i \, s_j} =  \see^{ab}_\delta \, \varpi_a (i) \varpi_b(j)
\eea
where the components $\see_\delta^{ab}$ are given by, 
\begin{align}
\label{4.see}
 \see_{\delta}^{11} &= \tfrac{1}{4} \phi_4 - \tfrac{3}{20} \mu_4 
 \,  \, = \tfrac{1}{4} \alpha_2 \beta_2 - \tfrac{3}{20} \mu_4
 \no \\
\see_\delta^{12} &=    \tfrac{1}{4}\phi_3 - \tfrac{9 \mu_3}{40}
 \,  =   \tfrac{1}{4}(\alpha_1 \beta_2 + \alpha_2 \beta_1) - \tfrac{9 \mu_3}{40}
\no \\
\see_\delta^{22} &= \tfrac{1}{4}\phi_2 - \tfrac{3}{20} \mu_2 
\, \,  = \tfrac{1}{4} \alpha_1 \beta_1 - \tfrac{3}{20} \mu_2 
\end{align}
and the holomorphic $(1,0)$-forms $\varpi_a(i)$ were defined in (\ref{defvarpis}).\footnote{It will be established in section \ref{sec:sum} that the sum  of $\see_\delta$ over all even spin structures against the unit measure, namely $\sum_\delta \see_\delta$, actually vanishes. This  property might alternatively have been used to justify the addition of the spin structure independent terms in (\ref{4.see}).} 
Expressed in terms of the components of $\see_\delta$, the function $\bL_\delta(i,j)$ takes the form,
\bea
\label{3.Ldel}
\bL_\delta (i,j) = \see_\delta^{11} - (x_i+x_j) \, \see_\delta^{12} + x_i x_j \, \see_\delta^{22}
\eea
Combining the transformation laws $\BL \varpi_1(z) =0$ and $\BL \varpi_2(z) = - \varpi_1(z)$ with those of $\see_\delta^{ab}$,
\bea
\BL \see ^{11}_\delta & = &  2\, \see _\delta ^{12} 
\no \\
\BL \see _\delta ^{12} & = & \see _\delta ^{22}
\no \\
\BL \see _\delta ^{22} & = &0
\eea
we readily verify that $\BL L_\delta (i,j)= \BL \bL_\delta (i,j)=0$. Using the transformation properties of $\phi_2, \phi_3, \phi_4$ and $\mu_m$ under $\cS$ given in (\ref{Tmuphi}) along with the transformation law $\cS x_i = -1/x_i$ of (\ref{SLTS}), we also verify that $L_\delta(i,j)$ is invariant under $\cS$. Since the combined invariance under $\cS$ and $\BL$ implies invariance under the full $SL(2,\CC)$ group,  we conclude that the symmetric bi-holomorphic form $L_\delta (i,j)$ in (\ref{Ldel2}) is $SL(2,\CC)$ invariant.

\sm

In particular this means that $ L_\delta (i,j)$ and $\bL_\delta (i,j)$ may be expressed as combinations of differences of the points $x_i,x_j$ and the branch points. For a spin structure $\delta$ corresponding to the partition $(u_{r_1}, u_{r_3}, u_{r_5} | u_{r_2}, u_{r_4}, u_{r_6})$, we may render this property explicit as follows, 
\bea
\bL_\delta(i,j) & = & - { 1 \over 80} 
\Big \{ \Big [ (x_i-u_{r_1})(x_j-u_{r_2})+(i \leftrightarrow j) \Big ] 
 (u_{r_3}-u_{r_4})(u_{r_5}-u_{r_6})  
\no \\ && \hskip 1in
+\hbox{cycl}(1,3,5) \Big \} 
+ \hbox{perm}(2,4,6)
\eea
where perm(2,4,6) stands for the five remaining permutations of $(2,4,6)$. The expression may be readily verified to agree with (\ref{3.Ldel}) using {\sc maple}. 

\sm 

The expansion of $L_\delta (i,j)$ in a basis of Abelian differentials $\omega_{I}(z_i) \omega_{J}(z_j)$ in the $Sp(4,\mathbb Z)$ frame will drive the translation of our results into $\tet$-functions \cite{DHS} and make contact with the spin structure dependent building blocks $P_{IJ}(\Omega_\delta)$ of $C_\delta$ considered in \cite{Tsuchiya:2022lqv}.

\subsection{Summary for the reduction of $Q_\delta(i|j)$ and $C_\delta(i,j)$}
\label{sec:2.3}

The result obtained earlier in this section  may be summarized by the following theorem.

{\thm
\label{thm:1}
The polynomial $Q_\delta(i|j)$ in (\ref{defQdelta}) and the cyclic product 
$C_\delta(i,j)$ admit the following $SL(2,\CC)$ covariant decompositions, 
\bea
Q_\delta(i|j)  & = & 4 x_{ij} x_{ji} \bL_\delta(i,j) + 2 Z(i,j)
\no \\
C_\delta(i,j) & = & L_\delta(i,j) + { Z(i,j) +  s_i s_j \over 2 x_{ij} x_{ji}}  { dx_i \, dx_j \over s_i \, s_j}
\eea
where the spin structure dependent symmetric bi-holomorphic $(1,0)$-form $L_\delta(i,j)$ was defined in (\ref{Ldel2}) while  the spin structure independent function $Z(i,j)$ is given by,
\bea
Z(i,j) & = & \mu_0 x_i^3 x_j^3 -{\mu_1 \over 2} (x_i^3x_j^2+x_i^2x_j^3) +{\mu_2 \over 5} (x_i x_j^3 + x_i^3 x_j + 3 x_i^2x_j^2) 
\label{defzij} \\ &&
-{\mu_3 \over 20} (x_i^3 + x_j^3 +9x_ix_j^2+9x_i^2x_j) +{\mu_4 \over 5} (x_i^2+x_j^2+3x_ix_j) -{\mu_5 \over 2} (x_i+x_j) + \mu_6
\no
\eea
An equivalent expression for $Z(i,j)$ may be given in terms of the rank six symmetric $SL(2,\CC)$-tensors $\CM_1$ and $\bX_{ij}$,
\bea
\label{Zij}
Z(i,j) = \CM_1 ^{a_1 \cdots a_6} \, \bX_{ij} ^{b_1 \cdots b_6} \, \ep_{a_1b_1} \cdots \, \ep_{a_6 b_6} 
\eea
where $\ep_{ab}$ is the anti-symmetric invariant tensor of $SL(2,\CC)$ normalized to $\ep_{12}=1$ and the components of the tensors $\CM_1$ and $\bX_{ij}$ are given by, }
\begin{align}
\label{MXcomps}
\CS^{111111} & = \mu_6 					&  \bX_{ij} ^{111111} & =  x_i^3 x_j^3
\no \\
\CS ^{111112} & =  \tfrac{1}{6} \mu_5		&  \bX_{ij} ^{111112} & =   \tfrac{1}{2} (x_i^3 x_j^2+x_i^2 x_j^3) 
\no \\
\CS^{111122} & = \tfrac{1}{15} \mu_4 		& \bX_{ij} ^{111122} & =  \tfrac{1}{5} ( x_i^3 x_j+x_i x_j^3+ 3 x_i^2 x_j^2)
\no \\
\CS ^{111222} & =  \tfrac{1}{20} \mu_3	&  \bX_{ij} ^{111222} & =  \tfrac{1}{20} ( x_i^3 +x_j^3+ 9 x_i^2 x_j+ 9 x_i x_j^2) 
\no \\
\CS^{112222} & = \tfrac{1}{15} \mu_2 		& \bX_{ij} ^{112222} & =  \tfrac{1}{5} (x_i^2 +x_j^2+ 3 x_i x_j)
\no \\
\CS ^{122222} & =  \tfrac{1}{6} \mu_1 		&  \bX_{ij} ^{122222} & =   \tfrac{1}{2} (x_i +x_j) 
\no \\
\CS^{222222} & = \mu_0 					&  \bX_{ij} ^{222222} & =  1
\end{align}

\sm

To prove the theorem, we have already done most of the work by identifying the bi-holomorphic form $L_\delta(i,j)$. It remains to show that $Q_\delta(i|j) - 4 x_{ij} x_{ji} \bL_\delta(i,j)$ indeed gives the expression for $Z(i,j)$. We carry out this analysis by expressing $Q_\delta$ in terms of $P_\delta$,  
\bea
&&
Q_\delta(i|j) - 4 x_{ij} x_{ji} \bL_\delta(i,j) 
\no \\ && \qquad \quad
=
P_\delta(i,j) +s_i^2 +s_j^2 - 4 x_{ij} x_{ji} \, \bL_\delta (i,j) 
 \\  && \qquad \quad
= -x_{ij}^2 \Big (X_{i,j}^2 + \mu_1 X_{i,j} Y_{i,j} + \mu_2 X_{i,j} +{3 \over 5}\mu_2 x_i x_j  +{ 9 \over 10} \mu_3 Y_{i,j}  + {3 \over 5} \mu_4 
 \Big )
+s_i^2 +s_j^2 
\no
\eea
with $X_{i,j}$ and $Y_{i,j}$ defined in (\ref{2.ratios}).
Using the relations $x_{ij} X_{i,j} = x_i^3-x_j^3$ and $x_{ij} Y_{i,j} = -x_i^2 +x_j^2$ one readily observes that the terms of degree 6, 5, and 4 in $x_i$ and $x_j$ cancel with the corresponding terms in $s_i^2+s_j^2$ to yield twice the expression for $Z(i,j)$  given in (\ref{defzij}). Recasting the expression in terms of the tensors $\CM_1$ and $\bX_{ij}$ is straightforward and completes the proof of the theorem.

\sm

We close this subsection by noting that the tensor $\CS$ corresponds to a unique (up to overall normalization) totally symmetric holomorphic $(1,0)^{\otimes 6}$ form given by, 
\bea
M_1(1,\cdots, 6) = \CS^{a_1  \cdots a_6} \, \varpi_{a_1}(1) \cdots \varpi_{a_6}(6)
\label{contractm1}
\eea
where $\varpi _a(z)$ are the familiar $(1,0)$-forms. In terms of the branch points, $M_1$ is given by,
\bea
M_1(1,\cdots , 6) = { 1 \over 6!} \sum _{\rho \in \mS_6} \, \prod _{j=1}^6 \big ( x_j -u_{\rho(j)}  \big ) { dx_j \over s_j}
\eea
Setting the six points $x_j$ equal to $x_1$, we have the relation $M_1(1,1,1,1,1,1) =  \psi(x_1)^2$, where $\psi(x)$ is the (unique up to an overall normalization) holomorphic $(3,0)$-form $dx^3/s^2$ whose six simple zeros are the  branch points.

\subsection{The function $P_\delta$ for the general case} 
\label{sec:2.4}
 
To isolate the spin structure dependence of the polynomials $Q_\delta(i_1, \cdots, i_m|j_1, \cdots, j_m)$ for arbitrary values of $m$, we introduce a suitable generalization of the function $P_\delta(i,j)$ in (\ref{defPtwo}) used for the case $m=1$. The number of points appearing in $P_\delta$ will be even, given by $n=2m$, and cyclically ordered as follows, $P_\delta(i_1, j_1, i_2, j_2, \cdots, i_m, j_m)$.  We lighten the notation by designating the points $i_1, j_1, i_2, j_2, \cdots, i_m, j_m$ in this order simply  by $1,2, \cdots, n$, namely arranged in the same cyclical order as the 
original cyclic product $C_\delta$. The generalized function $P_\delta(1,2,\cdots, n)$ is then defined by, 
\bea
\label{3.pdel}
P_\delta(1,\cdots, n) & = & \thalf \big (s_A(1)^2-s_A(2)^2 \big ) \big (s_B(2)^2-s_B(3)^2 \big ) \times \cdots 
\no \\ &&
\cdots  \times \big ( s_A(n{-}1)^2 - s_A(n)^2 \big ) \big (s_B(n)^2 - s_B(1)^2 \big ) + (A \leftrightarrow B)
\eea
In the same way as $P_\delta(i,j)$ in (\ref{defPtwo}) is automatically divisible by $x_{ij}^2$, a key advantage of its $n$-point generalization (\ref{3.pdel}) is its divisibility by the Parke-Taylor denominator $x_{12}x_{23}\ndots x_{n1}$ of $C_\delta(1,\cdots,n)$.
Note that the symmetrization in $A$ and $B$ will be essential, as this operation will guarantee that all the spin structure dependence of $P_\delta$ can be expressed in terms of the partially symmetric functions $\phi_2, \phi_3, \phi_4$ in (\ref{phi234}), as will be proven in Lemma~\ref{4.lem.1}. In the remainder of this subsection we shall relate $P_\delta$ to $Q_\delta$ while in the next subsection we shall state Lemma~\ref{4.lem.1}.

\sm

To relate $P_\delta$ to $Q_\delta$ we expand the products in (\ref{3.pdel}) and use the relation $s_A(x)s_B(x)=s(x)$, to express the function $P_\delta$ as a linear combination of the functions $Q_\delta$ with coefficients that are polynomials in $s_1, \cdots, s_n$. The simplest example  for $n=2$ was already given in (\ref{PQ2}). For the case $n=4$ we have, 
\bea
\label{PQ4}
P_\delta (1,2,3,4) & = &
Q_\delta (1,3|2,4) +  s_1^2s_3^2+s_2^2s_4^2
\no \\ &&
-\thalf \big\{ s_1^2 \big ( Q_\delta(2|3) + Q_\delta(3|4)- Q_\delta(2|4) \big ) + \hbox{cycl}(1,2,3,4) \big\}
\eea
where the cyclic permutations apply only to the last line. For the case $n=6$, we have,
\begin{align}
\label{PQ6a}
&P_\delta (1,2,3,4,5,6) =   Q_\delta (1,3,5|2,4,6)  - s_1^2 s_3^2 s_5^2 -  s_2^2 s_4^2 s_6^2
\no \\
&\quad +\Big \{ \tfrac{1}{2}  s_1^2 \big( Q_\delta(2,4|3,6)+Q_\delta(2,5|4,6) -Q_\delta(2,4|3,5)-Q_\delta(2,5|3,6)-Q_\delta(3,5|4,6) \big) 
\notag \\
&\quad \quad  + 
\tfrac{1}{4}  s_1^2 s_4^2  \big(  Q_\delta(2|5)+Q_\delta(3|6)-Q_\delta(2|6)-Q_\delta(3|5) \big)  
\no \\
&\quad \quad  + \tfrac{1}{2}  s_1^2 s_3^2 \big( Q_\delta(4|5)+Q_\delta(5|6)-Q_\delta(4|6) \big) +  \hbox{cycl}(1,2,3,4,5,6) \Big \} 
\end{align}
Equivalently, one may express $Q_\delta$ in terms of $P_\delta$ only by eliminating the $Q_\delta$ functions with fewer variables in terms of $P_\delta$ functions with fewer variables. For example,  the case $n=4$ reduces as follows,
\bea
\label{sping3.14}
Q_\delta(1,3|2,4) &= & 
P_\delta(1,2,3,4) 
- \tfrac{1}{2} P_\delta(1,3)(s_2^2 +s_4^2) - \tfrac{1}{2} P_\delta(2,4)(s_1^2 +s_3^2) 
\no \\ &&
+ \tfrac{1}{2} \big \{ P_\delta(1,2) (s_3^2 +s_4^2)  + {\rm cycl}(1,2,3,4) \big \} 
+ s_1^2 s_3^2 + s_2^2 s_4^2 
\eea
while the case $n=6$ becomes, 
\bea
\label{sping3.15}
Q_\delta(1,3,5|2,4,6) &= &
P_\delta(1,2,\ndots,6) 
+ \tfrac{1}{4} \big \{  P_\delta(1,2,4,5) (s_3^2 + s_6^2) + {\rm cycl}(1,2,\ndots,6)   \big \}
\no \\ &&
+  \tfrac{1}{2}\big \{ P_\delta(1,2,3,4) (s_5^2 + s_6^2) - P_\delta(1,2,3,5) (s_4^2 + s_6^2)
+ {\rm cycl}(1,2,\ndots,6) \big \}
\no \\ &&
+ \tfrac{1}{2} \big \{  P_\delta(1,2) (s_3^2 s_5^2 + s_4^2 s_6^2) - P_\delta(1,3) (s_2^2 s_5^2 + s_4^2 s_6^2) + {\rm cycl}(1,2,\ndots,6) \big \} 
\no \\ &&
+ \tfrac{1}{4} \big \{  P_\delta(1,4)(s_2^2 s_5^2 + s_3^2 s_6^2) + {\rm cycl}(1,2,\ndots,6)  \big \} 
+ s_1^2 s_3^2 s_5^2 + s_2^2 s_4^2 s_6^2 
\qquad\qquad
\eea
where the scope of the cyclic permutations is delimited by the respective braces. For the case of arbitrary values of $n=2m$, we have the following proposition, which was inferred by inspection of the {\sc mathematica} results for low values of $m$.

{\prop
The expression for $Q_\delta$ in terms of the functions $P_\delta$ for 
arbitrary $m$ is given by the following relations,
\begin{align}
&Q_\delta(1,3,\ndots,2m{-}1|2,4,\ndots,2m) = 
s_1^2 s_3^2\ndots s_{2m-1}^2 + s_2^2 s_4^2\ndots s_{2m}^2 + P_\delta(1,2,\ndots,2m) 
\notag \\
&\hskip 0.7in 
+ \frac{1}{2} \sum_{\ell=1}^{m-1} \sum_{1\leq i_1<\ndots  < i_{2\ell}\leq 2m}
(-1)^{\ell+i_1+i_2+\ndots+i_{2\ell}} P_{\delta}(i_1,i_2,\ndots,i_{2\ell}) \,
\Phi (j_1, \cdots, j_{2m-2\ell})
\label{QtoPdelta}
\end{align}
The $\delta$-independent function  $\Phi (j_1, \cdots, j_{2m-2\ell})$ may be expressed in terms of the ordered set of indices $1\leq j_1<j_2<\ndots<j_{2m-2\ell} \leq 2m$ obtained by removing $i_1,i_2,\ndots,i_{2\ell}$ from $\{1,2,\ndots,2m\}$,}
\bea
 \Phi (j_1, \cdots, j_{2m-2\ell}) = s_{j_1}^2 s_{j_3}^2 \ndots s_{j_{2m-2\ell -1}}^2 
 + s_{j_2}^2 s_{j_4}^2 \ndots s_{j_{2m-2\ell}}^2
\eea
We have proved the proposition using {\sc maple} for the values $m=2,3,4,5,6,7,8$. An analytical proof for arbitrary values of $m$ remains outstanding.

\subsection{Spin structure dependence of $P_\delta$}
\label{sec:2.5}

Having secured expressions for $Q_\delta$ in terms of the $P_\delta$ defined by
(\ref{3.pdel}) in the preceding subsection, we shall now reduce the spin structure dependence of $P_\delta$ to a standard form. By construction, $P_\delta$ is divisible by the Parke-Taylor denominator $x_{12}x_{23} \cdots x_{n1}$ in the expression (\ref{againdefC}) for $C_\delta$. Its quotient will be denoted by $\Pi_\delta$, i.e.
\bea
\label{PPi}
P_\delta(1,2, \cdots, n) = x_{12}x_{23} \cdots x_{n1} \, \Pi_\delta (1,2, \cdots, n)
\qquad\qquad n \ {\rm even}
\eea
Using the relations of (\ref{2.ratios}), we obtain an explicit expression for $\Pi_\delta$ in terms of the partially symmetric polynomials $\alpha_i,\beta_i$ in (\ref{albet}), which generalizes the case $n{=}2$ studied earlier in~(\ref{Pdelta2}), 
\bea
\label{Pifull}
\Pi_\delta (1, \cdots, n) & = & 
\big ( X_{1,2} + \a_1 Y_{1,2} + \a_2 \big ) \big ( X_{3,4} + \a_1 Y_{3,4} + \a_2 \big )
\cdots \big ( X_{n-1, n} + \a_1 Y_{n-1,n} + \a_2 \big )
\no \\ && \quad \times 
\big ( X_{2,3} + \b_1 Y_{2,3} + \b_2 \big )\big ( X_{4,5} + \b_1 Y_{4,5} + \b_2 \big )
\cdots \big ( X_{n, 1} + \b_1 Y_{n,1} + \b_2 \big )
\no \\ && + (\a \leftrightarrow \b)
\eea
Expanding the product on the first line above, we observe that the coefficient $\sigma _A^{a_1, a_2} (X,Y)$ of the monomial $\a_1^{a_1}  \a_2^{a_2} $ is a symmetric polynomial  in the variables $X_{1,2}, X_{3,4}, \cdots, X_{n-1,n}$ and $Y_{1,2}, Y_{3,4}, \cdots, Y_{n-1,n}$ which is
\begin{itemize}
\itemsep=-0.07in
\item homogeneous of degree $a_1$ in $Y$;
\item homogeneous of degree $n-a_1-a_2$ in $X$;
\item at most of combined degree 1 in $X_{i,j}$ and $Y_{i,j}$ at fixed $i,j$.
\end{itemize}
Proceeding similarly for the coefficient $\sigma_B^{b_1,b_2}(X,Y)$ of the monomial $\beta_1^{b_1} \beta_2^{b_2}$, the expression (\ref{Pifull}) for $\Pi_\delta$ at even $n=2m$ can reorganized into
 \bea
\Pi_\delta (1, \cdots, 2m) = \!\!\!
 \sum_{{ {a_1,a_2,b_1, b_2 \geq 0 \atop a_1+a_2 \leq m} \atop b_1+b_2 \leq m}}  \sigma_A^{a_1,a_2}(X,Y) \,
 \sigma_B^{b_1, b_2}(X,Y) \Big ( \a_1^{a_1} \a_2^{a_2} \b_1^{b_1} \b_2^{b_2} + \a_1^{b_1}  \a_2^{b_2} \b_1^{a_1} \b_2^{a_2}  \Big )
 \quad
  \label{pialphabeta}
 \eea
The coefficients $\sigma_A$ and $\sigma_B$ are manifestly independent of the spin structure. 
To narrow down the precise nature of the spin structure dependence, we appeal to the following lemma. 
 
 {\lem
 \label{4.lem.1} The spin structure dependence of $\Pi_\delta$ has the following properties.
 \begin{description}
 \itemsep=0in
\item (a) Every polynomial  in the variables $\a_1, \a_2, \b_1, \b_2$, which are defined in terms of the branch points in (\ref{albet}),  that is invariant under $(\a_1, \a_2) \leftrightarrow (\b_1, \b_2)$ may be expressed as a polynomial in the variables $\phi_2, \phi_3, \phi_4$ defined in (\ref{phi234}) with coefficients in $\CC[\mu_1, \cdots, \mu_6]$. 
\item (b)  Equivalently, every such polynomial  may be expressed as a polynomial in the variables $\see_\delta ^{ab}$ defined in (\ref{4.see})  with coefficients in $\CC[\mu_1, \cdots, \mu_6]$.
\item (c) The function $\Pi_\delta(1,\cdots, n)$ for even $n=2m$ is a polynomial in $\phi_2, \phi_3, \phi_4$,  whose allowed monomials  are subject to the following conditions, 
\bea
\phi_2^{\lambda _2} \, \phi_3 ^{\lambda_3} \, \phi_4 ^{\lambda _4}
\hskip 0.8in
0 \leq \lambda_2, \lambda_3, \lambda _4  
\hskip 0.8in 
\lambda_2 + \lambda _3 + \lambda _4 \leq m 
\eea
In particular, this condition requires that $\lambda _2, \lambda _3 , \lambda _4 \leq m$. 
\item (d) The same inequalities on $\lambda_2,\lambda_3, \lambda_4$ hold for the monomials,
\bea
(\see_\delta^{22})^{\lambda_2} (\see_\delta^{12}) ^{\lambda_3} (\see_\delta^{11})^{\lambda_4}
\hskip 0.7in
0 \leq \lambda_2, \lambda_3, \lambda _4  
\hskip 0.7in 
\lambda_2 + \lambda _3 + \lambda _4 \leq m 
\eea 
that can appear in the expansion of $\Pi_\delta$.
\end{description}
}

\sm

The lemma is proven in appendix \ref{sec:B} by explicit construction. The practical  significance of the lemma is the following:
 {\cor
 \label{allspinpi}
All the spin structure dependence of $\Pi_\delta$, and thus of $P_\delta,Q_\delta$ and $C_\delta$ is localized in a polynomial dependence on the functions $\phi_2, \phi_3, \phi_4$ or equivalently on the functions $\see_\delta ^{ab}$.}

\sm

An alternative, manifestly $SL(2,\mathbb C)$ covariant proof of Corollary \ref{allspinpi}  may be found in appendix \ref{sec:B1}. In the next section we shall show that this structure may be further reduced.

 \newpage

\section{The trilinear relations}
\label{sec:trilinear}
\setcounter{equation}{0}

In this section, we shall produce and prove three powerful related  theorems on the functions $\phi_2,\phi_3,\phi_4$ and their tensorial cousins $\see_\delta^{ab}$. In a nutshell, the theorems state that these functions obey a system of polynomial equations of degree 3, whose coefficients are elements of the polynomial ring $\CC[\mu_1, \cdots, \mu_6]$ and therefore independent on the spin structure $\delta$. We shall refer to these equations as the \textit{trilinear relations}. The trilinear relations will be essential in the simplification of the cyclic product of any number of Szeg\"o kernels and its spin sum, which is why we now proceed with their derivation. 

\sm

The existence of polynomial equations for the functions $\phi_2,\phi_3,\phi_4$ and their tensorial cousins $\see_\delta^{ab}$ is expected on general grounds.
Indeed, since the branch points $u_1, \cdots, u_6$ are the roots of the polynomial,
\bea
s(x)^2=x^6 - \mu_1x^5 + \mu_2 x^4 - \mu_3 x^3 + \mu_4 x^2-\mu_5 x +\mu_6
\eea 
they are  algebraic numbers over the ring $\CC[\mu_1, \cdots, \mu_6]$. Since the combinations $\a_1, \a_2, \b_1, \b_2$ and thus $\phi_2, \phi_3,\phi_4$ are polynomials in the branch points, they are also algebraic numbers over the ring $\CC[\mu_1, \cdots, \mu_6]$ and therefore must satisfy  polynomial equations whose coefficients belong to $\CC[\mu_1, \cdots, \mu_6]$. Our third and last theorem of this section will show that these trilinear relations have a simple and beautiful interpretation in terms of $SL(2,\CC)$-tensors.

\sm

We shall begin with an elementary derivation of two of the trilinear relations, and then use $SL(2,\CC)$ representation theory to construct all the trilinear relations. Some of the intermediate formulas get to be pretty lengthy and will be relegated to appendix \ref{sec:D1}.

\subsection{Elementary derivation of two trilinear relations}

The variables $\a_1, \a_2, \a_3, \b_1,\b_2,\b_3$ in (\ref{albet}) obey a set of equations that link them to $\phi_2,\phi_3,\phi_4$ in (\ref{phi234}) and the symmetric polynomials $\mu_m$ as follows,
\begin{align}
\label{abeq}
\a_1+ \b_1 & = \mu_1 & \a_1 \b_1 & = \phi_2 & \a_1 \b_2+\a_2\b_1 & = \phi_3
\no \\
\a_2+ \b_2 & = \mu_2-\phi_2 & \a_2 \b_2 & = \phi_4 & \a_2 \b_3+\a_3\b_2 & = \mu_5
\no \\
\a_3+ \b_3 & = \mu_3-\phi_3 & \a_3 \b_3 & = \mu_6 & \a_3 \b_1+\a_1\b_3 & = \mu_4 - \phi_4
\end{align}
Solutions to this system of nine equations for six unknowns $\a_1, \a_2, \a_3, \b_1, \b_2, \b_3$ will exist provided $\phi_2, \phi_3, \phi_4$ and $\mu_m$ satisfy certain relations that are obtained  by eliminating $\a_1,\a_2,\a_3,\b_1, \b_2, \b_3$ from the above system. Here we shall limit our attention to the derivation of only two equations that will suffice to establish all the trilinear relations. 

\sm

We solve for $\a_1, \b_1$ using the first and third relations on the top line in (\ref{abeq}) in terms of  $\mu_1,\phi_3$ and $\a_2,\b_2$, and similarly solve for $\a_3, \b_3$ using the first relation on the bottom line in (\ref{abeq}) and the third relation on the middle line in terms of  $\mu_3-\phi_3,\mu_5$ and $\a_2,\b_2$,
\begin{align}
\a_1 (\a_2-\b_2) & =  -  \phi_3 + \mu_1 \a_2  & \a_3 (\a_2-\b_2) & =  -  \mu_5 +  (\mu_3 - \phi_3) \a_2
\no \\
\b_1 (\a_2-\b_2) & =    \phi_3 - \mu_1 \b_2  & \b_3 (\a_2-\b_2) & =    \mu_5 -  (\mu_3-\phi_3) \b_2
\end{align}
Substituting the solution  for $\a_1,\b_1$ into the second relation on the top line of (\ref{abeq}), and similarly substituting the solution for $\a_3, \b_3$ into the second relation on the bottom line of (\ref{abeq}) leaves two equations involving $\a_2, \b_2$, 
\bea
(\a_2-\b_2)^2 \phi_2 & = & - (\phi_3 - \mu_1 \a_2)(\phi_3 - \mu_1 \b_2)
\no \\
(\a_2-\b_2)^2 \mu_6 & = & - (\mu_5 -  (\mu_3 - \phi_3) \a_2)(\mu_5 -  (\mu_3-\phi_3) \b_2)
\eea
Both relations are symmetric under swapping $\a_2$ and $\b_2$ so that we may eliminate $\a_2$ and $\b_2$ from them by using the first and second relations on the middle line of (\ref{abeq}),  to obtain,
\bea
  (\phi_2^2 - 2 \mu_2 \phi_2 + \mu_2^2 - 4 \phi_4)  \phi_2  & = & - \phi_3^2 + \mu_1\phi_3(\mu_2-\phi_2) - \mu_1^2 \phi_4
 \no \\
 (\phi_2^2 - 2 \mu_2 \phi_2 + \mu_2^2 - 4 \phi_4) \mu_6 & = & -\mu_5^2 + \mu_5 (\mu_3-\phi_3) (\mu_2-\phi_2) - (\mu_3-\phi_3)^2 \phi_4
\eea
We may recast these relations as giving $\phi_2^3$ and $\phi_3^2 \phi_4$, respectively,  in terms of a polynomial of degree two  in $\phi_2, \phi_3, \phi_4$ with spin structure independent coefficients in $\mathbb C[\mu_1,\cdots,\mu_6]$. Furthermore, we observe that all monomials may be rendered homogeneous of degree three in the combined set of variables $\mu$ and $\phi$ provided we insert appropriate powers of $\mu_0=1$. The resulting expressions are then trilinear in the variables $\phi$ and $\mu$ and will be referred to as \textit{trilinear relations}. They are given as follows, 
\bea
\label{phicubic}
\phi_2^3 & = & 
2 \mu_2 \phi_2^2 - \mu_1 \phi_2 \phi_3 + 4 \mu_0 \phi_2 \phi_4 - \mu_0 \phi_3^2 - \mu_2^2 \phi_2 
+\mu_1 \mu_2 \phi_3 -\mu_1^2 \phi_4
\no \\
\phi_3^2 \phi_4 & = &
{-} \mu_6 \phi_2^2 + \mu_5 \phi_2 \phi_3 + 2 \mu_3 \phi_3 \phi_4 
+ 2 \mu_2 \mu_6 \phi_2   - \mu_3 \mu_5 \phi_2 
- \mu_2 \mu_5 \phi_3 
\no \\ &&
 - \mu_3^2 \phi_4
+ 4 \mu_0 \mu_6 \phi_4 
-\mu_2^2 \mu_6 + \mu_2 \mu_3 \mu_5  - \mu_0 \mu_5^2 
\eea
The first relation is the first entry in the list of trilinear relations of Theorem \ref{4.tril} stated below, while the second relation is the second to last entry.

\subsection{The complete set of trilinear relations}

The trilinear relation for $\phi_2^3$ on the first line of (\ref{phicubic}) is seen to be invariant under the action of the translation generator $\BL$ of $SL(2,\CC)$ using the rules of (\ref{Tmuphi}). But it is not invariant under the action of the inversion $\cS$ and instead  maps it to a new relation,  
\bea
\label{phi4cubic}
\phi_4^3 = 
   4 \mu_6 \phi_2 \phi_4 - \mu_6 \phi_3^2
- \mu_5 \phi_3 \phi_4 + 2 \mu_4 \phi_4^2
 - \mu_5^2 \phi_2  
 + \mu_4 \mu_5 \phi_3 
 - \mu_4^2 \phi_4
\eea
which is the last entry in Theorem \ref{4.tril} stated below. Under a translation $\BL$, the relation (\ref{phi4cubic}) is mapped to  a descendant trilinear equation,
\bea
\label{phi3phi4}
\phi_3 \phi_4^2 & = &
%
 \mu_6 \phi_2 \phi_3 + \thalf \mu_5 \phi_2 \phi_4  - \thalf \mu_5 \phi_3^2 
  + \mu_4 \phi_3 \phi_4 + \mu_3 \phi_4^2
-\thalf \mu_4 \mu_5 \phi_2   + \thalf \mu_3 \mu_5 \phi_3 
\no \\ &&
 - \mu_2 \mu_6 \phi_3  
- \thalf \mu_2 \mu_5 \phi_4  - \mu_3 \mu_4 \phi_4  + 2 \mu_1 \mu_6 \phi_4 
+ \thalf \mu_2 \mu_4 \mu_5 - \thalf \mu_1 \mu_5^2  
\eea
which is the next-to-last relation in Theorem \ref{4.tril}. Further descendants are produced by successive application of $\BL$ in view of the relations, 
\begin{align}
\label{descend}
\BL^1  (\phi_4^3) & = 6 \, \phi_3 \phi_4^2 & 
\no \\
\BL^2 (\phi_4^3) & = 6 \, ( 4 \phi_3^2 \phi_4 + \phi_2\phi_4^2) + \cO(\phi^2) & 
\no \\
\BL ^3(\phi_4^3) & = 24 \, (3 \phi_2\phi_3\phi_4 + 2 \phi_3^3)  + \cO(\phi^2) & 
\no \\ 
\BL^4 (\phi_4^3) & = 72\, (4 \phi_2 \phi_3^2  +  \phi_2^2\phi_4)  + \cO(\phi^2) 
\no \\
\BL ^5(\phi_4^3) & = 720 \, \phi_2^2 \phi_3 + \cO(\phi^2) 
\no \\
\BL^6 (\phi_4^3) & =  720 \, \phi_2^3  + \cO(\phi^2) 
\no \\
\BL^7 (\phi_4^3) & = \cO(\phi^2)
\end{align}
where $\cO(\phi^2)$ refers to terms of second, first and zeroth order  in $\phi_2,\phi_3,\phi_4$.
We have retained only the contributions trilinear in $\phi$; the full relations will be deferred to Theorem \ref{4.tril}. Including the relation (\ref{phi4cubic}) from which we initiated the process of descent, we obtain a total of 7 trilinear relations. The last of those relation, namely corresponding to $\BL^6(\phi_4^3)$, is proportional to 
the first line in  (\ref{phicubic}). 

\sm

It is manifest from the list of descent equations that not all trilinear monomials in $\phi_2,\phi_3,\phi_4$ are being produced in the descent. 
For example, $\phi_3^2 \phi_4$ and $\phi_2 \phi_4^2$ are not produced independently, but only appear in the combination $4\phi_3^2 \phi_4+\phi_2 \phi_4^2$. However, the second relation obtained in (\ref{phicubic}) precisely fills this gap. With the expression for $\phi_3^2\phi_4$ included, we obtain three additional linearly independent trilinear relations. Since the combination $\phi_2\phi_4-\phi_3^2$ satisfies, 
\bea
\BL (\phi_2\phi_4-\phi_3^2) = 3 \mu_2 \phi_4 - 6 \mu_2 \phi_3 
\eea
and contains no terms bilinear in $\phi$, it will be convenient to initiate the descent for the additional trilinear relations from the combination $(\phi_2\phi_4-\phi_3^2) \phi_4$, which is linearly independent of the combination $4\phi_3^2 \phi_4+\phi_2 \phi_4^2$ that 
is produced by the descent (\ref{descend}) starting from $\phi_4^3$. We then obtain three additional trilinear relations, corresponding to the following cubic terms, 
\bea
(\phi_2\phi_4-\phi_3^2) \phi_m \hskip 1in m=2,3,4
\eea
These results establish Theorem \ref{4.tril} below, which encompasses all trilinear relations.

{\thm
\label{4.tril}
The functions $\phi_2, \phi_3, \phi_4$ obey the following trilinear relations,}
\bea
\phi_2^3 & = & 
2 \mu_2 \phi_2^2 - \mu_1 \phi_2 \phi_3 + 4 \mu_0 \phi_2 \phi_4 - \mu_0 \phi_3^2 - \mu_2^2 \phi_2 
+\mu_1 \mu_2 \phi_3 -\mu_1^2 \phi_4
\no \\
\phi_2^2 \, \phi_3 & = &
\mu_3 \phi_2^2 + \mu_2 \phi_2 \phi_3 + \thalf \mu_1 \phi_2 \phi_4  -\thalf \mu_1 \phi_3^2 + \mu_0 \phi_3 \phi_4
  - \thalf \mu_1 \mu_4 \phi_2   + 2 \mu_0 \mu_5 \phi_2
\no \\ &&
 - \mu_2 \mu_3 \phi_2   - \mu_0 \mu_4 \phi_3  + \thalf \mu_1 \mu_3 \phi_3 - \thalf \mu_1 \mu_2 \phi_4
  - \thalf \mu_1^2 \mu_5  + \thalf \mu_1 \mu_2 \mu_4
 \no \\
 \phi_2^2 \, \phi_4 & = & 
\mu_4 \phi_2^2
 - \mu_3 \phi_2 \phi_3
 +2 \mu_2 \phi_2 \phi_4 
+ \mu_2 \phi_3^2
  - 3 \mu_1 \phi_3 \phi_4 
+ 5 \mu_0 \phi_4^2
  \no \\ &&
  - 4 \mu_0 \mu_6 \phi_2 
+ \mu_1 \mu_5 \phi_2
- 2 \mu_2 \mu_4 \phi_2  
  + \mu_3^2 \phi_2
  + 2 \mu_0 \mu_5 \phi_3 
 + \mu_1 \mu_4 \phi_3
 - \mu_2 \mu_3 \phi_3
\no \\ &&
- 6 \mu_0 \mu_4 \phi_4 
 + 3 \mu_1 \mu_3 \phi_4
  - \mu_2^2 \phi_4
+ \mu_1^2 \mu_6 - \mu_1 \mu_2 \mu_5- \mu_1 \mu_3 \mu_4
+ \mu_2^2 \mu_4  + \mu_0 \mu_4^2 
 \no \\
 %
 \phi_2 \, \phi_3^2 & = &
  2 \mu_3 \phi_2 \phi_3
  + \mu_1 \phi_3 \phi_4
  - \mu_0 \phi_4^2
 + 4 \mu_0 \mu_6 \phi_2
  - \mu_3^2 \phi_2 
   -\mu_1 \mu_4 \phi_3 
 \no \\ &&
  + 2 \mu_0 \mu_4 \phi_4
  - \mu_1 \mu_3 \phi_4
  -\mu_1^2 \mu_6 
  + \mu_1 \mu_3 \mu_4 
  - \mu_0 \mu_4^2  
 \no \\
 \phi_2\phi_3\phi_4 & = & 
  \thalf \mu_5 \phi_2^2 
 +\tfrac{5}{4} \mu_3 \phi_2 \phi_4
 +\tfrac{1}{4} \mu_3 \phi_3^2 
 +\thalf \mu_1 \phi_4^2 
  -\tfrac{1}{4} \mu_3 \mu_4 \phi_2 
  - \thalf \mu_2 \mu_5 \phi_2
  + \thalf \mu_1 \mu_6 \phi_2
 \no \\ &&
  + \mu_0 \mu_6 \phi_3 
  - \tfrac{1}{4} \mu_3^2 \phi_3 
  + \thalf \mu_0 \mu_5 \phi_4 
  -\thalf \mu_1 \mu_4 \phi_4
  -\tfrac{1}{4} \mu_2 \mu_3 \phi_4 
 \no \\ &&
  -\thalf \mu_0 \mu_4 \mu_5 +\tfrac{1}{4} \mu_2 \mu_3 \mu_4 + \tfrac{1}{4} \mu_1 \mu_3 \mu_5 
 - \thalf \mu_1 \mu_2 \mu_6     
 \no \\
  \phi_3^3 & = &
  - \thalf \mu_5 \phi_2^2
   + \mu_4 \phi_2 \phi_3
  + \tfrac{7}{4} \mu_3 \phi_3^2
   -\tfrac{1}{4} \mu_3 \phi_2 \phi_4
   + \mu_2 \phi_3 \phi_4 
   -\thalf \mu_1 \phi_4^2
   + \thalf \mu_2 \mu_5 \phi_2
   \no \\ &&
  - \tfrac{3}{4} \mu_3 \mu_4 \phi_2
  +\tfrac{3}{2} \mu_1 \mu_6 \phi_2 
  + 3 \mu_0 \mu_6 \phi_3
   - \mu_1 \mu_5 \phi_3
    - \mu_2 \mu_4 \phi_3
  - \tfrac{3}{4} \mu_3^2 \phi_3    
    +\tfrac{3}{2} \mu_0 \mu_5 \phi_4 
   \no \\ &&
 + \thalf \mu_1 \mu_4 \phi_4
  - \tfrac{3}{4} \mu_2 \mu_3 \phi_4 
   - \tfrac{3}{2} \mu_0 \mu_4 \mu_5 + \tfrac{3}{4} \mu_2 \mu_3 \mu_4 + \tfrac{3}{4} \mu_1 \mu_3 \mu_5
   - \tfrac{3}{2} \mu_1 \mu_2 \mu_6    
   \no \\
 \phi_2 \phi_4^2 & = &
  5 \mu_6 \phi_2^2
 - 3 \mu_5 \phi_2 \phi_3 
 + 2 \mu_4 \phi_2 \phi_4 
 + \mu_4 \phi_3^2
 - \mu_3 \phi_3 \phi_4 
 + \mu_2 \phi_4^2 
  - 6 \mu_2 \mu_6 \phi_2 
 \no \\ &&
  + 3 \mu_3 \mu_5 \phi_2
  - \mu_4^2 \phi_2
 + 2 \mu_1 \mu_6 \phi_3 
 + \mu_2 \mu_5 \phi_3
  - \mu_3 \mu_4 \phi_3
  - 4 \mu_0 \mu_6 \phi_4 
 + \mu_1 \mu_5 \phi_4 
  \no \\ &&
   -2 \mu_2 \mu_4 \phi_4
 + \mu_3 ^2 \phi_4 
 -\mu_1 \mu_4 \mu_5 
 + \mu_2^2 \mu_6 
 - \mu_2 \mu_3 \mu_5 
 + \mu_2 \mu_4^2 
 + \mu_0 \mu_5^2 
  \no \\
\phi_3^2 \phi_4 & = &
 - \mu_6 \phi_2^2
 + \mu_5 \phi_2 \phi_3
 + 2 \mu_3 \phi_3 \phi_4 
 + 2 \mu_2 \mu_6 \phi_2
  - \mu_3 \mu_5 \phi_2 
\no \\ &&
 - \mu_2 \mu_5 \phi_3
 + 4 \mu_0 \mu_6 \phi_4 
 - \mu_3^2 \phi_4
-\mu_2^2 \mu_6 + \mu_2 \mu_3 \mu_5     
 - \mu_0 \mu_5^2 
\no \\
%
\phi_3 \phi_4^2 & = &
 \mu_6 \phi_2 \phi_3
+ \thalf \mu_5 \phi_2 \phi_4
- \thalf \mu_5 \phi_3^2
+ \mu_4 \phi_3 \phi_4
+ \mu_3 \phi_4^2
-\thalf \mu_4 \mu_5 \phi_2 
 - \mu_2 \mu_6 \phi_3
\no \\ &&
+ \thalf \mu_3 \mu_5 \phi_3  
+ 2 \mu_1 \mu_6 \phi_4 
- \thalf \mu_2 \mu_5 \phi_4
- \mu_3 \mu_4 \phi_4
 +\thalf \mu_2 \mu_4 \mu_5 - \thalf \mu_1 \mu_5^2    
\no \\ 
\phi_4^3 & = & 
4 \mu_6 \phi_2 \phi_4 - \mu_6 \phi_3^2
- \mu_5 \phi_3 \phi_4
+2 \mu_4 \phi_4^2  
 - \mu_5^2 \phi_2   
 + \mu_4 \mu_5 \phi_3 
- \mu_4^2 \phi_4 
\eea

The first relation coincides with the first line of (\ref{phicubic}); the last relation with (\ref{phi4cubic});  the next-to-last relation is (\ref{phi3phi4}); and the second to last relation is the second equation in (\ref{phicubic}).  All relations may be double-checked explicitly with {\sc maple} or {\sc mathematica}.

\sm

The use of the $SL(2,\CC)$ transformations of translation $\BL$ and inversion $\cS$ in the proof of Theorem \ref{4.tril} points towards a group theoretic underpinning of the system of trilinear relations, which is obscured in Theorem \ref{4.tril} but will be made manifest in the next subsection.

\subsection{$SL(2,\CC)$ structure of the trilinear relations}

While the trilinear relations of Theorem \ref{4.tril} were firmly established  in the preceding subsections,  their $SL(2,\CC)$ structure is not manifest. In this subsection, we shall show that the trilinear relations enjoy a simple and beautiful reformulation in terms of $SL(2,\CC)$ tensors. 

\sm

We begin by recasting the trilinear relations of Theorem \ref{4.tril} in terms of the components $\see_\delta ^{ab}$ of the $SL(2,\CC)$ tensor $\see_\delta$, using their expressions (\ref{4.see}) in terms of $\phi_m$ and $\mu_m$. The results are rather bulky and are relegated to Corollary \ref{sping3.55} of appendix \ref{sec:D1}. However, they are not yet much more illuminating than the relations of Theorem~\ref{4.tril} themselves.

\subsubsection{Decomposition of the trilinear tensor $\see_\delta \otimes \see_\delta \otimes\see_\delta$}

To realize the $SL(2,\CC)$ covariance in a manifest way, we consider the trilinear tensor product  $\see_\delta \otimes \see_\delta \otimes\see_\delta$ whose components,
\bea
\see_\delta ^{a_1 a_2} \, \see_\delta ^{a_3 a_4} \, \see_\delta ^{a_5 a_6}
\eea
precisely account for the terms in the trilinear equations that are trilinear in $\phi$. We recall that the symmetric rank-two tensor $\see_\delta$ itself transforms under the three-dimensional irreducible representation of $SL(2,\CC)$, denoted by ${\bf 3}$. The rank six tensor $\see_\delta \otimes \see_\delta \otimes\see_\delta$ of $SL(2,\CC)$  is symmetric under swapping the entries in each pair $(a_1,a_2)$, $(a_3,a_4)$, $(a_5,a_6)$ and symmetric under the permutations of the pairs.  One establishes, either by inspection of Corollary \ref{sping3.55}, or by direct counting, that the tensor $\see_\delta \otimes \see_\delta \otimes\see_\delta$ has ten independent components, and thus    transforms under a ten-dimensional representation of $SL(2,\CC)$. This counting precisely reproduces the number of trilinear relations listed in Theorem \ref{4.tril}, as expected.

\sm

The representation $\see_\delta \otimes \see_\delta \otimes\see_\delta$  is reducible as may be seen by constructing it out of the tensor product of three identical copies of the three-dimensional representation  ${\bf 3}$ of $SL(2,\CC)$ under which $\see_\delta$ transforms. The branching rules are as follows, see appendix \ref{sec:A.1} for further details,
\bea
\see_\delta \otimes \see_\delta  & = & {\bf 5} \oplus {\bf 1}
\no \\
 \see_\delta \otimes \see_\delta  \otimes \see_\delta & = & {\bf 7} \oplus {\bf 3}
 \eea
In the decomposition of the tensor product $\see_\delta \otimes \see_\delta$ the representation ${\bf 3}$, which would occur in the tensor product of distinct vectors, is absent here in  the tensor product of identical vectors. The representation ${\bf 5}$, and a second copy of the ${\bf 3}$, are absent in the tensor product $\see_\delta \otimes \see_\delta \otimes \see_\delta$ for the same reason. 

\sm

The remaining representation ${\bf 7} \oplus {\bf 3}$ accounts for the ten entries in Theorem \ref{4.tril} and Corollary \ref{sping3.55}. To see how this works, we start by rewriting the
singlet in $\see_\delta \otimes \see_\delta$ as follows
\bea
\det (\see_\delta) =  \tfrac{1}{2} \see_\delta ^{a_1a_2} \see_\delta ^{b_1b_2} \ep_{a_1b_1} \ep_{a_2b_2} 
= \see _\delta ^{11} \see_\delta^{22} - \see _\delta^{12} \see _\delta ^{12}
\eea
The representation ${\bf 3}$  in the decomposition of $\see_\delta \otimes \see_\delta  \otimes \see_\delta$ corresponds to  $\see _\delta ^{a_1 a_2} \, \det (\see_\delta)$, while the ${\bf 7}$ corresponds to the totally symmetrized tensor with components $\see_\delta ^{(a_1 a_2} \, \see_\delta ^{a_3 a_4} \, \see_\delta ^{a_5 a_6)}$. Since both of these tensors are irreducible, one may obtain the components of each tensor by successively applying the translation generator $\BL$ to the corresponding highest weight state, namely $\see_\delta ^{11} \, \det (\see_\delta)$ for the ${\bf 3}$ and $\see_\delta^{11} \, \see_\delta^{11} \,\see_\delta^{11} $ for the ${\bf 7}$, as illustrated in the table below, 
\begin{align}
\label{2.eeetrans}
\BL \big ( \see _\delta ^{11} \,  \see _\delta ^{11} \,  \see _\delta ^{11} \big ) & =  
 6 \, \see _\delta ^{12} \,  \see _\delta ^{11} \,  \see _\delta ^{11}
& 
\BL \big ( \see _\delta ^{12} \,  \see _\delta ^{11} \,  \see _\delta ^{11} \big ) & =  
 4 \, \see _\delta ^{12} \,  \see _\delta ^{12} \,  \see _\delta ^{11} 
+ \see _\delta ^{22} \,  \see _\delta ^{11} \,  \see _\delta ^{11}
\no \\
\BL \big ( \see _\delta ^{22} \,  \see _\delta ^{11} \,  \see _\delta ^{11} \big ) & =  
 4 \, \see _\delta ^{22} \,  \see _\delta ^{12} \,  \see _\delta ^{11}
& 
\BL \big ( \see _\delta ^{12} \,  \see _\delta ^{12} \,  \see _\delta ^{11} \big ) & =  
 2 \, \see _\delta ^{22} \,  \see _\delta ^{12} \,  \see _\delta ^{11} 
+ 2 \, \see _\delta ^{12} \,  \see _\delta ^{12} \,  \see _\delta ^{12}
\no \\
\BL \big ( \see _\delta ^{12} \,  \see _\delta ^{12} \,  \see _\delta ^{12} \big ) & =  
 3 \, \see _\delta ^{22} \,  \see _\delta ^{12} \,  \see _\delta ^{12}
&
\BL \big ( \see _\delta ^{22} \,  \see _\delta ^{12} \,  \see _\delta ^{11} \big ) & =  
 2 \, \see _\delta ^{22} \,  \see _\delta ^{12} \,  \see _\delta ^{12} 
+  \see _\delta ^{22} \,  \see _\delta ^{22} \,  \see _\delta ^{11}
\no \\
\BL \big ( \see _\delta ^{22} \,  \see _\delta ^{12} \,  \see _\delta ^{12} \big ) & =  
 2 \, \see _\delta ^{22} \,  \see _\delta ^{22} \,  \see _\delta ^{12}
&
\BL \big ( \see _\delta ^{22} \,  \see _\delta ^{22} \,  \see _\delta ^{11} \big ) & =  
 2 \, \see _\delta ^{22} \,  \see _\delta ^{22} \,  \see _\delta ^{12}
\no \\
\BL \big ( \see _\delta ^{22} \,  \see _\delta ^{22} \,  \see _\delta ^{12} \big ) & =  
  \see _\delta ^{22} \,  \see _\delta ^{22} \,  \see _\delta ^{22}
&
\BL \big ( \see _\delta ^{22} \,  \see _\delta ^{22} \,  \see _\delta ^{22} \big ) & =  
0
\end{align}
Similarly, under inversion, one has, 
\begin{align}
\cS \big ( \see _\delta ^{11} \,  \see _\delta ^{11} \,  \see _\delta ^{11} \big ) & =  
+  \mu_6^{-1} \, \see _\delta ^{22} \,  \see _\delta ^{22} \,  \see _\delta ^{22}
& 
\cS \big ( \see _\delta ^{12} \,  \see _\delta ^{11} \,  \see _\delta ^{11} \big ) & =  
-  \, \mu_6^{-1} \, \see _\delta ^{12} \,  \see _\delta ^{22} \,  \see _\delta ^{22} 
\no \\
\cS \big ( \see _\delta ^{22} \,  \see _\delta ^{11} \,  \see _\delta ^{11} \big ) & =  
+ \mu_6^{-1} \, \see _\delta ^{11} \,  \see _\delta ^{22} \,  \see _\delta ^{22}
& 
\cS \big ( \see _\delta ^{12} \,  \see _\delta ^{12} \,  \see _\delta ^{11} \big ) & =  
+ \mu_6^{-1} \, \see _\delta ^{12} \,  \see _\delta ^{12} \,  \see _\delta ^{22} 
\no \\
\cS \big ( \see _\delta ^{12} \,  \see _\delta ^{12} \,  \see _\delta ^{12} \big ) & =  
-  \, \mu_6^{-1} \, \see _\delta ^{12} \,  \see _\delta ^{12} \,  \see _\delta ^{12}
&
\cS \big ( \see _\delta ^{22} \,  \see _\delta ^{12} \,  \see _\delta ^{11} \big ) & =  
-  \, \mu_6^{-1} \, \see _\delta ^{11} \,  \see _\delta ^{12} \,  \see _\delta ^{22} 
\no \\
\cS \big ( \see _\delta ^{22} \,  \see _\delta ^{12} \,  \see _\delta ^{12} \big ) & =  
+ \mu_6^{-1} \, \see _\delta ^{11} \,  \see _\delta ^{12} \,  \see _\delta ^{12}
&
\cS \big ( \see _\delta ^{22} \,  \see _\delta ^{22} \,  \see _\delta ^{11} \big ) & =  
+ \mu_6^{-1} \, \see _\delta ^{11} \,  \see _\delta ^{11} \,  \see _\delta ^{22}
\no \\
\bS \big ( \see _\delta ^{22} \,  \see _\delta ^{22} \,  \see _\delta ^{12} \big ) & =  
-  \mu_6^{-1} \, \see _\delta ^{11} \,  \see _\delta ^{11} \,  \see _\delta ^{12}
&
\cS \big ( \see _\delta ^{22} \,  \see _\delta ^{22} \,  \see _\delta ^{22} \big ) & =  
+ \mu_6^{-1} \, \see _\delta ^{11} \,  \see _\delta ^{11} \,  \see _\delta ^{11}
\end{align}
One may verify by inspection that the trilinear relations for $(\see^{11}_\delta)^3$ and $(\see^{22}_\delta)^3$ 
indeed map into one another term by term under inversion. The corresponding behavior was already established for the trilinear equations expressed  in terms of $\phi_m$ variables in Theorem \ref{4.tril}.

\subsubsection{$SL(2,\CC)$-tensor formulation of the trilinear relations}

Having decomposed the trilinear tensor $\see_\delta \otimes \see_\delta \otimes\see_\delta$ in the trilinear relations, it now remains to find a similar representation for its contributions of homogeneity degree $2,1,0$ in $\see_\delta$.
To do so, we shall use the fact that the trilinear relations are homogeneous of combined degree 3 in  $\see_\delta$ and $\mu_m$ (inserting $\mu_0=1$  to achieve homogeneity, if needed). We shall also use the existence of the tensor $\CM_1$ which is homogeneous  in $\mu_m$ of degree 1, and whose components were given  in (\ref{MXcomps}). Combining these group-theoretic properties, the general structure of the trilinear relations thus takes the form, 
\bea
  \see_\delta \otimes \see_\delta  \otimes \see_\delta = 
\big [   \CS \otimes \see_\delta \otimes \see_\delta \big ]_{ {\bf 7} \oplus {\bf 3} } \, \oplus \,
\big [  \CS \otimes \CS \otimes \see _\delta \big ] _{{\bf 7} \oplus {\bf 3} } 
  \, \oplus \,  \big [ \CS \otimes \CS \otimes \CS \big ] _{{\bf 7} \oplus {\bf 3} } 
  \quad
  \eea
where $\big [ \cdots \big ] _{{\bf 7} \oplus {\bf 3} } $ indicates the projection onto the representation ${\bf 7} \oplus {\bf 3}$. We shall now obtain the projections onto the two irreducible representations ${\bf 3}$ and ${\bf 7}$ in turn.

\subsubsection{Projection onto the representation ${\bf 7}$}

The projection onto the ${\bf 7}$ is obtained by symmetrizing the trilinear part $\see_\delta ^{(a_1 a_2} \, \see_\delta ^{a_3 a_4} \, \see_\delta ^{a_5 a_6)}$. To obtain the components along the ${\bf 7}$ of the bilinear, linear, and $\see_\delta$-independent terms, we use the following decompositions,
\begin{align}
& \CS \otimes \see_\delta \otimes \see_\delta &&   {\bf 7} \otimes ( {\bf 5} \oplus {\bf 1} )
& \to {\bf 7} \oplus {\bf 7}
\no \\
& \CS \otimes \CS \otimes \see _\delta &&  {\bf 7} \otimes ({\bf 9} \oplus {\bf 7} \oplus  {\bf 5} )  
& \to {\bf 7} \oplus {\bf 7}
\no \\
& \CS \otimes \CS \otimes \CS &&  {\bf 7}  \otimes ({\bf 13} \oplus {\bf 9} \oplus {\bf 5} \oplus {\bf 1} )  & \to {\bf 7} \oplus {\bf 7}
\end{align}
where the right-most entry gives the multiplicity of the representation ${\bf 7}$ in the decomposition. The corresponding decomposition 
in components is readily obtained, and we have, 
\bea
\label{ell3}
\see_\delta ^{(a_1 a_2} \see_\delta ^{a_3 a_4} \see_\delta ^{a_5 a_6)} 
& = &
C_1 \, \CS ^{b_1b_2 (a_1 a_2 a_3 a_4} \, \see_\delta ^{a_5 a_6)} \see _\delta ^{b_3b_4} \ep _{b_1b_3 } \ep _{b_2b_4}
\no \\ &&
+ C_2 \,  \CS ^{b_1 b_2 (a_1 a_2 a_3 a_4 } \, \see _\delta ^{a_5| b_3} \see_\delta ^{a_6) b_4} \ep_{b_1b_3} \ep _{b_2b_4}
\no \\ &&
+ C_3 \,  \CM_2^{(a_1 a_2 a_3 a_4} \see_\delta^{a_5a_6)} 
+C_4 \, \CM_2 ^{a_1 \cdots a_6 b_1b_2} \see_\delta ^{b_3b_4} \, \ep_{b_1b_3} \ep_{b_2b_4} 
\no \\ &&
+ C_5 \, \CM_2 \,  \CS^{a_1 a_2 a_3 a_4 a_5 a_6}
 + C_6 \,    \CM_3^{a_1 a_2 a_3 a_4 a_5 a_6}
\eea
where $C_1, \cdots, C_6$ are coefficients that are not determined by $SL(2,\CC)$ group theory, and the vertical bar in the superscript of $ \CS^{b_1 b_2 (a_1 a_2 a_3 a_4 } \, \see _\delta ^{a_5| b_3} \see_\delta ^{a_6) b_4}$
instructs to exclude $b_3$ from the symmetrization in $a_1,\ndots,a_6$. 
The tensors $\CM_2$ of ranks 0, 4, and 8 are defined as follows,
\bea 
\label{5.M2}
\CM_2 & = &  \thalf \, \CS ^{b_1 \cdots b_6} \, \CS^{c_1 \cdots c_6} \, \ep_{b_1 c_1} \cdots \ep _{b_6 c_6}
\no \\
\CM_2^{a_1\cdots  a_4} & = &  \thalf \,  \CS ^{b_1 b_2 b_3 b_4 (a_1a_2} \, \CS^{a_3a_4) c_1 c_2 c_3 c_4} \, 
\ep_{b_1 c_1} \cdots \ep _{b_4 c_4}
\no \\
\CM_2^{a_1\cdots a_8} & = &  \thalf \,  \CS ^{b_1 b_2 ( a_1 \cdots a_4 } \, \CS^{a_5 \cdots a_8) c_1 c_2 } \, 
\ep_{b_1 c_1} \ep _{b_2 c_2}
\eea
The projections onto the ${\bf 3} $ and ${\bf 7}$ of the  tensor product $\CS \otimes \CS \otimes \CS$
are given as follows in components (the projections onto the remaining representations will not be needed),  
\bea
\label{5.M3}
\CM_3 ^{a_1 a_2} & = & \CS^{a_1 a_2 b_1 b_2 b_3 b_4} \, \CM_2 ^{c_1 c_2 c_3 c_4} \, 
\ep_{b_1 c_1} \cdots \ep _{b_4 c_4}
\no \\
\CM_3 ^{a_1 \cdots a_6} & = & 
\CS^{b_1 b_2(a_1 a_2 a_3 a_4} \, \CM_2 ^{a_5 a_6)  c_1 c_2} \, \ep_{b_1 c_1} \, \ep _{b_2 c_2} \, 
\eea
The components of these tensors are given explicitly  in terms of $\mu_m$ in appendix \ref{sec:D2},
and their generalizations $\CM_w$ to  degree $4\leq w\leq 6$ in the $\mu_m$ are introduced in appendix~\ref{sec:D2higher}. A representation-theoretic method for counting the number of independent components of the $\CM_w^{a_1\ndots a_r}$-tensors at various ranks $r$ and degrees of homogeneity $w$ in $\mu_m$ is explained in appendix \ref{sec:cl.1}. Our normalization conventions for {(anti-)}symmetrizing $k$ indices include a prefactor $\frac{1}{k!}$ to ensure overall weight one, for instance,
\bea
\see_\delta ^{(a_1 a_2} \see_\delta ^{a_3 a_4)}
= \frac{1}{3} ( \see_\delta ^{a_1 a_2} \see_\delta ^{a_3 a_4}
+\see_\delta ^{a_1 a_3} \see_\delta ^{a_2 a_4}
+\see_\delta ^{a_1 a_4} \see_\delta ^{a_2 a_3})
\eea
To evaluate the coefficients $C_1, \cdots, C_6$ in (\ref{ell3}), it suffices to examine the highest weight component in the representation ${\bf 7}$ of the tensorial equation, as all other components may be deduced from it by successive application of the translation generator. To this end we set the free indices to $a_1=\cdots = a_6=1$ and evaluate (\ref{5.M3}) accordingly on this component,
\bea
(\see_\delta ^{11})^3  
& = &
C_1 \, \see_\delta ^{11} \Big ( \CS ^{111122} \, \see_\delta ^{11} - 2 \CS^{111112} \see_\delta^{12} + \CS^{111111} \see_\delta^{22} \Big )
\no \\ &&
+ C_2 \Big ( 
 \CS ^{111111 } \, (\see _\delta ^{1 2})^2 
-  2\CS ^{1 1111 2} \, \see _\delta ^{1 2} \see_\delta ^{1 1} 
 +    \CS ^{111122 } \, (\see _\delta ^{1 1})^2  \Big ) 
 \no \\ &&
+C_4 \Big (
\CM_2 ^{111111 11} \see_\delta ^{22} 
-2 \CM_2 ^{111111 12} \see_\delta ^{12} 
+\CM_2 ^{111111 22} \see_\delta ^{11}  \Big ) 
\no \\ &&
+ C_3 \,  \CM_2^{1111} \see_\delta^{11} + C_5 \, \CM_2 \,  \CS^{111111} + C_6 \,    \CM_3^{111111}
\eea
We now match this expression  with the first relation in Corollary \ref{sping3.55} by converting the components of $\CS$ into $\mu_m$ using (\ref{MXcomps}).   Identifying the terms proportional to $\see_\delta^{11} \see_\delta ^{22}$ and $(\see_\delta^{11})^2$ readily gives $C_1=1$ and $C_2 = -\tfrac{1}{4}$. All other  terms bilinear in $\see_\delta$ then automatically match. Identifying the terms linear in $\see_\delta$ gives $C_4 = \tfrac{9}{4}$ and $C_3 =\tfrac{99}{56}$. Finally, identifying the terms independent of $\see_\delta$ gives $C_5=\tfrac{9}{32}$ and $C_6 = - \tfrac{27}{16}$. 

\sm

We note that the decomposition (\ref{ell3}) may be recast in an alternative basis of tensors using
one or both of the following relations,
\bea
 \label{clean.31} 
\CM_1^{b_1b_2(a_1 a_2 a_3 a_4} \see_\delta^{a_5 |c_1} \see_\delta^{a_6)c_2}
  \varepsilon_{b_1 c_1}  \varepsilon_{b_2c_2} 
& = & 
\CM_1^{b_1 b_2(a_1 a_2 a_3 a_4}  \see_\delta^{a_5 a_6)}  \see_\delta^{c_1c_2}
 \varepsilon_{b_1 c_1}  \varepsilon_{b_2 c_2}  
 \no \\ &&
 {-} ( \det \see_\delta) \CM_1^{a_1 a_2\ndots a_6}
\no \\
\CM_1^{b_1 b_2 b_3(a_1 a_2 a_3} \CM_1^{a_4 a_5 a_6)c_1 c_2 c_3} \see_\delta^{gh}
    \varepsilon_{b_1 c_1 }  \varepsilon_{b_2 c_2}  \varepsilon_{b_3 g}  \varepsilon_{c_3h}  
& = & 
2 \CM_2^{a_1 \ndots a_6 b_1 b_2} 
    \see_\delta^{c_1 c_2}  \varepsilon_{b_1 c_1}  \varepsilon_{b_2 c_2} 
    \no \\ &&
     - \frac{3}{7} \CM_2^{(a_1 \ndots a_4} \see_\delta^{a_5 a_6)}
\eea

\subsubsection{Projection onto the representation ${\bf 3}$}

In addition to the totally symmetric combination in the ${\bf 7}$, the tensor product  $\see_\delta \otimes \see_\delta \otimes \see_\delta$ also contains the ${\bf 3}$ which is obtained by anti-symmetrizing one pair of indices, 
\bea
\see_\delta ^{a_1 a_2} \see_\delta ^{a_3 a_4} \see_\delta ^{a_5 a_6} \ep_{a_4a_6}
=  (\det \see _\delta) \, \see_\delta ^{a_1 a_2} \,  \ep^{a_3 a_5}
\eea
where $ (\det \see _\delta)$ is a singlet and our sign conventions for $\ep^{ab}$ are fixed
by $\ep^{12}=\ep_{12}=1$, leading to $\ep_{ab} \ep^{cd}=\delta_a^c \delta_b^d - 
\delta_a^d \delta_b^c$ and hence $\ep_{ae} \ep^{ed} = - \delta_a^d$. 
The component decomposition is as follows,
\bea
(\det \see _\delta) \, \see_\delta ^{a_1 a_2} & = & 
D_1 \, \CM_1 ^{a_1 a_2 b_1 \cdots b_4} \, \see_\delta^ {c_1c_2} \, \see_\delta^ {c_3c_4} \, \ep_{b_1 c_1} \cdots \ep_{b_4 c_4}
+ D_2 \, \CM_2 \, \see _\delta ^{a_1 a_2}
\no \\ &&
 + D_3 \, \CM_2 ^{a_1a_2b_1b_2} \see_\delta^{c_1c_2} \ep_{b_1c_1} \ep _{b_2 c_2}
+ D_4 \, \CM_3^{a_1 a_2}
\eea 
where the coefficients $D_1, \cdots, D_4$ are not  determined by $SL(2,\CC)$ group theory alone.
Identifying the terms bilinear in $\see_\delta$ we find $D_1=\tfrac{3}{2}$; those  linear in $\see_\delta^{12}$ and $\see_\delta^{22}$ give $D_3=-\tfrac{3}{4}$; those linear in $\see_\delta^{11}$ give $D_2=-\tfrac{3}{8}$; and those independent of $\see_\delta$ give  $D_4=-\tfrac{9}{16}$.

\subsubsection{Summary of the tensorial representation of the trilinear relations}
 
 The results of the preceding subsections on the $SL(2,\CC)$ tensorial structure of the trilinear relations may be collected in the following theorem which is equivalent to Theorem \ref{4.tril}.   
{\thm 
\label{thm:6}
The component of the trilinear relations transforming under the {\bf 7} of $SL(2,\CC)$ is given as follows, 
\bea
\label{ell3a}
\see_\delta ^{(a_1 a_2} \see_\delta ^{a_3 a_4} \see_\delta ^{a_5 a_6)} 
& = &
\CS ^{b_1b_2 (a_1 \cdots a_4} \, \see_\delta ^{a_5 a_6)} \see _\delta ^{b_3b_4} \ep _{b_1b_3 } \ep _{b_2b_4}
\no \\ &&
- \tfrac{1}{4}  \,  \CS ^{b_1 b_2 (a_1 \cdots a_4 } \, 
\see _\delta ^{a_5 | b_3 } \see_\delta ^{a_6) b_4} \ep_{b_1b_3} \ep _{b_2b_4}
\no \\ &&
+ \tfrac{99}{56} \,  \CM_2^{(a_1 \cdots a_4} \, \see_\delta^{a_5a_6)} 
+ \tfrac{9}{4}  \, \CM_2 ^{a_1 \cdots a_6 b_1b_2} \, \see_\delta ^{b_3b_4} \, \ep_{b_1b_3} \ep_{b_2b_4} 
\no \\ &&
+ \tfrac{9}{32}  \, \CM_2 \,  \CS^{a_1 \cdots a_6}
- \tfrac{27}{16} \,    \CM_3^{a_1 \cdots a_6}
\eea
while its component transforming under the {\bf 3} of $SL(2,\CC)$ is given as follows, 
\bea
\label{ell3b}
(\det \see _\delta) \, \see_\delta ^{a_1 a_2} & = & 
\tfrac{3}{2} \, \CM_1 ^{a_1 a_2 b_1 \cdots b_4} \, \see_\delta^ {c_1c_2} \, \see_\delta^ {c_3c_4} \, \ep_{b_1 c_1} \cdots \ep_{b_4 c_4}
-\tfrac{3}{8}  \, \CM_2 \, \see _\delta ^{a_1 a_2}
\no \\ &&
 -\tfrac{3}{4} \, \CM_2 ^{a_1a_2b_1b_2} \see_\delta^{c_1c_2} \ep_{b_1c_1} \ep _{b_2 c_2}
-\tfrac{9}{16} \, \CM_3^{a_1 a_2}
\eea }
The proof of both was given in the derivations of the preceding subsections. The tensor relation of (\ref{ell3a}) is equivalent to the seven linear combinations of the trilinear relations of Corollary \ref{sping3.55} that descend from $(\see_\delta^{11})^3$, while the tensor relation of (\ref{ell3b}) is equivalent to the trilinear relation (\ref{delapp}) for $  (\det \see_\delta) \, \see_\delta^{11}$ and its two descendants under $SL(2,\CC)$. The combined system of relations in (\ref{ell3a}) and (\ref{ell3b}) is equivalent to the system of trilinear equations in Theorem \ref{4.tril} or equivalently in Corollary \ref{sping3.55}.

{\cor
\label{cor:noprod}
The trilinear relation for $\see _\delta ^{a_1 a_2} \see _\delta ^{b_1b_2} \see _\delta ^{c_1 c_2}$, 
considered without any additional  (anti-)symmetrization prescription for the indices $a_1,a_2,b_1,b_2,c_1,c_2$, transforms under the reducible representation
${\bf 7} \oplus {\bf 3}$ of $SL(2,\mathbb C)$ and can be assembled from (\ref{ell3a}) and (\ref{ell3b}),
\begin{align}
&\see_\delta^{a_1 a_2} \see_\delta^{b_1 b_2} \see_\delta^{c_1 c_2}
= \Big\{ \tfrac{ 4}{3}  \CM_1^{a_1 a_2b_1 b_2 ef} \see_\delta^{c_1 c_2} \see_\delta^{gh} \varepsilon_{eg} \varepsilon_{fh} -\tfrac{1}{12}  \CM_1^{a_1 a_2b_1 b_2 ef} \see_\delta^{c_1 g} \see_\delta^{ c_2 h} \varepsilon_{eg} \varepsilon_{fh}
\notag \\
&\quad \quad \quad
 -\tfrac{1}{4} \Big ( \CM_1^{a_1 a_2b_1 c_1 ef} \see_\delta^{b_2 c_2 } 
 +\CM_1^{a_1 a_2b_1 c_2 ef} \see_\delta^{b_2 c_1 } 
 +\CM_1^{a_1 a_2b_2 c_1 ef} \see_\delta^{b_1 c_2 } 
 +\CM_1^{a_1 a_2b_2 c_2 ef} \see_\delta^{b_1 c_1 } \Big ) \see_\delta^{ g h} \varepsilon_{eg} \varepsilon_{fh} 
\notag \\
&\quad \quad \quad
+  \tfrac{15}{56} \Big (
\CM_2^{a_1 a_2 b_1 c_1} \see_\delta^{b_2 c_2}
+\CM_2^{a_1 a_2 b_1 c_2} \see_\delta^{b_2 c_1}
+\CM_2^{a_1 a_2 b_2 c_1} \see_\delta^{b_1 c_2}
+\CM_2^{a_1 a_2 b_2 c_2} \see_\delta^{b_1 c_1}
\Big )
\notag \\
&\quad \quad \quad
 -\tfrac{27}{56}  \CM_2^{a_1 a_2 b_1 b_2} \see_\delta^{c_1 c_2} 
 -\tfrac{3}{40}  \CM_2 \see_\delta^{a_1 a_2} \big ( \varepsilon^{b_1 c_1} \varepsilon^{b_2 c_2}
+ \varepsilon^{b_1 c_2}  \varepsilon^{b_2 c_1} \big )  
\notag \\
&\quad \quad \quad
+ \tfrac{3}{4} \CM_2^{a_1 a_2 b_1 b_2 c_1 c_2 d_1 d_2} \see_\delta^{e_1 e_2} \varepsilon_{d_1 e_1}  \varepsilon_{d_2 e_2}   -\tfrac{9}{16} \CM_3^{a_1 a_2 b_1 b_2 c_1 c_2} + \tfrac{3}{32} \CM_2 \CM_1^{a_1 a_2 b_1 b_2 c_1 c_2}
\notag \\
&\quad \quad \quad
-\tfrac{9}{80}  \CM_3^{a_1 a_2} \big ( \varepsilon^{b_1 c_1} \varepsilon^{b_2 c_2}
+ \varepsilon^{b_1 c_2}  \varepsilon^{b_2 c_1} \big )
+ {\rm cycl}(a_1a_2,b_1b_2,c_1c_2) \Big\} \label{sdel.18}
\end{align}
As indicated through the curly brackets, the cyclic symmetrization w.r.t.\ the three pairs $a_1a_2,b_1b_2,c_1c_2$ of indices
applies to the entire right-hand side (for instance adding the
images $ -\tfrac{27}{56}  \CM_2^{ b_1 b_2 c_1 c_2} \see_\delta^{a_1 a_2} $
and $ -\tfrac{27}{56}  \CM_2^{c_1 c_2 a_1 a_2} \see_\delta^{b_1 b_2} $ of the term
$ -\tfrac{27}{56}  \CM_2^{a_1 a_2 b_1 b_2} \see_\delta^{c_1 c_2} $ on the fourth line).}

 \newpage

\section{Isolating spin structure dependence for arbitrary $n$}
\label{sec:low-n}
\setcounter{equation}{0}

In this section we shall use the results of section \ref{sec:cyclic}
on the generalized $Q_\delta$ and $P_\delta$ functions in (\ref{defQdelta}) and (\ref{3.pdel}) to extend the construction of the case $n=2$ in subsection \ref{sec:2.3} to higher values of $n$, starting with the low values $n=3,4,5,6$. Explicit formulas for simplified cyclic products $C_\delta(z_1,\ndots,z_n)$ at $n>6$ will be relegated to appendix \ref{apponCdelta}.  In each case we shall obtain a decomposition of the cyclic product of Szeg\"o kernels $C_\delta$ in terms of a polynomial in the universal spin structure dependent symmetric bi-holomorphic $(1,0)$-form $L_\delta$ in (\ref{Ldel2}). The coefficients are spin structure independent and built out of the function $Z(i,j)$ introduced in Theorem \ref{thm:1}, as well as simple combinations of $x_{ij}^{-1}$ and $s_i$. We shall show that each result simply and naturally matches the pole structure of $C_\delta$. 

\sm

According to the discussion in section \ref{sec:qdel}, the entire
spin structure dependence of $C_\delta$ can be expressed in terms
of the polynomials $Q_\delta$ in (\ref{defQdelta}). We present a conjectural
all-multiplicity formula in section \ref{sec:conj} decomposing the $Q_\delta$
into cyclic products of $x_{ij}$ and polynomials in two-point building 
blocks $L_\delta(i,j)$, $Z(i,j)$. These expressions are the key ingredient
for the analogous decompositions of $n$-point $C_\delta$.

\sm

As will be detailed in appendix \ref{sec:group}, cyclic products $C_\delta(z_1, \cdots, z_n)$
simplify further upon symmetrization in the insertion points $z_1,\cdots, z_n$. Symmetrized
cyclic products for an arbitrary number of $n\geq 4$ points are 
holomorphic in all of $z_1,\cdots, z_n$, see
(\ref{symcyc.4}), (\ref{symcyc.6}), (\ref{symmszg.05}) and (\ref{symmszg.06})
for their simplified form at $n=4,6,8$ and $10$.

\subsection{The case $n=2$ recalled}

For completeness we begin by recalling the solution for $n=2$ and cast it in a language that will be used for $n >2$. The results given by Theorem \ref{thm:1} may be re-expressed as follows,
\bea
\label{5.C2}
Q_\delta(i | j) & = & - 4 x_{ij}^2 \, \bL_\delta(i,j) + 2 \, Z(i,j)
\no \\
C_\delta(i,j) & = & L_\delta(i,j) + { W_2^+(i,j) \over 2 x_{ij} x_{ji}}
\eea 
where the bi-holomorphic form $L_\delta(i,j)$ and the polynomial $\bL_\delta(i,j)$ were given in (\ref{Ldel2}) and (\ref{4.see}), respectively. The spin structure
independent combination $W_2^+(i,j)$ is defined by,
\bea
\label{5.W2}
W^\pm_2(i,j) = \big (  s_i s_j\pm Z(i,j)   \big ) {dx_i \, dx_j \over s_i \, s_j} 
\eea
with $Z(i,j)$ given in Theorem \ref{thm:1}. We have additionally introduced the variant $W_2^-(i,j)$ 
with opposite relative sign for later convenience, where both of $W_2^+$ and $W_2^-$ obey 
the symmetry $W^\pm_2(i,j)=  W^\pm_2(j,i)$. Recall that $L_\delta(i,j)$ is a symmetric 
bi-holomorphic $(1,0)$ form while $W_2^+(i,j)$ has a double zero as $(x_j, s_j) \to (x_i, -s_i)$ 
so that $C_\delta(i,j)$ is regular in this limit. Finally, the double pole of (\ref{5.C2})
as $(x_j, s_j) \to (x_i, s_i)$ has residue $-1$, as expected from the definition of $C_\delta(i,j)$.

\subsection{The case $n=3$}

For $n=3$, the numerator function $N_\delta$ is given in terms of $Q_\delta(i|j)$ by the second line in (\ref{NQ}). Using the decomposition of $Q_\delta(i|j)$ in terms of $L_\delta(i,j)$ and $Z(i,j)$ in (\ref{5.C2}), we readily derive the expression for $N_\delta$ in terms of the latter objects, 
\bea
N_\delta (i,j,k) & = &
 - 4 s_i x_{jk}^2 \, \bL_\delta(j,k) 
  - 4 s_j x_{ki}^2 \, \bL_\delta(k,i) 
   - 4 s_k x_{ij}^2 \, \bL_\delta(i,j) 
   \no \\ &&
+2 s_i Z(j,k)    +2 s_j Z(k,i)    +2 s_k Z(i,j)    +2 s_i s_j s_k
\eea
As  a result, we have,
\bea
C_\delta(i,j,k) & = & 
 - { x_{jk}  \, dx_i \over 2 x_{ij}  x_{ki}} L_\delta(j,k) 
 - { x_{ki}  \, dx_j \over 2 x_{jk}  x_{ij}} L_\delta(k,i) 
 - { x_{ij}  \, dx_k \over 2 x_{ki}  x_{jk}} L_\delta(i,j)
+ {W^+_3(i,j,k) \over 4 x_{ij} x_{jk} x_{ki}}
\qquad \notag \\
&= &\bigg \{ \frac{ dx_i \, L_\delta(j,k) + dx_j   \, L_\delta(i,k)}{2 x_{ij}} + {\rm cycl}(i,j,k) \bigg \}
+ {W^+_3(i,j,k) \over 4 x_{ij} x_{jk} x_{ki}}
 \label{5.C-three}
\eea
where the form $W_3^+(i,j,k)$  (along with its variant $W_3^-(i,j|k)$ to be encountered below) is defined as follows,
\begin{align}
W^+_3 (i,j,k)  &= 
 \Big ( s_i Z(j,k)   +  s_j Z(k,i) +  s_k Z(i,j)    + s_i s_j s_k \Big ) { dx_i \, dx_j \, dx_k \over s_i \, s_j \, s_k} \label{5.W3} \\
 W^-_3 (i , j| k)  &= 
 \Big (  s_k Z(i,j) - s_i Z(j,k)  - s_j Z(k,i)  + s_i s_j s_k \Big ) { dx_i \, dx_j \, dx_k \over s_i \, s_j \, s_k}
 \notag
\end{align}
Note that the combination $W_3^+(i,j,k)$ is invariant under all permutations of $i,j,k$, while $W_3^-(i , j| k)$ is invariant only under swapping $j,k$ as indicated by the vertical bar.

\subsubsection{Pole structure}

The denominators in (\ref{5.C-three}) produce simple poles as $(x_j, s_j) \to (x_i, \pm s_i)$ in individual terms of the expression (\ref{5.C-three}) for $C_\delta(i,j,k)$. The residues of these poles may be evaluated with the help of the following limits, 
\bea
\label{5.limits2}
\lim_{(x_j, s_j) \to (x_i, \pm s_i)} L_\delta(j,k) & = & \pm L_\delta(i,k)
\no \\
\lim_{(x_j, s_j) \to (x_i, \pm s_i)} Z(i,j) & = & s_i^2
\no \\
\lim_{(x_j, s_j) \to (x_i, \pm s_i)} Z(j,k) & = & Z(i,k)
\eea 
The $\pm$ sign on the right side of the first line arises from the limit of the Abelian differential $dx_j/s_j \to \pm dx_i/s_i$.
Using these limits, we readily establish that the poles as $(x_j, s_j) \to (x_i,  s_i)$ produce the expected residues $C_\delta(i,k)$.  The poles as $(x_j, s_j) \to (x_i, - s_i)$ do not occur in the original definition of $C_\delta(i,j,k) $ and must cancel in the sum of all terms in (\ref{5.C-three}). To verify this fact, we use the limits of (\ref{5.limits2}) again and observe that 
all poles in the first three terms of (\ref{5.C-three}) cancel one another.  In $W^+_3$ the sum of the first two terms of the numerator cancel one another using the last line in (\ref{5.limits2}) and the fact that $s_j \to -s_i$, while the last two terms cancel one another using the second line in (\ref{5.limits2}). In view of the cyclic symmetry of $C_\delta(i,j,k)$ all other spurious poles such as $(x_j, s_j) \to (x_k, - s_k)$ also cancel in (\ref{5.C-three}).

\subsection{The case $n=4$}

Inspection of the relation between $N_\delta$ and $Q_\delta$ given in (\ref{NQ}) for the case $n=4$ reveals that only the 2- and 4-point combinations $Q_\delta(1|2)$ and $Q_\delta(1,3|2,4)$ are needed. The expression for $Q_\delta(1|2)$ in terms of $\bL_\delta(1,2)$ and $Z(1,2)$ was already given in (\ref{5.C2}), while the analogous expression for $Q_\delta(1,3|2,4)$ is given by the following lemma.
{\lem
\label{lem:Q4}
The polynomial $Q_\delta(1,3|2,4)$ is given in terms of $ \bL_\delta(i,j)$ and  $Z(i,j)$ as follows,
\bea
\label{5.Q4}
Q_\delta(1,3|2,4) & = & 
8 x_{12} x_{23} x_{34} x_{41} \Big ( \bL_\delta(1,2) \bL_\delta(3,4) +  \bL_\delta(1,4) \bL_\delta(2,3) \Big ) 
\no \\ &&
+ \Big \{ 2  \, x_{24}^2  Z(1,3) \bL_\delta(2,4)  - 4  \, x_{34}^2 Z(1,2) \bL_\delta(3,4) 
 + {\rm cycl}(1,2,3,4) \Big  \} 
\no \\ &&
+ 2 \, Z_4(1,2,3,4)
\eea
where the combination $Z_4$ is defined by,
\begin{align}
Z_4(a,b,c,d) = Z(a,b) Z(c,d)  + Z(b,c) Z(d,a)  - Z(a,c) Z(b,d) 
\label{z4def}
\end{align}
} 
The proof of the lemma is relegated to appendix \ref{sec:E.1}.

\sm

Substituting the expression for $Q_\delta(1|2)$ given in (\ref{5.C2}) and $Q_\delta(1,3|2,4)$ given in (\ref{5.Q4}) into the relations (\ref{NQ}) for $n=4$ we obtain the following expression, after some simplifications, 
{\small \bea
\label{Cdel4}
C_\delta(1,2,3,4) & = &
\half   L_\delta(1,2) L_\delta(3,4) + \half L_\delta(1,4) L_\delta(2,3) 
\no \\ &&
+ \bigg \{ 
 \frac{W_2^+(2,3) L_\delta(4,1) +  W_2^+(4,1) L_\delta(2,3) +  W^-_2(1,3) L_\delta(2,4) 
  + W^-_2(2,4) L_\delta(1,3)}{8 x_{12}x_{34}} 
 \no \\  && \hskip 0.2in
+\frac{W_2^+(1,2)  L_\delta(3,4) +  W_2^+(2,3) L_\delta(4,1) +  W^-_2(1,3)L_\delta(2,4)}{4 x_{12}x_{23}} 
 +{\rm cycl}(1,2,3,4) \bigg \}
\no \\ &&
+ \frac{  W_2^+(1,2) W_2^+(3,4)  +W_2^+(2,3) W_2^+(4,1) -  W_2^-(1,3)W_2^-(2,4)}{ 8 x_{12}x_{23}x_{34} x_{41}}
\eea}
see (\ref{5.W2}) for the definition of both $W_2^+(i,j)$ and $W_2^-(i,j)$.

\subsubsection{Pole structure}

The singularity structure of the cyclic product $C_\delta(1,2,3,4)$ may be read off directly from its expression in terms of the Szeg\"o kernel $S_\delta(z_i, z_j)$ which has a single pole at $z_i=z_j$.  The only singularities of $C_\delta(1,2,3,4)$ are simple poles when two neighboring points in the cyclic product come together, and the residue is the corresponding three-point cyclic product, 
\bea
\label{5.pole4}
C_\delta(1,2,3,4) = { C_\delta(2,3,4)  \over z_1-z_2} + \hbox{regular in $z_{1}{-}z_2$}
\eea
In particular, there are no poles when two non-neighboring points come together, a property that is manifestly borne out  by formula (\ref{Cdel4}). The poles produced by the Parke-Taylor factor between neighboring points $z_i, z_j$  as $(x_j, s_j) \to (x_i, s_i)$ are simple and their residues precisely match the expected 3-point value of (\ref{5.pole4}). The poles of the Parke-Taylor factor between neighboring points $z_i, z_j$  as $(x_j, s_j) \to (x_i, -s_i)$ are absent from (\ref{5.pole4}) and must cancel in the sum (\ref{Cdel4}). To see this, we use the limits of (\ref{5.limits2}) which imply,
\begin{align}
\lim _{(x_j, s_j) \to (x_i, - s_i)} W_2^+(i,j) &=  0
\notag \\
\lim _{(x_j, s_j) \to (x_i, - s_i)} W_2^- (i,j) &=   2 dx_i^2 
\label{wpwmlim} \\
\lim _{(x_j, s_j) \to (x_i, - s_i)} W_2^\pm (j,k) &=   W^\mp _2(i,k)
\notag
\end{align}
The spurious poles all cancel one another in (\ref{Cdel4}), as expected, since
(\ref{wpwmlim}) leads to cancellations such as $W_2^+(2,3) L_\delta(4,1) + W_2^-(1,3) L_\delta(2,4)
\rightarrow 0$ under $(x_2, s_2) \to (x_1, - s_1)$.

\subsection{The case $n=5$}

Inspection of the relation between $N_\delta$ and $Q_\delta$ given in (\ref{NQ}) for the case $n=5$ reveals that again only the  combinations $Q_\delta(1|2)$ and $Q_\delta(1,3|2,4)$ are needed. The expression for $Q_\delta(1|2)$ in terms of $\bL_\delta(1,2)$ was already given in (\ref{5.C2}), while $Q_\delta(1,3|2,4)$ was obtained in Lemma \ref{lem:Q4}.  Substituting these expressions into $C_\delta$, we obtain, 
\bea
\label{C.del5}
C_\delta (1,2,3,4,5)& = & 
\bigg\{\frac{ x_{41} \, dx_5}{4  x_{45} x_{51} } \, \Big[ L_\delta(1,2) L_\delta(3,4) +  L_\delta(1,4) L_\delta(2,3) \Big]
 \\ && \quad
- \frac{ x_{12}^2  L_\delta (1,2)  W_3^+ (3,4,5) + x_{13}^2  L_\delta(1,3)  W_3 ^-(4,5| 2) }{ 8\,   x_{12} x_{23} x_{34} x_{45} x_{51} } 
\no \\ &&  \quad
+ \frac{W_2^+ (1,2) W_3^+(3,4,5) - W_2 ^-(1,3)  W_3^-(4,5| 2)}{32 \, x_{12}x_{23}x_{34}x_{45}x_{51} }
\no \\ &&  \quad
+ \frac{ dx_2 \, dx_4 \, dx_5 W^-_2(1,3) }{16 \, x_{12}x_{23}x_{34}x_{45}x_{51} }
 +\hbox{cycl} (1,2,3,4,5) \bigg\}
 - \frac{  dx_1 dx_2 dx_3 dx_4 dx_5 }{ 4 \, x_{12}x_{23}x_{34}x_{45}x_{51} } 
 \no
\eea
where $W_2^\pm$ and $W_3^\pm$ were introduced in (\ref{5.W2}) and (\ref{5.W3}), respectively.
By rearranging all numerator factors $ x_{ij}^2$ to cancel the denominators, we can manifest
the pole structure of (\ref{C.del5}) to be of the form
\begin{align}
C_\delta (1,2,3,4,5) &=
\bigg\{   \frac{ {\cal N}^{(5)}_\delta[12] }{4x_{12} }
+ \frac{  {\cal N}^{(5)}_\delta[1234] }{8x_{12}x_{23}x_{34} }
+  \frac{  {\cal N}^{(5)}_\delta[123,45] }{8x_{12}x_{23}x_{45} }
  + {\rm cycl}(1,2,3,4,5)  \bigg\}   \notag \\
&\quad   +  \frac{  {\cal N}^{(5)}[12345] }{16x_{12}x_{23}x_{34}x_{45}x_{51} }
  \label{cdel5.1}
\end{align}
with numerators
\begin{align}
\label{cdel5.2}
{\cal N}^{(5)}_{\delta}[12]  &= dx_1 \big( L_\delta(2,3) L_\delta(4,5) +  L_\delta(2,5) L_\delta(3,4) \big)
\no \\ & \quad
+dx_2 \big( L_\delta(1,3) L_\delta(4,5) +  L_\delta(1,5) L_\delta(3,4) \big)
\notag \\
{\cal N}^{(5)}_\delta[1234] &=
L_\delta(4,5) W_3^+(1,2,3)
{+} L_\delta(3,5) W_3^-(1,2 | 4)
\no \\ & \quad \,
{+} L_\delta(2,5) W_3^-(3,4 | 1)
{+} L_\delta(1,5) W_3^+(2,3,4)
 \\
{\cal N}^{(5)}_\delta[123,45]  &= 
L_\delta(3,4) W_3^+(5,1,2) +L_\delta(2,4) W_3^-(5,1| 3)  + L_\delta(1,4) W_3^-(2,3 | 5) 
\no  \\
&\quad + L_\delta(3,5) W_3^-(1,2 | 4)+ L_\delta(2,5) W_3^-(3,4 | 1)+ L_\delta(1,5) W_3^+(2,3,4)
\notag \\
{\cal N}^{(5)}[12345] &= \Big \{  \tfrac{1}{2} W_2^+(1,2) W_3^+(3,4,5) 
- \tfrac{1}{2} W_2^-(1,3) W_3^-(4,5 | 2) 
\no \\
&\qquad 
+ dx_2 dx_4 dx_5 W_2^-(1,3)  + {\rm cycl}(1,2,3,4,5) \Big \}   - 4 dx_1 dx_2 dx_3 dx_4 dx_5 
\no
\end{align}
Note that the last numerator ${\cal N}^{(5)}[12345]$ associated with five simultaneous poles is independent on $\delta$, and the subtraction in the last line of (\ref{cdel5.2}) ensures that the overall coefficient of $dx_1 \ndots dx_5$ in ${\cal N}^{(5)}[12345] $ is $1$ rather than 5.

\subsection{The case $n=6$}

The six-point relation between $N_\delta$ and $Q_\delta$ given in (\ref{NQ}) involves all of $Q_\delta(1|2)$, $Q_\delta(1,3|2,4)$ and $Q_\delta(1,3,5|2,4,6)$.  The expressions for $Q_\delta(1|2)$ and $Q_\delta(1,3|2,4)$ in terms of $\bL_\delta(i,j)$ and $Z(i,j)$ were given in (\ref{5.C2}) and Lemma \ref{lem:Q4}, while the analogous expression for $Q_\delta(1,3,5|2,4,6)$ is given by the following lemma, the proof of which is relegated to appendix \ref{sec:E.2}.

{\lem
\label{lem:Q6}
The polynomial $Q_\delta(1,3,5|2,4,6)$ is given in terms of the functions $ \bL_\delta(i,j)$ and  $Z(i,j)$ as well as
polynomials in $x_{ij}$ by the following formula,
{\small 
\begin{align}
 \label{Qdel.13} 
& Q_\delta(1,3,5|2,4,6) =32 \eecyc_\delta(1,2,3,4,5,6) 
\no \\
&\quad \!\!\!
+ 8 \Big \{ Z(1,2) \eecyc_\delta(3,4,5,6)
- Z(1,3) \eecyc_\delta(2,4,5,6)
+\tfrac{1}{2} Z(1,4) \eecyc_\delta(2,3,5,6)
+{\rm cycl}(1,\ndots,6) \Big \} \notag \\
&\quad  \!\!\!
+ 4  \Big \{  Z_4(1,2,3,4) \eecyc_\delta(5,6)
- Z_4(1,2,3,5) \eecyc_\delta(4,6)
+\tfrac{1}{2} Z_4(1,2,4,5) \eecyc_\delta(3,6)
+{\rm cycl}(1,\ndots,6) \Big \}
 \notag \\
&\quad \!\!\!
+ 2 Z_6(1,2,3,4,5,6)
\end{align}}
where we employ the following combinations for the spin structure dependent terms,
\begin{align}
\label{Qdel.05}
\eecyc_\delta(1,2) &= x_{12} x_{21}\bL_\delta(1,2)
\\
\eecyc_\delta(1,2,3,4) &=
x_{12}x_{23}x_{34} x_{41} \Big[
\bL_\delta(1,2) \bL_\delta(3,4)
+ \bL_\delta(2,3) \bL_\delta(4,1) \Big]
 \no \\
\eecyc_\delta(1,2,3,4,5,6) &= x_{12}x_{23}x_{34}x_{45}x_{56}x_{61}
\Big[ \bL_\delta(1,2) \bL_\delta(3,4) \bL_\delta(5,6)
+ \bL_\delta(2,3) \bL_\delta(4,5) \bL_\delta(6,1) \Big]
\notag
\end{align}
The bilinears in $Z(i,j)$ have been regrouped into the form $Z_4$ defined in (\ref{z4def}),
and we have furthermore introduced
\begin{align}
Z_6(1,2,3,4,5,6) &= Z(1,2) Z_4(3,4,5,6) - Z(1,3) Z_4(2,4,5,6) + Z(1,4) Z_4(2,3,5,6)
\notag \\
&\quad
- Z(1,5) Z_4(2,3,4,6)
+ Z(1,6) Z_4(2,3,4,5)
 \label{Qdel.10}
\end{align}
}

Based on the expressions for $Q_\delta(1|2)$, $Q_\delta(1,3|2,4)$ and
$Q_\delta(1,3,5|2,4,6)$ in (\ref{5.C2}), (\ref{5.Q4}) and (\ref{Qdel.13}), the
numerator for $n=6$ points in (\ref{NQ}) yields,
\begin{align}
C_\delta(1,\cdots,6) &=
\frac{1}{2} \, L_\delta(1,2) L_\delta(3,4) L_\delta(5,6)
+ \frac{1}{2}\, L_\delta(2,3) L_\delta(4,5) L_\delta(6,1)
\notag \\
&\quad +
\bigg\{ \frac{ {\cal N}^{(6)}_\delta[123] }{8x_{12}x_{23} }
+ \frac{ {\cal N}^{(6)}_\delta[12,34] }{8x_{12}x_{34} }
+ \frac{ {\cal N}^{(6)}_\delta[12,45] }{16x_{12}x_{45} }
+ \frac{ {\cal N}^{(6)}_\delta[12345] }{16 x_{12}x_{23} x_{34}x_{45} } \label{cdel6.1} \\
&\qquad \quad
+ \frac{ {\cal N}^{(6)}_\delta[1234,56] }{16 x_{12}x_{23} x_{34}x_{56} }
+ \frac{ {\cal N}^{(6)}_\delta[123,456] }{32 x_{12}x_{23} x_{45}x_{56} }
+ {\rm cycl}(1,\cdots,6) \bigg\} \notag  \\
&\quad + \frac{  {\cal N}^{(6)}[123456] }{32 
 x_{12}x_{23} x_{34}x_{45}  x_{56}x_{61} } 
 \notag
\end{align}
The numerators multiplying two simultaneous poles $(x_{ab} x_{cd})^{-1}$
are bilinear in $L_\delta(i,j)$ and may be most conveniently written as follows,
\begin{align}
{\cal N}^{(6)}_\delta[123] &=
 L_{\delta}(3, 4,5,6) W_2^+(1, 2) 
+ L_{\delta}(4, 5,6, 1)  W_2^+(2, 3) 
+ L_{\delta}(2, 4,5, 6) W_2^-(1, 3)
\notag \\
{\cal N}^{(6)}_\delta[12,34] &=
L_{\delta}(2, 3,5, 6)  W_2^+(1, 4) 
+ L_{\delta}(4, 5,6, 1)   W_2^+(2, 3)  \notag \\
&\quad
+  L_{\delta}(2, 4,5, 6)   W_2^-(1, 3) 
+ L_{\delta}(1, 3,5, 6)   W_2^-(2, 4)
\label{cdel6.2} \\
{\cal N}^{(6)}_\delta[12,45]&=
 L_{\delta}(2, 3,5, 6)  W_2^+(1, 4) 
+  L_{\delta}(3, 4,6, 1)  W_2^+(2, 5)  \notag \\
&\quad
+ L_{\delta}(1, 3,5, 6)  W_2^-(2, 4) 
+ L_{\delta}(2, 3,4, 6)  W_2^-(1, 5)
\notag
\end{align}
in terms of the bilinear combination,
\bea
L_\delta(a,b,c,d) =  L_\delta(a,b)L_\delta(c,d) + L_\delta(b,c)L_\delta(d,a)
\label{cdel6.3}
\eea
The numerators multiplying four simultaneous poles 
$(x_{ab}x_{cd}x_{ef}x_{gh} )^{-1}$ in (\ref{cdel6.1}) are linear
in $L_\delta(i,j)$ and take the following form,
\begin{align}
{\cal N}^{(6)}_\delta[12345] &= L_{\delta}(5, 6) W_4^+(1, 2, 3, 4) + L_{\delta}(4, 6) W_4^-(1, 2, 3 | 5) 
+     L_{\delta}(3, 6) W_4^-(1, 2 | 4, 5) \notag \\
&\quad +  L_{\delta}(2, 6) W_4^-(3, 4, 5 | 1) +  L_{\delta}(1, 6) W_4^+(2, 3, 4, 5)
\notag \\
{\cal N}^{(6)}_\delta[1234,56] &=
L_{\delta}(4, 5) W_4^+(6, 1, 2, 3) + L_{\delta}(4, 6) W_4^-(1, 2, 3 | 5) + 
 L_{\delta}(3, 5) W_4^-(6, 1, 2 | 4) \notag \\
 &\quad +   L_{\delta}(3, 6) W_4^-(1, 2 | 4, 5) + 
 L_{\delta}(2, 5) W_4^-(3, 4 | 6, 1) +   L_{\delta}(2, 6) W_4^-(3, 4, 5 | 1)  \notag \\
 &\quad + 
 L_{\delta}(1, 5) W_4^-(2, 3, 4 | 6) +    L_{\delta}(1, 6) W_4^+(2, 3, 4, 5)
 \label{cdel6.4}
 \\
{\cal N}^{(6)}_\delta[123,456] &=    L_{\delta}(3, 4) W_4^+(5, 6, 1, 2) + L_{\delta}(3, 5) W_4^-(6, 1, 2 | 4) + 
L_{\delta}(3, 6) W_4^-(1, 2 | 4, 5)  \notag \\
&\quad +  L_{\delta}(2, 4) W_4^-(5, 6, 1 | 3) + 
L_{\delta}(2, 5) W_4^-(6, 1 | 3, 4) + L_{\delta}(2, 6) W_4^-(3, 4, 5 | 1)  \notag \\
&\quad + 
L_{\delta}(1, 4) W_4^-(5, 6 | 2, 3) +  L_{\delta}(1, 5) W_4^-(2, 3, 4 | 6) + 
L_{\delta}(1, 6) W_4^+(2, 3, 4, 5)   \notag 
\end{align}
in terms of the following combinations,
\begin{align}
W_4^+(a,b,c,d) &=  W_2^+(a,b) W_2^+(c,d)  + W_2^+(a,d) W_2^+(b,c)  - W_2^-(a,c) W_2^-(b,d)  
\notag \\
W_4^-(a,b,c | d) &=  W_2^+(a,b) W_2^-(c,d)  + W_2^-(a,d) W_2^+(b,c)  - W_2^-(a,c) W_2^+(b,d) 
\label{cdel6.5} \\
W_4^-(a,b| c,d) &=  W_2^+(a,b) W_2^+(c,d)  + W_2^-(a,d) W_2^-(b,c)  - W_2^+(a,c) W_2^+(b,d) 
\notag \\
W_4^-(a|b| c|d) &=  W_2^-(a,b) W_2^-(c,d)  + W_2^-(a,d) W_2^-(b,c)  - W_2^-(a,c) W_2^-(b,d)
\notag
\end{align}
In contrast to the lower-point analogues $W_2^{\pm}$ in (\ref{5.W2}) and $W_3^{\pm}$ in 
(\ref{5.W3}), the four-point objects $W_4^{\pm}$ come in four different variants that differ
in relative signs. The first variant $W_4^+(a,b,c,d) $ also captures the
contribution $C_\delta(1,2,3,4) = \ndots +  W_4^+(1,2,3,4) / (8 x_{12}x_{23}x_{34}x_{41})$
to the four-cycle with four simultaneous poles in the last line of (\ref{Cdel4}). 
The last variant $W_4^-(a|b| c|d) $ does not yet enter the six-point building blocks
(\ref{cdel6.4}) but will appear in the eight-point cycle in appendix \ref{app.Cdel8}.

\sm 

Finally, the numerator multiplying six simultaneous poles $( x_{12}x_{23} x_{34}x_{45}  x_{56}x_{61})^{-1}$ in (\ref{cdel6.1}) is independent on $\delta$ and given by,
\begin{align}
{\cal N}^{(6)}[123456]  &=
\Big \{ \tfrac{1}{2} W_2^+(1,4) W_2^+(2,3) W_2^+(5,6)
+ \tfrac{1}{2} W_2^+(1,4) W_2^-(2,6) W_2^-(3,5) \notag \\
&\quad \quad \quad
- W_2^+(1,2) W_2^-(3,5) W_2^-(4,6)
+{\rm cycl}(1,\ndots,6) \Big \} \label{cdel6.6} \\
&\quad  
+ W_2^+(1,2) W_2^+(3,4) W_2^+(5,6)
+ W_2^+(2,3) W_2^+(4,5) W_2^+(6,1) \notag \\
&\quad
-  W_2^+(1,4) W_2^+(2,5) W_2^+(3,6)  \notag
\end{align}
It can be formally obtained from the expression 
(\ref{Qdel.10}) for $Z_6(1,2,\cdots,6)$ written as a trilinear in $Z(i,j)$
by promoting each $Z(i,j)$ to $W_2^+(i,j)$ if $i{-}j$ is odd and to $W_2^-(i,j)$ 
if $i{-}j$ is even.

\subsection{Higher multiplicity}
\label{sec:conj}

The analogous decompositions of the cyclic product $C_\delta(1,\ndots,n)$ for the cases of $n=7,8$ points are given explicitly in appendix \ref{apponCdelta}.

\sm

The examples above illustrate that the bottleneck in the simplification of cyclic products $C_\delta(1,\ndots,n)$ for higher multiplicity $n$ stems from the $Q_\delta$-functions with the highest numbers of points in (\ref{lastterms}).
This subsection is dedicated to obtaining a conjectural expression for $Q_\delta(1,3,\ndots,n{-}1|2,4,\ndots,n)$ at arbitrary $ n$ that generalizes the expressions (\ref{5.Q4}) and (\ref{Qdel.05})  at $n=4,6$ and makes the simplification of $C_\delta$ accessible at all multiplicities.

\subsubsection{Towards higher-point building blocks}

As a first step, we rewrite the examples at $n=2,4$ in
terms of the quantities $\eecyc_\delta$ in (\ref{Qdel.05}),
\begin{align}
Q_\delta(1|2) &= 4\eecyc_\delta(1,2) + 2 Z(1,2) 
\notag\\
 Q_\delta(1,3|2,4) &=8 \eecyc_\delta(1,2,3,4)
 -4 \eecyc_\delta(1,3) Z(2,4) - 4 \eecyc_\delta(2,4) Z(1,3)  \label{Qdel.11}  \\
 &\quad
+ 4 \big \{  \eecyc_\delta(1,2) Z(3,4) + {\rm cycl}(1,2,3,4) \big \}
 + 2 Z_4(1,2,3,4)
 \notag
\end{align}
which closely follows the structure of $Q_\delta(1,3,5|2,4,6)$ in (\ref{Qdel.13}) and guides the extrapolation to higher multiplicity. We furthermore note that both $Z_4$ and $Z_6$ in (\ref{z4def}) and (\ref{Qdel.10}) can be identified as Pfaffians of anti-symmetric matrices with entries $\pm Z(i,j)$, for instance
\begin{align}
Z_4(1,2,3,4) &= 
{\rm Pf}
\left( \begin{array}{cccc} 0 &Z(1,2) &Z(1,3) &Z(1,4) \\
-Z(1,2) &0 &Z(2,3) &Z(2,4) \\
-Z(1,3) &-Z(2,3) &0 &Z(3,4) \\
-Z(1,4) &-Z(2,4) &-Z(3,4) & 0 \end{array} \right)
\label{Qdel.07}
\end{align}
The recursive structure of $Z_4$ and $Z_6$ in (\ref{z4def}) and (\ref{Qdel.10})
can be straightforwardly uplifted to define higher-point objects $Z_k$ at arbitrary even
$k$ of homogeneity degree $\frac{k}{2}$ in $Z(i,j)$,
\bea
Z_k(a_1,\ndots ,a_k) = \sum_{j=2}^{k} (-1)^{j} Z(a_1,a_j) Z_{k-2}(a_2,a_3,\ndots,\widehat{a_j},\ndots,a_k)
 \label{Qdel.09}
\eea
where the notation $\widehat{a_j}$ on the right-hand side instructs to omit the entry $a_j$.
One can equivalently define $Z_k$ as the Pfaffian of the $k\times k$ matrix ${\cal Z}_k$ with the
following entries,
\bea
Z_k(a_1,\ndots ,a_k) = {\rm Pf} \, {\cal Z}_k 
\ \ \ \ \ \ \ \ 
({\cal Z}_k)_{ij} = \left\{ \begin{array}{rl} 
Z(a_i,a_j) &: \ 1 \leq i<j\leq k \\
-Z(a_i,a_j) &: \ 1 \leq j<i\leq k \\
0&: \ i=j \\
\end{array} \right.
 \label{Qdel.08}
\eea
reducing to (\ref{Qdel.07}) at $k=4$.
As a higher-multiplicity uplift of the spin structure dependent building blocks $\eecyc_\delta$ in (\ref{Qdel.05}), we define
(assuming even $k\geq 4$),
\begin{align}
&\eecyc_\delta(1,2,\ndots,k) =
x_{12}x_{23}\ndots x_{k-1,k} x_{k1} \label{Qdel.06}\\
&\ \ \ \ \times \big[
 \bL_\delta(1,2) \bL_\delta(3,4)\ndots  \bL_\delta(k{-}1,k)
+  \bL_\delta(2,3)  \bL_\delta(4,5) \ndots   \bL_\delta(k,1) \big]
\notag
\end{align}
Just like the spin structure independent $Z_k$, the ${\cal L}_\delta$ are cyclically invariant at 
any multiplicity,
\bea
\eecyc_\delta(1,2,\ndots,k) & = & \eecyc_\delta(2,3,\ndots,k,1) 
\no \\
Z_k(1,2,\ndots,k) & = & Z_k(2,3,\ndots,k,1) 
\label{cycinvs}
\eea
 
\subsubsection{Results at $n=6,8,10$ points} 

By reorganizing the cyclic orbits of the six-point expression (\ref{Qdel.13}), we can bring it into the following suggestive form associating all 
ordered subsets $(a_1,a_2,\ndots, a_k)$ 
of $\{1,2,3,4,5,6\}$ with even $k$ to a factor of $Z_k$
\begin{align}
\label{Qdel.12} 
& Q_\delta(1,3,5|2,4,6) =32 \eecyc_\delta(1,2,3,4,5,6) +2 Z_6(1,2,3,4,5,6) 
 \\
& \quad \quad
 - 8 \sum_{1\leq a_1 <a_2}^6  (-1)^{a_1+a_2} Z(a_1,a_2) 
\eecyc_\delta(1,\ndots,\widehat{a_1},\ndots,\widehat{a_2},\ndots,6) \notag \\
&\quad \quad
+ 4 \sum_{1\leq a_1 <a_2<a_3 <a_4}^6  (-1)^{a_1+a_2+a_3 +a_4} Z_4(a_1,a_2,a_3,a_4)
\eecyc_\delta(1,\ndots,\widehat{a_1}, ~\ndots ~,
\widehat{a_4},\ndots,6) \no
\end{align}
Based on the $k$-point building blocks $Z_k$ and $\eecyc_\delta$ in
(\ref{Qdel.09}) and (\ref{Qdel.06}), this can be generalized to a natural guess for the 
analogous eight- and ten-point expressions
\begin{align}
\label{Qdel.14}
& Q_\delta(1,3,5,7|2,4,6,8) = 128 \, \eecyc_\delta(1,2,\ndots,8) +2 Z_8(1,2,\ndots,8) \\
& \quad \quad
- 32 \! \! \sum_{1\leq a_1 <a_2}^8 \! \!  (-1)^{a_1+a_2} Z(a_1,a_2) 
\eecyc_\delta(1,\ndots,\widehat{a_1},\ndots,\widehat{a_2},\ndots,8) \notag \\
&\quad \quad
+ 8 \sum_{1\leq a_1 <a_2<a_3 <a_4}^8  (-1)^{a_1+a_2+a_3 +a_4} Z_4(a_1,a_2,a_3,a_4)
\eecyc_\delta(1,\ndots,\widehat{a_1}, ~\ndots~,\widehat{a_4},\ndots,8) \notag \\
&\quad \quad
- 4 \sum_{1\leq a_1 <a_2<\ndots<a_6}^8  (-1)^{a_1+a_2+\ndots+a_6} Z_6(a_1,\ndots,a_6)
\eecyc_\delta(1,\ndots,\widehat{a_1}, ~\ndots~,\widehat{a_6},\ndots,8) \notag 
\end{align}
and
 \begin{align}
\label{Qdel.15}
& Q_\delta(1,3,5,7,9|2,4,6,8,10) = 512 \, \eecyc_\delta(1,2,\ndots,10) +2 Z_{10}(1,2,\ndots,10)   \\
&\quad \quad
- 128 \sum_{1\leq a_1 <a_2}^{10}  (-1)^{a_1+a_2} Z(a_1,a_2) 
\eecyc_\delta(1,\ndots,\widehat{a_1},\ndots,\widehat{a_2},\ndots,10) \notag \\
&\quad \quad
+ 32 \sum_{1\leq a_1 <a_2<a_3 <a_4}^{10}  (-1)^{a_1+a_2+a_3 +a_4} Z_4(a_1,a_2,a_3,a_4)
\eecyc_\delta(1,\ndots,\widehat{a_1},~\ndots ~,\widehat{a_4},\ndots,10) \notag \\
&\quad \quad
- 8 \sum_{1\leq a_1 <\ndots < a_6}^{10}  (-1)^{a_1+a_2+\ndots+a_6} Z_6(a_1,\ndots,a_6)
\eecyc_\delta(1,\ndots,\widehat{a_1}, ~\ndots ~,\widehat{a_6},\ndots,10) \notag \\
&\quad \quad
+ 4 \sum_{1\leq a_1 <\ndots < a_8}^{10}  (-1)^{a_1+a_2+\ndots+a_8} Z_8(a_1,\ndots,a_8)
\eecyc_\delta(1,\ndots,\widehat{a_1},~ \ndots~,\widehat{a_8},\ndots,10) \notag 
\end{align}
In each term, the subsets of vertex points associated with $Z_k$ and ${\cal L}_\delta$ are ordered according to the cycle $1,2,\ndots,n$, and the accompanying powers of $\pm 2$ are fixed by analogy with (\ref{Qdel.12}). Both (\ref{Qdel.14}) and (\ref{Qdel.15}) are numerically verified to reproduce the polynomials in the definition (\ref{defQdelta}) of $Q_\delta$.

\subsubsection{All-multiplicity conjecture} 

The examples of $Q_\delta$ at six, eight and ten points in (\ref{Qdel.12}), 
(\ref{Qdel.14}) and (\ref{Qdel.15}) motivate the following all-multiplicity 
conjecture for $k\geq 2$,
\begin{align}
& Q_\delta(1,3,\ndots,2k{-}1|2,4,\ndots,2k) = 2^{2k-1} \eecyc_\delta(1,2,\ndots,2k) +2 Z_{2k}(1,2,\ndots,2k)
 \label{Qdel.16}  \\
&\quad \quad
- 2^{2k-3} \sum_{1\leq a_1 <a_2}^{2k}  (-1)^{a_1+a_2} Z(a_1,a_2) 
\eecyc_\delta(1,\ndots,\widehat{a_1},\ndots,\widehat{a_2},\ndots,2k) \notag \\
&\quad \quad
+ 2^{2k-5} \sum_{1\leq a_1 <a_2<a_3 <a_4}^{2k}  (-1)^{a_1+a_2+a_3 +a_4} Z_4(a_1,a_2,a_3,a_4)
\eecyc_\delta(1,\ndots,\widehat{a_1},~\ndots~,\widehat{a_4},\ndots,2k) 
\notag \\
&\quad \quad +\cdots  \notag \\
&\quad \quad
+(-1)^k 8 \sum_{1\leq a_1 <\ndots < a_{2k-4}}^{2k}  
(-1)^{a_1+a_2+\ndots+a_{2k-4}} Z_{2k-4}(a_1,a_2,\ndots,a_{2k-4}) \notag \\
&\quad \quad\quad \quad\quad \quad\quad \quad \quad \quad \quad \quad\times
\eecyc_\delta(1,\ndots,\widehat{a_1},~\ndots~,\widehat{a_{2k-4}},\ndots,2k) \notag \\
&\quad \quad
- (-1)^k 4 \sum_{1\leq a_1 < \ndots < a_{2k-2}}^{2k}  (-1)^{a_1+a_2+\ndots+a_{2k-2}} Z_{2k-2}(a_1,a_2,\ndots,a_{2k-2}) \notag \\
&\quad \quad\quad \quad\quad \quad\quad \quad \quad \quad \quad \quad\times
\eecyc_\delta(1,\ndots,\widehat{a_1},~\ndots~,\widehat{a_{2k-2}},\ndots,2k)  \notag
\end{align}
More precisely, the terms in the ellipsis in the fourth line can be spelt out
through the following alternating sum over $m$ (where again $k\geq 2$),
\begin{align}
&Q_\delta(1,3,\ndots,2k{-}1|2,4,\ndots,2k) = 2^{2k-1} \eecyc_\delta(1,2,\ndots,2k)
+2 Z_{2k}(1,2,\ndots,2k)
  \notag \\
&\quad + \sum_{m=1}^{k-2} (-1)^m 2^{2k-1-2m}
\sum_{1\leq a_1 < a_2 < \ndots < a_{2m}}^{2k}
(-1)^{a_1+a_2+\ndots+a_{2m}} Z_{2m}(a_1,a_2,\ndots,a_{2m})
\notag \\
&\quad \quad\quad \quad\quad \quad\quad \quad \quad \quad \quad \quad\times
\eecyc_\delta(1,\ndots,\widehat{a_1},\ndots,\widehat{a_2}
,\ndots,\ndots,\widehat{a_{2m}},\ndots,2k) \label{Qdel.17} \\
&\quad  
- (-1)^k 4 \sum_{1\leq a_1 <a_2<\ndots < a_{2k-2}}^{2k}  (-1)^{a_1+a_2+\ndots+a_{2k-2}} Z_{2k-2}(a_1,a_2,\ndots,a_{2k-2}) \notag \\
&\quad \quad\quad \quad\quad \quad\quad \quad \quad \quad \quad \quad\times
\eecyc_\delta(1,\ndots,\widehat{a_1},\ndots,\widehat{a_2}
,\ndots,\ndots,\widehat{a_{2k-2}},\ndots,2k) \notag 
\end{align}
The term in the last two lines has an irregular prefactor $-(-1)^k 4$ and thereby could not
be absorbed into an extension of the sum over $m$
to $m=k{-}1$: Such an extension would give $(-1)^m 2^{2k-1-2m} \rightarrow -(-1)^k 2$
instead of the desired factor of $-(-1)^k 4$.

\sm

It would be interesting to prove the above formulas for $Q_\delta(1,3,\ndots,2k{-}1|2,4,\ndots,2k)$, for instance via induction or the $SL(2,\mathbb C)$ covariant techniques of appendix \ref{sec:B1}.

\newpage

\section{Evaluating the sums over spin structures}
\label{sec:sum}
\setcounter{equation}{0}

While the previous sections were dedicated to studying the cyclic products $C_\delta$ of Szeg\"o kernels  for a given spin structure $\delta$, we shall now investigate the summation of $C_\delta$ over the ten even spin structures $\delta$. In particular, this sum will provide key parts of the parity-even contribution to the genus-two chiral amplitudes for all perturbative superstring theories. 

\sm

Our earlier results dramatically simplify these spin structure sums since Corollary \ref{allspinpi} and Theorem \ref{thm:6} reduce all the spin structure dependence of the cyclic product $C_\delta$ of an arbitrary number $n$ of Szeg\"o kernels to a linear combination of $1, \see^{ab}_\delta$ and $\see^{ab}_\delta \see^{cd}_\delta$ with spin structure-independent coefficients. Thus, the summation over spin structures of $C_\delta$ has been reduced to the problem of the summation over spin structures of the three basic ingredients $1, \see^{ab}_\delta$ and $\see^{ab}_\delta \see^{cd}_\delta$. 
In practice, it will be convenient to decompose the bilinear combination into irreducible representations of $SL(2,\CC)$, as was already done in section \ref{sec:trilinear}, namely the representations  ${\bf 1}$  for $(\det \see_\delta)$ and ${\bf 7}$ for the symmetrized tensor product with components $\see_\delta^{(ab} \see_\delta ^{cd)}$, where symmetrization of the indices is indicated by the parentheses in the superscript.

\subsection{Summation measures}

From a mathematical point of view it may be natural to carry out the summation over the even spin structures with \textit{unit measure}. The result is given by the following theorem, whose proof may be obtained using the hyper-elliptic representation of $\see_\delta^{ab}$ and by summing over all permutations of the branch points.

{\thm 
\label{thm:6.0}
The spin structure sums with unit measure may be obtained by reducing all spin structure dependence to expressions bilinear, linear, and independent of $\see_\delta ^{ab}$ combined with the following summation identities,
\bea
\sum_{\delta \ \rm{even}}  (1,\see_\delta ^{ab}) =(10,0)   \hskip 0.5in 
\sum_{\delta \ \rm{even}}  \det \see_\delta = { 9 \over 4} \, \CM_2 \hskip 0.5in
\sum_{\delta \ \rm{even}}  \see_\delta^{(ab} \see_\delta ^{cd)} = { 45 \over 4} \, \CM_2^{abcd}
\eea
where the $SL(2,\CC)$ scalar $\CM_2$ and the rank-4 symmetric tensor $\CM_2^{abcd}$ were  defined in (\ref{5.M2}), and their components are given in  (\ref{sping3.65}) and (\ref{sping3.64})  in terms of symmetric polynomials $\mu_m$.}

\sm

From a physics point of view, however, the measure against which the summation over spin structures is to be carried out is not the unit measure. To obtain the five critical superstring theories in $\RR^{10}$, or any flat toroidal compactification thereof, three different measures are to be considered. The spin structure dependent parts of these measures on the chiral amplitudes are given as follows,
\bea
\label{spinsums.01}
\Upsilon_8[\delta] \hskip 0.55in & & \textrm{supersymmetric sector} 
\no \\
\tet[\delta](0)^{16} \hskip 0.4in & & \textrm{Heterotic Spin$(32)/\mathbb Z_2$ sector } 
\no \\
\tet[\delta_1](0)^8 \, \tet[\delta_2](0)^8 \hskip 0.1in & & \textrm{Heterotic} \ E_8\times E_8 \textrm{ sector}
\eea
 The definitions and properties of Riemann $\tet$-constants, $\Psi_{10}$ and other Igusa modular forms,  as well as the composite form $\Upsilon_8[\delta]$ are briefly reviewed in appendix~\ref{sec:F}. Their hyper-elliptic representations  may be obtained using the Thomae formulas of (\ref{sping2.8}) and may be found in (\ref{thetfourth}), (\ref{sqrteq}), and (\ref{F.Ups}), respectively. For each sector, the full chiral measure is obtained by dividing the above expressions by the Igusa cusp form $\Psi_{10}$.

\sm

The left- and right-moving sectors of Type II theories are both supersymmetric, while for the Heterotic theories, only the right-moving sector is supersymmetric. The left-moving sector of Heterotic strings, in the worldsheet fermion representation of the gauge degrees of freedom, corresponds to the gauge group under consideration. The GSO projection assigns independent spin structures to the left- and right-moving sectors of all the closed superstring theories, and then sums over these spin structures independently of one another. For the $E_8 \times E_8$ Heterotic theory, the even spin structures $\delta_1$ and $\delta_2$ in (\ref{spinsums.01}) are summed independently.  The measure factor  $\Upsilon_8[\delta]$ is the top component of the chiral measure on supermoduli space \cite{DP2, DP3, DP4}. The corresponding bottom component also enters into the calculation, but involves linear chains rather than cyclic products of Szeg\"o kernels. The evaluation of the spin structure sums of its contribution to $n$-point amplitudes is relegated to  future work.

\subsection{Spin structure sums in the supersymmetric sector}
\label{sec:sum.1} 

In the supersymmetric sector, the spin structure summation is carried out against the measure $\Upsilon_8[\delta]$ in (\ref{defUps}), and the results are given by the following theorem, whose proof may be obtained using the hyper-elliptic representation of $\see_\delta^{ab}$ and $\Upsilon_8[\delta]$, derived in appendix \ref{sec:F}, and by summing over all permutations of the branch points.

{\thm 
\label{thm:6.1}
The spin structure sums with  the supersymmetric measure $\Upsilon_8[\delta]$ in (\ref{F.Ups}) may be obtained by reducing all spin structure dependence to expressions bilinear, linear, and independent of $\see_\delta ^{ab}$ combined with the following summation identities,
\begin{align}
 \label{spinsums.02}
& \sum_{\delta \ {\rm even}}  \Upsilon _8 [\delta]  =
\sum_{\delta \ {\rm even}} \Upsilon _8 [\delta] \,  \see^{ab}_\delta  = 0
\no \\
& \sum_{\delta \ {\rm even}}  { \Upsilon _8 [\delta] \over \Psi_{10}} \, 
 \see_\delta^{ab} \, \see_\delta^{cd}
= \frac{1}{4} (\det \sigma)^2 \,  (\varepsilon^{ac} \varepsilon^{bd} + \varepsilon^{ad} \varepsilon^{bc} ) 
\end{align}
where the matrix $\sigma^a{}_I$ was defined in (\ref{2.sigma}).  Equivalently, by decomposing the tensor $ \see_\delta^{ab} \, \see_\delta^{cd}$ into irreducible representations of $SL(2,\CC)$, the spin sum (\ref{spinsums.02}) may be expressed as follows,
\bea
 \sum_{\delta \ {\rm even}}  \Upsilon _8 [\delta] \,  \see_\delta^{(ab} \, \see_\delta^{cd)}
= 0 
\hskip 1in
  \sum_{\delta \ {\rm even}}  {\Upsilon _8 [\delta] \over \Psi_{10} }  \,  \det \see_\delta
= \frac{3}{4} (\det \sigma)^2 
 \label{spinsums.08a}
\eea}
A few comments are in order. 
Firstly, the vanishing relations on the first line of (\ref{spinsums.02}) feed into the non-renormalization theorems for the genus-two amplitudes with 0, 1, 2, and 3 external massless states in Type I, Type II, or Heterotic strings.
Secondly,  one readily verifies that both sides of the second equation in (\ref{spinsums.08a})  transform as Siegel modular forms of weight $-2$ since $\Upsilon_8$ has weight 8, $\Psi_{10}$ has weight 10, and $\det \sigma$ has weight $-1$.  Finally, we have the following corollary, which immediately follows from Theorem \ref{thm:6.1}.

{\cor
\label{cor:6.1}
Expressed in terms of the differential forms $L_\delta(i,j) $ defined  in (\ref{Ldel2}), the spin structure sums of Theorem \ref{thm:6.1} take the form \cite{DP6}, 
\begin{align}
& \sum_{\delta \ {\rm even}}   \Upsilon _8 [\delta] =  \sum_{\delta \ {\rm even}}   \Upsilon _8 [\delta] L_\delta(1,2)  =0
\no \\
& \sum_{\delta \ {\rm even}}  { \Upsilon _8 [\delta] \over \Psi_{10}} L_\delta(1,2)  L_\delta(3,4) 
= \frac{1}{4}  \, \Big ( \Delta(1,3) \Delta(2,4) + \Delta(1,4) \Delta(2,3) \Big )
 \label{spinsums.04}
\end{align}
The anti-symmetric bi-holomorphic form $\Delta$ was defined in (\ref{defdelta}).  On the second line, both sides transform as Siegel modular forms of weight $-2$ as $\Delta$ is a modular form of weight $-1$.}

\subsubsection{Summation over higher powers of $\see_\delta$}
\label{sec:SUSY}

The basic spin structure sums of Theorem \ref{thm:6.1} may be generalized to summands that contain higher powers of $\see_\delta$, which will contribute to the spin structure sums  for higher-point cyclic products. These spin structure sums follow from reducing the spin structure dependence of the summands with the help of the trilinear relations in Theorem \ref{thm:6} to those powers whose spin structure sums are given by Theorem \ref{thm:6.1}. Organizing the spin structure sums by the degree of homogeneity of the summand, we have the following three corollaries, which will suffice to evaluate all the spin structure sums in the supersymmetric sector for up to the eleven-point cyclic products. The tensors $\CM_1$, $\CM_2$ and $\CM_3$ on the right-hand sides below can be found in (\ref{MXcomps}), (\ref{5.M2}) and (\ref{5.M3}), respectively.

{\cor
\label{cor:6.2}
The spin structure sums of trilinear combinations of $\see_\delta$ are given by, 
\begin{align}
 \sum_{\delta \ {\rm even}}   \Upsilon _8 [\delta] 
\, \big ( \det \see_\delta \big ) \, \see^{a_1 a_2}_\delta &= 0 \label{proj.01} 
\no \\
 \sum_{\delta \ {\rm even}}  { \Upsilon _8 [\delta] \over \Psi_{10}}
 \, \see^{(a_1 a_2}_\delta \see^{a_3 a_4}_\delta \see^{a_5 a_6)}_\delta 
 &=   \frac{9}{16} \, (\det \sigma)^2 \,  \CM_1^{a_1  \ndots a_6}
\end{align}
These equations show that the spin structure sum of trilinears in $\see_\delta$ projects the triple tensor product of identical tensors $\see_\delta^{\otimes 3} = {\bf 3} \oplus {\bf 7}$ onto the  ${\bf 7}$ of $SL(2,\CC)$.}
{\cor
\label{cor:6.3}
The spin structure sums of quadri-linear combinations of $\see_\delta$ are given by, 
\begin{align}
\label{proj.02}
\sum_{\delta \ {\rm even}}  { \Upsilon _8 [\delta] \over \Psi_{10}}
\, (\det \see_\delta)^2 &=   \frac{9}{16}\, (\det \sigma)^2 \, \CM_2
 \notag \\
\sum_{\delta \ {\rm even}}  { \Upsilon _8 [\delta] \over \Psi_{10}}
\, \big ( \det \see_\delta \big ) \, \see^{(a_1 a_2}_\delta  \see^{a_3 a_4)}_\delta  
&=  \frac{21}{16}\, (\det \sigma)^2 \,  \CM_2^{a_1 a_2 a_3 a_4}
\no \\
 \sum_{\delta \ {\rm even}} { \Upsilon _8 [\delta] \over \Psi_{10}}
 \, \see^{(a_1 a_2}_\delta \see^{a_3 a_4}_\delta \see^{a_5 a_6}_\delta 
 \see^{a_7 a_8)}_\delta &= \frac{63}{32} \, (\det \sigma)^2 \, \CM_2^{a_1  \ndots a_8}
\end{align}
which shows that the spin structure sum of quadri-linears realizes all the irreducible representations
of $SL(2,\CC)$ on the right-hand side of $\see_\delta^{\otimes 4} = {\bf 1} \oplus {\bf 5} \oplus {\bf 9}$, associated with the tensors $\CM_2^{a_1 \ndots a_r}$ of rank $r=0,4,8$.}
{\cor
\label{cor:6.4}
The spin structure sums of penta-linear combinations of $\see_\delta$ are given by, 
\begin{align}
\label{proj.03}
\sum_{\delta \ {\rm even}} { \Upsilon _8 [\delta] \over \Psi_{10}}
\, (\det \see_\delta)^2 \see^{a_1 a_2}_\delta &=  \frac{99}{64}\, (\det \sigma)^2 \, \CM_3^{a_1 a_2} 
\notag \\
\sum_{\delta \ {\rm even}} { \Upsilon _8 [\delta] \over \Psi_{10}}
\, \big ( \det \see_\delta \big ) \, \see^{(a_1 a_2}_\delta  \see^{a_3 a_4}_\delta \see^{a_5 a_6)}_\delta  
&=    \frac{81}{128}\, (\det \sigma)^2 \, \CM_2 \CM_1^{a_1\ndots a_6}
- \frac{9}{32}\, (\det \sigma)^2 \, \CM_3^{a_1   \ndots a_6} 
\no \\
 \sum_{\delta \ {\rm even}} { \Upsilon _8 [\delta] \over \Psi_{10}}
 \, \see^{(a_1 a_2}_\delta \see^{a_3 a_4}_\delta \see^{a_5 a_6}_\delta 
 \see^{a_7 a_8}_\delta \see^{a_9 a_{10})}_\delta &=   \frac{705}{256} \, (\det \sigma)^2 \, 
 \CM_2^{(a_1 a_2 a_3a_4} \CM_1^{a_5  \ndots a_{10})}
\end{align}
The spin structure sum again realizes all the $SL(2,\CC)$ irreducible representations on the right-hand side of
$\see_\delta^{\otimes 5} = {\bf 3} \oplus {\bf 7} \oplus {\bf 11}$.}

\sm

Based on (\ref{spinsums.02}) and (\ref{proj.01}) to (\ref{proj.03}), 
the spin structure sums of cyclic products $C_\delta$ of up to 
eleven Szeg\"o kernels are readily available in a simplified form. The analogous 
spin structure sums over six powers of $\see_\delta^{ab}$ can be found in appendix \ref{highM.1}
which apply to cyclic products of Szeg\"o kernels up to thirteen points. 
We list spin structure sums over higher powers of 
$\see_\delta^{ab}$ in appendix \ref{highM.2}.

\sm

All of the expressions in this section and appendix \ref{highM} may be proven using the basic spin structure sums (\ref{spinsums.02}) and the trilinear relations in Theorem \ref{thm:6}. In practice, we often employ {\sc mathematica} and {\sc maple} to compare the hyper-elliptic representation of the spin structure sums with an Ansatz for modular tensors of suitable degree of homogeneity in the branch points and corresponding $SL(2,\mathbb C)$ representation properties. The multiplicities of the various representations in the respective Ans\"atze are derived from the methods in appendix \ref{sec:cl.1} and can be read off from table \ref{specirrep}. Matching high-degree polynomials in the branch points (including the cancellation of denominators $(u_i{-}u_j)^{-1}$ required by the presence of $\Psi_{10}^{-1}$) is facilitated by fixing three branch points, say  $u_4,u_5,u_6$,  at numerical values using $SL(2,\CC)$ covariance. Matching the Ansatz as a function of arbitrary values for the remaining branch points $u_1, u_2, u_3$ then provides an analytical proof of the Corollaries by {\sc mathematica} or {\sc maple}.

\subsubsection{Summation over three and four powers of $L_\delta$}

A major goal of this work is to set the stage for evaluating higher-point genus-two amplitudes of Type I, Type II or Heterotic strings, extending the five-point computations in \cite{DHoker:2021kks}. Here we shall specialize to the summation measure appropriate for the supersymmetric chiral sector, and spell out the explicit form of spin structure sums with three and four insertions of the tensor $L_\delta(i,j)$ in (\ref{Ldel2}) which follows from the Corollaries \ref{cor:6.1}---\ref{cor:6.4}. In this way, the spin structure sums of cyclic products $C_\delta$ of up to  eight Szeg\"o kernels are readily available in a simplified form. 

\sm

The results for the spin structure sums involve a quadri-holomorphic form $ \Delta(i,j|k,l)$ that generalizes the bi-holomorphic form $\Delta(z_i,z_j)$ in (\ref{defdelta}), and is defined as follows, 
\bea
 \Delta(i,j|k,l)  = \Delta(i,k) \Delta(j,l) + \Delta(i,l) \Delta(j,k) 
   \label{quadriholo}
 \eea
which is easily checked to obey the following identities, 
\bea
\Delta(i,j|k,l)=  \Delta(j,i|k,l) =  \Delta(k,l |i,j)
\no \\
\Delta(i,j|k,l) +  \Delta(i,k|l,j) +  \Delta(i,l|j,k) =0
\eea 
We shall also use the following notation for the contractions of the tensors $\CM_2^{a_1\ndots a_r}$ of  (\ref{5.M2}) of rank $r=4,8$ with the basis of holomorphic $(1,0)$ forms $\varpi _a$, 
\bea
M_2(1,\ndots,r) = \CM_2^{a_1 \ndots a_r} \varpi_{a_1}(1) \ndots \varpi_{a_r}(r) 
   \label{m2contract}
 \eea
In terms of these quantities, we have the following Corollaries.

\sm

{\cor 
\label{cor:6.6}
The spin structure sums with the supersymmetric measure $ \Upsilon _8 [\delta]$ for the product of three factors of $L_\delta$ is given as follows, 
\begin{align}
\sum_{\delta \ {\rm even}}  { \Upsilon _8 [\delta] \over \Psi_{10}} \,
\bee_\delta(1,2) \bee_\delta(3,4) \bee_\delta(5,6) &= \frac{9}{16}  \,  (\det \sigma)^2 \, M_1(1,2,3,4,5,6)
\label{sping3.58}
\end{align} 
where the contraction $M_1(1,\ndots,6)$ of $\CM_1^{a_1\ndots a_6}$ with $\varpi_{a_1}(1)\ndots \varpi_{a_6}(6)$ is defined in (\ref{contractm1}). The corresponding spin structure sum for the product of four factors of $L_\delta$ is given by, 
 \begin{align}
  \label{sping3.81}
\sum_{\delta \ {\rm even}}   { \Upsilon _8 [\delta] \over \Psi _{10}} \, 
\bee_\delta(1,2) & \bee_\delta(3,4)  \bee_\delta(5,6) \bee_\delta(7,8)= 
\frac{63}{32} \,  (\det \sigma)^2 M_2(1,2,3,4,5,6,7,8)
\no \\
&
+ \frac{3}{16} \,  \Big\{ 
\Delta(1, 2 | 3, 4) M_2(5, 6, 7, 8) + 
\Delta(1, 2 | 5, 6) M_2(3, 4, 7, 8) 
    \no \\ & \qquad ~
+ \Delta(1, 2 | 7, 8) M_2(3, 4, 5, 6) 
+  \Delta(3, 4 | 5, 6) M_2(1, 2, 7, 8) 
\no \\ & \qquad ~
+ \Delta(3, 4 | 7, 8) M_2(1, 2, 5, 6) + 
    \Delta(5, 6 | 7, 8) M_2(1, 2, 3, 4) \Big\}  \notag \\
& +
 \frac{3 \CM_2}{80 (\det \sigma)^2} \,  \Big\{ \Delta(1, 2 | 3, 4) \Delta(5, 6 | 7, 8)  
  +  \Delta(1, 2 | 5, 6) \Delta(3, 4 | 7, 8)  
    \no \\ & \qquad \qquad
   + \Delta(1, 2 | 7, 8) \Delta(3, 4 | 5, 6) \Big\}\, 
\end{align}}
 The identity (\ref{sping3.58}) will play a key role in the evaluation of the genus-two six-point amplitude.

\subsection{Supersymmetric spin structure summations for $C_\delta$}
 
Using these summation formulas for $\see_\delta$ and $L_\delta$, we may readily collect the results for the summation of the full cyclic products $C_\delta$ of Szeg\"o kernels, and products thereof, against the supersymmetric chiral measure. We group the results in terms of the following theorems.

\sm

{\thm 
\label{thm:6.3}
By combining the expressions in (\ref{5.C2}) and (\ref{5.C-three}) with (\ref{spinsums.02}) we obtain the vanishing of the cyclic product for zero, two, and three points upon spin structure sum, 
\begin{align}
\sum_{\delta \ {\rm even}}  \Upsilon _8 [\delta]  =
\sum_{\delta \ {\rm even}}  \Upsilon _8 [\delta]  \, C_\delta(1,2) =
\sum_{\delta \ {\rm even}} \Upsilon _8 [\delta] \,  C_\delta(1,2,3)  &= 0
 \label{Csums.1}
\end{align}
By combining (\ref{Cdel4}) with (\ref{spinsums.04}), we obtain the spin structure sum for the 4-point function~\cite{DP6}, 
\bea
\label{Cdel4sum}
\sum _{\delta \ {\rm even}} { \Upsilon_8[\delta] \over \Psi_{10}} \, C_\delta(1,2,3,4) 
& = &
-\frac{1}{8} \, \Delta(1,3|2,4)
=
\frac{1}{8} \,  \Big (  \Delta(1,4) \Delta(2,3)  + \Delta(1,2) \Delta(4,3) \Big )
\qquad
\eea
By combining (\ref{C.del5})  with (\ref{spinsums.04}) we obtain the spin structure sum for the 5-point function~\cite{DHoker:2021kks}, 
\begin{align}
\label{C.del5else}
\sum _{\delta \ {\rm even}} { \Upsilon_8[\delta] \over \Psi_{10}}  \, C_\delta (1,\cdots,5) 
=  -   \frac{ dx_1 \, \Delta(2,4|3,5) + d x_2 \, \Delta(1,4|3,5) }{ 16 \, x_{12} } 
  +{\rm cycl} (1,\cdots,5) 
\end{align}
Both sides of the equations transform as Siegel modular forms under $Sp(4,\ZZ)$ of weight $-2$.}

\sm

{\thm
\label{thm:6.7}
The six-point function in (\ref{cdel6.1}) introduces two types of non-vanishing spin structure sums,
both (\ref{spinsums.04}) over two powers of $L_\delta(i,j)$ and (\ref{sping3.58}) over three powers,
\bea
\label{6ptsum}
\sum _{\delta \ {\rm even}} \frac{\Upsilon_8[\delta] }{\Psi_{10}} \, C_\delta (1,\cdots,6)
& = & \frac{9}{16}\,(\det \sigma)^2 \, M_1(1,\cdots,6) 
 - \frac{1}{32} \bigg \{ 
\frac{\mN^{(6)}[123] }{x_{12}\, x_{23}}  
\no \\ &&
+ \frac{\mN^{(6)} [12,34] }{x_{12} \, x_{34} }  
 + \frac{\mN^{(6)} [12,45] }{2\, x_{12}\, x_{45} } 
 + {\rm cycl}(1,\cdots,6) \bigg\} 
\eea
where the numerators multiplying two simultaneous poles $(x_{ab}x_{cd})^{-1}$ are given by
\bea
\mN^{(6)}[123] & = &  \Delta(3,5|4,6) W_2^+(1,2)
+  \Delta(1,5|4,6) W_2^+(2,3) 
\no \\ && 
+  \Delta(2,5|4,6) W_2^-(1,3)
\no \\
\mN^{(6)} [12,34] & = & 
\Delta(2,5|3,6) W_2^+(1,4) +  \Delta(1,5|4,6) W_2^+(2,3)
\no \\ && 
+  \Delta(2,5|4,6) W_2^-(1,3) +  \Delta(1,5|3,6) W_2^-(2,4)
\no \\
\mN^{(6)} [12,45] & = & 
\Delta(2,5|3,6) W_2^+(1,4) +  \Delta(1,4|3,6) W_2^+(2,5)
\no \\ &&
+  \Delta(2,4|3,6) W_2^-(1,5) +  \Delta(1,5|3,6) W_2^-(2,4)
\eea
and the functions $W_2^{\pm}(i,j)$ are defined in (\ref{5.W2}).}

\sm

The residues of the poles in $x_{12}x_{34}$ or $x_{12}x_{45}$ entirely stem from the limits 
$W_2^{+}(i,j) \rightarrow s_i s_j$, while the linear part in $Z(i,j)$ vanishes as $x_1\rightarrow x_2$ and $x_3\rightarrow x_4$ or $x_4\rightarrow x_5$. The residues of the overlapping poles in $x_{12}x_{23}$, however, also receive 
contributions from the $Z(i,j)$, for instance $\Delta(3,5|4,6) Z(1,2)+  \Delta(1,5|4,6) Z(2,3) -  \Delta(2,5|4,6) Z(1,3) $ reduces to $\Delta(3,5|4,6)s_2^2$ as $z_1\rightarrow z_2$.

\sm

{\thm
\label{thm:6.8}
The spin structure sum of the seven-point function may be obtained from the representation (\ref{cdel7.1}) of $C_\delta(1,\ndots,7)$
and again receives non-vanishing contributions from both (\ref{spinsums.04})  and (\ref{sping3.58}),
\begin{align}
\label{7ptsum}
\sum _{\delta \ {\rm even}} \frac{\Upsilon_8[\delta] }{\Psi_{10}} \, C_\delta(1,\cdots,7) 
= & ~ \frac{ 9 \, \mathfrak{N}^{(7)}[12] }{32 \, x_{12}}
-  \frac{    \mathfrak{N}^{(7)}[1234] }{64 \, x_{12} \, x_{23} \, x_{34} }
-  \frac{    \mathfrak{N}^{(7)}[123,45] }{64 \, x_{12} \, x_{23} \, x_{45} }
\no \\ &
-  \frac{    \mathfrak{N}^{(7)}[123,56] }{64 \, x_{12} \, x_{23} \, x_{56} }
-  \frac{   \mathfrak{N}^{(7)}[123,67] }{64 \, x_{12}\, x_{23} \, x_{67} }
-  \frac{   \mathfrak{N}^{(7)}[12,34,56] }{64 \, x_{12} \, x_{34} \, x_{56} }  
\no \\ &
+ {\rm cycl}(1,\cdots,7)
 \end{align}
The instruction to add cyclic permutations applies to the entire right side of the formula.

\sm

The numerators are given by the following expressions, 
\begin{align}
\label{7ptsum.2}
 \mathfrak{N}^{(7)}[12] &= (\det \sigma)^2 \big[  dx_1 M_1(2,3,4,5,6,7) + dx_2 M_1(3,4,5,6,7,1) \big] 
\end{align}
multiplying a single pole $x_{12}^{-1}$ and the following cyclically inequivalent
cases for three simultaneous poles $(x_{ab}x_{cd}x_{ef})^{-1}$,
\begin{align}
 \mathfrak{N}^{(7)}[1234] &=
 \Delta(4,6|5,7) W_3^+(1,2,3)   +   \Delta(3,6|5,7)W_3^-( 1,2|4)  
\no \\ & \quad 
  +  \Delta(2,6|5,7)W_3^-( 3,4| 1) +  \Delta(1,6|5,7)W_3^+(2, 3,4) 
\notag \\
 \mathfrak{N}^{(7)}[123,45] &=
  \Delta(3,6|5,7)W_3^-(1,2 | 4)  + \Delta(3,6|4,7)W_3^+(1,2,5)
  \no \\ & \quad
 +  \Delta(2,6|5,7)W_3^-(3,4 |1)  +  \Delta(2,6|4,7)W_3^-(1,5 | 3) 
 \no \\ & \quad
 +  \Delta(1,6|5,7)W_3^+(2,3,4)  +  \Delta(1,6|4,7)W_3^-(2,3 | 5) 
 \notag \\
 \mathfrak{N}^{(7)}[123,56] &=
  \Delta(3,6|4,7)W_3^+(1,2,5) +   \Delta(3,5|4,7)W_3^-(1,2 |6)
  \no \\ & \quad
+  \Delta(2,6|4,7)W_3^-(1,5 | 3) +   \Delta(2,5|4,7)W_3^-(3,6 | 1) 
\no \\ & \quad
+ \Delta(1,6|4,7)W_3^-(2,3 | 5) + \Delta(1,5|4,7)W_3^+(2,3,6)
 \notag \\
 \mathfrak{N}^{(7)}[123,67] &=
  \Delta(3,5|4,7)W_3^-(1,2 |6)  +  \Delta(3,5|4,6)W_3^+(7,1,2) 
  \no \\ & \quad
+  \Delta(2,5|4,7) W_3^-(3,6 | 1) +   \Delta(2,5|4,6)W_3^-(7,1 |3) 
\no \\ & \quad
+   \Delta(1,5|4,7) W_3^+(2,3,6) +  \Delta(1,5|4,6)W_3^-(2,3 |7)  
\no \\
 \mathfrak{N}^{(7)}[12,34,56]    &=
  \Delta(2,6|4,7) W_3^-(1,5 |3) +   \Delta(2,5|4,7) W_3^-(3,6 | 1) 
  \no \\ & \quad
+  \Delta(2,6|3,7)W_3^+(1,4,5)  +  \Delta(2,5|3,7) W_3^-(1,4 | 6) 
\no \\ & \quad
+  \Delta(1,6|4,7)W_3^-(2,3 |5) +   \Delta(1,5|4,7)W_3^+(2,3,6) 
 \notag \\
 &\quad
+ 
  \Delta(1,6|3,7)W_3^-(4,5 |2) 
+  
  \Delta(1,5|3,7)W_3^-(2,6 |4) 
\end{align}
As for six points, some of the $Z(i,j)$-contributions to $W_3^{\pm}$ in (\ref{5.W3}) may cancel at the residues of individual poles.}

\newpage

{\thm
\label{thm:6.9}
The spin structure sum of the cyclic product of Szeg\"o kernels for eight points is given by, 
\begin{align}
\label{cdel8.A}
\sum _{\delta \ {\rm even}}  \frac{\Upsilon_8[\delta] }{\Psi_{10}} \,  C_\delta & (1,  \ndots,8) = 
 \frac{63}{32}\, (\det \sigma)^2  \, M_2(1,\cdots,8) 
 \\
& \quad +
\bigg\{
\frac{3}{32} \,  \Big (  \Delta(1, 2 | 3, 4)\,  M_2(5,6,7,8)
+ \frac{1}{2} \Delta(1, 2 | 5,6)\,  M_2(3,4,7,8) \Big )
\notag \\
&\qquad ~
+ \frac{3\, \CM_2}{320 \, (\det \sigma)^2}  \, \Big ( 
\Delta(1, 2 | 3, 4) \Delta(5, 6 | 7, 8) 
 + \frac{1}{2} \Delta(1, 2 | 5, 6) \Delta(3, 4 | 7, 8)  \Big )  \notag \\
&\qquad ~
+ (\det \sigma)^2 \bigg[\frac{ 9 \mathfrak{N}^{(8)}[123] }{64 \, x_{12} \, x_{23} }
+  \frac{ 9 \mathfrak{N}^{(8)}[12,34] }{64 \, x_{12} \, x_{34} }
+  \frac{ 9 \mathfrak{N}^{(8)}[12,45] }{64 \, x_{12} \, x_{45} }
+  \frac{ 9 \mathfrak{N}^{(8)}[12,56] }{128 \, x_{12} \, x_{56} } \bigg] 
\notag \\
&\qquad ~
-  \frac{  \mathfrak{N}^{(8)}[12345] }{128 \, x_{12} \, x_{23}  \, x_{34} \, x_{45} }
-  \frac{  \mathfrak{N}^{(8)}[1234,56] }{128 \, x_{12} \, x_{23}  \, x_{34} \, x_{56} }
-  \frac{  \mathfrak{N}^{(8)}[1234,67] }{128 \, x_{12} \, x_{23}  \, x_{34} \, x_{67} }
\notag \\
&\qquad ~
-  \frac{  \mathfrak{N}^{(8)}[1234,78] }{128 \, x_{12} \, x_{23}  \, x_{34} \, x_{78} }
-  \frac{  \mathfrak{N}^{(8)}[123,45,78] }{128 \, x_{12} \, x_{23}  \, x_{45} \, x_{78} }
-  \frac{  \mathfrak{N}^{(8)}[123,56,78] }{128 \, x_{12} x_{23}  \, x_{56} \, x_{78} }
\notag \\
&\qquad ~
-  \frac{  \mathfrak{N}^{(8)}[123,45,67] }{128 \, x_{12} \, x_{23}  \, x_{45} \, x_{67} }
-  \frac{  \mathfrak{N}^{(8)}[123,456] }{128 \, x_{12} \, x_{23}  \, x_{45} \, x_{56} } 
-  \frac{  \mathfrak{N}^{(8)}[123,567] }{256 \, x_{12} \, x_{23}  \, x_{56} \, x_{67} }
\no \\ & \qquad ~
-  \frac{  \mathfrak{N}^{(8)}[12,34,56,78] }{512 \, x_{12} \, x_{34}  \, x_{56} \, x_{78} } 
+ {\rm cycl}(1,2,\ndots,8) \bigg\}
\no
\end{align} 
The expressions for the numerators $\mathfrak{N}^{(8)}$ in the last five lines are lengthy and given in appendix \ref{sec:spinsum8}}.

\subsubsection{Spin structure sums of products of cyclic products}
\label{sec:prodcyc}

The spin structure sums over products of $C_\delta$ at four and five points can be readily derived from the
representations (\ref{5.C2}) and (\ref{5.C-three}) of the cyclic products $C_\delta$ of length two and three as well as the spin structure sum (\ref{spinsums.04}), and are given by the following corollary.
{\cor 
\label{cor:6.15}
The spin structure sums, with the supersymmetric measure, of the products of $C_\delta(1,2)$ with the cyclic product of two and three Szeg\"o kernels are given by \cite{DP6, DHoker:2021kks}, 
\begin{align}
\sum _{\delta \ {\rm even}} \frac{\Upsilon_8[\delta] }{\Psi_{10}} \, C_\delta(1,2) C_\delta(3,4) &=
\frac{1}{4}\, \Delta(1,2|3,4) \label{prods.01} \\
 \sum _{\delta \ {\rm even}} \frac{\Upsilon_8[\delta] }{\Psi_{10}} \, C_\delta(1,2,3) C_\delta(4,5) &=
 \frac{1}{8} \, \bigg\{ \frac{dx_1\, \Delta(2,3|4,5) + dx_2\, \Delta(1,3|4,5) }{x_{12}}
 + {\rm cycl}(1,2,3) \bigg\}
 \notag
\end{align}}
There are three distinct factorized cyclic products contributing to the six-point function, which may be evaluated  
using the above methods along with (\ref{Cdel4}) and (\ref{sping3.58}). The results are given by the corollary below. 
{\cor
\label{cor:6.16}
The spin structure sums of factorized cyclic products of Szeg\"o kernels contributing to the six-point function are given by,
\begin{align}
\sum _{\delta \ {\rm even}} & \frac{\Upsilon_8[\delta] }  {\Psi_{10}} \,  C_\delta(1,2) C_\delta(3,4) C_\delta(5,6) 
 = \frac{9}{16} \, (\det \sigma)^2 \, M_1(1,2,\ndots,6) 
 \\ &
 - \frac{1}{8} \bigg\{ \Delta(1,2|3,4) \, \frac{ W_2^+(5,6) }{x_{56}^2}
+\Delta(1,2|5,6) \, \frac{ W_2^+(3,4) }{x_{34}^2}
+\Delta(3,4|5,6) \, \frac{ W_2^+(1,2) }{x_{12}^2} \bigg\} 
\no
\end{align}
as well as,
\begin{align}
\label{prods.02}
\sum _{\delta \ {\rm even}} \frac{\Upsilon_8[\delta] }{\Psi_{10}} \,  C_\delta(1,2,3,4) C_\delta(5,6) 
&=  \frac{9}{16} \, (\det \sigma)^2 \, M_1(1,2,\ndots,6)  +   \frac{  \Delta(1,3|2,4) W_2^+(5,6) }{16 x_{56}^2}
\no \\
&\quad + \bigg\{
\frac{ W_2^+(2,3) \Delta(1,4|5,6) }{32 x_{12}x_{34}}
+ \frac{W_2^+(4,1) \Delta(2,3|5,6) }{32 x_{12}x_{34}}
\no \\ & \qquad
+ \frac{W_2^-(2,4) \Delta(1,3|5,6) }{32 x_{12}x_{34}}
+ \frac{W_2^-(1,3) \Delta(2,4|5,6) }{32 x_{12}x_{34}} 
\no \\ & \qquad
+ \frac{ W_2^+(1,2) \Delta(3,4|5,6) }{16 x_{12}x_{23}}
+ \frac{W_2^+(2,3) \Delta(1,4|5,6) }{16 x_{12}x_{23}}
\notag \\
&\qquad 
+ \frac{W_2^-(1,3) \Delta(2,4|5,6) }{16 x_{12}x_{23}}
+{\rm cycl}(1,2,3,4) \bigg\} 
\end{align}
and finally, 
\begin{align}
\sum _{\delta \ {\rm even}} \frac{\Upsilon_8[\delta] }{\Psi_{10}} \, C_\delta(1,2,3) C_\delta(4,5,6) 
&= 
\frac{  \mathfrak{N}^{(6)}[12|45]}{16 x_{12} x_{45}}
+\frac{  \mathfrak{N}^{(6)}[12|56]}{16 x_{12} x_{56}}
\no \\ 
& \quad
+\frac{  \mathfrak{N}^{(6)}[12|64]}{16 x_{12} x_{64}} + {\rm cycl}(1,2,3)
\end{align}
where the instruction to add cyclic permutations applies to all terms on the right side of the formula and we have used the following definition,
\begin{align}
\mathfrak{N}^{(6)}[12|45] &=
dx_1 \, dx_4 \, \Delta(2,3|5,6)
+ dx_1 \, dx_5 \, \Delta(2,3|4,6)  \label{prods.03}\\
&\quad
+ dx_2 \, dx_4 \, \Delta(1,3|5,6)
+ dx_2 \, dx_5 \, \Delta(1,3|4,6) \notag
\end{align}}
All other components of these expressions have been defined earlier.

\newpage

\subsection{Spin structure sums in the $E_8\times E_8$ sector}
\label{sec:sum.2} 

In the $E_8\times E_8$ Heterotic string the 32 Majorana-Weyl fermions are partitioned into two 16-element sets, 
with all fermions in the first set of 16 carrying the same spin structure $\delta_1$ and all fermions in the second set of 16 carrying the same spin structure $\delta_2$, where $\delta_1$ and $\delta_2$ are independent of one another. The spin structure sum then consists of summing independently over  $\delta_1$ and $\delta_2$. Thus, the entire spin structure sum required for the $E_8\times E_8$ Heterotic string factorizes into two independent spin structure sums for 16 fermions, producing the measure factor $\tet [\delta](0)^8$ noted in (\ref{spinsums.01}). The basic spin structure sums in a single $E_8$ sector are then given by the following theorem.
 
{\thm 
\label{thm:6.11}
The basic spin structure sums for a single $E_8$ sector in the  Heterotic $E_8\times E_8$ string are given by,
{\small \begin{align}
(\det \sigma)^4  \sum_{\delta \ {\rm even}} \tet[\delta](0)^8  
&
= 720 \Big ( 75 \, \CM_4  - 4 \, \CM_2^2 \Big ) 
 \label{spinsums.05}
 \\
(\det \sigma)^4  \sum_{\delta \ {\rm even}} \tet[\delta](0)^8   \see_\delta^{ab} &= 
 1350 \Big (15 \, \CM_5^{ab} - 4 \, \CM_2 \, \CM_3^{ab} \Big )
 \notag \\
 (\det \sigma)^4  \sum_{\delta \ {\rm even}} \tet[\delta](0)^8  ( \det \see_\delta)&=
 81 \Big ( 24 \, \CM_2^3 - 300 \, \CM_2  \, \CM_4 - 125 \, \CM_6 \Big )
\no \\ %
(\det \sigma)^4  \sum_{\delta \ {\rm even}} \tet[\delta](0)^8   \see_\delta^{(ab} \see_\delta^{cd)} &=
  \frac{405}{2} \Big (25 \, \CM_3^{(ab} \, \CM_3^{cd)} 
  {+}   \big ( 200 \, \CM_4 - 16 \CM_2^2 \big )  \CM_2^{abcd} 
   {-}   40 \, \CM_2 \, \CM_4^{abcd}  \Big )
   \no
\end{align}} 
where the tensors $\CM_2, \CM_3$ were defined in (\ref{5.M2}), (\ref{5.M3}),
and the tensors $ \CM_4, \CM_5, \CM_6$ can be found in appendix \ref{sec:D2higher}.}

The proof of Theorem \ref{thm:6.11} is obtained in the same manner as the proofs of Theorem \ref{thm:6.1} and Corollaries \ref{cor:6.2}---\ref{cor:6.4}. We note that the combination $75 \, \CM_4  - 4 \, \CM_2^2$ is proportional to the Siegel modular form $\Psi_4$ defined in (\ref{defpsifour}).

\subsection{Spin structure sums in the Spin$(32)/\mathbb Z_2$ sector}
\label{sec:sum.1.3} 

In the Spin$(32)/\mathbb Z_2$ Heterotic string, 32 Majorana-Weyl fermions are all in the same 
spin structure so that the chiral measure includes the factor of $\tet[\delta](0)^{16}$ in (\ref{spinsums.01}), and 
produces the basic spin structure sums given in the following theorem.
{\thm 
\label{thm:6.10}
The spin structure sums of the Heterotic Spin$(32)/\mathbb Z_2$ string are given by, 
\begin{align}
\label{spinsums.08}
(\det \sigma)^8  \sum_{\delta \ {\rm even}} \tet[\delta](0)^{16}   &=  129600 
\Big ( 75 \, \CM_4  - 4 \, \CM_2^2 \Big )^2
 \\
(\det \sigma)^8  \sum_{\delta \ {\rm even}} \tet[\delta](0)^{16} \,   \see_\delta^{ab} &= 
243000 \Big ( 75 \, \CM_4  - 4 \, \CM_2^2 \Big ) \Big ( 15 \CM_5^{ab} - 4  \CM_2 \CM_3^{ab} \Big )
\notag \\
(\det \sigma)^8  \sum_{\delta \ {\rm even}} \tet[\delta](0)^{16}   \see_\delta^{(ab}  \see_\delta^{cd)} &=
12150 \Big( ( 64 \CM_2^4  - 
   2000 \CM_2^2 \CM_4 + 
   15000 \CM_4^2 ) \CM_2^{abcd} \notag \\
   &\quad + 
  ( 300 \CM_2^2  {+} 
   1875 \CM_4 )  \CM_3^{(ab} \CM_3^{cd)}  - 
   3000 \CM_2\CM_3^{(ab} \CM_5^{cd)}  
   \no \\
   &\quad  + 
   5625  \CM_5^{(ab} \CM_5^{cd)}  + 
  40 \CM_2 ( 4 \CM_2^2 {-}  75   \CM_4 ) 
    \CM_4^{abcd} \Big)
\notag \\
(\det \sigma)^8  \sum_{\delta \ {\rm even}} \tet[\delta](0)^{16}  ( \det \see_\delta) &= 
14580 \big (  75 \CM_4 - 4 \CM_2^2\big ) \big ( 24 \CM_2^3 -  300 \CM_2\CM_4 - 125 \CM_6 \big )  
\notag \\
   &\quad - \tfrac{3}{8} (\det \sigma)^{10} \Psi_{10} 
\notag
\end{align}
These results remain expressible in terms of the tensors $ \CM_{w=2,3}$ in (\ref{5.M2}), (\ref{5.M3})
and $ \CM_{w=4,5,6}$ in appendix \ref{sec:D2higher} as well as the Igusa 
cusp form $\Psi_{10}$ in (\ref{defpsifour}).}

The proof of Theorem \ref{thm:6.10} is obtained in the same manner as the proofs of Theorem \ref{thm:6.1}, Corollaries \ref{cor:6.2}---\ref{cor:6.4}, and Theorem \ref{thm:6.11} above.

 \newpage
 
\section{Conclusion and outlook}
\label{sec:concl}
\setcounter{equation}{0}

In this work, we have described and implemented a procedure that organizes the evaluation of cyclic products $C_\delta$ of an arbitrary number of Szeg\"o kernels on a  genus two Riemann surface with arbitrary even spin structure $\delta$. The procedure drastically simplifies the dependence on $\delta$ and reduces all spin structure dependence to a degree-two polynomial in the components $\see_\delta^{ab}$ of a symmetric rank-two tensor $\see_\delta$ under $SL(2,\CC)$. The tensor $\see_\delta$  depends on the branch points of the surface in the hyper-elliptic description  but not on the insertion points of the Szeg\"o kernels.  The dependence of $C_\delta$ on the insertion points of the Szeg\"o kernels is  organized in simple  $SL(2,\mathbb C)$ invariant building blocks that are identified from the cyclic product of two Szeg\"o kernels and suffice to evaluate cyclic products with an arbitrary number of Szeg\"o kernels. The explicit form of this reduction can be assembled from the all-multiplicity conjecture (\ref{Qdel.17}) combined with the trilinear relations among the variables $\see_\delta^{ab}$ in Theorem~\ref{thm:6}.

\sm

Our results dramatically simplify the summation of cyclic products of Szeg\"o kernels over even spin structures that arise in genus-two amplitudes of Type I, Type II and Heterotic strings. For a given string theory, only the spin structure sums over the ten monomials in $\see_\delta^{ab}$ of degree $0,1$ and 2 are needed and given in terms of branch-point dependent $SL(2,\mathbb C)$-tensors in (\ref{spinsums.02}), (\ref{spinsums.05}) and (\ref{spinsums.08}).  In future work  we will apply these advances to the evaluation of genus-two superstring amplitudes involving six or more massless  external NS states which are currently uncharted territory in both the RNS and the pure-spinor formulation. The construction of these higher-point amplitudes, in the formulation of \cite{DP5,DP6}, will also require similar reductions of the spin structure dependence for products  of Szeg\"o kernels that form linear chains rather than closed cycles. The simplifications of such linear 
chains may be obtained by methods similar to the ones used here, including the trilinear relations, and they will greatly streamline the challenging steps in the assembly of the amplitudes. In a more immediate follow-up work \cite{DHS}, the building blocks of our results will be translated into the language of  genus-two theta functions and modular tensors of $Sp(4,\mathbb Z)$.

\sm 

It is expected that, in the long run, the ideas developed here will lead to further developments related to (i) string and field-theory scattering amplitudes; (ii) tests of the S-duality of the low energy effective action of Type IIB string theory and its relation to modular forms of higher genus; (iii) mathematical questions on iterated integrals on genus-two surfaces; (iv) improving the understanding of the cohomology of chiral amplitudes at fixed spin structure:
\begin{itemize}
\item[(i)] The organization of supersymmetric chiral amplitudes at genus two as degree-two polynomials in $\see_\delta^{ab}$ can be viewed as a space-time supersymmetry decomposition of the particle content in the loops. In a string-theory context, combinations of $\see_\delta^{ab}$ that drop out from maximally supersymmetric spin sums
may contribute to genus-two amplitudes in K3, Calabi-Yau, and orbifold compactifications with reduced supersymmetry \cite{Aoki:2003sy,DHoker:2013sqy}, see also \cite{Bianchi:2006nf, Berg:2016wux} for analogous studies of Szeg\"o kernels at genus one. In a field-theory context, each monomial in $\see_\delta^{ab}$ can be associated with a different combination of massless particle species in the two loops of the Feynman graphs that arise in the low energy limit. This mechanism has also been studied in detail from the viewpoint of ambi-twistor strings at genus one \cite{Geyer:2015bja, Geyer:2015jch, He:2017spx} and genus two \cite{Geyer:2016wjx, Geyer:2018xwu, Geyer:2019hnn}.
\item[(ii)] The availability of explicit expressions for various amplitudes and their low energy expansions has resulted in extensive checks on the S-duality of Type IIB superstring contributions to the effective interactions, based on amplitudes with four external states in \cite{DP6,DHoker:2005jhf,Gomez:2013sla, DHoker:2014oxd}, and five external states in \cite{Richards:2008jg, Green:2013bza, Gomez:2015uha, DHoker:2020tcq}. The coefficients of these genus-two low energy effective interactions may be reformulated in terms of non-holomorphic genus-two modular graph forms \cite{DHoker:2017pvk,DHoker:2018mys}, which satisfy a wealth of novel identities as functions on Torelli space \cite{DHoker:2020uid}. Availability of the genus-two 6-point amplitude will, for sure, add much insight into the structure of both the low energy effective interactions and their significance to S-duality, as well as for the understanding of genus-two modular graph functions.
\item[(iii)] 
At genus one, cyclic products of Szeg\"o kernels and their spin structure sums generate coefficients of the Kronecker-Eisenstein series \cite{Broedel:2014vla} which can be used to construct elliptic polylogarithms and homotopy-invariant iterated integrals on a torus \cite{BrownLev}. Not withstanding recent progress in \cite{EZpinoeer}, the construction of iterated integrals on Riemann surfaces of genus two and beyond is largely an open problem, which is of relevance to both mathematicians and physicists. Our simple spin-structure independent functions of the insertion points, encountered  in the cyclic products of Szeg\"o kernels at genus two, are expected to provide guidance for the construction of integration kernels on Riemann surfaces beyond genus one. In particular, it would be interesting to extract higher-genus generalizations of Kronecker-Eisenstein coefficients from Szeg\"o kernels and to make contact with the proposal for their generating series in \cite{Tsuchiya:2017joo, Tsuchiya:2022lqv}.
\item[(iv)] The relation between super-holomorphicity and holomorphicity of chiral superstring $n$-point functions for NS bosons on a genus-two Riemann surface was shown to be encoded in a hybrid of both de Rham and Dolbeault cohomologies in   \cite{DP7}. The reduction procedure developed in the present paper is expected to drastically simplify the constructive algorithm provided in \cite{DP7} for the chiral amplitude cohomology classes and their representatives. Investigations of these questions are relegated to future work.
\end{itemize}

 \newpage
 
\appendix

\section{Synopsis of $SL(2,\CC)$  representations}
\label{sec:A}
\setcounter{equation}{0}

In this appendix we collect some simple basic formulas on tensor products and tensor powers of irreducible representations of $SL(2,\CC)$. 

\sm

We denote the irreducible finite-dimensional representations of $SL(2,\CC)$ alternatively by the standard symbol $D_j$ or by its dimension written in bold face $\bd=2j+1$ for half positive integer $j$. In this notation, the defining representation is denoted alternatively either by $D_\half$ or by ${\bf 2}$, while the vector representation is denoted either by $D_1$ or by ${\bf 3}$. The latter is actually the representation under which the vectors $\see_\delta$ transform, while the tensor $\CM_1$ transforms under the representation denoted by either $D_3$ or ${\bf 7}$.  The branching rule for the tensor product of two arbitrary representations $D_{j_1}$ and $D_{j_2}$ is given by the standard formula,
\bea
D_{j_1} \otimes D_{j_2} = \bigoplus_{k =  |j_1-j_2 |} ^{j_1+j_2} D_k
\eea
where the sum proceeds in integer steps.  For example, $D_1 \otimes D_j = D_{j-1} \oplus D_j \oplus D_{j+1}$.

\subsection{Tensor powers of irreducible representations}
\label{sec:A.1}

We shall often need the tensor power of an irreducible representation, such as the tensor power of $\see_\delta$, the tensor power of $\CM_1$, and the tensor product of $n$ identical copies of an irreducible representation $\bd$. We shall use the following notation for these tensor powers, 
\bea
\see_\delta ^{\otimes n} & = & \see_\delta \otimes \cdots \otimes \see_\delta
\no \\
\CM_1^{\otimes n} & = & \CM_1 \otimes \cdots \otimes \CM_1
\no \\
\bd ^{\otimes n} & = & \bd \otimes \cdots \otimes \bd
\eea
where each tensor product on the right contains $n$ factors. The dimension of the representation $\bd^{\otimes n}$  is generally smaller than $d^n$ because the tensor power involves \textit{identical} tensors. For example, the general rule gives ${\bf 3} \otimes {\bf 3} = {\bf 5} \oplus {\bf 3} \oplus {\bf 1}$. However, when the two factors in the tensor product are identical tensors (as is always the case in a tensor power), the ${\bf 3}$ in the direct-sum decomposition is absent, and the dimension is reduced to 6. The following lemma gives the dimension of the $n$-th tensor power $\bd^{\otimes n}$ of an irreducible representation $\bd$.

{\lem
\label{tenpow}
The dimension $D(d,n)$ of the tensor power $\bd^{\otimes n}$ of an irreducible representation $\bd$ of $SL(2,\CC)$ is given by,
\bea
D(d,n)= \dim \left ( \bd^{\otimes n} \right ) = \binom{d+n-1}{d-1}
\eea}

\sm

To prove the lemma, we denote the components of the tensor by $\bd = (\xi_1, \cdots, \xi_d)$. The components of the tensor power $\bd^{\otimes n}$ are then given by the monomials $\xi_{i_1} \cdots \xi_{i_n}$ for all possible distinct orderings $1\leq i_1 \leq i_2 \leq \cdots \leq i_{n-1} \leq i_n\leq d$. To list all such monomials, we begin with those whose first entry is $\xi_1$, of which there are $D(d,n{-}1)$.  The remaining monomials cannot contain the component $\xi_1$ any more in view of the ordering we have adopted, so their number is $D(d{-}1, n)$, which gives the double recursion relation, 
\bea
D(d,n)=D(d,n{-}1) + D(d{-}1,n)
\eea
with the initial conditions $D(d,0)=D(1,n)=1$. The solution is given by Pascal's triangle, which completes the proof of the lemma. The lemma implies the following corollary.
 {\lem
 The decomposition of  $\see_\delta ^{\otimes n}$ into  irreducible representations is given by, 
 \bea 
 \see_\delta ^{\otimes n} = D_n \oplus D_{n-2} \oplus \cdots \oplus D_\star
 \eea
 each representation appearing with multiplicity 1.  The representation $D_\star$ equals $D_0$ when $n$ is even and $D_1$ when $n$ is odd.}
 
 \sm
 
To prove this lemma we proceed as follows. From inspection of the contractions with $\ep_{ab}$ it is clear that we must  have the following inclusion,
\bea
D_n \oplus D_{n-2} \oplus \cdots \oplus D_\star \subset  \see_\delta ^{\otimes n}
 \eea
each representation appearing with multiplicity 1.  Thus we must have,
\bea
\label{dimineq}
 \dim (D_n) + \dim (D_{n-2}) + \cdots + \dim (D_\star) \leq \dim (\see_\delta ^{\otimes n}) = \thalf (n+1)(n+2)
\eea
where the last equality on the right side uses the result of Lemma \ref{tenpow}. However, computing the dimension of $D_n \oplus D_{n-2} \oplus \cdots \oplus D_\star$ shows that the inequality in (\ref{dimineq}) is actually an equality which proves the lemma.

\sm

Explicit results for low values of $n$ are given as follows,
\begin{align} 
\see_\delta^{\otimes 1} & = {\bf 3} &
\see_\delta^{\otimes 5} & = {\bf 11} \oplus {\bf 7} \oplus {\bf 3}
\no \\
\see_\delta^{\otimes 2} & = {\bf 5} \oplus {\bf 1} &
\see_\delta^{\otimes 6} & = {\bf 13} \oplus {\bf 9} \oplus {\bf 5} \oplus {\bf 1}
\no \\
\see_\delta^{\otimes 3} & = {\bf 7} \oplus {\bf 3} &
\see_\delta^{\otimes 7} & = {\bf 15} \oplus {\bf 11} \oplus {\bf 7} \oplus {\bf 3}
\no \\
\see_\delta^{\otimes 4} & = {\bf 9} \oplus {\bf 5} \oplus {\bf 1} &
\see_\delta^{\otimes 8} & = {\bf 17 } \oplus {\bf 13} \oplus {\bf 9} \oplus {\bf 5} \oplus {\bf 1}
\end{align}

\subsection{Representation theory of $\CM_n$ tensors}
\label{sec:cl.1} 

In order to count the modular tensor with degree $n$ in $\mu_j$ and components of homogeneity degree $h$ in $u_j$, we enumerate the monomials $\mu_{j_1} \mu_{j_2}\ndots \mu_{j_n}$ with $h=\sum_{i=1}^n j_i$ and $0\leq j_i\leq 6$. 
Since the spectrum of such monomials is symmetric under $\mu_j \leftrightarrow \mu_{6-j}$, the distribution
of homogeneity degrees in $u_j$ at given degree $n$ is symmetric around $h=3n$. The counting is spelled out for $n\leq 6$ in table \ref{homdegree} with the symmetry
point $h=3n$ highlighted in red.\footnote{From a representation-theoretic viewpoint,
the more natural quantity is $h\rightarrow h{-}3n$. In this way, the ``average value'' of 
$h$ is mapped to 0 and the spectrum of $h$ becomes symmetric under 
the ``inversion'' $h \rightarrow - h$ of the shifted $h$.}

\begin{table}[h]
\begin{center}
\begin{tabular}{|c||c|c|c|c|c|c|c|c|c|c|c|c|c|c|c|c|c|c|c|}
\hline
$n\backslash h$ &0&1&2&3
& 4 &5&6 &7&8&9 &10 &11 &12 &13 &14 &15 &16 &17 &18 \\ \hline\hline
1 &1 &1 &1 &\textcolor{red}{1} &1 &1 &1 &&&&&&&&&&&& \\ \hline
2 &1& 1& 2& 2& 3& 3& \textcolor{red}{4}& 3& 3& 2& 2& 1& 1 &&&&&& \\ \hline
3 &1& 1& 2& 3& 4& 5& 7& 7& 8& \textcolor{red}{8}& 8& 7& 7& 5& 4& 3& 2& 1& 1  \\ \hline
4 &1& 1& 2& 3& 5& 6& 9& 10& 13& 14& 16& 16& \textcolor{red}{18}& 16& 16& 14& 13& 10& 9  \\ \hline
5 &1& 1& 2& 3& 5& 7& 10& 12& 16& 19& 23& 25& 29& 30& 32& \textcolor{red}{32}& 32& 30 &29  \\ \hline
6 &1& 1& 2& 3& 5& 7& 11& 13& 18& 22& 28& 32& 39& 42& 48& 51& 55& 55&\textcolor{red}{58} \\ \hline
\end{tabular}
\end{center}
\caption{The numbers of monomials $\mu_{j_1} \mu_{j_2}\ndots \mu_{j_n}$ with $h=\sum_{i=1}^n j_i$.}
\label{homdegree}
\end{table}

The next step is to organize these monomials in $\mu_j$ into irreducible representations (irreps) of $SL(2,\mathbb C)$. The $\mu_j$ themselves at $n=1$ form the tensor $\CM_1$ in (\ref{MXcomps}) in the seven-dimensional
representation {\bf 7}. At $n\geq 2$, the entirety of degree-$n$ modular tensors furnishes the symmetric $n$-fold tensor product ${\bf 7}^{\otimes_s n}$ which can be conveniently organized into irreps of $SL(2,\mathbb C)$ by means of
the entries of table \ref{homdegree}:
\begin{itemize}
\item[(i)] At given $n$, start from the average value of $h=3n$ and combine all the neighboring entries at $h=3n{\pm}1, \ h=3n{\pm}2,\ndots$ with the same counting of $\mu_j$-monomials into one irrep. If there are $m_1$ adjacent cells in table \ref{homdegree} with the same counting, this irrep is the $m_1$-dimensional one $\bf{m_1}$.
\item[(ii)] Remove the monomials in the above $\bf{m_1}$ from the table (i.e.\ subtract one from the relevant entries) and once more identify the largest sequence of identical entries centered around $h=3n$. The length
$m_2$ of that sequence determines the next irrep $\bf{m_2}$.
\item[(iii)] Repeat the process of removing the entries of the irrep $\bf{m_k}$ identified in the
previous step from the table and combining the leftover entries with the same counting
centered around $h=3n$ into the next irrep $\bf{m_{k+1}}$. The process terminates if the removal of $\bf{m_k}$ reduces the leftover entries of the table to zero.
\end{itemize}
At $n=2$ for instance, the entries $1,1,2,2,3,3,\textcolor{red}{4},3,3,2,2,1,1$ of  table \ref{homdegree} lead to identify a {\bf 1} in step (i) which leaves 5 neighboring entries  3 after lowering the central entry 4 highlighted in red to 3. Step (ii) then identifies a {\bf 5} whose removal from the table leads to 9 neighboring entries 2 which form a {\bf 9}.
By also removing the latter from the table, one is finally left with a {\bf 13} and
we reproduce the decomposition 
\bea
{\bf 7}^{\otimes_s 2}= {\bf 13} \oplus {\bf 9}
\oplus {\bf 5} \oplus {\bf 1}
\label{clean.02}
\eea
The spectrum of irreps ${\bf m}$ at given degree $n$ in $\mu_j$ can be found in table \ref{specirrep}. Note that the total number of degree-$n$ monomials is ${n+6 \choose n}$, i.e.\ $7, 28, 84, 210, 462, 924, 1716, 3003,\ndots$ at $n=1,2,3,4,5,6,7,8,\ndots$, as indeed predicted by Lemma \ref{tenpow} for the case $d=7$. 

\begin{table}[h]
\begin{center}
\begin{tabular}{|c||c|c|c|c|c|c|c|c|c|c|c|c|c|c|c|c|}
\hline
$n\backslash \bd $ &\phantom{,}{\bf 1}\phantom{,}&\phantom{,}{\bf 3}\phantom{,}&\phantom{,}{\bf 5}\phantom{,}&\phantom{,}{\bf 7}\phantom{,}&\phantom{,}{\bf 9}\phantom{,}& {\bf 11} & {\bf 13} & {\bf 15} & {\bf 17} 
& {\bf 19} & {\bf 21}  & {\bf 23}  & {\bf 25}  & {\bf 27}  & {\bf 29}  & {\bf 31}  \\ \hline\hline
1 &0 &0 &0 &1 &&&&&&&&&&&&  \\ \hline
2 &1& 0& 1& 0& 1& 0& 1 &&&&&&&&& \\\hline
3 &0& 1& 0& 2& \textcolor{red}{1}& 1& \textcolor{red}{1}& 1& 0& 1 &&&&&&  \\ \hline
4 &2& 0& 2& \textcolor{red}{1}& 3& \textcolor{red}{1}& 3& \textcolor{red}{1}& 2& \textcolor{red}{1}& 1& 0& 1 
&&& \\ \hline
5 &0& 2& \textcolor{red}{1}& 4& \textcolor{red}{2}& 4& \textcolor{red}{3}& 4& \textcolor{red}{2}& 3& \textcolor{red}{2}& 2& \textcolor{red}{1}& 1& 0& 1  \\\hline
6
&3& 0& 4& \textcolor{red}{3}& 6& \textcolor{red}{3}& 7& \textcolor{red}{4}& 6& \textcolor{red}{4}& 5& \textcolor{red}{2}& 4& \textcolor{red}{2}& 2& \textcolor{red}{1} \\\hline
7
&0& 4& \textcolor{red}{2}& 7& \textcolor{red}{5}& 8& \textcolor{red}{7}& 9& \textcolor{red}{6}& 9& \textcolor{red}{6}& 7& \textcolor{red}{5}& 5& \textcolor{red}{3}& 4 \\\hline
8
&4& \textcolor{red}{1}& 7& \textcolor{red}{5}& 11& \textcolor{red}{7}& 13& \textcolor{red}{9}& 13& \textcolor{red}{10}& 12& \textcolor{red}{8}& 11& \textcolor{red}{7}& 8& \textcolor{red}{5}
\\ \hline
\end{tabular}
\end{center}
\caption{Multiplicities of the decomposition of the tensor power ${\bf 7}^{\otimes_s n}$ into irreducible representations  $\bd$ of $SL(2,\CC)$. For tensor powers $n\geq 6$, the multiplicities
of $(d\geq 33)$-dimensional irreps are beyond the
space limitations of this table.}
\label{specirrep}
\end{table}

All the irreps in table \ref{specirrep} correspond to modular tensors obtained from contracting the $n$-fold outer product of $\CM_1^{a_1 \cdots a_6}$ with $\varepsilon$-symbols. However, the spin sums and trilinear relations of $\see_\delta^{ab}$ we shall be interested in always involve an even number of $\varepsilon$-contractions.
Tensors with an odd number of $\varepsilon$-contractions appear in the red entries of table \ref{specirrep} with odd $n+\frac{d-1}{2}$. These tensors are not expected to play any role in organizing cyclic products of genus-two Szeg\"o kernels. At $n=3$, for instance, table \ref{specirrep} lists the irreps,
\bea
{\bf 7}^{\otimes_s 3}= {\bf 19} \oplus {\bf 15}
\oplus {\bf 13} \oplus {\bf 11}
\oplus {\bf 9} \oplus {\bf 7} \oplus {\bf 7} \oplus {\bf 3}
\label{clean.03}
\eea
but only the ${\bf 19} \oplus {\bf 15} \oplus {\bf 11}  \oplus {\bf 7} \oplus {\bf 7} \oplus {\bf 3}$ representations
are expected to enter the Szeg\"o-kernel discussion. Indeed, the ${\bf 9} $ and ${\bf 13} $ correspond to 8- and 12-index tensors that follow from $\CM_1^{a_1 a_2 \ndots a_6}\CM_1^{b_1 b_2 \ndots b_6} \CM_1^{c_1 c_2 \ndots c_6}$ 
via contractions with an odd number of $\varepsilon$-symbols.

\newpage

\section{Proof of Lemma \ref{4.lem.1}}
\label{sec:B}
\setcounter{equation}{0}

To prove part (a) of Lemma \ref{4.lem.1}, it suffices to prove the statement of the lemma for arbitrary symmetrized monomials out of which an arbitrary $\alpha \leftrightarrow \b$ symmetric polynomial may be built.  We shall produce a simple proof by explicit construction and  parametrize the  symmetrized monomials in the expression (\ref{pialphabeta}) for $\Pi_\delta(1,\cdots,n)$ as follows, 
\bea
\a_1^{a_1} \, \a_2^{a_2} \, \b_1^{b_1} \, \b_2^{b_2} +  \a_1^{b_1} \, \a_2^{b_2} \, \b_1^{a_1} \, \b_2^{a_2} 
\eea
with integers $a_1, a_2, b_1, b_2 \geq 0$ and $\alpha_i,\beta_i$ defined in (\ref{albet}). 
Without loss of generality, we may assume that $a_2 \geq b_2$, and factor out $\phi_4^{b_2}$, see (\ref{phi234}) for the definition of the $\phi_k$. It remains to prove that the lemma holds for symmetrized monomials of the form,  
\bea
\label{4.reda}
\a_1^{a_1} \, \a_2^{c_2} \, \b_1^{b_1} +  \a_1^{b_1}  \, \b_1^{a_1} \, \b_2^{c_2} 
\eea
with $c_2=a_2-b_2 \geq 0$. We proceed by distinguishing the cases $c_2=0$ and $c_2\not=0$. 

\sm

$\bullet$ For $c_2=0$ we set $c_1 = |a_1-b_1|$ and  the symmetrized monomial (\ref{4.reda}) reduces to, 
\bea
\label{4.redb}
\a_1^{a_1} \b_1^{b_1}  + \b_1^{a_1}  \a_1^{b_1} & = & \phi_2^{\min(a_1,b_1)} \big ( \a_1^{c_1} + \b_1^{c_1} \big )
\eea
If $c_1=0$, then the reduction is complete, while if $c_1=1$ then the sum in the parentheses equals $\mu_1$ so that the reduction is again complete. If $c_1 \geq 2$ we use the following two-step recursion relation,
\bea
\a_1^{c_1} + \b_1^{c_1} & = & \mu_1 (\a_1^{c_1-1} + \b_1^{c_1-1} ) -\phi_2 (\a_1^{c_1-2} + \b_1^{c_1-2} )
\eea
which may be solved in terms of a polynomial in $\mu_1$ and $\phi_2$. Thus,  for $c_2=0$, the monomial (\ref{4.reda}) may be expressed as a polynomial in $\phi_2, \phi_3, \phi_4$ with coefficients in $\CC[\mu_1, \cdots, \mu_6]$.

\sm

$\bullet$ For $c_2>0$ the reduction of the symmetrized monomial (\ref{4.reda}) is more involved. For $c_2\geq 2$ we use the following recursion relation,
\bea
\label{4.redc}
\a_1^{a_1} \a_2^{c_2} \b_1^{b_1}   + \b_1^{a_1}   \b_2^{c_2} \a_1^{b_1}
& = & 
\phi_3 \Big ( \a_1^{a_1}  \a_2^{c_2-1} \b_1^{b_1-1} + \b_1^{a_1}   \b_2^{c_2-1} \a_1^{b_1-1} \Big ) 
\no \\ &&
- \mu_1 \phi_4 \Big ( \a_1^{a_1} \a_2^{c_2-2}  \b_1^{b_1-1}  + \b_1^{a_1}  \b_2^{c_2-2}  \a_1^{b_1-1} \Big ) 
\no \\ &&
+\phi_4 \Big (  \a_1^{a_1} \a_2^{c_2-2} \b_1^{b_1}   + \b_1^{a_1}  \b_2^{c_2-2}  \a_1^{b_1} \Big )
\eea
\indent
$\star$ If $b_1 \geq c_2$ then all three lines on the right side will terminate after at most $c_2$ iterations in a term of the reduced form $ \a_1^{a_1} \b_1^{b_1} + \b_1^{a_1}  \a_1^{b_1} $ when $c_2$ is even plus a term of the reduced form $ \a_1^{a_1} \b_1^{b_1} \a_2 + \b_1^{a_1}  \a_1^{b_1} \b_2$ when $c_2$ is odd, which includes the case $c_2=1$. The first reduced form satisfies the recursion relation already given in (\ref{4.redb}). The second reduced form satisfies the following recursion relations for $c_1=a_1 - b_1\geq 0$,
 \bea
 \a_1^{a_1} \b_1^{b_1} \a_2 + \b_1^{a_1}  \a_1^{b_1} \b_2 & = &
\phi_2^{b_1} \Big ( \a_1^{c_1} \a_2 + \b_1^{c_1} \b_2 \Big )
\eea
For $c_1\geq 2$ the combination inside the parentheses satisfies the following recursion relation,
\bea
\label{4.rede}
 \a_1^{c_1} \a_2 + \b_1^{c_1} \b_2 & = & \mu_1 \Big ( \a_1^{c_1-1} \a_2 + \b_1^{c_1-1} \b_2 \Big ) 
 +\phi_2 \Big ( \a_1^{c_1-2} \a_2 + \b_1^{c_1-2} \b_2 \Big )
 \eea
 while for $c_1=1$, we have $ \a_1 \a_2 +\b_1 \b_2 = \mu_1(\mu_2-\phi_2) - \phi_3$.
 
 \indent
$\star$ If $b_1 <c_2$ then the second and third lines in (\ref{4.redc}) will terminate as in the case $b_1 \geq c_2$, but the first line will terminate after $b_1$ iterations in a term of the  form $ \a_1^{a_1} \a_2^{c_2-b_1}  + \b_1^{a_1}  \b_2^{c_2-b_1}$ with $d_2=c_2-b_1>0$, which obeys the following recursion relation for $d_2\geq 2$, 
 \bea
 \a_1^{a_1} \a_2^{d_2}  + \b_1^{a_1}  \b_2^{d_2}
 & = & (\mu_2-\phi_2) \Big (  \a_1^{a_1} \a_2^{d_2-1}  + \b_1^{a_1}  \b_2^{d_2-1} \Big )
 \no \\ &&
 - \phi_4 \Big ( \a_1^{a_1} \a_2^{d_2-2}  + \b_1^{a_1}  \b_2^{d_2-2} \Big ) 
 \eea
 and
 \bea
 \a_1^{a_1} \a_2  + \b_1^{a_1}  \b_2
 & = & (\mu_2-\phi_2) \Big (  \a_1^{a_1}   + \b_1^{a_1}   \Big ) - \a_1^{a_1} \b_2 -  \b_1^{a_1} \a_2
 \eea
 It remains to exhibit the reduction of the last two terms on the right side, $\a_1^{a_1} \b_2 + \b_1^{a_1} \a_2$.
 For $a_1=1$, this is just $\phi_3$. For $a_1 \geq 2$, we use the following recursion relation, 
 \bea
 \a_1^{a_1} \b_2 + \b_1^{a_1} \a_2 & = & 
 \phi_3( \a_1^{a_1-1} + \b _1 ^{a_1-1}) - \phi_2 ( \a_1^{a_1-2} \a_2 + \b_1^{a_1-2} \b_2)
 \eea
 The reduction of the parentheses in the first term on the right was given in the second line of (\ref{4.redb}) while the reduction of the parentheses in the second term was given in (\ref{4.rede}). 
 
 \sm
 
The proof of part (b) of the lemma follows immediately from part (a), using the relations of (\ref{4.see}) between $\phi_2, \phi_3, \phi_4$ and $\see_\delta^{ab}$. 
 
\sm

To prove part (c) of the  lemma we use the fact that $\Pi_\delta$ is a polynomial in $\phi_2, \phi_3, \phi_4$ as guaranteed by part (a). The combination $\Pi_\delta$ also explicitly depends on the symmetric polynomials $\mu_1, \mu_2$. To prove the formula on the total degree of a given monomial, we reason by analyzing the degree of homogeneity in $\a_1, \b_1$ on the one hand and the degree of homogeneity in $\a_2, \b_2$ on the other hand. Schematically, we have, 
\bea
\phi_4 & \to & \a_2^2 , \, \a_2 \b_2 , \, \b_2^2
\no \\
 \phi_3 & \to & \a_1\b_2 , \, \b_1 \a_2
 \no \\
 \phi_2 & \to & \a_1^2 , \, \a_1 \b_1 , \, \b_1^2
 \eea
 The maximum value that $\lambda _4 $ can take is when all its $\a_2$ and $\b_2$ are assembled into $\phi_4$,
 giving the maximum value $\left \lfloor \tfrac{a_2+b_2}{2} \right \rfloor $. However, this number may be decreased by a number $0\leq j \leq \left \lfloor \tfrac{a_2+b_2}{2} \right \rfloor$ when some of the $\a_2$ and $\b_2$ are collected into $\phi_3$. It will be convenient to split up into the case where $a_2+b_2$ is even or odd. 
 
 \sm
 
 $\bullet$ For even $a_2+b_2$ we have,
  \bea
 \lambda _4  \leq   \tfrac{a_2+b_2}{2} -j
\hskip 0.8in
 \lambda_3  \leq  2j 
\hskip 0.8in
 \lambda _2  \leq  \left \lfloor \tfrac{a_1+b_1}{2} \right \rfloor -j
 \eea
 The inequality for $\lambda_2$ is obtained by using the fact that the presence of $2j$ factors of $\phi_3$ takes up $2j$ factors of $\a_1$ and $\b_1$ combined, which amounts to $j$ factors of $\phi_2$. 
 Adding up these lines gives the following upper bound, 
 \bea
 \lambda_2 + \lambda _3 + \lambda _4 \leq \tfrac{a_2+b_2}{2} + \left \lfloor \tfrac{a_1+b_1}{2} \right \rfloor  
 \leq \tfrac{ a_1+a_2+b_2+b_2}{2} \leq m
 \eea
 where we have used the bounds $a_1+a_2 \leq m$ and $b_1+b_2 \leq m$. 
 
 \sm
 
$\bullet $ For odd $a_2+b_2$ we have instead, 
\bea
 \lambda _4  \leq   \tfrac{a_2+b_2-1}{2} -j
\hskip 0.7in
 \lambda_3  \leq  2j +1
\hskip 0.7in
 \lambda _2  \leq  \left \lfloor \tfrac{a_1+b_1}{2} \right  \rfloor -j
 \eea
For $\lambda_3$ we have added 1 in view of the fact that $a_2+b_2$ is odd and the only way a term of first order in $\a_2$ and $\b_2$ can be produced in the iterative process outlined earlier is through one factor of $\phi_3$. 
 Adding up these lines gives the following upper bound, 
 \bea
 \lambda_2 + \lambda _3 + \lambda _4 \leq \tfrac{a_2+b_2-1}{2} + 1 + \left \lfloor \tfrac{a_1+b_1}{2} \right \rfloor  
 \leq \tfrac{ a_1+a_2+b_2+b_2}{2} + \thalf 
 \eea
Using again the bounds $a_1+a_2 \leq m$ and $b_1+b_2 \leq m$, we obtain $ \lambda_2 + \lambda _3 + \lambda _4 \leq m + \thalf $. Since $\lambda_2, \lambda_3, \lambda_4$ and $m$ are integers, this implies $ \lambda_2 + \lambda _3 + \lambda _4 \leq m$ which completes the proof of part (c) of the lemma. 

\sm

Part (d) of the lemma follows immediately from part of (c) by using the relations of (\ref{4.see}) between $\phi_2, \phi_3, \phi_4$ and $\see_\delta^{ab}$. This completes the proof of the lemma.

\newpage

\section{Manifestly $SL(2,\CC)$ invariant reduction of $Q_\delta$}
\label{sec:B1}
\setcounter{equation}{0}

In section \ref{sec:cyclic} we developed a systematic and efficient procedure by which all spin structure dependence of the cyclic product $C_\delta$ of an arbitrary number of Szeg\"o kernels is reduced to polynomials $Q_\delta$ which, in turn, are expressed in terms of further reduced polynomials $P_\delta$.
In appendix \ref{sec:E} the polynomials $P_\delta$ are shown to play a crucial role for 
the reformulations in section \ref{sec:low-n} of the cyclic products of Szeg\"o kernels solely in terms of the fundamental building blocks $L_\delta(i,j)$,  $Z(i,j)$, and standard Parke-Taylor factors built from $x_{ij}$.

\sm

Efficient as the polynomials $P_\delta$ may be in carrying through the reduction process reviewed above,  their use is responsible for giving up \textit{manifest invariance} under $SL(2,\CC)$, i.e.\ invariance term by term, at intermediate stages of the reduction.  Indeed, the factors $(s_A(i)^2-s_A(j)^2)$ and $(s_B(i)^2-s_B(j)^2)$, out of which $P_\delta$ is constructed, do not enjoy tensorial $SL(2,\CC)$ transformation properties. However, the final expressions we obtain in terms of $L_\delta(i,j)$,  $Z(i,j)$, and Parke-Taylor factors are perfectly $SL(2,\CC)$ invariant term by term.

\sm

This situation naturally raises the question as to whether a reduction procedure exists that is \textit{manifestly} $SL(2,\CC)$ covariant at all stages of the reduction process, and avoids the intermediate polynomials $P_\delta$ altogether. In this appendix, we shall outline precisely such a construction which reduces $Q_\delta(1,3,\ndots,2m{-}1|2,4,\ndots,2m)$ to a degree-$m$ polynomial in tensor components $\see_\delta^{ab}$ with $\delta$-independent coefficients and thereby constitutes an alternative proof of Corollary \ref{allspinpi}. In practice, it remains to be seen whether the substantial group-theoretic calculations required to complete this process are competitive with the use of $P_\delta$, whose efficiency has been established beyond a doubt in section \ref{sec:low-n}
and appendix \ref{sec:E}.

\subsection{The tensors $\bA, \bB, \bX$}

To obtain good $SL(2,\CC)$ tensorial expressions for $Q_\delta$ we begin by rendering the tensorial properties of $s_A(i)^2$ and $s_B(i)^2$ manifest.  We represent the spin structure $\delta$ by a partition $A\cup B$ of the branch points, where $A=\{ u_{r_1}, u_{r_3}, u_{r_5} \}$ and $B= \{ u_{r_2}, u_{r_4}, u_{r_6} \}$ and $A \cap B = \emptyset$. The partially symmetric polynomials $\a_1, \a_2, \a_3$ and $\b_1, \b_2,\b_3$,  defined in terms of the branch points in (\ref{albet}), and powers of the points $x_i$ may be arranged in terms of totally symmetric rank-3 tensors $\bA, \bB$, and $\bX_i$, respectively. Their components are given as follows,
\begin{align}
\bA^{111} & =  \a_3 & \bB^{111} & =  \b_3 & \bX_i ^{111} & =  x_i^3
\no \\
\bA^{112} & = \tfrac{1}{3}  \a_2  & \bB^{112} & = \tfrac{1}{3} \b_2 & \bX_i ^{112} & = x_i^2
\no \\
\bA^{122} & =  \tfrac{1}{3} \a_1  & \bB^{122} & =  \tfrac{1}{3} \b_1  & \bX_i ^{122} & =  x_i
\no \\
\bA^{222} & = 1 & \bB^{222} & = 1 & \bX_i ^{222} & = 1
\end{align} 
The combinatorial denominators are equal to the number of monomials in the sum. Each one of these symmetric rank-3 tensors  transforms in the ${\bf 4}$ of $SL(2,\CC)$. This property  may also be inferred from their representation as the following totally symmetrized tensor products of the $SL(2,\CC)$ doublets $\bx ,\bu_r$ in (\ref{doubleteq}),
\bea
\bA= \bu_{r_1} \otimes \bu_{r_3} \otimes \bu_{r_5} \Big |_{\bf 4} 
\hskip 0.7in
\bB= \bu_{r_2} \otimes \bu_{r_4} \otimes \bu_{r_6} \Big |_{\bf 4} 
\hskip 0.7in
\bX_i = \bx_i ^{ \otimes 3}\Big |_{\bf 4}
\eea
The prescription $ | _{\bf 4}$ stands for the projection onto the ${\bf 4}$, which is obtained by symmetrization.  The pairings of $\bA$ and $\bB$ with $\bX$ give the polynomials $s_A(i)^2$ and $s_B(i)^2$ in (\ref{sabpoly}),
\bea
\bX_i^{a_1a_2a_3}  \bA^{b_1 b_2 b_3} \ep_{a_1 b_1} \ep_{a_2 b_2} \ep_{a_3 b_3}   
= \bX_i ^{a_1a_2a_3} \bA_{a_1a_2a_3}  = \bX_i \cdot \bA = s_A(i)^2
\no \\
\bX_i^{a_1a_2a_3}  \bB^{b_1 b_2 b_3} \ep_{a_1 b_1} \ep_{a_2 b_2} \ep_{a_3 b_3}  
=  \bX_i ^{a_1a_2a_3} \bB_{a_1a_2a_3}   = \bX_i \cdot \bB =s_B(i)^2
\eea
Note that the pairing of two tensors of even rank is symmetric, while the pairing of two tensors of odd rank is anti-symmetric, so that we have $\bX_i \cdot \bA = - \bA \cdot \bX_i$.

\subsection{Expressing $Q_\delta$ in terms of $\bA, \bB, \bX_i$}

In terms of the tensors $\bA, \bB$ and $\bX_i$ the polynomial $Q_\delta(i_1, \cdots, i_m| j_1 , \cdots , j_m)$ defined by (\ref{defQdelta})  takes the form,
\bea
Q_\delta(i_1, \cdots, i_m| j_1 , \cdots , j_m) 
& = &
\big ( \bX_{i_1} \cdot \bA \big )  \, \big ( \bX_{j_1} \cdot \bB \big )  \cdots \big ( \bX_{i_m} \cdot \bA \big )  \, \big (  \bX_{j_m} \cdot \bB \big ) + (\bA \leftrightarrow \bB)
\qquad
\label{qdelcov}
\eea
More abstractly, but equivalently, one may view this object as an $SL(2,\CC)$ singlet formed out of the following tensor products, 
\bea
Q_\delta(i_1, \cdots, i_m| j_1 , \cdots , j_m) 
=  \Big ( \bX_{i_1}  \otimes \cdots \otimes \bX_{i_m} \otimes \bX_{j_1}
 \otimes \cdots \otimes \bX_{j_m} \Big ) 
\cdot ( \bA^{\otimes m} \otimes \bB^{\otimes m} \Big ) \Big |_{{\bf 1}}
\eea
suitably symmetrized in $\bA$ and $\bB$.  The components in the decomposition of the tensor products $\bA^{\otimes m} \otimes \bB^{\otimes m}$ and 
$ \bX_{i_1}  {\otimes} \! \cdots \! {\otimes} \bX_{i_m} {\otimes} \bX_{j_1}
 {\otimes} \! \cdots \! {\otimes} \bX_{j_m}$
 into irreducible representations may then be reorganized in terms of the functions $L_\delta(i,j), \, Z(i,j)$ and products of powers of $x_{ij}$.

\subsection{Tensor product decomposition of $\bA\otimes \bB$}

Since $\bA$ and $ \bB$ transform under the ${\bf 4}$ of $SL(2,\CC)$, the tensor product  $\bA \otimes \bB$ decomposes into the direct sum of the following  irreducible representations ${\bf 7} \oplus {\bf 5} \oplus {\bf 3} \oplus {\bf 1}$. In this decomposition, the ${\bf 7}$ component is obtained by complete symmetrization, and is given by the spin structure independent tensor $\CS$ introduced earlier  in (\ref{MXcomps}), 
\bea
\CS^{a_1a_2a_3a_4a_5a_6} = \bA^{(a_1 a_2 a_3} \bB^{a_4a_5a_6)}  
\eea
This relation may be readily verified by identifying the highest weight components, namely $\CS^{111111}= \mu_6=\a_3\b_3= \bA^{111} \bB^{111}$  on both sides, and then using the translation operator $\cT$ to construct the full multiplets. 

\sm

The remaining components ${\bf 5} \oplus {\bf 3} \oplus {\bf 1}$ of $\bA \otimes \bB$ result from one, two, and three anti-symmetrizations, followed by total symmetrization in the remaining indices, and may be defined as follows,
\bea
\bN_\delta^{a_1a_2a_3a_4} & = & \thalf  \bA^{b (a_1 a_2} \bB^{a_3a_4) c} \, \ep_{bc}
\no \\
\bN_\delta^{a_1a_2} & = & \tfrac{1}{4} \bA^{b_1 b_2 (a_1} \bB^{a_2) c_1 c_2} \, \ep_{b_1c_1} \ep_{b_2c_2} 
\no \\
\bN_\delta& = & \tfrac{1}{8}  \bA^{b_1 b_2 b_3} \bB^{c_1 c_2 c_3} \, \ep_{b_1c_1} \ep_{b_2c_2} \ep_{b_3c_3}
\eea
Also, we use the same letter $\bN_\delta$ to designate the tensors of ranks $4, 2$ and 0, following the familiar notation introduced earlier for the tensors  $\bM_w$. The multiplets of these irreducible representations are completely determined by their highest weight components, which are given as follows,
\bea
\bN_\delta^{1111}  & = & 
\thalf (\bA^{111} \bB^{112} - \bA^{112} \bB^{111})  =  \tfrac{1}{6} (\a_3 \b_2 - \a_2 \b_3)
\no \\
\bN_\delta^{11}  & = & 
\tfrac{1}{4} ( \bA^{111} \bB^{122} + \bA^{122} \bB^{111} -2 \bA^{112} \bB^{112})
= \tfrac{1}{36} ( 3 \a_3 \b_1 + 3 \a_1 \b_3 - 2 \a_2 \b_2) 
\no \\
\bN_\delta  & = & 
\tfrac{1}{8} \big ( \bA^{111} \bB^{222} -3 \bA^{112} \bB^{122} + 3 \bA^{122} \bB^{122} - \bA^{222} \bB^{111} \big )
\no \\ & = &
\tfrac{1}{24} \big ( 3 \a_3 - 3\b_3 +\a_1  \b_2 - \a_2 \b_1 \big )
\eea
Comparison with the expression (\ref{4.see}) for $\see_\delta^{11}$ then gives the following relation,
\bea
\bN_\delta^{ab}  = - { 5 \over 9} \, \see_\delta ^{ab}
\label{ntoell}
\eea
We note that $\CS^{a_1 \cdots a_6}$ and $ \bN_\delta^{ab}$ are symmetric under the interchange of $\bA$ and $\bB$, while $\bN_\delta^{a_1 a_2a_3a_4}$ and $\bN_\delta$ are anti-symmetric. We may summarize the decomposition as follows, 
\bea
\bA^{a_1a_2a_3} \bB^{b_1b_2b_3} & = & \CS^{a_1a_2a_3b_1b_2b_3} 
+ \Big \{ 3 \ep^{a_1b_1} \bN_\delta ^{a_2a_3b_2b_3} 
+{18 \over 5} \ep^{a_1b_1} \ep ^{a_2 b_2} \bN_\delta ^{a_3 b_3} 
 \label{fierzAB} \\ &&
+ 2 \ep^{a_1b_1} \ep^{a_2 b_2} \ep^{a_3b_3} \bN_\delta^{} 
\quad \hbox{ symmetrized in } (a_1,a_2,a_3) \hbox{ and } (b_1, b_2, b_3) \Big \}
\no
\eea
where the symmetrization is to be carried out independently on the triplets of indices $(a_1,a_2,a_3)$ and $(b_1, b_2, b_3)$.

\sm

Finally, symmetrization in $\bA$ and $\bB$ cancels all terms in $Q_\delta$ that contain an odd combined number of factors of $\bN_\delta ^{a_1 a_2 a_3 a_4}$ and $\bN_\delta ^{}$. For this reason, it will be useful to express bilinear tensor products of $\bN_\delta ^{a_1 a_2 a_3 a_4}$ and $\bN_\delta ^{}$ in terms of the purely even tensors $\CS$ and $\bN_\delta ^{a_1a_2}$, which are given as follows,
\bea
\bN_\delta ^2& = & - \tfrac{1}{16} \bM_2 - \tfrac{5}{18} \det \see_\delta
\no \\
\bN_\delta ^{} \bN_\delta ^{a_1 a_2 a_3 a_4} & = & - \tfrac{1}{4} \bM_2^{a_1 a_2 a_3 a_4} 
+\tfrac{1}{6} \CS^{a_1a_2a_3a_4 b_1b_2} \see_\delta ^{c_1 c_2} \ep_{b_1c_1} \ep _{b_2 c_2} 
+ \tfrac{2}{9} \see_\delta ^{(a_1a_2} \see_\delta ^{a_3a_4)} 
\no \\
\bN_\delta ^{a_1 a_2 a_3 a_4} \, \bN_\delta ^{b_1 b_2 b_3 b_4} & = &
- \bM_2^{a_1 a_2 a_3 a_4b_1 b_2 b_3 b_4}
+ \tfrac{2}{3}  \CS^{a_1 a_2 a_3 a_4b_1 b_2} \see_\delta ^{b_3b_4} 
+ \tfrac{2}{3}  \CS^{b_1 b_2 b_3 b_4a_1 a_2} \see_\delta ^{a_3a_4} 
\no \\ &&
- \tfrac{16}{9}  \CS^{a_1 a_2a_3b_1 b_2 b_3} \see_\delta ^{a_4b_4} 
+ \tfrac{2}{7} \ep^{a_1b_1} \ep^{a_2b_2} \bM_2^{a_3a_4b_3b_4} 
- \tfrac{8}{9} \ep^{a_1 b_1} \ep^{a_2 b_2} \see_\delta ^{(a_3 b_3} \see_\delta ^{a_4 b_4)}
\no \\ &&
- \ep^{a_1 b_1} \ep^{a_2 b_2} \ep^{a_3 b_3} \ep^{a_4 b_4}
\Big (  \tfrac{4}{27} \det \see_\delta + \tfrac{1}{30} \bM_2 \Big ) \quad \hbox{symmetrized}
\label{ndelreduce}
\eea
where the symmetrization in the last entry above is to be carried out independently on the quadruplets 
 $(a_1,a_2,a_3, a_4)$ and $(b_1, b_2, b_3, b_4)$, and the
tensors $\bM_2^{a_1 \ndots a_r}$ of rank $r=8,4,0$ are defined in (\ref{5.M2}).

\subsection{Tensor product decomposition of $\bX_i \otimes \bX_j$}

Since $\bX_i$ and $\bX_j$ transform under the ${\bf 4}$ of $SL(2,\CC)$, the tensor product $\bX_i \otimes \bX_j$ decomposes into ${\bf 7} \oplus {\bf 5} \oplus {\bf 3} \oplus {\bf 1}$. The ${\bf 7}$ component  is obtained by total symmetrization of $\bX_i \otimes \bX_j$ and coincides with the rank-six tensor $\bX_{ij}$ introduced in (\ref{MXcomps}), and we have,
\bea
\bX_{ij} ^{a_1 \cdots a_6 } & = & \bX_i ^{(a_1 a_2 a_3} \bX_j^{a_4a_5a_6)} 
\eea
The remaining ${\bf 5} \oplus {\bf 3} \oplus {\bf 1}$ corresponds to tensors of ranks 4, 2 and 0, respectively, obtained from (partial) anti-symmetrization and will be denoted by $\bY_{ij}$ with components,
\bea
\bY_{ij} ^{a_1 a_2 a_3 a_4} & = & \thalf \bX_i ^{b (a_1 a_2} \bX_j ^{a_3a_4) c} \ep_{bc}
\no \\
\bY_{ij} ^{a_1 a_2} & = & \tfrac{1}{4} \bX_i ^{b_1 b_2 (a_1} \bX_j ^{a_2) c_1 c_2} \ep_{b_1c_1} \ep_{b_2c_2}
\no \\
\bY_{ij} & = & \tfrac{1}{8} \bX_i ^{b_1 b_2 b_3} \bX_j ^{c_1 c_2 c_3} \ep_{b_1c_1} \ep_{b_2c_2} \ep_{b_3c_3}
\label{idYij}
\eea
The multiplets of these irreducible representations are completely determined by their highest weight components, which are given as follows,
\bea
\bY_{ij}^{1111} & = & \thalf ( \bX_i ^{111} \bX_j^{112}- \bX_i^{112} \bX_j ^{111}) =  \thalf x_{ij} x_i^2 x_j^2
\no \\
\bY_{ij}^{11} & = & \tfrac{1}{4} ( \bX_i^{111} \bX_j ^{122} + \bX_i^{122} \bX_j ^{111}  - 2 \bX_i^{112} \bX_j ^{112} \big )
= \tfrac{1}{4}  x_{ij} ^2 x_i x_j
\no \\
\bY_{ij} & = & 
\tfrac{1}{8} \big ( \bX_i^{111} \bX_j ^{222} - 3 \bX_i^{112} \bX_j ^{122} + 3 \bX_i^{122} \bX_j^{112} - \bX_i^{222} \bX_j^{111} 
\big ) = \tfrac{1}{8}  x_{ij}^3
\eea
We see that each anti-symmetrization in $\bX_i \otimes \bX_j$ produces a factor of $x_{ij}$. Such  factors are indeed encountered in the decomposition of $Q_\delta$. 

\subsection{Manifestly $SL(2,\mathbb C)$ invariant decomposition of $Q_\delta(1|2)$}

The above tensor-product decompositions greatly facilitate a manifestly 
$SL(2,\mathbb C)$ invariant derivation of the expression $Q_\delta(i|j)  =  
4 x_{ij} x_{ji} \bL_\delta(i,j) + 2 Z(i,j)$ in Theorem \ref{thm:1}. The symmetrization
over ${\bf A} \leftrightarrow {\bf B}$ directly projects the $m=1$ instance of (\ref{qdelcov}) to the ${\bf 7} \oplus {\bf 3} $ parts of $\bA^{c_1 c_2 c_3} \bB^{d_1 d_2 d_3}$ in (\ref{fierzAB}),
\begin{align}
Q_\delta(i|j) &= \big ( \bX_{i} \cdot \bA \big )  \, \big ( \bX_{j} \cdot \bB \big )
+\big ( \bX_{i} \cdot \bB \big )\, \big ( \bX_{j} \cdot \bA \big )  \notag \\
&= \bX_i^{a_1 a_2 a_3} \bX_j^{b_1 b_2 b_3} 
\varepsilon_{a_1 c_1} \varepsilon_{a_2 c_2} \varepsilon_{a_3 c_3} 
\varepsilon_{b_1 d_1} \varepsilon_{b_2 d_2} \varepsilon_{b_3 d_3}
\bigg( 2 \, \CM_1^{c_1 c_2 c_3 d_1 d_2 d_3} + \frac{36}{5} \,
\varepsilon^{c_1 d_1} \varepsilon^{c_2 d_2} \bN_\delta ^{c_3 d_3} \bigg) \notag \\
&= 2 Z(i,j) - 16  \bY_{ij}^{ab}   \varepsilon_{ac} \varepsilon_{bd} \see_\delta^{cd}
= 2 Z(i,j) - 4 x_{ij}^2 \bL_\delta (i,j) 
\notag 
\end{align}
We have identified $\bY_{ij}^{ab}$ and $\see_\delta^{cd}$ via (\ref{idYij}) and (\ref{ntoell}) in passing to the last line, and the last equality follows from the rewriting of (\ref{3.Ldel}) as
\bea
 x_{ij}^2 \bL_\delta (i,j) = 4 \bY_{ij}^{a_1a_2}\see_\delta^{b_1 b_2} \ep_{a_1b_1} \ep_{a_2b_2}
\eea

\subsection{Manifestly $SL(2,\mathbb C)$ invariant decomposition of arbitrary $Q_\delta$}

A key step in the above rewriting of $Q_\delta(i|j)$ is the tensor-product decomposition
(\ref{fierzAB}) of $\bA^{c_1 c_2 c_3} \bB^{d_1 d_2 d_3}$ and its projection to the even part under
${\bf A}\leftrightarrow {\bf B}$. The analogous treatment of higher-point $Q_\delta$ will require an 
iterative use of (\ref{fierzAB}) followed
by a projection to terms with an even combined number of factors $\bN_\delta ^{a_1 a_2 a_3 a_4}$ 
and $\bN_\delta$. At four points, for instance, this will bring the spin structure dependence of (\ref{qdelcov}) into the form of
\begin{align}
&\bA^{a_1 a_2 a_3} \bB^{b_1 b_2 b_3} \bA^{c_1 c_2 c_3} \bB^{d_1 d_2 d_3}
+ (\bA \leftrightarrow \bB) \label{ABABdec} \\
&=
2 \Big(  \CM_1^{a_1 a_2 a_3 b_1 b_2 b_3} -2 \,
\varepsilon^{a_1 b_1} \varepsilon^{a_2 b_2} \see_\delta ^{a_3 b_3} \Big)
\Big(  \CM_1^{c_1 c_2 c_3 d_1 d_2 d_3} -2 \,
\varepsilon^{c_1 d_1} \varepsilon^{c_2 d_2} \see_\delta ^{c_3 d_3} \Big)
\notag \\
&\quad + 2 \Big( 3 \varepsilon^{a_1 b_1}  \bN_\delta ^{a_2 a_3 b_2 b_3} 
+ 2 \varepsilon^{a_1 b_1} \varepsilon^{a_2 b_2} \varepsilon^{a_3 b_3}  \bN_\delta \Big)
 \Big( 3 \varepsilon^{c_1 d_1}  \bN_\delta ^{c_2 c_3 d_2 d_3} 
+ 2 \varepsilon^{c_1 d_1} \varepsilon^{c_2 d_2} \varepsilon^{c_3 d_3}  \bN_\delta \Big)
\notag
\end{align}
where independent symmetrizations in the triplets of $a_i,b_i,c_i,d_i$ are understood throughout.
On the one hand, a detailed analysis of the $x_i$-dependence in the follow-up steps of (\ref{ABABdec}) and its generalization to higher points is beyond the scope of this work. On the other hand, it is easy to explain via (\ref{ABABdec}) that the entire $\delta$-dependence 
of $Q_\delta(i_1,\ndots,i_m|j_1,\ndots,j_m)$ can be arranged in a degree-$m$ polynomial in $\see_\delta^{ab}$.

\sm

Based on the identities (\ref{ndelreduce}) for bilinears in $ \bN_\delta$ and $ \bN_\delta^{abcd}$, the third line of (\ref{ABABdec}) boils down to a degree-two-polynomial in $\see_\delta^{ab}$ with $\delta$-independent coefficients. The same kind of conversion can be found for the symmetrization of $\bA^{\otimes m} \otimes \bB^{\otimes m}$ in ${\bf A}\leftrightarrow {\bf B}$ at higher points $m\geq 3$ since
each summand will have an even combined number of factors $\bN_\delta ^{a_1 a_2 a_3 a_4}$ 
and $\bN_\delta$. All of them can be grouped into pairs of rank $0,4$ or $8$ which reduce to degree-two polynomials in $\see_\delta^{ab}$ with $\delta$-independent coefficients by virtue of (\ref{ndelreduce}). Contributions to $Q_\delta(i_1,\ndots,i_m|j_1,\ndots,j_m)$ from $k$ such pairs are multiplied by a $(m{-}2k)$ fold
tensor power of $\CM_1^{a_1 a_2 \ndots a_6} -2 
\varepsilon^{a_1 a_2} \varepsilon^{a_3 a_4} \see_\delta ^{a_5 a_6}$. This casts the entire dependence on the spin structure into a polynomial in $\see_\delta^{ab}$ of total degree $m$, separately for any number $k=1,2,\ndots,\lfloor m/2\rfloor$ of pairs in (\ref{ndelreduce}), and concludes a manifestly $SL(2,\mathbb C)$ invariant proof of Corollary \ref{allspinpi}.

\sm

It would be interesting to prove the all-multiplicity conjecture
(\ref{Qdel.17}) for the decomposition of $Q_\delta$ into $Z (i,j) $
and $\bL_\delta (i,j) $ from the methods of this appendix. Already at four points, 
it will require a variety of tensor rearrangements to derive the form (\ref{5.Q4}) 
of $Q_\delta(1,3|2,4)$ from the contraction of (\ref{ABABdec}) with the rank-three tensors $\bX_1,\bX_2,\bX_3,\bX_4$. 
A major challenge in the higher-point
proof is to obtain the Parke-Taylor numerator $x_{ij} x_{jk} \ndots x_{mi}$ in
the building blocks ${\cal L}_\delta$ of $Q_\delta$ defined in (\ref{Qdel.06}) from
the rank-three tensors $\bX_i^{a_1 a_2 a_3}$ in (\ref{qdelcov}).
 
\newpage

\section{List of trilinear relations}
\label{sec:D1}
\setcounter{equation}{0}

In this appendix, we rewrite the trilinear relations of the $\phi_m$ in Theorem \ref{4.tril} in terms of the $SL(2,\mathbb C)$-tensor components $\see_\delta ^{ab}$.

{\cor
The variables $\see_\delta ^{ab}$ in (\ref{4.see}) obey the following trilinear relations.}
 \begin{subequations}
\label{sping3.55}
{\small \begin{align}
(\see_{\delta}^{11})^3 &=  \frac{ \mu_4 (\see_{\delta}^{11})^2}{20} 
-   \frac{ \mu_5 \see_{\delta}^{11} \see_{\delta}^{12}}{4}  
+ \mu_6 \see_{\delta}^{11} \see_{\delta}^{22} - \frac{ \mu_6 (\see_{\delta}^{12})^2}{  4}
+ \frac{ \mu_4^2 \see_{\delta}^{11}}{50} 
- \frac{ 9 \mu_3 \mu_5 \see_{\delta}^{11}}{160} + \frac{ 3 \mu_2 \mu_6 \see_{\delta}^{11}}{ 20} 
 + \frac{ \mu_4 \mu_5 \see_{\delta}^{12}}{40} 
\notag \\
&\ \ 
- \frac{   9 \mu_3 \mu_6 \see_{\delta}^{12}}{80} - \frac{ \mu_5^2 \see_{\delta}^{22}}{16} 
+ \frac{ 3 \mu_4 \mu_6 \see_{\delta}^{22}}{20} 
 - \frac{ 3 \mu_4^3}{2000} + \frac{ 9 \mu_3 \mu_4 \mu_5}{1600} 
- \frac{   3 \mu_2 \mu_5^2}{320} - \frac{ 81 \mu_3^2 \mu_6}{6400} 
+ \frac{ 9 \mu_2 \mu_4 \mu_6}{400} 
  \notag \\
(\see_{\delta}^{11})^2 \see_{\delta}^{12} &=   \frac{\mu_3 (\see_{\delta}^{11})^2}{40}
- \frac{  \mu_4 \see_{\delta}^{11}\see_{\delta}^{12}}{20} 
- \frac{\mu_5 (\see_{\delta}^{12})^2}{8} 
 + \frac{\mu_5 \see_{\delta}^{11} \see_{\delta}^{22}}{8} 
 + \frac{  \mu_6 \see_{\delta}^{12}\see_{\delta}^{22}}{4} \notag \\
 &\ \  + \frac{\mu_3 \mu_4 \see_{\delta}^{11}}{800} 
 - \frac{\mu_2 \mu_5 \see_{\delta}^{11}}{  80} + \frac{\mu_1 \mu_6 \see_{\delta}^{11}}{8} 
 + \frac{3 \mu_4^2 \see_{\delta}^{12}}{ 200} 
 - \frac{\mu_3 \mu_5 \see_{\delta}^{12}}{40} - \frac{\mu_2 \mu_6 \see_{\delta}^{12}}{40} 
 - \frac{\mu_4 \mu_5 \see_{\delta}^{22}}{  80}
 \notag \\
 & \ \ 
 + \frac{9 \mu_3 \mu_6 \see_{\delta}^{22}}{160}  
 - \frac{ 3 \mu_3 \mu_4^2}{8000} 
+ \frac{9 \mu_3^2 \mu_5}{12800} 
+ \frac{  \mu_2 \mu_4 \mu_5}{800} 
- \frac{\mu_1 \mu_5^2}{128} 
- \frac{9 \mu_2 \mu_3 \mu_6}{1600} 
+  \frac{ 3 \mu_1 \mu_4 \mu_6}{160} \notag \\
\see_{\delta}^{11} (\see_{\delta}^{12})^2 &=
 \frac{\mu_3 \see_{\delta}^{11}\see_{\delta}^{12}}{20} 
- \frac{  3 \mu_4 (\see_{\delta}^{12})^2}{20}
+ \frac{\mu_5 \see_{\delta}^{12}\see_{\delta}^{22}}{4} 
- \frac{\mu_6 (\see_{\delta}^{22})^2}{4}  \notag \\
&\ \ - \frac{\mu_3^2 \see_{\delta}^{11}}{  1600} 
+ \frac{\mu_6 \see_{\delta}^{11}}{4} + \frac{3 \mu_3 \mu_4 \see_{\delta}^{12}}{400} 
- \frac{  \mu_2 \mu_5 \see_{\delta}^{12}}{40}  - \frac{\mu_3 \mu_5 \see_{\delta}^{22}}{160} 
+ \frac{\mu_2 \mu_6 \see_{\delta}^{22}}{20} \no \\
&\ \ - \frac{ 3 \mu_3^2 \mu_4}{32000} 
+ \frac{\mu_2 \mu_3 \mu_5}{1600} - \frac{ \mu_5^2}{64} - \frac{\mu_2^2 \mu_6}{400} 
+ \frac{3 \mu_4 \mu_6}{80}  \notag  \\
(\see_{\delta}^{11})^2 \see_{\delta}^{22} &=
 \frac{\mu_2 (\see_{\delta}^{11})^2}{10}
- \frac{\mu_3 \see_{\delta}^{11}\see_{\delta}^{12}}{4} + \frac{  \mu_4 (\see_{\delta}^{12})^2}{4} 
+ \frac{\mu_4 \see_{\delta}^{11}\see_{\delta}^{22}}{5} 
- \frac{3 \mu_5 \see_{\delta}^{12}\see_{\delta}^{22}}{4} 
+ \frac{ 5 \mu_6 (\see_{\delta}^{22})^2}{4}
+ \frac{\mu_3^2 \see_{\delta}^{11}}{  160}  \notag \\
& \ \  - \frac{\mu_2 \mu_4 \see_{\delta}^{11}}{50} 
+ \frac{\mu_1 \mu_5 \see_{\delta}^{11}}{16} - \frac{\mu_6 \see_{\delta}^{11}}{  4} 
 + \frac{\mu_3 \mu_4 \see_{\delta}^{12}}{80} 
- \frac{\mu_2 \mu_5 \see_{\delta}^{12}}{  20} 
+ \frac{\mu_1 \mu_6 \see_{\delta}^{12}}{8} 
- \frac{\mu_4^2 \see_{\delta}^{22}}{100}   \notag \\
&\ \ 
+ \frac{3 \mu_3 \mu_5 \see_{\delta}^{22}}{  160}
-\frac{3 \mu_3^2 \mu_4}{6400} 
+ \frac{\mu_2 \mu_4^2}{1000} + \frac{  \mu_2 \mu_3 \mu_5}{800} 
- \frac{\mu_1 \mu_4 \mu_5}{160} + \frac{ \mu_5^2}{64} - \frac{\mu_2^2 \mu_6}{  80} 
+ \frac{9 \mu_1 \mu_3 \mu_6}{320} - \frac{3 \mu_4 \mu_6}{80} 
\notag \\
(\see_{\delta}^{12})^3 &= 
{-} \frac{\mu_1 (\see_{\delta}^{11})^2}{8}
+ \frac{\mu_2 \see_{\delta}^{11}\see_{\delta}^{12}}{4} 
- \frac{  19 \mu_3 (\see_{\delta}^{12})^2}{80} 
- \frac{\mu_3 \see_{\delta}^{11}\see_{\delta}^{22}}{16} 
+ \frac{  \mu_4 \see_{\delta}^{12}\see_{\delta}^{22}}{4} 
- \frac{\mu_5 (\see_{\delta}^{22})^2}{8} 
\notag \\
& \ \ 
- \frac{  \mu_1 \mu_4 \see_{\delta}^{11}}{160} + \frac{3 \mu_5 \see_{\delta}^{11}}{32} 
 - \frac{  3 \mu_3^2 \see_{\delta}^{12}}{1600} 
+ \frac{\mu_2 \mu_4 \see_{\delta}^{12}}{80} - \frac{\mu_1 \mu_5 \see_{\delta}^{12}}{16} 
+ \frac{3 \mu_6 \see_{\delta}^{12}}{16} - \frac{\mu_2 \mu_5 \see_{\delta}^{22}}{160} 
+ \frac{  3 \mu_1 \mu_6 \see_{\delta}^{22}}{32}  \notag \\
& \ \ +\frac{ 27 \mu_3^3}{128000} - \frac{3 \mu_2 \mu_3 \mu_4}{3200} 
+ \frac{  3 \mu_1 \mu_4^2}{1600} + \frac{3 \mu_2^2 \mu_5}{1600} 
- \frac{3 \mu_1 \mu_3 \mu_5}{1280} - \frac{  3 \mu_4 \mu_5}{320} 
- \frac{3 \mu_1 \mu_2 \mu_6}{320} + \frac{27 \mu_3 \mu_6}{640} 
\notag \\
\see_{\delta}^{11} \see_{\delta}^{12} \see_{\delta}^{22} &=
 \frac{\mu_1 (\see_{\delta}^{11})^2}{8}
- \frac{3 \mu_2 \see_{\delta}^{11}\see_{\delta}^{12}}{20} 
+ \frac{  \mu_3 (\see_{\delta}^{12})^2}{16}
+ \frac{7 \mu_3 \see_{\delta}^{11}\see_{\delta}^{22}}{80} 
- \frac{  3 \mu_4 \see_{\delta}^{12}\see_{\delta}^{22}}{20} 
+ \frac{\mu_5 (\see_{\delta}^{22})^2}{8} 
- \frac{  \mu_2 \mu_3 \see_{\delta}^{11}}{400}
\no \\
& \ \
 + \frac{\mu_1 \mu_4 \see_{\delta}^{11}}{160} 
+ \frac{\mu_5 \see_{\delta}^{11}}{ 32}  
+ \frac{\mu_3^2 \see_{\delta}^{12}}{80} - \frac{9 \mu_2 \mu_4 \see_{\delta}^{12}}{ 400} 
+ \frac{\mu_6 \see_{\delta}^{12}}{16}  - \frac{\mu_3 \mu_4 \see_{\delta}^{22}}{400} 
+ \frac{\mu_2 \mu_5 \see_{\delta}^{22}}{160} + \frac{\mu_1 \mu_6 \see_{\delta}^{22}}{32} 
\no  \\
&\ \
- \frac{ 9 \mu_3^3}{25600} 
+ \frac{19 \mu_2 \mu_3 \mu_4}{16000} - \frac{3 \mu_1 \mu_4^2}{1600} 
- \frac{3 \mu_2^2 \mu_5}{1600} + \frac{\mu_1 \mu_3 \mu_5}{ 256} 
- \frac{\mu_4 \mu_5}{320} - \frac{\mu_1 \mu_2 \mu_6}{320} + \frac{9 \mu_3 \mu_6}{640} 
\label{alltril.1}
\end{align}}
{\small  \begin{align}  
(\see_{\delta}^{12})^2 \see_{\delta}^{22} &=
- \frac{(\see_{\delta}^{11})^2}{4} 
+ \frac{\mu_1 \see_{\delta}^{11}\see_{\delta}^{12}}{4} 
- \frac{3 \mu_2 (\see_{\delta}^{12})^2}{  20} 
+ \frac{  \mu_3 \see_{\delta}^{12}\see_{\delta}^{22}}{20} \notag \\
&\ \ - \frac{\mu_1 \mu_3 \see_{\delta}^{11}}{160} 
+ \frac{\mu_4 \see_{\delta}^{11}}{20} 
+ \frac{3 \mu_2 \mu_3 \see_{\delta}^{12}}{400} - \frac{ \mu_1 \mu_4 \see_{\delta}^{12}}{40} 
- \frac{\mu_3^2 \see_{\delta}^{22}}{1600} + \frac{\mu_6 \see_{\delta}^{22}}{4}  \notag \\
& \ \ - \frac{ 3 \mu_2 \mu_3^2}{32000} 
+ \frac{\mu_1 \mu_3 \mu_4}{1600} - \frac{ \mu_4^2}{400} - \frac{\mu_1^2 \mu_6}{64} 
+ \frac{3 \mu_2 \mu_6}{80} 
 \notag  \\
\see_{\delta}^{11} (\see_{\delta}^{22})^2 &=
 \frac{5 (\see_{\delta}^{11})^2}{4}
- \frac{  3 \mu_1 \see_{\delta}^{11}\see_{\delta}^{12}}{4} 
 + \frac{\mu_2 (\see_{\delta}^{12})^2}{4} 
+ \frac{\mu_2 \see_{\delta}^{11}\see_{\delta}^{22}}{5} 
 - \frac{\mu_3 \see_{\delta}^{12}\see_{\delta}^{22}}{4} 
 + \frac{  \mu_4 (\see_{\delta}^{22})^2}{10}  
 - \frac{\mu_2^2 \see_{\delta}^{11}}{100} 
\notag \\
&\ \ + 
 \frac{ 3 \mu_1 \mu_3 \see_{\delta}^{11}}{160}  
 + \frac{\mu_2 \mu_3 \see_{\delta}^{12}}{ 80} - \frac{\mu_1 \mu_4 \see_{\delta}^{12}}{20} 
 + \frac{\mu_5 \see_{\delta}^{12}}{8} + \frac{\mu_3^2 \see_{\delta}^{22}}{ 160} 
 - \frac{\mu_2 \mu_4 \see_{\delta}^{22}}{50} + \frac{\mu_1 \mu_5 \see_{\delta}^{22}}{16} 
 \notag \\
& \ \ 
- \frac{\mu_6 \see_{\delta}^{22}}{  4} 
-\frac{3 \mu_2 \mu_3^2}{6400} 
+ \frac{\mu_2^2 \mu_4}{1000} + \frac{  \mu_1 \mu_3 \mu_4}{800} - \frac{ \mu_4^2}{80} 
- \frac{\mu_1 \mu_2 \mu_5}{160} + \frac{9 \mu_3 \mu_5}{ 320} + \frac{\mu_1^2 \mu_6}{64} 
- \frac{3 \mu_2 \mu_6}{80}  \no \\
\see_{\delta}^{12} (\see_{\delta}^{22})^2 &=
 \frac{ \see_{\delta}^{11}\see_{\delta}^{12}}{4} - \frac{\mu_1 (\see_{\delta}^{12})^2}{8} 
+ \frac{\mu_1 \see_{\delta}^{11}\see_{\delta}^{22}}{  8} 
- \frac{\mu_2 \see_{\delta}^{12} \see_{\delta}^{22}}{20} 
+ \frac{\mu_3 (\see_{\delta}^{22})^2}{40}
- \frac{\mu_1 \mu_2 \see_{\delta}^{11}}{80} 
\notag \\
&\ \
+ \frac{9 \mu_3 \see_{\delta}^{11}}{160} + \frac{  3 \mu_2^2 \see_{\delta}^{12}}{200} 
- \frac{\mu_1 \mu_3 \see_{\delta}^{12}}{40} - \frac{\mu_4 \see_{\delta}^{12}}{40} 
+ \frac{\mu_2 \mu_3 \see_{\delta}^{22}}{800} - \frac{  \mu_1 \mu_4 \see_{\delta}^{22}}{80} 
\label{alltril.2} \\
&\ \
+ \frac{\mu_5 \see_{\delta}^{22}}{8} 
- \frac{3 \mu_2^2 \mu_3}{8000} 
+ \frac{9 \mu_1 \mu_3^2}{12800} + \frac{  \mu_1 \mu_2 \mu_4}{800} 
- \frac{9 \mu_3 \mu_4}{1600} - \frac{\mu_1^2 \mu_5}{128} + 
\frac{  3 \mu_2 \mu_5}{160}  \notag \\
(\see_{\delta}^{22})^3 &= 
 \see_{\delta}^{11}\see_{\delta}^{22}
- \frac{(\see_{\delta}^{12})^2}{4}
 - \frac{ \mu_1 \see_{\delta}^{12}\see_{\delta}^{22}}{4} 
+ \frac{\mu_2 (\see_{\delta}^{22})^2}{20} 
- \frac{\mu_1^2 \see_{\delta}^{11}}{  16} + \frac{3 \mu_2 \see_{\delta}^{11}}{20} 
+ \frac{\mu_1 \mu_2 \see_{\delta}^{12}}{40} - \frac{9 \mu_3 \see_{\delta}^{12}}{  80} 
\notag \\
&\ \ 
 + \frac{\mu_2^2 \see_{\delta}^{22}}{50} 
- \frac{9 \mu_1 \mu_3 \see_{\delta}^{22}}{160} + \frac{3 \mu_4 \see_{\delta}^{22}}{20}  - \frac{3 \mu_2^3}{2000} + \frac{9 \mu_1 \mu_2 \mu_3}{1600} 
- \frac{81 \mu_3^2}{  6400} - \frac{3 \mu_1^2 \mu_4}{320} 
+ \frac{9 \mu_2 \mu_4}{400} 
 \notag
\end{align}
}
\end{subequations}

\noindent
The following combination constructed from $\det \see_\delta$ will also be useful,
\bea
\label{delapp}
(\det \see_\delta) \see_\delta^{11} & = & 
{ \mu_2 (\see_\delta^{11})^2 \over 10}
-{ 3 \mu_3 \see_\delta^{11} \see_\delta ^{12}  \over 10} 
+{ \mu_4 \see_\delta ^{11} \see_\delta ^{22} \over 5}
+{2 \mu_4 (\see_\delta^{12})^2 \over 5}
- \mu_5 \see_\delta ^{12} \see_\delta^{22}
+{3 \mu_6 (\see_\delta^{22})^2 \over 2}
\no \\ &&
-{\mu_6 \see_\delta^{11} \over 2} 
+{ \mu_1 \mu_5 \see_\delta ^{11} \over 16}
-{ \mu_2 \mu_4 \see_\delta ^{11} \over 50}
+{11 \mu_3^2 \see_\delta^{11} \over 1600}
+{\mu_1 \mu_6 \see_\delta^{12} \over 8}
-{ \mu_2 \mu_5 \see_\delta^{12} \over 40}
\no \\ &&
+{ \mu_3 \mu_4 \see_\delta^{12} \over 200}
-{ \mu_2 \mu_6 \see_\delta^{22} \over 20}
+{ \mu_3 \mu_5 \see_\delta^{22} \over 40}
-{\mu_4^2 \see_\delta^{22} \over 100} 
 -{3 \mu_4 \mu_6 \over 40}
 +{\mu_5^2 \over 32}
 \\ && 
 -{\mu_1 \mu_4 \mu_5 \over 160}
 +{9 \mu_1 \mu_3 \mu_6 \over 320}
 -{\mu_2^2 \mu_6 \over 100} 
+{\mu_2 \mu_3 \mu_5 \over 1600}
+{\mu_2 \mu_4^2 \over 1000}
-{ 3 \mu_3^2 \mu_4 \over 8000}
\no
\eea

\newpage

\section{Modular tensors $\CM_w$}
\label{sec:D2meta}
\setcounter{equation}{0}

This appendix gathers definitions and components of the modular tensors $\CM_w$
constructed from symmetric polynomials $\mu_m$ in the branch points $u_i$.
The outer products of $\CM_1$ and the $\CM_{w\leq 6}$ tensors in
this appendix span the irreducible representations of $SL(2,\mathbb C)$ in the 
decomposition of ${\bf 7}^{\otimes_s \leq 6}$ noted in the black entries of table \ref{specirrep}.

\subsection{List of components of tensors $\CM_2$ and $\CM_3$}
\label{sec:D2}

In this appendix we provide the expressions in terms of polynomials in $\mu_m$, for the components of the scalar $\CM_2$; of the rank 4 and rank 8 totally symmetric tensors $\CM_2$; and of the rank 2 and rank 6 totally symmetric tensors $\CM_3$, defined in (\ref{5.M2}) and (\ref{5.M3}), respectively.

\sm

The scalar $\CM_2$ is given by,
\bea
\CM_2 = \mu_0 \mu_6 - \tfrac{\mu_1 \mu_5}{6} + \tfrac{\mu_2 \mu_4}{15}  -\tfrac{ \mu_3^2}{40} 
\label{sping3.65}
\eea
The tensor $\CM_2^{a_1 \cdots a_4}$ transforms under the ${\bf 5}$ of $SL(2,\CC)$, and its components are given by,
\begin{align}
\label{sping3.64}
\CM_2^{1111} &=  \tfrac{\mu_2 \mu_6}{15}   - \tfrac{\mu_3 \mu_5}{30}   + \tfrac{\mu_4^2 }{75}  &
\CM_2^{2222} &=   \tfrac{\mu_0 \mu_4}{15}  -\tfrac{\mu_1  \mu_3}{30}  + \tfrac{\mu_2^2}{75} 
\notag \\
\CM_2^{1112} &=   \tfrac{\mu_1 \mu_6 }{12}  - \tfrac{\mu_2  \mu_5}{60}  + \tfrac{ \mu_3 \mu_4}{300}  &
\CM_2^{1222} &=  \tfrac{\mu_0 \mu_5}{12}  - \tfrac{\mu_1 \mu_4}{60}   + \tfrac{\mu_2 \mu_3}{300}   
\notag \\
\CM_2^{1122} &=  \tfrac{ \mu_3^2}{300} - \tfrac{ \mu_2 \mu_4}{150} + \tfrac{ \mu_0 \mu_6}{6}
\end{align}
The components of $\CM_2^{a_1 \cdots a_8}$  are given as follows,
\begin{align}
\CM_2^{1 \cdots 1} & = \tfrac{\mu_4 \mu_6 }{ 15} - \tfrac{\mu_5^2 }{ 36} &
\CM_2^{2 \cdots 2} & = \tfrac{\mu_0 \mu_2 }{ 15} -\tfrac{\mu_1^2 }{ 36} 
\no \\
\CM_2^{1 \cdots 12} & = \tfrac{\mu_3 \mu_6 }{ 40} -\tfrac{\mu_4 \mu_5 }{ 180} &
\CM_2^{1 2 \cdots 2} & =  \tfrac{\mu_0 \mu_3 }{ 40} -\tfrac{\mu_1 \mu_2 }{ 180} 
\no \\
\CM_2^{1 \cdots 122} & = \tfrac{ \mu_2 \mu_6 }{ 70} +\tfrac{\mu_3 \mu_5 }{ 840} -\tfrac{\mu_4^2 }{ 630} &
\CM_2^{11 2 \cdots 2} & = \tfrac{\mu_0 \mu_4 }{ 70} +\tfrac{\mu_1 \mu_3 }{ 840} -\tfrac{\mu_2^2 }{ 630}
\no \\
\CM_2^{1 \cdots 1222} & = \tfrac{\mu_1 \mu_6 }{ 84} +\tfrac{ \mu_2 \mu_5 }{ 315} -\tfrac{\mu_3 \mu_4 }{ 840} &
\CM_2^{111 2 \cdots 2} & = \tfrac{\mu_0 \mu_5 }{ 84} +\tfrac{\mu_1 \mu_4 }{ 315} -\tfrac{\mu_2 \mu_3 }{ 840}  
\no \\
\CM_2^{11112222} & = \tfrac{\mu_0 \mu_6 }{ 70} + \tfrac{\mu_1 \mu_5 }{ 180} +\tfrac{\mu_2 \mu_4 }{ 3150}  -  \tfrac{\mu_3^2 }{ 1400}
\end{align}
The components of the tensor $\CM_3^{a_1 a_2}$ are given by,
\bea
\CM_3^{11} & = & 
\tfrac{2 \mu_0 \mu_4 \mu_6 }{ 15} -\tfrac{ \mu_1 \mu_3 \mu_6 }{ 20} +\tfrac{ 4 \mu_2^2 \mu_6 }{ 225}
-\tfrac{ \mu_0 \mu_5^2  }{ 18} +\tfrac{\mu_1 \mu_4 \mu_5 }{ 90} -\tfrac{ \mu_2 \mu_3 \mu_5 }{ 900} 
-\tfrac{ 2 \mu_2 \mu_4^2 }{ 1125} +\tfrac{ \mu_3 ^2 \mu_4 }{ 1500}
\no \\
\CM_3^{12} & = & 
 \tfrac{ \mu_0 \mu_3 \mu_6 }{ 20} - \tfrac{ \mu_1 \mu_2 \mu_6 }{ 90} - \tfrac{\mu_0  \mu_4 \mu_5 }{ 90} -\tfrac{\mu_1 \mu_3 \mu_5 }{ 90} 
+\tfrac{  \mu_2^2 \mu_5  }{ 150} + \tfrac{\mu_1 \mu_4^2 }{ 150} - \tfrac{ 17 \mu_2 \mu_3 \mu_4 }{ 4500} 
+\tfrac{  \mu_3^3 }{ 1000}  
\no \\
\CM_3^{22} & = & 
\tfrac{2 \mu_0 \mu_2 \mu_6 }{ 15} -\tfrac{ \mu_0 \mu_3 \mu_5  }{ 20} +\tfrac{ 4 \mu_0 \mu_4^2  }{ 225}
-\tfrac{  \mu_1^2 \mu_6 }{ 18} +\tfrac{\mu_1 \mu_2 \mu_5 }{ 90} -\tfrac{ \mu_1 \mu_3 \mu_4 }{ 900} 
-\tfrac{ 2 \mu_2^2 \mu_4  }{ 1125} +\tfrac{  \mu_2 \mu_3 ^2 }{ 1500}
\qquad
\eea
Finally, the components of $\CM_3^{a_1 \cdots a_6}$  work out as follows,
\label{5.M30}
\bea
\CM_3^{111111} & = & 
\tfrac{\mu_0 \mu_6^2 }{ 6} -\tfrac{\mu_1 \mu_5 \mu_6 }{ 36}-\tfrac{\mu_2 \mu_4 \mu_6 }{ 450} +\tfrac{\mu_3^2 \mu_6 }{ 300}
+\tfrac{\mu_2 \mu_5^2 }{ 180} -\tfrac{\mu_3 \mu_4 \mu_5 }{ 300} +\tfrac{\mu_4^3 }{ 1125}
\no \\
\CM_3^{111112} & = & 
  \tfrac{\mu_0 \mu_5 \mu_6 }{ 36}  +\tfrac{\mu_2 \mu_3 \mu_6 }{ 300}
-\tfrac{\mu_3^2 \mu_5 }{ 900} +\tfrac{\mu_3 \mu_4^2 }{ 4500}
-\tfrac{\mu_1 \mu_4 \mu_6 }{ 90} +\tfrac{\mu_2 \mu_4 \mu_5 }{ 900}  
\no \\
\CM_3^{111122} & = & 
- \tfrac{\mu_0 \mu_4 \mu_6 }{ 450} +\tfrac{\mu_0 \mu_5^2 }{ 180}-\tfrac{\mu_2 \mu_3 \mu_5 }{ 2250} -\tfrac{\mu_1 \mu_3 \mu_6 }{ 300}
+\tfrac{\mu_2^2 \mu_6 }{ 375} -\tfrac{\mu_3^2 \mu_4}{ 5625} +\tfrac{7 \mu_2 \mu_4^2 }{ 11250}
-\tfrac{\mu_1 \mu_4 \mu_5 }{ 900}
\no \\
\CM_3^{111222} & = & 
-  \tfrac{\mu_0 \mu_3 \mu_6 }{ 150}  + \tfrac{\mu_0 \mu_4 \mu_5 }{ 300}
+ \tfrac{2\mu_2  \mu_3 \mu_4 }{ 5625} - \tfrac{\mu_1 \mu_3 \mu_5 }{ 450}
+\tfrac{\mu_2^2  \mu_5 }{ 4500} +\tfrac{\mu_1 \mu_2 \mu_6 }{ 300} 
-\tfrac{\mu_3^3 }{ 7500} +\tfrac{\mu_1 \mu_4^2 }{ 4500} 
\no \\
\CM_3^{112222} & = & 
- \tfrac{\mu_0 \mu_2 \mu_6 }{ 450} +\tfrac{\mu_1^2 \mu_6 }{ 180}-\tfrac{\mu_1 \mu_3 \mu_4 }{ 2250} -\tfrac{\mu_0 \mu_3 \mu_5 }{ 300}
+\tfrac{\mu_0 \mu_4^2 }{ 375} -\tfrac{\mu_2 \mu_3^2 }{ 5625} +\tfrac{7 \mu_2^2 \mu_4 }{ 11250}
-\tfrac{\mu_1 \mu_2 \mu_5 }{ 900}
\no \\
\CM_3^{122222} & = & 
  \tfrac{\mu_0 \mu_1 \mu_6 }{ 36}  +\tfrac{\mu_0 \mu_3 \mu_4 }{ 300}
-\tfrac{\mu_1 \mu_3^2 }{ 900} + \tfrac{\mu_2^2 \mu_3  }{ 4500}
- \tfrac{\mu_0 \mu_2 \mu_5  }{ 90} + \tfrac{\mu_1 \mu_2 \mu_4  }{ 900}  
\no \\
\CM_3^{222222} & = & 
\tfrac{\mu_0^2 \mu_6 }{ 6} -\tfrac{\mu_0 \mu_1 \mu_5  }{ 36}-\tfrac{\mu_0 \mu_2 \mu_4  }{ 450} +\tfrac{\mu_0 \mu_3^2 }{ 300}
+\tfrac{\mu_1^2 \mu_4  }{ 180} -\tfrac{\mu_1 \mu_2 \mu_3 }{ 300} +\tfrac{\mu_2^3 }{ 1125}
\eea
We note in every multiplet the invariance under inversion which swaps the top and bottom component along with the other pairs. 

\subsection{Higher-rank tensors $\CM_w$ at $4\leq w\leq 6$}
\label{sec:D2higher}

Similar to the modular tensors $\CM_{w=2,3}$ introduced in (\ref{5.M2}) and (\ref{5.M3}),
we shall here introduce shorthands $\CM_{w=4,5,6}$ that are used in higher-order 
computations such as section \ref{sec:sum} or (\ref{symmszg.06}). 
At degree four in the $\mu_m$, we have,
 \begin{align}
 \CM_4 &= \tfrac{1}{2} \CM_2^{a_1a_2a_3a_4} \CM_2^{b_1 b_2 b_3 b_4}
\varepsilon_{a_1 b_1}\varepsilon_{a_2 b_2}\varepsilon_{a_3 b_3}\varepsilon_{a_4 b_4}
 \notag \\
\CM_4^{a_1a_2a_3a_4} &= \tfrac{1}{2} \CM_2^{b_1b_2(a_1a_2} \CM_2^{a_3a_4)c_1c_2}
\varepsilon_{b_1 c_1} \varepsilon_{b_2 c_2} 
\label{defm4ten}
 \end{align}
with highest weight components,
\begin{align}
\CM_4 &= \tfrac{ \mu_3^4}{30000} -  \tfrac{\mu_2 \mu_3^2 \mu_4}{5625} + 
 \tfrac{ 7 \mu_2^2 \mu_4^2}{22500} -  \tfrac{\mu_1 \mu_3 \mu_4^2}{4500} + 
 \tfrac{ \mu_0 \mu_4^3}{1125} -  \tfrac{\mu_2^2 \mu_3 \mu_5}{4500} + 
  \tfrac{\mu_1 \mu_3^2 \mu_5}{900}   \notag \\
  &\quad -  \tfrac{\mu_1 \mu_2 \mu_4 \mu_5}{900}  - 
  \tfrac{\mu_0 \mu_3 \mu_4 \mu_5}{300}  +  \tfrac{\mu_0 \mu_2 \mu_5^2}{180}  + 
 \tfrac{ \mu_2^3 \mu_6}{1125} -  \tfrac{\mu_1 \mu_2 \mu_3 \mu_6}{300}   \notag \\
 &\quad + 
  \tfrac{\mu_0 \mu_3^2 \mu_6}{300}  +  \tfrac{\mu_1^2 \mu_4 \mu_6}{180}  - 
  \tfrac{\mu_0 \mu_2 \mu_4 \mu_6}{450}  -  \tfrac{ \mu_0 \mu_1 \mu_5 \mu_6}{36} + 
  \tfrac{\mu_0^2 \mu_6^2}{12} 
\label{clean.12} \\
\CM_4^{1111} &=  
  \tfrac{\mu_3^2 \mu_4^2}{30000} -  \tfrac{\mu_2 \mu_4^3}{11250} - 
 \tfrac{ \mu_3^3 \mu_5}{9000} +  \tfrac{\mu_2 \mu_3 \mu_4 \mu_5}{3000} - 
 \tfrac{ \mu_2^2 \mu_5^2}{3600} +  \tfrac{\mu_2 \mu_3^2 \mu_6}{4500} - 
 \tfrac{ \mu_2^2 \mu_4 \mu_6}{2250} 
 \notag \\
 &\quad -  \tfrac{\mu_1 \mu_3 \mu_4 \mu_6}{1800}  + 
  \tfrac{\mu_0 \mu_4^2 \mu_6}{450}  +  \tfrac{\mu_1 \mu_2 \mu_5 \mu_6}{360}  - 
  \tfrac{\mu_0 \mu_3 \mu_5 \mu_6}{180}  -  \tfrac{\mu_1^2 \mu_6^2}{144}  + 
  \tfrac{\mu_0 \mu_2 \mu_6^2}{90} 
  \notag
\end{align}
We furthermore employ a two-tensor at degree five and a
scalar of degree six, 
\begin{align}
\CM_5^{a_1 a_2} &= \CM_2^{a_1 a_2 b_1 b_2} \CM_3^{c_1 c_2}
\varepsilon_{b_1 c_1}\varepsilon_{b_2 c_2}
\notag \\
\CM_6 &= \tfrac{1}{2} \CM_3^{a_1 a_2} \CM_3^{b_1 b_2}
 \varepsilon_{a_1 b_1}  \varepsilon_{a_2 b_2}
\label{clean.23} 
\end{align}
with highest weight components,
\begin{align}
 \label{clean.17}
\CM_5^{11} &= 
-  \tfrac{\mu_3^4 \mu_4}{225000} +  \tfrac{2 \mu_2 \mu_3^2 \mu_4^2}{84375} - 
 \tfrac{ \mu_2^2 \mu_4^3}{84375} -  \tfrac{\mu_1 \mu_3 \mu_4^3}{16875} + 
 \tfrac{ 4 \mu_0 \mu_4^4}{16875} +  \tfrac{\mu_2 \mu_3^3 \mu_5}{135000} - 
 \tfrac{ 7 \mu_2^2 \mu_3 \mu_4 \mu_5}{67500}  \notag \\
 &\quad + 
 \tfrac{ \mu_1 \mu_3^2 \mu_4 \mu_5}{6750} + 
 \tfrac{ \mu_1 \mu_2 \mu_4^2 \mu_5}{3375} - 
 \tfrac{ 4 \mu_0 \mu_3 \mu_4^2 \mu_5}{3375} +  \tfrac{\mu_2^3 \mu_5^2}{4500} 
 -  \tfrac{ \mu_1 \mu_2 \mu_3 \mu_5^2}{1350} +  \tfrac{\mu_0 \mu_3^2 \mu_5^2}{675}  
 \notag \\
 &\quad + 
 \tfrac{ 7 \mu_2^2 \mu_3^2 \mu_6}{67500} -  \tfrac{\mu_1 \mu_3^3 \mu_6}{3000} - 
 \tfrac{ 4 \mu_2^3 \mu_4 \mu_6}{16875} +  \tfrac{ 13 \mu_1 \mu_2 \mu_3 \mu_4 \mu_6}{13500} 
 +  \tfrac{ \mu_0 \mu_3^2 \mu_4 \mu_6}{4500} -  \tfrac{\mu_1^2 \mu_4^2 \mu_6 }{540} 
  \no \\
 &\quad +  \tfrac{ 2 \mu_0 \mu_2 \mu_4^2 \mu_6}{1125} -  \tfrac{
 \mu_1 \mu_2^2 \mu_5 \mu_6}{1350} +  \tfrac{\mu_1^2 \mu_3 \mu_5 \mu_6}{270}  - 
 \tfrac{ 17 \mu_0 \mu_2 \mu_3 \mu_5 \mu_6}{2700} + 
  \tfrac{\mu_0 \mu_1 \mu_4 \mu_5 \mu_6}{270}  
  \notag \\
  &\quad -  \tfrac{\mu_0^2 \mu_5^2 \mu_6}{108}  - 
  \tfrac{\mu_1^2 \mu_2 \mu_6^2}{540}  +  \tfrac{8 \, \mu_0 \mu_2^2 \mu_6^2}{675}  - 
  \tfrac{\mu_0 \mu_1 \mu_3 \mu_6^2}{60}  +  \tfrac{\mu_0^2 \mu_4 \mu_6^2}{45} 
\end{align}
as well as 
\begin{align}
\CM_6 &= - \tfrac{ \mu_3^6}{1000000} + \tfrac{\mu_2 \mu_3^4 \mu_4}{125000}
 - \tfrac{ 337 \mu_2^2 \mu_3^2 \mu_4^2}{20250000} - 
\tfrac{ 19 \mu_1 \mu_3^3 \mu_4^2}{1350000} + \tfrac{4 \mu_2^3 \mu_4^3}{1265625} 
+ \tfrac{ 53 \mu_1 \mu_2 \mu_3 \mu_4^3}{1012500} + 
\tfrac{ \mu_0 \mu_3^2 \mu_4^3}{84375} \notag \\
&\quad - \tfrac{\mu_1^2 \mu_4^4}{22500 }
- \tfrac{ 8 \mu_0 \mu_2 \mu_4^4}{253125} - 
\tfrac{ 19 \mu_2^2 \mu_3^3 \mu_5}{1350000} 
+ \tfrac{\mu_1 \mu_3^4 \mu_5}{45000} + 
\tfrac{ 53 \mu_2^3 \mu_3 \mu_4 \mu_5}{1012500} - 
\tfrac{ 11 \mu_1 \mu_2 \mu_3^2 \mu_4 \mu_5}{162000} 
- \tfrac{ \mu_0 \mu_3^3 \mu_4 \mu_5}{90000} \notag \\
&\quad - 
\tfrac{ 13 \mu_1 \mu_2^2 \mu_4^2 \mu_5}{101250} 
+ \tfrac{ 11 \mu_1^2 \mu_3 \mu_4^2 \mu_5}{81000} - 
\tfrac{ \mu_0 \mu_2 \mu_3 \mu_4^2 \mu_5}{67500} + 
\tfrac{ 7 \mu_0 \mu_1 \mu_4^3 \mu_5}{20250} - \tfrac{\mu_2^4 \mu_5^2}{22500} 
+ \tfrac{ 11 \mu_1 \mu_2^2 \mu_3 \mu_5^2}{81000} \notag \\
&\quad - 
\tfrac{ \mu_1^2 \mu_3^2 \mu_5^2}{8100} + 
\tfrac{ \mu_0 \mu_2 \mu_3^2 \mu_5^2}{54000} + 
\tfrac{ \mu_1^2 \mu_2 \mu_4 \mu_5^2}{8100} + 
\tfrac{ \mu_0 \mu_2^2 \mu_4 \mu_5^2}{4050} -
\tfrac{ \mu_0 \mu_1 \mu_3 \mu_4 \mu_5^2}{1350} - 
 \tfrac{ \mu_0^2 \mu_4^2 \mu_5^2 }{900} - \tfrac{\mu_0 \mu_1 \mu_2 \mu_5^3}{1620} \notag \\
&\quad+ 
 \tfrac{ \mu_0^2 \mu_3 \mu_5^3}{360} + \tfrac{\mu_2^3 \mu_3^2 \mu_6}{84375 }- 
\tfrac{ \mu_1 \mu_2 \mu_3^3 \mu_6}{90000} - \tfrac{\mu_0 \mu_3^4 \mu_6}{10000} - 
\tfrac{ 8 \mu_2^4 \mu_4 \mu_6}{253125} - 
\tfrac{ \mu_1 \mu_2^2 \mu_3 \mu_4 \mu_6}{67500} + 
\tfrac{ \mu_1^2 \mu_3^2 \mu_4 \mu_6}{54000}  \label{m6compon} \\
&\quad+ 
\tfrac{ \mu_0 \mu_2 \mu_3^2 \mu_4 \mu_6}{1800} + 
\tfrac{ \mu_1^2 \mu_2 \mu_4^2 \mu_6}{4050} - 
\tfrac{ 8 \mu_0 \mu_2^2 \mu_4^2 \mu_6}{50625} - 
\tfrac{ 23 \mu_0 \mu_1 \mu_3 \mu_4^2 \mu_6}{13500} + 
\tfrac{ 8 \mu_0^2 \mu_4^3 \mu_6}{3375} + 
\tfrac{ 7 \mu_1 \mu_2^3 \mu_5 \mu_6}{20250} \notag \\
&\quad - 
\tfrac{ \mu_1^2 \mu_2 \mu_3 \mu_5 \mu_6}{1350} - 
\tfrac{ 23 \mu_0 \mu_2^2 \mu_3 \mu_5 \mu_6}{13500} + 
\tfrac{ 13 \mu_0 \mu_1 \mu_3^2 \mu_5 \mu_6}{3600} - 
\tfrac{ \mu_1^3 \mu_4 \mu_5 \mu_6}{1620} + 
\tfrac{ 11 \mu_0 \mu_1 \mu_2 \mu_4 \mu_5 \mu_6}{4050} \notag \\
&\quad - 
 \tfrac{ \mu_0^2 \mu_3 \mu_4 \mu_5 \mu_6}{180} + 
\tfrac{ \mu_0 \mu_1^2 \mu_5^2 \mu_6 }{324} - 
\tfrac{ \mu_0^2 \mu_2 \mu_5^2 \mu_6}{135} -\tfrac{ \mu_1^2 \mu_2^2 \mu_6^2}{900} 
+ \tfrac{ 8 \mu_0 \mu_2^3 \mu_6^2}{3375} + \tfrac{ \mu_1^3 \mu_3 \mu_6^2 }{360 } \notag \\
&\quad- 
 \tfrac{ \mu_0 \mu_1 \mu_2 \mu_3 \mu_6^2 }{180} - 
 \tfrac{  \mu_0^2 \mu_3^2 \mu_6^2 }{400} - \tfrac{ \mu_0 \mu_1^2 \mu_4 \mu_6^2 }{135} + 
\tfrac{ 4 \mu_0^2 \mu_2 \mu_4 \mu_6^2 }{225}
\notag
\end{align} 

\newpage

\section{Proof of the lemmas for $Q_\delta$}
\label{sec:E}
\setcounter{equation}{0}

In this appendix, we provide the proofs for the Lemmas \ref{lem:Q4} and \ref{lem:Q6} that reduce $Q_\delta(1,3|2,4)$ and $Q_\delta(1,3,5|2,4,6)$ to the spin structure dependent polynomial $\bL_\delta(i,j)$ and the spin structure independent polynomial $Z(i,j)$.

\subsection{Proof of Lemma \ref{lem:Q4}}
\label{sec:E.1}

Lemma \ref{lem:Q4} provides an expression for $Q_\delta(1,3|2,4)$ in terms of $\bL_\delta(i,j)$ and $Z(i,j)$, which we prove in this appendix. The starting point consists of the expression (\ref{PQ4}) for $Q_\delta(1,3|2,4)$ in terms of $P_\delta(1,2,3,4)$ and $Q_\delta(i|j)$ for various values of $i,j$; the relation between $P_\delta$ and $\Pi_\delta$ in (\ref{PPi}); and the expression for $\Pi_\delta$  in terms of the partially symmetric polynomials $\a_1, \a_2,\b_1, \b_2$ given in (\ref{Pifull}). Combining these relations,  we obtain the following expression for $Q_\delta(1,3|2,4)$ in terms of $\Pi_\delta(1,2,3,4)$ and $Q_\delta(i|j)$, 
\bea
\label{Qdelta4}
Q_\delta(1,3|2,4) & = & x_{12} x_{23}x_{34} x_{41}\, \Pi_\delta(1,2,3,4) -s_1^2 s_3^2 - s_2^2 s_4^2
\no \\ &&
+ \Big ( \thalf s_1^2 \big (Q_\delta(2|3) + Q_\delta(3|4)  -Q_\delta(2|4) \big ) + \hbox{cycl}(1,2,3,4) \Big )
\eea
where $\Pi_\delta$ is given in terms of $\a_1, \a_2,\b_1, \b_2$ by,
\bea
\label{Pi4}
\Pi_\delta(1,2,3,4) & = & \half \Big ( X_{1,2} +\a_1 Y_{1,2} +\a_2 \Big ) \Big ( X_{2,3} +\b_1 Y_{2,3} +\b_2 \Big ) 
\Big ( X_{3,4} +\a_1 Y_{3,4} +\a_2 \Big ) 
\no \\ && \quad \times 
\Big ( X_{4,1} +\b_1 Y_{4,1} +\b_2 \Big )
+ (\a \leftrightarrow \b)
\eea
The polynomials $Q_\delta(i|j)$ on the second line of (\ref{Qdelta4}) may be readily expressed in terms of $\bL_\delta(i,j)$ and $Z(i,j)$ using Theorem \ref{thm:1}. Furthermore, Lemma \ref{4.lem.1} prescribes that the spin structure dependence of $\Pi_\delta(1,2,3,4)$ may be expressed as a polynomial in $\see_\delta^{ab}$ whose total degree is 2. Combining these observations shows that all spin structure dependence of $Q_\delta(1,3|2,4)$ may be reduced to a degree-two polynomial in $\see_\delta^{ab}$. Below we shall extract the terms bilinear, linear, and independent in $\see_\delta$.

\subsubsection{Terms bilinear in ${\see_\delta}$}

The contributions to (\ref{Qdelta4}) bilinear in $\see_\delta$ can arise only from $\Pi_\delta$ and are readily extracted from (\ref{Pi4}). They may be regrouped entirely in terms of the functions $\bL_\delta$ of Theorem \ref{thm:1},
\bea
Q_\delta(1,3|2,4) \Big |_{\see_\delta ^2} & = & 8 x_{12} x_{23} x_{34} x_{41} 
\Big ( \bL_\delta(1,2) \bL_\delta(3,4) +  \bL_\delta(1,4) \bL_\delta(2,3) \Big ) 
\eea

\subsubsection{Terms linear in $\see_\delta$}

The contributions to (\ref{Qdelta4}) linear in $\see_\delta$ may be parametrized by a symmetric rank-two $SL(2,\CC)$ tensor $\bH^{ab}$,  which is independent of the spin structure $\delta$, 
\bea
Q_\delta(1,3|2,4) \Big |_{\see_\delta^1 } & = & 
 \bH^{a_1a_2} (1,2,3,4) \, \see _\delta ^{b_1 b_2} \, \ep_{a_1b_1} \ep _{a_2 b_2}
\eea
The coefficients of $\see_\delta$ in $Q_\delta(i|j)$ and  $\Pi_\delta(1,2,3,4)$  are homogeneous of degree 1 in the symmetric polynomials $\mu_m$ (including $\mu_0=1$ for homogeneity where needed), so that the tensor $\bH$ must be homogeneous of degree 1  in $\mu_m$ or equivalently, using (\ref{MXcomps}), in the tensor $\CM_1$.

\sm

Homogeneity of the tensor $\bH$ in $\CM_1$   allows us to parametrize $\bH^{a_1a_2}$ as follows,
\bea
\bH^{a_1 a_2} (1,2,3,4) & = & \CM_1^{b_1 \cdots b_6} \, \bH_1^{a_1 a_2 | c_1 \cdots c_6} (1,2,3,4)  \, \ep_{b_1 c_1}\ndots \ep_{b_6 c_6}
\eea
where the tensor  $\bH_1$ depends on $x_1,x_2,x_3,x_4$ but is independent of $\mu_m$.  To determine the components of $\bH$ and $\bH_1$, we proceed as follows.  The symmetric rank-two tensor $\bH$  transforms under the ${\bf 3}$ of $SL(2,\CC)$. As a result $\bH$ is  determined completely by the lowest weight component $\bH^{22}$ or equivalently by $\bH_1^{22|c_1 \cdots c_6}$. The components of $\bH_1^{22|c_1 \cdots c_6}$ may be determined in turn by successively applying the translation generator $\cT$ to its component $\bH^{22|111111}_1$.  As usual, the remaining components of $\bH_1$ may be determined by applying the inversion generator  $\cS$.  Finally, the components $\bH^{22|111111}_1$ may be obtained from $Q_\delta(1,3|2,4)$ in (\ref{Qdelta4}) by extracting the terms linear in $\see_\delta$ and proportional to $\mu_0$,\footnote{The tensor $\bH_1$ is symmetric in its first two indices and separately symmetric in its last six indices but it is not  symmetric in all its indices. Therefore, it corresponds to a reducible representation of $SL(2,\CC)$, whose dimension is 21, and which  is readily identified as  ${\bf 9} \oplus {\bf 7} \oplus {\bf 5}$ of $SL(2,\CC)$. One  verifies that $\cT \, \bH_1^{22|222222}=0$ as expected. One also readily verifies  that the tensor $\bH_1$ is not totally symmetric by evaluating for example the component $\bH_1^{22|222221}-\bH_1^{12|222222}= 2(x_1+x_3-x_2-x_4)(x_1+x_2-x_3-x_4)(x_1+x_4-x_2-x_3)\not =0$.}
\bea
\bH^{22|111111}_1 & = & 2 x_1^3 x_3^3 x_{24}^2-4 x_1^3 x_2^3x_{34}^2 + {\rm cycl}(1,2,3,4) 
\no \\
\bH^{22|211111}_1 & = & (x_1^3 x_3^2+x_1^2 x_3^3)x_{24}^2 - 2(x_1^3 x_2^2 + x_1^2 x_2^3)x_{34}^2+ {\rm cycl}(1,2,3,4) 
\no \\
\bH^{22|221111}_1 & = &  \tfrac{2}{5} (x_1^3 x_3+x_1 x_3^3+ 3 x_1^2 x_3^2)x_{24}^2 - \tfrac{4}{5} (x_1^3 x_2 + x_1 x_2^3+ 3 x_1^2 x_2^2 )x_{34}^2 + {\rm cycl}(1,2,3,4) 
\no \\
\bH^{22|222111}_1 & = & \tfrac{1}{10} (x_1^3 + x_3^3 + 9 x_1^2 x_3 + 9 x_1 x_3^2) x_{24}^2 + {\rm cycl}(1,2,3,4)
\no \\ &&
- \tfrac{1}{5} (x_1^3 + x_2^3 + 9 x_1^2 x_2 + 9 x_1 x_2^2) x_{34}^2 + {\rm cycl}(1,2,3,4) 
\no \\
\bH^{22|222211}_1 & = & \tfrac{2}{5} (x_1^2 + x_3^2 + 3 x_1 x_3) x_{24}^2
- \tfrac{4}{5} (x_1^2 + x_2^2 + 3 x_1 x_2) x_{34}^2 + {\rm cycl}(1,2,3,4) 
\no \\
\bH^{22|222221}_1 & = & (x_1+x_3)x_{24}^2 - 2(x_1+x_2)x_{34}^2+ {\rm cycl}(1,2,3,4) 
\no \\
\bH^{22|222222} _1& = & 2 x_{24}^2-4 x_{34}^2 + {\rm cycl}(1,2,3,4) 
\eea
Appealing to the components of the tensor $\bX_{ij}$ which were defined in (\ref{MXcomps}),  we may recast the above expressions and its corresponding expressions for $\bH_1^{12|c_1 \cdots c_6}$ and $\bH_1^{11|c_1 \cdots c_6}$ as follows,
\bea
\bH^{22|c_1 \cdots c_6}_1 & = & 2 \bX^{c_1 \cdots c_6}_{13} x_{24}^2-4 \bX^{c_1 \cdots c_6}_{12} x_{34}^2 + {\rm cycl}(1,2,3,4) 
\no \\ 
\bH^{12|c_1 \cdots c_6 }_1 & = &  \bX^{c_1 \cdots c_6} _{13}  (x_2+x_4) x_{24}^2 - 2 \bX^{c_1 \cdots c_6}_{12} (x_3+x_4) x_{34}^2 + {\rm cycl}(1,2,3,4) 
\no \\
\bH^{11|c_1 \cdots c_6}_1 & = & 2 \bX^{c_1 \cdots c_6} _{13}  x_2x_4 x_{24}^2-4 \bX^{c_1 \cdots c_6}_{12} x_3x_4 x_{34}^2 + {\rm cycl}(1,2,3,4) 
\eea
The coefficients of $x_{24}^2 \bX_{13}$ and $x_{34}^2 \bX_{12}$ contract  with the tensor $\see_\delta$ to form the combinations $\bL_\delta$, and contracting with the tensor $\CM_1$ to produce $\bH$, we obtain, 
\bea
&&
\bH^{a_1 a_2} (1,2,3,4) \, \see _\delta ^{b_1 b_2} \, \ep_{a_1b_1} \ep _{a_2b_2}
\no \\ && \qquad =  
\CM_1^{a_1 \cdots a_6}  \Big ( 2  \bX_{13}^{b_1 \cdots b_6} x_{24}^2  \bL_\delta(2,4)  
- 4 \bX_{12}^{b_1 \cdots b_6} x_{34}^2 \bL_\delta(3,4) \Big )  \ep_{a_1b_1}\ndots
\ep_{a_6b_6}  + {\rm cycl}(1,2,3,4) 
\no \\ && \qquad =
2  Z(1,3) x_{24}^2  \bL_\delta(2,4)   - 4 Z(1,2) x_{34}^2 \bL_\delta(3,4)    + {\rm cycl}(1,2,3,4) 
\eea
where we have used the definition of $Z(i,j)$ in (\ref{Zij}) to present the result in its final from on the last line above.  
In summary, we have, 
\bea
Q_\delta(1,3|2,4) \Big |_{\see_\delta^1 } & = & 
2  Z(1,3) \, x_{24}^2 \,  \bL_\delta(2,4)   - 4 Z(1,2) \,  x_{34}^2 \, \bL_\delta(3,4)    + {\rm cycl}(1,2,3,4) 
\eea

\subsubsection{Terms independent of $\see_\delta$}

Extracting the terms independent of $\see_\delta$ from the expression for $Q_\delta(1,3|2,4)$ given in (\ref{Qdelta4}), we observe that they are  homogeneous of degree 2 in $\mu_m$ (including $\mu_0=1$ for homogeneity where needed)  and thus are also homogeneous of degree 2 in $\CM_1$. Collecting all terms and expressing the result in terms of the tensors $\CM_1$ and $\bX_{ij}$, we find, 
\bea
Q_\delta(1,3|2,4) \Big |_{\see_\delta^0 }  & = & 
2 (\CM_1 \cdot \bX_{12}) (\CM_1 \cdot \bX_{34}) 
+2 (\CM_1 \cdot \bX_{23}) (\CM_1 \cdot \bX_{41}) 
\no \\ &&
-2 (\CM_1 \cdot \bX_{13}) (\CM_1 \cdot \bX_{24}) 
\eea
Expressing the contractions $\CM_1 \cdot \bX_{ij} =\CM_1^{a_1\cdots a_6}   \bX_{ij}^{b_1\cdots b_6} \ep_{a_1b_1}\cdots \ep_{a_6b_6}$ in terms of $Z(i,j)$ using (\ref{Zij}), we obtain our final expression for the $\see_\delta$-independent contribution $Z_4$ in (\ref{z4def}),
\bea
Q_\delta(1,3|2,4) \Big |_{\see_\delta^0 }  & = &    2 \, Z(1,2) Z(3,4) + 2 \, Z(2,3)Z(4,1) - 2 \, Z(1,3) Z(2,4)
\eea
Assembling all contributions to $Q_\delta(1,3|2,4)$, we recover the result of (\ref{5.Q4}) and thereby have completed the proof of Lemma \ref{lem:Q4}.

\subsection{Proof of Lemma \ref{lem:Q6}}
\label{sec:E.2}

The function $Q_\delta(1,3,5|2,4,6)$ is given in terms of  $\Pi_\delta(1,\cdots,6)=\Pi_\delta(1,2,3,4,5,6)$, as well as $Q_\delta(i,j|k,l)$ and $Q_\delta(i|j)$ by (\ref{PQ6a}) and (\ref{PPi}) for  $n=6$, and takes the following form, 
\begin{align}
&Q_\delta (1,3,5|2,4,6)  =  x_{12} x_{23} x_{34} x_{45} x_{56} x_{61} \, \Pi_\delta (1,\cdots,6) 
 + s_1^2 s_3^2 s_5^2 + s_2^2 s_4^2 s_6^2
\no \\
&\quad -\Big( \tfrac{1}{2}  s_1^2 \big( Q_\delta(2,4|3,6)+Q_\delta(2,5|4,6) -Q_\delta(2,4|3,5)-Q_\delta(2,5|3,6)-Q_\delta(3,5|4,6) \big) 
\notag \\
&\quad \quad  + 
\tfrac{1}{4}  s_1^2 s_4^2  \big(  Q_\delta(2|5)+Q_\delta(3|6)-Q_\delta(2|6)-Q_\delta(3|5) \big)  
\label{QPi6}\\
&\quad \quad  + \tfrac{1}{2}  s_1^2 s_3^2 \big( Q_\delta(4|5)+Q_\delta(5|6)-Q_\delta(4|6) \big) +  \hbox{cycl}(1,2,3,4,5,6) \Big) \notag
\end{align}
Since the functions $Q_\delta(i|j)$ and $Q_\delta(i,j|k,l)$ were already expressed in terms of $\bL_\delta(i,j)$ and $Z(i,j)$ in equations (\ref{5.C2}) and (\ref{5.Q4}), it remains only to show that $\Pi_\delta(1,\cdots,6)$ may also be decomposed in terms of these functions. The starting point is formula (\ref{Pifull}) for the case $n=6$, 
\bea
\label{E.Pi}
\Pi_\delta (1, \cdots, 6) & = & 
\big ( X_{1,2} + \a_1 Y_{1,2} +  \a_2 \big ) \big ( X_{3,4} + \a_1 Y_{3,4} + \a_2 \big )
 \big ( X_{5,6} + \a_1 Y_{5,6} + \a_2 \big )
\no \\ && \, \times 
\big ( X_{2,3} + \b_1 Y_{2,3} + \b_2 \big )\big ( X_{4,5} + \b_1 Y_{4,5} + \b_2 \big )
\big ( X_{6, 1} + \b_1 Y_{6,1} + \b_2 \big )
\no \\ && + (\a \leftrightarrow \b)
\eea
By Lemma \ref{4.lem.1} all spin structure dependence of $\Pi_\delta(1, \cdots, 6)$ resides in a polynomial in $\see_\delta ^{ab}$ of degree 3, containing trilinear, bilinear, linear and $\see_\delta$-independent terms. The reduction of the dependence of $\Pi_\delta$ on $\a_1,\a_2,\b_1, \b_2$ to its expression in terms of $\see_\delta$ may be carried out with the algorithm used in appendix \ref{sec:B}, and is best performed using {\sc maple} or {\sc mathematica}.

\subsubsection{Terms trilinear in $\see_\delta$}

From the outset the case $n=6$ involves an important new twist, whose presence will persist to all higher orders $n\geq 3$. Indeed, the trilinear relations of Theorem \ref{thm:6}, written out in components in Corollary \ref{sping3.55} of appendix \ref{sec:D1}, guarantee that all trilinear dependence on $\see_\delta$ may be reduced to a polynomial in $\see_\delta$ of degree 2. Thus, we are faced with a choice as to how the final expressions for $\Pi_\delta(1,\cdots, 6)$ and $Q_\delta(1,3,5|2,4,6)$ should be presented. A propitious choice turns out to be one that naturally generalizes the structure of $Q_\delta(i|j)$ and $Q_\delta(i,j|k,l)$, namely whose top power is represented by a sum of cyclic products of $\bL_\delta$ functions. As no trilinear contribution to $Q_\delta(1,3,5|2,4,6)$ arise from $Q_\delta(i|j)$ and $Q_\delta(i,j|k,l)$ all trilinear terms arise from $\Pi_\delta(1,\cdots, 6)$ and may be brought into the following form, 
\begin{align}
Q_\delta&(1,3,5|2,4,6) \Big |_{\see_\delta ^3} 
 = 32 \, x_{12} \, x_{23} \, x_{34} \, x_{45} \, x_{56} \, x_{61} 
\label{E.Q6-3} \\ &
\times \big (
\bL_\delta(1,2) \bL_\delta(3,4) \bL_\delta(5,6) 
+ \bL_\delta (2,3)  \bL_\delta(4,5) \bL_\delta(6,1) \big )  
\notag
\end{align}
as seen the first line of (\ref{Qdel.13}).
Actually, the straightforward reduction of the product in (\ref{E.Pi}), following the algorithm of appendix \ref{sec:B},  will produce, in addition to the trilinear contribution of (\ref{E.Q6-3}), also a single term proportional to $(\see_\delta ^{22})^3$. The precise form of this extra term will, in general, depend on exactly how the reduction is carried out as different reduction algorithms may dissimulate the trilinear relations in different ways. Whatever the extra terms may be, it will be our choice to convert all trilinear terms, other than those collected in (\ref{E.Q6-3}), into a degree two polynomial in $\see_\delta$.   In the sequel, we shall assume that these operations have been carried out.

\subsubsection{Terms bilinear in $\see_\delta$}

Having eliminated the $(\see_\delta^{22})^3$ term using the trilinear relations, as explained above, we isolate the contributions bilinear in $\see_\delta$. This process yields a unique answer.  The terms bilinear in $\see_\delta$ transform under the representation ${\bf 3} \otimes {\bf 3}$ of $SL(2,\CC)$ with identical ${\bf 3}$-vectors. This representation reduces to ${\bf 1} \oplus {\bf 5}$, namely the singlet $\det (\see_\delta)$ and the totally symmetrized combination. Accordingly, the terms bilinear in $\see_\delta$ may be parametrized as follows,
\bea
Q_\delta(1,3,5|2,4,6) \Big |_{\see_\delta^2} 
& = & 
\bK _{\bf 1} \, \det (\see_\delta) + \bK_{\bf 5} ^{a_1 a_2 a_3 a_4} \, \see_\delta ^{b_1 b_2} \see _\delta ^{ b_3 b_4} \ep_{a_1b_1}\ndots  \ep_{a_4b_4}
\label{qdelquad}
\eea
where the $SL(2,\CC)$ scalar  $\bK_{\bf 1}$ and the symmetric rank-four tensor $\bK_{\bf 5}$  depend only on $x_i$ and $\mu_m$. Inspection of the terms in $Q_\delta(i|j), Q_\delta(i,j|k,l)$ and $\Pi_\delta(1,\cdots, 6)$ shows that their dependence on $\mu_m$ is via a homogeneous polynomial of degree 1, so that they must be linear in the tensor $\CM_1$.

\subsubsection*{$\bullet$ {\sl Evaluating the contribution of $\bK_{\bf 5}$}}

To determine the symmetric tensor $\bK_{\bf 5}$ in (\ref{qdelquad}), it suffices to start from its lowest weight vector $\bK_{\bf 5}^{2222}$ which we obtain using {\sc maple} analysis, and immediately re-express in terms of the functions $Z(i,j)$ as follows,
\bea
\bK_{\bf 5}^{2222} & = & 
16 \, Z(1,2)  \, x_{34} \, x_{45} \, x_{56} \, x_{63} 
- 16 \, Z(1,3)  \, x_{24} \, x_{45} \, x_{56} \, x_{62} 
\no \\ && 
+ 8 \, Z (1,4) \, x_{23} \, x_{35} \, x_{56} \, x_{62}  
+\hbox{cycl} (1,\cdots,6)
\eea
Cyclic permutations $\hbox{cycl} (1,\cdots,6)=\hbox{cycl} (1,2,3,4,5,6)$ are to be applied to both lines. 
To obtain the other components of the symmetric tensor $\bK_{\bf 5}$, we use the familiar methods of group theory: we obtain $\bK_{\bf 5}^{1111} $ from $\bK_{\bf 5}^{2222} $ by applying the inversion generator $\cS$,
\bea
\bK_{\bf 5}^{1111} & = & 
16 \, Z(1,2)  \, x_{34} \, x_{45} \, x_{56} \, x_{63} \, x_3 \, x_4 \, x_5 \, x_6
- 16 \, Z(1,3)  \, x_{24} \, x_{45} \, x_{56} \, x_{62} \, x_2 \, x_4 \, x_5 \, x_6
\no \\ && 
+ 8 \, Z(1,4) \, x_{23} \, x_{35} \, x_{56} \, x_{62}  \, x_2 \, x_3 \, x_5 \, x_6 
+\hbox{cycl} (1,\cdots,6)
\eea
and then use the $\cT$ generator to obtain the remaining descendant components,
\bea
\bK_{\bf 5}^{1112} & = & 
 4 \, Z(1,2)  \, x_{34} \, x_{45} \, x_{56} \, x_{63} \, (x_4 \, x_5 \, x_6+ x_3 \, x_5 \, x_6+ x_3 \, x_4  \, x_6+ x_3 \, x_4 \, x_5 )
\no \\ && 
-4 \, Z(1,3)  \, x_{24} \, x_{45} \, x_{56} \, x_{62} \, (x_4 \, x_5 \, x_6+ x_2 \, x_5 \, x_6+ x_2 \, x_4  \, x_6+ x_2 \, x_4 \, x_5 )
\no \\ && 
+2 \, Z (1,4) \, x_{23} \, x_{35} \, x_{56} \, x_{62}  \, (x_2 \, x_5 \, x_6+ x_3 \, x_5 \, x_6+ x_3 \, x_2  \, x_6+ x_3 \, x_2 \, x_5 )
\no \\ && 
+\hbox{cycl} (1,\cdots,6)
\no \\
3\, \bK_{\bf 5}^{1122} & = &
8 \, Z(1,2)  \, x_{34} \, x_{45} \, x_{56} \, x_{63} \, (x_3x_4+ x_3 x_5 + x_3 x_6+x_4 x_5 + x_4 x_6 + x_5 x_6) 
\no \\ && 
- 8 \, Z(1,3)  \, x_{24} \, x_{45} \, x_{56} \, x_{62} \, (x_2x_4+ x_2 x_5 + x_2 x_6+x_4 x_5 + x_4 x_6 + x_5 x_6) 
\no \\ && 
+ 4 \, Z (1,4) \, x_{23} \, x_{35} \, x_{56} \, x_{62}  \, (x_3x_2+ x_3 x_5 + x_3 x_6+x_2 x_5 +  x_2 x_6 + x_5 x_6) 
\no \\ && 
+\hbox{cycl} (1,\cdots,6)
\no \\
\bK_{\bf 5}^{1222} & = &
4 \, Z(1,2)  \, x_{34} \, x_{45} \, x_{56} \, x_{63} \, (x_3+x_4+x_5+x_6)
\no \\ && 
- 4 \, Z(1,3)  \, x_{24} \, x_{45} \, x_{56} \, x_{62} \, (x_2+x_4+x_5+x_6)
\no \\ && 
+ 2 \, Z (1,4) \, x_{23} \, x_{35} \, x_{56} \, x_{62}  \, (x_3+x_2+x_5+x_6)
\no \\ &&
+\hbox{cycl} (1,\cdots,6)
\eea
It is readily verified that $\cT \, \bK_{\bf 5}^{1222}=\bK_{\bf 5}^{2222}$ from this last explicit expression, as expected. 
Expressing the result in terms of the functions $\bL_\delta(i,j)$ and $Z(i,j)$, we obtain,
\bea
\bK_{\bf 5} ^{a_1 a_2 a_3 a_4} \, \see_\delta ^{b_1 b_2} \see _\delta ^{ b_3 b_4} \ep_{a_1b_1}\ndots \ep_{a_4b_4}
& = &  
\tfrac{16}{3} \, x_{34}  x_{45}  x_{56}  x_{63} \, Z(1,2)   \Big \{ \bL_\delta(3,4) \bL_\delta (5,6) + {\rm cycl}(4,5,6) \Big \}
\no \\ && 
-  \tfrac{16}{3} \, x_{24}  x_{45}  x_{56}  x_{62} \, Z(1,3)  \Big \{ \bL_\delta(2,4) \bL_\delta (5,6) + {\rm cycl}(4,5,6) \Big \}
\no \\ &&
+  \tfrac{8}{3}  \, x_{23}  x_{35}  x_{56}  x_{62}  \, Z (1,4)  \Big \{ \bL_\delta(2,3) \bL_\delta (5,6) + {\rm cycl}(3,5,6) \Big \}
\no \\ && 
+\hbox{cycl} (1,\cdots,6)
\label{k5terms}
\eea
The instruction to add cyclic permutations applies to the entire expression on the right side.

\subsubsection*{$\bullet$ {\sl Evaluating the contribution of $\bK_{\bf 1}$}}

The singlet contribution $\bK _{\bf 1}$ is homogeneous of degree one in $\mu_m$ and may be expressed as the contraction of $\CM_1$ with a tensor $\bK_0$ which has rank 6 and only depends on $x_i$, 
\bea
\bK_{\bf 1} = \CM_1 ^{a_1 \cdots a_6} \bK_0^{b_1 \cdots b_6} \ep_{a_1b_1 }\ndots \ep_{a_6b_6}
\eea
The lowest weight component $\bK_0^{222222}$ of the tensor $\bK_0$ may be obtained using {\sc maple},
\bea
\bK^{222222}_0 & = & 
-\tfrac{8}{3} \, x_{12}^2 \, x_{23} \, x_{34}^2 \, x_{41} 
-\tfrac{8}{3} \, x_{12} \, x_{23}^2 \, x_{34} \, x_{41}^2 
-\tfrac{4}{3} \, x_{12}^2 \, x_{24} \, x_{45}^2 \, x_{51} 
\no \\ &&
+\tfrac{8}{3}  \, x_{12}^2 \, x_{23} \, x_{35}^2 \, x_{51}  
+\tfrac{8}{3} \, x_{12} \, x_{23}^2 \, x_{35} \, x_{51}^2 
-\tfrac{4}{3} \, x_{12} \, x_{24}^2 \, x_{45} \, x_{51}^2 
\no \\ &&
+ {\rm cycl}(1,\cdots,6)
\eea
Since it is a sum of products of differences $x_{ij}$, we manifestly have $\cT \, \bK^{222222}_0=0$ as expected. 
By inversion, we get,
\bea
\bK^{111111}_0 & = & 
- \tfrac{8}{3} \, x_{12}^2 \, x_{23} \, x_{34}^2 \, x_{41}\, x_5^3 x_6^3 
- \tfrac{8}{3} \, x_{12} \, x_{23}^2 \, x_{34} \, x_{41}^2 \, x_5^3 x_6^3 
- \tfrac{4}{3} \, x_{12}^2 \, x_{24} \, x_{45}^2 \, x_{51} \, x_3^3 x_6^3 
\no \\ &&
+ \tfrac{8}{3} \, x_{12}^2 \, x_{23} \, x_{35}^2 \, x_{51}  \, x_4^3 x_6^3 
+ \tfrac{8}{3} \, x_{12} \, x_{23}^2 \, x_{35} \, x_{51}^2 \, x_4^3 x_6^3 
- \tfrac{4}{3} \, x_{12} \, x_{24}^2 \, x_{45} \, x_{51}^2 \, x_3^3 x_6^3 
\no \\ &&
+ {\rm cycl}(1,\cdots,6)
\eea
The descent to obtain the remaining components of $\bK_0$ by successively applying the generator $\cT$ follows the pattern of the descent in the expression for $Z(i,j)$ and may be verified by a direct {\sc maple} calculation, resulting in
\bea
\bK_{\bf 1} & = &
- \tfrac{8}{3} \, x_{12} \, x_{23} \, x_{34} \, x_{41} (x_{12} \,  x_{34} + x_{23} \, x_{41} )\, Z(5,6)
- \tfrac{4}{3} \, x_{12} \, x_{24} \, x_{45} \, x_{51}  ( x_{12} \, x_{45} + x_{24} \, x_{51}  )\, Z(3,6) 
\no \\ &&
+ \tfrac{8}{3} \, x_{12} \, x_{23} \, x_{35} \, x_{51}  ( x_{12} \, x_{35}  + x_{23} \, x_{51})\, Z(4,6) 
+ {\rm cycl}(1,\cdots,6)
\label{bfk1}
\eea
Upon multiplication with $\det (\see_\delta)$, the contributions $(x_{ij} \,  x_{kl} + x_{jk} \, x_{li} )$ in the round brackets may be combined with the determinant via
\bea
 x_{i\ell} \, x_{jk} \, \det (\see_\delta)  = \bL_\delta(i,j) \bL_\delta(k,\ell) - \bL_\delta(i,k) \bL_\delta(j,\ell)
\eea
and we obtain,
%
\bea
\bK_{\bf 1} \, \det \see_\delta & = &
\tfrac{8}{3} \, x_{12}  x_{23}  x_{34}  x_{41} Z(5,6) 
\Big (  \bL_\delta(1,4) \bL_\delta(2,3)  + \bL_\delta(1,2) \bL_\delta(3,4) -2 \bL_\delta(1,3) \bL_\delta(2,4) \Big ) 
\no \\ &&
+ \tfrac{4}{3} \,x_{12}  x_{24}  x_{45}  x_{51}  Z(3,6) 
\Big (  \bL_\delta(1,5) \bL_\delta(2,4) + \bL_\delta(1,2) \bL_\delta(4,5) - 2 \bL_\delta(1,4) \bL_\delta(2,5)  \Big ) 
\no \\ &&
- \tfrac{8}{3} \,x_{12}  x_{23}  x_{35}  x_{51}   Z(4,6) 
\Big (  \bL_\delta(1,5) \bL_\delta(2,3)  + \bL_\delta(1,2) \bL_\delta(3,5) - 2 \bL_\delta(1,3) \bL_\delta(2,5) \Big ) 
\no \\ &&
+ {\rm cycl}(1,\cdots,6)
\eea
To combine this expression with the contribution arising from $\bK_{\bf 5}$, we cyclically permute the arguments so that the functions $Z$ are evaluated at $Z(1,2)$, $Z(1,3)$, and $Z(1,4)$ only, 
\bea
\bK_{\bf 1} \, \det \see_\delta & = &
\tfrac{8}{3} \,  x_{34}  x_{45}  x_{56}  x_{63} Z(1,2) 
\Big (  \bL_\delta(3,6) \bL_\delta(4,5)  + \bL_\delta(3,4) \bL_\delta(5,6) -2 \bL_\delta(3,5) \bL_\delta(4,6) \Big ) 
\no \\ &&
+ \tfrac{4}{3} \, x_{56}  x_{62}  x_{23}  x_{35}  Z(1,4) 
\Big (  \bL_\delta(3,5) \bL_\delta(2,6) + \bL_\delta(5,6) \bL_\delta(2,3) - 2 \bL_\delta(2,5) \bL_\delta(3,6)  \Big ) 
\no \\ &&
- \tfrac{8}{3} \,  x_{45}  x_{56}  x_{62}  x_{24}   Z(1,3) 
\Big (  \bL_\delta(2,4) \bL_\delta(5,6)  + \bL_\delta(4,5) \bL_\delta(2,6) - 2 \bL_\delta(4,6) \bL_\delta(2,5) \Big ) 
\no \\ &&
+ {\rm cycl}(1,\cdots,6)
\label{k1terms}
\eea

\subsubsection*{$\bullet$ {\sl Assembling the terms bilinear in $\see_\delta$}}

Combining the contributions from $\bK_{\bf 1}$ in (\ref{k1terms}) and $\bK_{\bf 5}$ in (\ref{k5terms}), we have,
\bea
Q_\delta(1,3,5|2,4,6) \Big |_{\see_\delta^2}  &= 8 \big[ Z(1,2) \eecyc_\delta(3,4,5,6)
- Z(1,3) \eecyc_\delta(2,4,5,6) \notag \\
&\quad 
+\tfrac{1}{2} Z(1,4) \eecyc_\delta(2,3,5,6)
+{\rm cycl}(1,2,\ndots,6) \big] 
\label{qdellin}
\eea
where $\eecyc_\delta(a,b,c,d)$ is defined in (\ref{Qdel.05}).

\subsubsection{Terms linear in $\see_\delta$}

The terms linear in $\bL_\delta$ may be read off directly from the {\sc maple} calculation, and we have, 
\bea
Q_\delta(1,3,5|2,4,6) \Big |_{\see_\delta^1} 
& = & 
-4\, x_{12}^2 \, \bL_\delta(1,2) \Big ( Z(3,4)Z(5,6)+Z(3,6)Z(4,5)-Z(3,5)Z(4,6) \Big )
\no \\ &&
+4\, x_{13}^2 \, \bL_\delta(1,3) \Big ( Z(2,4)Z(5,6)+Z(2,6)Z(4,5)-Z(2,5)Z(4,6) \Big )
\no \\ &&
-2\, x_{14}^2 \, \bL_\delta(1,4) \Big ( Z(2,3)Z(5,6)+Z(3,5)Z(2,6)-Z(3,6)Z(2,5) \Big )
\no \\ &&
+ {\rm cycl}(1,\cdots,6)
\eea
where the instruction to add cyclic permutations applies to all three lines. These
contributions are represented via permutations of $\pm 4 Z_4(1,2,3,4) {\cal L}_\delta(5,6)$
in (\ref{Qdel.13}).

\subsubsection{Contributions to $Q_\delta(1,3,5|2,4,6)$ independent of $\see_\delta$}

The terms independent of $\see_\delta$ may be read off directly from the {\sc maple} calculation, 
\bea
Q_\delta(1,3,5|2,4,6) \Big |_{\see_\delta^0} 
& = & 
\tfrac{2}{3} \, Z(1,2) Z(3,4)Z(5,6)  - \tfrac{1}{3} \,Z(1,4)Z(2,5)Z(3,6)
\no \\ &&
+Z(1,2)Z(3,6)Z(4,5) + Z(1,3)Z(2,5) Z(4,6)
\no \\ && 
 - 2 \, Z(1,2)Z(3,5)Z(4,6) + {\rm cycl}(1,\cdots,6) 
\eea
which matches twice the expression (\ref{Qdel.10}) for $Z_6(1,2,3,4,5,6)$.

\newpage

\section{Simplified cyclic products $C_\delta$ for 7 and 8 points}
\label{apponCdelta}
\setcounter{equation}{0}

In this appendix, we generalize the simplified results of (\ref{5.C2}), (\ref{5.C-three}), (\ref{Cdel4}), (\ref{cdel5.1}) and (\ref{cdel6.1}) for the cyclic products $C_\delta (1,\cdots, n) = C_\delta(z_1,\ndots,z_n)$ to higher multiplicity $n$, with
$n=7$ in appendix \ref{app.Cdel7} and $n=8$ in appendix \ref{app.Cdel8}.  Furthermore, we gather the numerators $\mathfrak{N}^{(8)}$ of the eight-point spin structure sum (\ref{cdel8.A}) in appendix \ref{sec:spinsum8}.
Throughout the points $z_i$ are represented in the hyper-elliptic representation by $z_i = (x_i, s_i)$. 

\subsection{Seven points}
\label{app.Cdel7}

The spin structure dependence of the cyclic product of seven Szeg\"o kernels may be isolated based on the expressions for $Q_\delta(1|2)$, $Q_\delta(1,3|2,4)$ and $Q_\delta(1,3,5|2,4,6)$ provided in (\ref{5.C2}), (\ref{5.Q4}) and (\ref{Qdel.13}), respectively, leading to,
\begin{align}
 \label{cdel7.1}  
C_\delta(1,\cdots,7) &= \bigg\{ \frac{  {\cal N}^{(7)}_\delta[12] }{4 x_{12}}
+  \frac{   {\cal N}^{(7)}_\delta[1234] }{16 x_{12}x_{23}x_{34} }
+  \frac{   {\cal N}^{(7)}_\delta[123,45] }{16 x_{12}x_{23}x_{45} }
+  \frac{   {\cal N}^{(7)}_\delta[123,56] }{16 x_{12}x_{23}x_{56} }
\no \\ & \qquad
+  \frac{  {\cal N}^{(7)}_\delta[123,67] }{16 x_{12}x_{23}x_{67} }
+  \frac{   {\cal N}^{(7)}_\delta[12,34,56] }{16 x_{12}x_{34}x_{56} } 
+  \frac{   {\cal N}^{(7)}_\delta[123456] }{32 x_{12}x_{23}x_{34} x_{45}x_{56} }
\no \\ & \qquad
+  \frac{   {\cal N}^{(7)}_\delta[12345,67] }{32 x_{12}x_{23}x_{34} x_{45}x_{67} }
+  \frac{   {\cal N}^{(7)}_\delta[1234,567] }{32 x_{12}x_{23}x_{34} x_{56}x_{67} }
+ {\rm cycl}(1,\cdots,7) \bigg\}  
\no \\ & \quad 
+ \frac{  {\cal N}^{(7)}[1234567] }{64 x_{12}x_{23} x_{34}x_{45}x_{56}x_{67} x_{71}}
\end{align}
The numerator function on one simultaneous pole is given by,
\begin{align}
\label{cdel7.3}
 {\cal N}^{(7)}_\delta[12] &= dx_1 \big( L_\delta(2,3)   L_\delta(4,5)   L_\delta(6,7) 
+  L_\delta(2,7)   L_\delta(3,4)   L_\delta(5,6)  \big) 
 \\
&\quad + dx_2 \big( L_\delta(1,3)   L_\delta(4,5)   L_\delta(6,7) 
+  L_\delta(1,7)   L_\delta(3,4)   L_\delta(5,6)  \big)
\no
\end{align}
The numerators on 3 simultaneous poles are conveniently expressed in terms of
$L_\delta(a,b,c,d)$ and $W^\pm_3$ defined in (\ref{cdel6.3}) and (\ref{5.W3}),
{\small \begin{align}
 {\cal N}^{(7)}_\delta[1234] &=
 L_\delta(4,5,6,7)W_3^+(1,2,3)  +  L_\delta(3,5,6,7)W_3^-( 1,2|4 )  
 \notag \\ &\quad
+  L_\delta(2,5,6,7)W_3^-( 3,4 |1)+  L_\delta(1,5,6,7)W_3^+(2, 3,4) 
    \notag \\
 {\cal N}^{(7)}_\delta[123,45] &=
 L_\delta(3,5,6,7) W_3^-(1,2 | 4) + L_\delta(3,4,6,7) W_3^+(1,2,5)
 \no \\ & \quad
 + L_\delta(2,5,6,7) W_3^-(3,4 | 1)   +  L_\delta(2,4,6,7) W_3^-(1,5 | 3) 
 \no \\ & \quad
 + L_\delta(1,5,6,7)  W_3^+(2,3,4)  + L_\delta(1,4,6,7)  W_3^-(2,3 | 5) 
 \notag  \\
  {\cal N}^{(7)}_\delta[123,56] &=
L_\delta(3,4,6,7) W_3^+(1,2,5)+ L_\delta(3,4,5,7) W_3^-(1,2 |6)
\no \\ & \quad
+ L_\delta(2,4,6,7) W_3^-(1,5 | 3)+ L_\delta(2,4,5,7) W_3^-(3,6 | 1)
\no \\ & \quad
+ L_\delta(1,4,6,7) W_3^-(2,3 | 5)+ L_\delta(1,4,5,7) W_3^+(2,3,6)    
\label{cdel7.5} 
\end{align}}
as well as,
{\small \begin{align} 
 {\cal N}^{(7)}_\delta[123,67] &=
L_\delta(3,4,5,7) W_3^-(1,2 |6) + L_\delta(3,4,5,6) W_3^+(7,1,2)  
\no \\ & \quad
+ L_\delta(2,4,5,7) W_3^-(3,6 | 1)  + L_\delta(2,4,5,6) W_3^-(7,1 | 3) 
\no \\ & \quad 
+ L_\delta(1,4,5,7) W_3^+(2,3,6)  +  L_\delta(1,4,5,6) W_3^-(2,3 |7)  
\no \\
 {\cal N}^{(7)}_\delta[12,34,56]    &=
L_\delta(2,4,6,7) W_3^-(1,5| 3)  + L_\delta(2,4,5,7)  W_3^-(3,6 | 1)  
\no \\ & \quad
+ L_\delta(2,3,6,7) W_3^+(1,4,5) + L_\delta(2,3,5,7) W_3^-(1,4 | 6) 
\no \\ & \quad
+ L_\delta(1,4,6,7) W_3^-(2,3 |5) + L_\delta(1,4,5,7) W_3^+(2,3,6) 
 \notag \\
 &\quad
+ L_\delta(1,3,6,7) W_3^-(4,5|2) +  L_\delta(1,3,5,7)  W_3^-(2,6 |4)       
\label{cdel7.5more} 
\end{align}}
For the next set of numerators, on 5 simultaneous poles,
{\small \begin{align}
 {\cal N}^{(7)}_\delta[123456] &=
L_\delta(6, 7) W_5^+(1, 2, 3, 4, 5) 
+ L_\delta(5, 7) W_5^-(1, 2, 3, 4 | 6) 
\no \\ & \quad
+  L_\delta(4, 7) W_5^-(1, 2, 3 | 5, 6)  
+  L_\delta(3, 7) W_5^-(4, 5, 6 | 1, 2) 
\no \\ & \quad
+  L_\delta(2, 7) W_5^-(3, 4, 5, 6 | 1) 
+ L_\delta(1, 7) W_5^+(2, 3, 4, 5, 6) 
\notag \\
 {\cal N}^{(7)}_\delta[12345,67] &=
 L_\delta(5, 7) W_5^-(1, 2, 3, 4 | 6) 
 + L_\delta(5, 6) W_5^+(7, 1, 2, 3, 4) 
 \no \\ & \quad
 +  L_\delta(4, 7) W_5^-(1, 2, 3 | 5, 6) 
 + L_\delta(4, 6) W_5^-(7, 1, 2, 3 | 5) 
 \no \\ & \quad
 +  L_\delta(3, 7) W_5^-(4, 5, 6 | 1, 2) 
 + L_\delta(3, 6) W_5^-(7, 1, 2 | 4, 5)  
 \notag \\
 &\quad+ 
 L_\delta(2, 7) W_5^-(3, 4, 5, 6 | 1) + 
 L_\delta(2, 6) W_5^-(3, 4, 5 | 7, 1)  
 \notag \\
 &\quad+ L_\delta(1, 7) W_5^+(2, 3, 4, 5, 6) +
  L_\delta(1, 6) W_5^-(2, 3, 4, 5 | 7)
\notag \\
 {\cal N}^{(7)}_\delta[1234,567]  &= 
  L_\delta(4, 7) W_5^-(1, 2, 3 | 5, 6) + 
 L_\delta(4, 6) W_5^-(7, 1, 2, 3 | 5) 
 \no \\ & \quad
 + L_\delta(4, 5) W_5^+(6, 7, 1, 2, 3)  
 + L_\delta(3, 7) W_5^-(4, 5, 6 | 1, 2) 
 \no \\ & \quad
 + L_\delta(3, 6) W_5^-(7, 1, 2 | 4, 5) 
 + L_\delta(3, 5) W_5^-(6, 7, 1, 2 | 4)  
 \notag \\
 &\quad+ 
 L_\delta(2, 7) W_5^-(3, 4, 5, 6 | 1) + 
 L_\delta(2, 6) W_5^-(3, 4, 5 | 7, 1) 
 \no \\ & \quad
 +  L_\delta(2, 5) W_5^-(6, 7, 1 | 3, 4)  
 + L_\delta(1, 7) W_5^+(2, 3, 4, 5, 6) 
 \no \\ & \quad
 + L_\delta(1, 6) W_5^-(2, 3, 4, 5 | 7) + 
 L_\delta(1, 5) W_5^-(2, 3, 4 | 6, 7)
\label{cdel7.6} 
 \end{align}}
we need new building blocks $W_5^{\pm}$ that generalize the $W_4^{\pm}$ in (\ref{cdel6.5}). They are defined by, 
{\small  \begin{align}
 \label{cdel7.7a}
W_5^+(a,b,c,d,e) &=  
\frac{1}{2} \Big\{
W_2^+(a, b) W_3^+(c, d, e) + W_2^+(b, c) W_3^+(d, e, a) 
\no \\ & \qquad~
+ W_2^+(c, d) W_3^+(e, a, b)   + W_2^+(d, e) W_3^+(a, b, c) 
+  W_2^+(e, a) W_3^+(b, c, d) \Big\} \notag \\
    &\quad 
-  \frac{1}{2} \Big\{
W_2^-(a, c) W_3^-(d, e | b) + W_2^-(b, d) W_3^-(e, a | c) 
\no \\ & \qquad
+ W_2^-(c, e) W_3^-(a, b | d)  
+ W_2^-(a, d) W_3^-(b, c | e) 
+ W_2^-(b, e) W_3^-(c, d | a)\Big\}  
\no \\  &\quad ~ 
+  dx_a dx_c dx_d W_2^-(b, e) + dx_b dx_d dx_e W_2^-(a, c) + 
    dx_a dx_c dx_e W_2^-(b, d)  \notag \\
    &\quad ~  + dx_a dx_b dx_d W_2^-(c, e) + 
    dx_b dx_c dx_e W_2^-(a, d) - 4 dx_a dx_b dx_c dx_d dx_e
    \qquad
\end{align}}
as well as
{\small \begin{align}
W_5^-(a,b,c,d | e) &=  
\frac{1}{2} \Big\{W_2^+(a, b) W_3^-(c, d | e) + W_2^+(b, c) W_3^-(d, a | e) + 
    W_2^+(c, d) W_3^-(a, b | e)  \notag \\
    &\quad \quad \quad+ W_2^-(d, e) W_3^+(a, b, c) + 
    W_2^-(e, a) W_3^+(b, c, d)\Big\}  \notag \\
    &\quad  
    -  \frac{1}{2} \Big\{W_2^-(a, c) W_3^-(b, e | d) + 
    W_2^-(b, d) W_3^-(e, c | a) + W_2^+(c, e) W_3^-(a, b | d)
    \notag \\
    &\quad \quad \quad + 
    W_2^-(a, d) W_3^+(b, c, e) + W_2^+(b, e) W_3^-(c, d | a)\Big\} 
    \no \\
    &\quad   + 
 dx_a dx_c dx_d W_2^+(b, e) + dx_b dx_d dx_e W_2^-(a, c) + 
    dx_a dx_c dx_e W_2^-(b, d) 
    \notag \\
    &\quad  + dx_a dx_b dx_d W_2^+(c, e) + 
    dx_b dx_c dx_e W_2^-(a, d) - 4 dx_a dx_b dx_c dx_d dx_e
   \notag  \\
W_5^-(a,b, c|d,e) &=  
\frac{1}{2} \Big\{ W_2^+(a, b) W_3^-(d, e | c) + W_2^+(b, c) W_3^-(d, e | a) + 
    W_2^-(c, d) W_3^-(a, b | e) \notag \\
    &\quad \quad \quad + W_2^+(d, e) W_3^+(a, b, c) + 
    W_2^-(e, a) W_3^-(b, c | d)\Big\} \notag \\
    &\quad  
    - \frac{1}{2} \Big\{W_2^-(a, c) W_3^+(b, d, e) + W_2^+(b, d) W_3^-(e, c | a) + 
    W_2^+(c, e) W_3^+(a, b, d)  \notag \\
    &\quad \quad \quad+ W_2^+(a, d) W_3^+(b, c, e) + 
    W_2^+(b, e) W_3^-(a, d | c)\Big\} \no \\
    &\quad   + 
 dx_a dx_c dx_d W_2^+(b, e) + dx_b dx_d dx_e W_2^-(a, c) + 
    dx_a dx_c dx_e W_2^+(b, d)  \notag \\
    &\quad  + dx_a dx_b dx_d W_2^+(c, e) + 
    dx_b dx_c dx_e W_2^+(a, d)  - 4 dx_a dx_b dx_c dx_d dx_e
    \qquad
    \label{cdel7.7b} 
\end{align}}
Finally, the numerator on 7 simultaneous poles is given by
  \begin{align}
 {\cal N}^{(7)}[1234567] &= \bigg( \prod_{j=1}^7 \frac{ dx_j }{s_j} \bigg) \bigg ( s_1 s_2 \ndots s_7  
   + \sum_{1\leq i < j}^7 Z(i,j) s_1\ndots \widehat{s_i} \ndots \widehat{s_j} \ndots s_7  \notag \\
&\qquad\qquad + \sum_{1\leq i<j<k<l}^7 Z_4(i,j,k,l) s_1\ndots \widehat{s_i} \ndots \widehat{s_j} \ndots
 \widehat{s_k} \ndots \widehat{s_l} \ndots s_7 
\label{cdel7.8}  \\
&\qquad\qquad
+ \Big \{  s_1  Z_6(2,3,4,5,6,7) + {\rm cycl}(1,\cdots,7) \Big \}  \bigg )
\notag
   \end{align}
where the notation $\widehat{s_i}$ in the first two lines instructs to omit
the factor of $s_i$, and $Z_4$, $Z_6$ are defined in (\ref{z4def}), (\ref{Qdel.10}).

\subsection{Eight points}
\label{app.Cdel8}

The spin structure dependence of the cyclic product of eight Szeg\"o kernels is isolated using the expressions for $Q_\delta(1|2)$, $Q_\delta(1,3|2,4)$,  $Q_\delta(1,3,5|2,4,6)$ and $Q_\delta(1,3,5,7|2,4,6,8)$ provided in (\ref{5.C2}), (\ref{5.Q4}), (\ref{Qdel.13}) and  (\ref{Qdel.14}), respectively, leading to the following expression,
\begin{align}
C_\delta(1,\ndots,8) &=  \frac{1}{2} L_\delta(1,2) L_\delta(3,4) L_\delta(5,6) L_\delta(7,8)
+ \frac{1}{2} L_\delta(2,3) L_\delta(4,5) L_\delta(6,7) L_\delta(8,1) \notag \\
&\ +
\bigg\{ \frac{  {\cal N}^{(8)}_\delta[123] }{8 x_{12} x_{23} }
+  \frac{  {\cal N}^{(8)}_\delta[12,34] }{8 x_{12} x_{34} }
+  \frac{  {\cal N}^{(8)}_\delta[12,45] }{8 x_{12} x_{45} }
+  \frac{  {\cal N}^{(8)}_\delta[12,56] }{16 x_{12} x_{56} } 
+  \frac{  {\cal N}^{(8)}_\delta[12345] }{32 x_{12} x_{23}  x_{34} x_{45} }
\notag \\ & 
\quad \ \ +  \frac{  {\cal N}^{(8)}_\delta[1234,56] }{32 x_{12} x_{23}  x_{34} x_{56} }
+  \frac{  {\cal N}^{(8)}_\delta[1234,67] }{32 x_{12} x_{23}  x_{34} x_{67} }
+  \frac{  {\cal N}^{(8)}_\delta[1234,78] }{32 x_{12} x_{23}  x_{34} x_{78} }
+  \frac{  {\cal N}^{(8)}_\delta[123,45,78] }{32 x_{12} x_{23}  x_{45} x_{78} }
\no \\ & 
\quad\ \ +  \frac{  {\cal N}^{(8)}_\delta[123,56,78] }{32 x_{12} x_{23}  x_{56} x_{78} }
+  \frac{  {\cal N}^{(8)}_\delta[123,45,67] }{32 x_{12} x_{23}  x_{45} x_{67} }
+  \frac{  {\cal N}^{(8)}_\delta[123,456] }{32 x_{12} x_{23}  x_{45} x_{56} } 
+  \frac{  {\cal N}^{(8)}_\delta[123,567] }{64 x_{12} x_{23}  x_{56} x_{67} }
\no \\ & 
\quad\ \ +  \frac{  {\cal N}^{(8)}_\delta[12,34,56,78] }{128 x_{12} x_{34}  x_{56} x_{78} }
+  \frac{  {\cal N}^{(8)}_\delta[1234567] }{64 x_{12} x_{23} x_{34} x_{45}  x_{56} x_{67} }
+  \frac{  {\cal N}^{(8)}_\delta[123456,78] }{64 x_{12} x_{23} x_{34} x_{45}  x_{56} x_{78} }
\notag \\ & 
\quad\ \ +  \frac{  {\cal N}^{(8)}_\delta[12345,678] }{64 x_{12} x_{23} x_{34} x_{45}  x_{67} x_{78} }
+  \frac{  {\cal N}^{(8)}_\delta[1234,5678] }{128 x_{12} x_{23} x_{34} x_{56}  x_{67} x_{78} }+ {\rm cycl}(1,2,\ndots,8) \bigg\}   
\no \\ & 
\ + \frac{  {\cal N}^{(8)}[12345678] }{128 x_{12}x_{23} x_{34}x_{45}x_{56}x_{67} x_{78} x_{81} }
 \label{cdel8.1} 
\end{align}
The numerators along with two simultaneous poles are,
{\small \begin{align}
\label{cdel8.2}
 {\cal N}^{(8)}_\delta[123] &= W_2^+(1, 2) L_\delta(3, 4, 5, 6, 7, 8) + 
 W_2^-(1, 3) L_\delta(2, 4, 5, 6, 7, 8) 
 \no \\ & \quad 
 + W_2^+(2, 3) L_\delta(4, 5, 6, 7, 8, 1)
 \notag \\
 {\cal N}^{(8)}_\delta[12,34] &=  W_2^+(1, 4) L_\delta(2, 3, 5, 6, 7, 8) + 
 W_2^-(1, 3) L_\delta(2, 4, 5, 6, 7, 8)  \notag \\
 &\quad+ 
 W_2^+(2, 3) L_\delta(4, 5, 6, 7, 8, 1) + 
 W_2^-(2, 4) L_\delta(1, 3, 5, 6, 7, 8)
 \notag \\
 {\cal N}^{(8)}_\delta[12,45] &= W_2^+(1, 4) L_\delta(2, 3, 5, 6, 7, 8) + 
 W_2^-(1, 5) L_\delta(2, 3, 4, 6, 7, 8) \no  \\
 &\quad + 
 W_2^+(2, 5) L_\delta(1, 3, 4, 6, 7, 8) + 
 W_2^-(2, 4) L_\delta(1, 3, 5, 6, 7, 8)
 \notag \\
 {\cal N}^{(8)}_\delta[12,56] &=  W_2^+(1, 6) L_\delta(2, 3, 4, 5, 7, 8) + 
 W_2^-(1, 5) L_\delta(2, 3, 4, 6, 7, 8)  \notag \\
 &\quad+ 
 W_2^+(2, 5) L_\delta(1, 3, 4, 6, 7, 8) + 
 W_2^-(2, 6) L_\delta(1, 3, 4, 5, 7, 8)
\end{align}}
{where $W_2^{\pm}$ are defined in (\ref{5.W2}) and we introduced,
\begin{align}
L_\delta(a,b,c,d,e,f) &=
L_\delta(a,b) L_\delta(c,d) L_\delta(e,f) 
+ L_\delta(b,c) L_\delta(d,e) L_\delta(f,a) 
 \label{cdel8.3} 
\end{align}
see (\ref{cdel6.3}) and (\ref{cdel6.5}) for the analogous definition of $L_\delta(a,b,c,d)$
and $W_4^{\pm}$. The numerators along with four simultaneous poles are given by,}
{\small \begin{align}
 \label{cdel8.21}
{\cal N}^{(8)}_\delta[12345] &= 
W_4^+(1, 2, 3, 4) L_\delta(5, 6, 7, 8) + 
 W_4^-(1, 2, 3 | 5) L_\delta(4, 6, 7, 8) 
 \no \\ & \quad
 + W_4^-(1, 2 | 4, 5) L_\delta(3, 6, 7, 8) +  W_4^-(3, 4, 5 | 1) L_\delta(2, 6, 7, 8) 
 \notag \\
 &\quad + 
 W_4^+(2, 3, 4, 5) L_\delta(6, 7, 8, 1)
 \notag \\
 {\cal N}^{(8)}_\delta[1234,56] &= 
 W_4^-(1, 2, 3 | 5) L_\delta(4, 6, 7, 8) + 
 W_4^-(1, 2 | 4, 5) L_\delta(3, 6, 7, 8) 
 \notag \\
 &\quad+ 
 W_4^-(3, 4, 5 | 1) L_\delta(2, 6, 7, 8)  + 
 W_4^+(2, 3, 4, 5) L_\delta(6, 7, 8, 1) 
 \notag \\
 &\quad+ 
 W_4^+(1, 2, 3, 6) L_\delta(4, 5, 7, 8) + 
 W_4^-(6, 1, 2 | 4) L_\delta(3, 5, 7, 8)  
 \notag \\
 &\quad+ 
 W_4^-(3, 4 | 6, 1) L_\delta(2, 5, 7, 8) + 
 W_4^-(2, 3, 4 | 6) L_\delta(5, 7, 8, 1)
 \notag \\
 {\cal N}^{(8)}_\delta[1234,67] &= 
 W_4^+(1, 2, 3, 6) L_\delta(4, 5, 7, 8) +  W_4^-(6, 1, 2 | 4) L_\delta(3, 5, 7, 8) 
 \notag \\
 &\quad+ 
 W_4^-(3, 4 | 6, 1) L_\delta(2, 5, 7, 8) +  W_4^-(2, 3, 4 | 6) L_\delta(5, 7, 8, 1) 
 \notag \\
 &\quad + 
 W_4^-(1, 2, 3 | 7) L_\delta(4, 5, 6, 8) + 
 W_4^-(1, 2 | 4, 7) L_\delta(3, 5, 6, 8)  \notag \\
 &\quad+ 
 W_4^-(3, 4, 7 | 1) L_\delta(2, 5, 6, 8) + 
 W_4^+(2, 3, 4, 7) L_\delta(5, 6, 8, 1)
 \notag \\
  {\cal N}^{(8)}_\delta[1234,78] &=  
  W_4^-(1, 2, 3 | 7) L_\delta(4, 5, 6, 8) + 
 W_4^-(1, 2 | 4, 7) L_\delta(3, 5, 6, 8) 
 \notag \\
 &\quad+ 
 W_4^-(3, 4, 7 | 1) L_\delta(2, 5, 6, 8)  + 
 W_4^+(2, 3, 4, 7) L_\delta(5, 6, 8, 1) 
 \notag \\
 &\quad+ 
 W_4^+(8, 1, 2, 3) L_\delta(4, 5, 6, 7) + 
 W_4^-(8, 1, 2 | 4) L_\delta(3, 5, 6, 7)  \notag \\
 &\quad+ 
 W_4^-(3, 4 | 8, 1) L_\delta(2, 5, 6, 7) + 
 W_4^-(2, 3, 4 | 8) L_\delta(5, 6, 7, 1)
\no \\
  {\cal N}^{(8)}_\delta[123,45,78] &= 
    W_4^-(1, 2 | 4, 7) L_\delta(3, 5, 6, 8) +  W_4^-(8, 1, 2 | 4) L_\delta(3, 5, 6, 7) 
 \notag \\ & \quad 
 +  W_4^-(1, 2, 5 | 7) L_\delta(3, 4, 6, 8) +  W_4^+(1, 2, 5, 8) L_\delta(3, 4, 6, 7) 
 \no \\ & \quad 
 +  W_4^-(3, 4, 7 | 1) L_\delta(2, 5, 6, 8) +  W_4^-(8, 1 | 3, 4) L_\delta(2, 5, 6, 7) 
  \notag \\ & \quad
 +  W_4^-(1|3|5| 7)  L_\delta(2, 4, 6, 8) +  W_4^-(5, 8, 1 | 3) L_\delta(2, 4, 6, 7) 
 \notag \\ & \quad 
 +  W_4^+(2, 3, 4, 7) L_\delta(1, 5, 6, 8)   +  W_4^-(2, 3, 4 | 8) L_\delta(1, 5, 6, 7) 
 \no \\ & \quad + 
 W_4^-(7, 2, 3 | 5) L_\delta(1, 4, 6, 8) +   W_4^-(2, 3 | 5, 8) L_\delta(1, 4, 6, 7)
 \notag \\
     {\cal N}^{(8)}_\delta[123,56,78] &= 
 W_4^-(1, 2, 5 | 7) L_\delta(3, 4, 6, 8) +  W_4^-(1|3|5| 7)  L_\delta(2, 4, 6, 8) 
 \notag \\ & \quad
 +  W_4^-(7, 2, 3 | 5) L_\delta(1, 4, 6, 8)  +  W_4^+(8, 1, 2, 5) L_\delta(3, 4, 6, 7) 
 \no \\ & \quad
 +  W_4^-(5, 8, 1 | 3) L_\delta(2, 4, 6, 7) +  W_4^-(2, 3 | 5, 8) L_\delta(1, 4, 6, 7)  
 \notag \\ & \quad
+  W_4^-(1, 2 | 6, 7) L_\delta(3, 4, 5, 8) +  W_4^-(3, 6, 7 | 1) L_\delta(2, 4, 5, 8) 
\no \\ & \quad 
+  W_4^+(2, 3, 6, 7) L_\delta(1, 4, 5, 8)  +  W_4^-(8, 1, 2 | 6) L_\delta(3, 4, 5, 7) 
\no \\ & \quad 
+  W_4^-(8, 1 | 3, 6) L_\delta(2, 4, 5, 7) +  W_4^-(2, 3, 6 | 8) L_\delta(1, 4, 5, 7)
 \notag \\
     {\cal N}^{(8)}_\delta[123,45,67] &= 
 W_4^-(6, 1, 2 | 4) L_\delta(3, 5, 7, 8) + 
 W_4^-(6, 1 | 3, 4) L_\delta(2, 5, 7, 8) 
 \no \\ & \quad 
 +  W_4^-(2, 3, 4 | 6) L_\delta(1, 5, 7, 8)  +  W_4^-(1, 2 | 4, 7) L_\delta(3, 5, 6, 8) 
 \no \\ & \quad
 +  W_4^-(3, 4, 7 | 1) L_\delta(2, 5, 6, 8) +  W_4^+(2, 3, 4, 7) L_\delta(1, 5, 6, 8) 
 \notag \\
 &\quad + 
 W_4^+(1, 2, 5, 6) L_\delta(3, 4, 7, 8) + 
 W_4^-(5, 6, 1 | 3) L_\delta(2, 4, 7, 8) 
 \no \\ & \quad 
 + W_4^-(2, 3 | 5, 6) L_\delta(1, 4, 7, 8)  + W_4^-(1, 2, 5 | 7) L_\delta(3, 4, 6, 8) 
 \no \\ & \quad
 + W_4^-(1|3|5| 7) L_\delta(2, 4, 6, 8) +  W_4^-(7, 2, 3 | 5) L_\delta(1, 4, 6, 8)
\end{align}}
as well as,
{\small  \begin{align}
  {\cal N}^{(8)}_\delta[123,456] &= 
W_4^-(1, 2 | 4, 5) L_\delta(3, 6, 7, 8) +  W_4^-(6, 1, 2 | 4) L_\delta(3, 5, 7, 8) 
\no \\ & \quad
+  W_4^+(1, 2, 5, 6) L_\delta(3, 4, 7, 8)  +  W_4^-(3, 4, 5 | 1) L_\delta(2, 6, 7, 8) 
 \notag \\
 &\quad+ 
 W_4^-(6, 1 | 3, 4) L_\delta(2, 5, 7, 8) +  W_4^-(5, 6, 1 | 3) L_\delta(2, 4, 7, 8)  
 \notag \\
 &\quad+ 
 W_4^+(2, 3, 4, 5) L_\delta(1, 6, 7, 8) + 
 W_4^-(2, 3, 4 | 6) L_\delta(1, 5, 7, 8) 
 \no \\ & \quad 
 + W_4^-(2, 3 | 5, 6) L_\delta(1, 4, 7, 8)
 \notag \\
{\cal N}^{(8)}_\delta[123,567]&= 
W_4^+(1, 2, 5, 6) L_\delta(3, 4, 7, 8) +  W_4^-(5, 6, 1 | 3) L_\delta(2, 4, 7, 8) 
 \notag \\
 &\quad+ 
 W_4^-(2, 3 | 5, 6) L_\delta(1, 4, 7, 8)  +  W_4^-(1, 2, 5 | 7) L_\delta(3, 4, 6, 8) 
 \notag \\
 &\quad+ 
 W_4^-(1|3|5| 7) L_\delta(2, 4, 6, 8) +  W_4^-(7, 2, 3 | 5) L_\delta(1, 4, 6, 8)  
 \notag \\
 &\quad+ 
 W_4^-(1, 2 | 6, 7) L_\delta(3, 4, 5, 8) + 
 W_4^-(3, 6, 7 | 1) L_\delta(2, 4, 5, 8) 
 \notag \\
 &\quad+ 
 W_4^+(2, 3, 6, 7) L_\delta(1, 4, 5, 8)
\no \\
         {\cal N}^{(8)}_\delta[12,34,56,78]&= 
 W_4^-(1|3|5| 7) L_\delta(2, 4, 6, 8) +  W_4^-(5, 8, 1 | 3) L_\delta(2, 4, 6, 7) 
 \notag \\
 &\quad
 +  W_4^-(3, 6, 7 | 1) L_\delta(2, 4, 5, 8)  + 
 W_4^-(8, 1 | 3, 6) L_\delta(2, 4, 5, 7) 
 \notag \\
 &\quad+ 
 W_4^-(1, 4, 5 | 7) L_\delta(2, 3, 6, 8) + 
 W_4^+(1, 4, 5, 8) L_\delta(2, 3, 6, 7)  
 \notag \\
 &\quad+ 
 W_4^-(6, 7 | 1, 4) L_\delta(2, 3, 5, 8) + 
 W_4^-(8, 1, 4 | 6) L_\delta(2, 3, 5, 7) 
 \notag \\
 &\quad+ 
 W_4^-(7, 2, 3 | 5) L_\delta(1, 4, 6, 8)  + 
 W_4^-(5, 8 | 2, 3) L_\delta(1, 4, 6, 7) 
  \notag \\
 &\quad+ 
 W_4^+(2, 3, 6, 7) L_\delta(1, 4, 5, 8) +  W_4^-(2, 3, 6 | 8) L_\delta(1, 4, 5, 7)  
 \notag \\
 &\quad+ 
 W_4^-(4, 5 | 7, 2) L_\delta(1, 3, 6, 8) + 
 W_4^-(4, 5, 8 | 2) L_\delta(1, 3, 6, 7)  \notag \\
 &\quad+ 
 W_4^-(6, 7, 2 | 4) L_\delta(1, 3, 5, 8) + 
 W_4^-(2 | 4 | 6 | 8) L_\delta(1, 3, 5, 7) \label{cdel8.22}
 \end{align}}
Finally, the numerators on six simultaneous poles are,
{\small \begin{align}
{\cal N}^{(8)}_\delta[1234567]&=  W_6^+(1, 2, 3, 4, 5, 6) L_\delta(7, 8) + 
 W_6^-(1, 2, 3, 4, 5 | 7) L_\delta(6, 8) \notag \\
 &\quad + 
 W_6^-(1, 2, 3, 4 | 6, 7) L_\delta(5, 8) + 
 W_6^-(1, 2, 3 | 5, 6, 7) L_\delta(4, 8) \notag \\
 &\quad + 
 W_6^-(4, 5, 6, 7 | 1, 2) L_\delta(3, 8) + 
 W_6^-(3, 4, 5, 6, 7 | 1) L_\delta(2, 8)  \notag \\
 &\quad+ 
 W_6^+(2, 3, 4, 5, 6, 7) L_\delta(1, 8)
\no \\
 {\cal N}^{(8)}_\delta[123456,78]&= W_6^-(1, 2, 3, 4, 5 | 7) L_\delta(6, 8) + 
 W_6^-(1, 2, 3, 4 | 6, 7) L_\delta(5, 8)  \notag \\
 &\quad+ 
 W_6^-(1, 2, 3 | 5, 6, 7) L_\delta(4, 8) + 
 W_6^-(4, 5, 6, 7 | 1, 2) L_\delta(3, 8)  \notag \\
 &\quad+ 
 W_6^-(3, 4, 5, 6, 7 | 1) L_\delta(2, 8) + 
 W_6^+(2, 3, 4, 5, 6, 7) L_\delta(1, 8) \notag \\
 &\quad + 
 W_6^+(8, 1, 2, 3, 4, 5) L_\delta(6, 7) + 
 W_6^-(8, 1, 2, 3, 4 | 6) L_\delta(5, 7)  \notag \\
 &\quad+ 
 W_6^-(8, 1, 2, 3 | 5, 6) L_\delta(4, 7) + 
 W_6^-(8, 1, 2 | 4, 5, 6) L_\delta(3, 7)  \notag \\
 &\quad+ 
 W_6^-(3, 4, 5, 6 | 8, 1) L_\delta(2, 7) + 
 W_6^-(2, 3, 4, 5, 6 | 8) L_\delta(1, 7)
 \label{cdel8.31}
 \end{align}}
 as well as
{\small  \begin{align}
 {\cal N}^{(8)}_\delta[12345,678] &= W_6^-(1, 2, 3, 4 | 6, 7) L_\delta(5, 8) + 
 W_6^-(1, 2, 3 | 5, 6, 7) L_\delta(4, 8)  \notag \\
 &\quad+ 
 W_6^-(4, 5, 6, 7 | 1, 2) L_\delta(3, 8) + 
 W_6^-(3, 4, 5, 6, 7 | 1) L_\delta(2, 8) \notag \\
 &\quad + 
 W_6^+(2, 3, 4, 5, 6, 7) L_\delta(1, 8) + 
 W_6^-(8, 1, 2, 3, 4 | 6) L_\delta(5, 7) \notag \\
 &\quad + 
 W_6^-(8, 1, 2, 3 | 5, 6) L_\delta(4, 7) + 
 W_6^-(8, 1, 2 | 4, 5, 6) L_\delta(3, 7)  \notag \\
 &\quad+ 
 W_6^-(3, 4, 5, 6 | 8, 1) L_\delta(2, 7) + 
 W_6^-(2, 3, 4, 5, 6 | 8) L_\delta(1, 7)  \notag \\
 &\quad+ 
 W_6^+(7, 8, 1, 2, 3, 4) L_\delta(5, 6) + 
 W_6^-(7, 8, 1, 2, 3 | 5) L_\delta(4, 6)  \notag \\
 &\quad+ 
 W_6^-(7, 8, 1, 2 | 4, 5) L_\delta(3, 6) + 
 W_6^-(3, 4, 5 | 7, 8, 1) L_\delta(2, 6) \notag \\
 &\quad + 
 W_6^-(2, 3, 4, 5 | 7, 8) L_\delta(1, 6)
\no \\
 {\cal N}^{(8)}_\delta[1234,5678]&= W_6^-(1, 2, 3 | 5, 6, 7) L_\delta(4, 8) + 
 W_6^-(4, 5, 6, 7 | 1, 2) L_\delta(3, 8) \notag \\
 &\quad + 
 W_6^-(3, 4, 5, 6, 7 | 1) L_\delta(2, 8) + 
 W_6^+(2, 3, 4, 5, 6, 7) L_\delta(1, 8) \notag \\
 &\quad + 
 W_6^-(8, 1, 2, 3 | 5, 6) L_\delta(4, 7) + 
 W_6^-(8, 1, 2 | 4, 5, 6) L_\delta(3, 7) \notag \\
 &\quad + 
 W_6^-(3, 4, 5, 6 | 8, 1) L_\delta(2, 7) + 
 W_6^-(2, 3, 4, 5, 6 | 8) L_\delta(1, 7) \notag \\
 &\quad + 
 W_6^-(7, 8, 1, 2, 3 | 5) L_\delta(4, 6) + 
 W_6^-(7, 8, 1, 2 | 4, 5) L_\delta(3, 6)  \notag \\
 &\quad+ 
 W_6^-(3, 4, 5 | 7, 8, 1) L_\delta(2, 6) + 
 W_6^-(2, 3, 4, 5 | 7, 8) L_\delta(1, 6)  \notag \\
 &\quad+ 
 W_6^+(6, 7, 8, 1, 2, 3) L_\delta(4, 5) + 
 W_6^-(6, 7, 8, 1, 2 | 4) L_\delta(3, 5)  \notag \\
 &\quad+ 
 W_6^-(6, 7, 8, 1 | 3, 4) L_\delta(2, 5) + 
 W_6^-(2, 3, 4 | 6, 7, 8) L_\delta(1, 5)
 \label{cdel8.32}
\end{align}}
Apart from the building block of the six-point cyclic product in (\ref{cdel6.6}),
 {\small \begin{align}
W_6^+(1,2,3,4,5,6) &= W_2^+(1,2) W_2^+(3,4) W_2^+(5,6)
+ W_2^+(2,3) W_2^+(4,5) W_2^+(6,1) \notag \\
&\quad - W_2^+(1,2) W_2^-(3,5) W_2^-(4,6)
- W_2^+(2,3) W_2^-(4,6) W_2^-(1,5)\notag \\
&\quad - W_2^+(3,4) W_2^-(1,5) W_2^-(2,6)
- W_2^+(4,5) W_2^-(2,6) W_2^-(1,3)
\notag \\
&\quad - W_2^+(5,6) W_2^-(1,3) W_2^-(2,4)
- W_2^+(1,6) W_2^-(2,4) W_2^-(3,5)  \notag \\
&\quad + W_2^+(1,4) W_2^+(2,3) W_2^+(5,6)
 + W_2^+(2,5) W_2^+(3,4) W_2^+(1,6) \notag \\
&\quad + W_2^+(3,6) W_2^+(4,5) W_2^+(1,2)
+ W_2^+(1,4) W_2^-(2,6) W_2^-(3,5) \notag \\
&\quad + W_2^+(2,5) W_2^-(1,3) W_2^-(4,6)
+ W_2^+(3,6) W_2^-(2,4) W_2^-(1,5) \notag \\
&\quad -  W_2^+(1,4) W_2^+(2,5) W_2^+(3,6) 
 \label{cdel8.33} 
\end{align}}
we employed three additional variants in (\ref{cdel8.31}) and (\ref{cdel8.32}), namely,
{\small \begin{align}
W_6^-(1,2,3,4,5|6) &= W_2^+(1,2) W_2^+(3,4) W_2^-(5,6)
+ W_2^+(2,3) W_2^+(4,5) W_2^-(6,1) \notag \\
&\quad - W_2^+(1,2) W_2^-(3,5) W_2^+(4,6)
- W_2^+(2,3) W_2^+(4,6) W_2^-(1,5)\notag \\
&\quad - W_2^+(3,4) W_2^-(1,5) W_2^+(2,6)
- W_2^+(4,5) W_2^+(2,6) W_2^-(1,3)
\notag \\
&\quad - W_2^-(5,6) W_2^-(1,3) W_2^-(2,4)
- W_2^-(1,6) W_2^-(2,4) W_2^-(3,5)  \notag \\
&\quad + W_2^+(1,4) W_2^+(2,3) W_2^-(5,6)
 + W_2^+(2,5) W_2^+(3,4) W_2^-(1,6) \notag \\
&\quad + W_2^-(3,6) W_2^+(4,5) W_2^+(1,2)
+ W_2^+(1,4) W_2^+(2,6) W_2^-(3,5) \notag \\
&\quad + W_2^+(2,5) W_2^-(1,3) W_2^+(4,6)
+ W_2^-(3,6) W_2^-(2,4) W_2^-(1,5) \notag \\
&\quad -  W_2^+(1,4) W_2^+(2,5) W_2^-(3,6) 
\notag \\
W_6^-(1,2,3,4|5,6) &= W_2^+(1,2) W_2^+(3,4) W_2^+(5,6)
+ W_2^+(2,3) W_2^-(4,5) W_2^-(6,1) \notag \\
&\quad - W_2^+(1,2) W_2^+(3,5) W_2^+(4,6)
- W_2^+(2,3) W_2^+(4,6) W_2^+(1,5)\notag \\
&\quad - W_2^+(3,4) W_2^+(1,5) W_2^+(2,6)
- W_2^-(4,5) W_2^+(2,6) W_2^-(1,3)
\notag \\
&\quad - W_2^+(5,6) W_2^-(1,3) W_2^-(2,4)
- W_2^-(1,6) W_2^-(2,4) W_2^+(3,5)  \notag \\
&\quad + W_2^+(1,4) W_2^+(2,3) W_2^+(5,6)
 + W_2^-(2,5) W_2^+(3,4) W_2^-(1,6) \notag \\
&\quad + W_2^-(3,6) W_2^-(4,5) W_2^+(1,2)
+ W_2^+(1,4) W_2^+(2,6) W_2^+(3,5) \notag \\
&\quad + W_2^-(2,5) W_2^-(1,3) W_2^+(4,6)
+ W_2^-(3,6) W_2^-(2,4) W_2^+(1,5) \notag \\
&\quad -  W_2^+(1,4) W_2^-(2,5) W_2^-(3,6)
\notag \\
W_6^-(1,2,3|4,5,6) &= W_2^+(1,2) W_2^-(3,4) W_2^+(5,6)
+ W_2^+(2,3) W_2^+(4,5) W_2^-(6,1) \notag \\
&\quad - W_2^+(1,2) W_2^+(3,5) W_2^-(4,6)
- W_2^+(2,3) W_2^-(4,6) W_2^+(1,5)\notag \\
&\quad - W_2^-(3,4) W_2^+(1,5) W_2^+(2,6)
- W_2^+(4,5) W_2^+(2,6) W_2^-(1,3)
\notag \\
&\quad - W_2^+(5,6) W_2^-(1,3) W_2^+(2,4)
- W_2^-(1,6) W_2^+(2,4) W_2^+(3,5)  \notag \\
&\quad + W_2^-(1,4) W_2^+(2,3) W_2^+(5,6)
 + W_2^-(2,5) W_2^-(3,4) W_2^-(1,6) \notag \\
&\quad + W_2^-(3,6) W_2^+(4,5) W_2^+(1,2)
+ W_2^-(1,4) W_2^+(2,6) W_2^+(3,5) \notag \\
&\quad + W_2^-(2,5) W_2^-(1,3) W_2^-(4,6)
+ W_2^-(3,6) W_2^+(2,4) W_2^+(1,5) \notag \\
&\quad -  W_2^-(1,4) W_2^-(2,5) W_2^-(3,6) 
 \label{cdel8.36}
\end{align}}
Finally, the $\delta$-independent numerator of the Parke-Taylor factor in (\ref{cdel8.1}) is given by
\begin{align}
{\cal N}^{(8)}[12345678]  &=  \bigg( \prod_{j=1}^8 \frac{ dx_j }{s_j} \bigg) \bigg\{ s_1 s_2 \ndots s_8  
   + \sum_{1\leq i < j}^8 Z(i,j) s_1\ndots \widehat{s_i} \ndots \widehat{s_j} \ndots s_8  \notag \\
&\quad + \sum_{1\leq i<j<k<l}^8 Z_4(i,j,k,l) s_1\ndots \widehat{s_i} \ndots \widehat{s_j} \ndots
\widehat{s_k} \ndots \widehat{s_l} \ndots s_8
\label{cdel8.0}  \\
&\quad
+  \sum_{1\leq i < j}^8 Z_6(1,\ndots, \widehat{i},\ndots,\widehat{ j},\ndots,8) s_i s_j
+ Z_8(1,2,\ndots,8)
 \bigg\} \notag
\end{align}

\subsection{The spin structure sum of the eight-point $C_\delta$}
\label{sec:spinsum8}

This appendix gathers the numerators $\mathfrak{N}^{(8)}$ in the eight-point
spin structure sum (\ref{cdel8.A}).

\sm 

The numerator functions with two simultaneous poles are given by, 
{\small \begin{align}
 \mathfrak{N}^{(8)}[123] &= W_2^+(1, 2) M_1(3, 4, 5, 6, 7, 8) + 
 W_2^-(1, 3) M_1(2, 4, 5, 6, 7, 8) 
 \no \\ & \quad + 
 W_2^+(2, 3)M_1(4, 5, 6, 7, 8, 1)
 \notag \\
 \mathfrak{N}^{(8)}[12,34] &=  W_2^+(1, 4) M_1(2, 3, 5, 6, 7, 8) + 
 W_2^-(1, 3) M_1(2, 4, 5, 6, 7, 8)  \notag \\
 &\quad+ 
 W_2^+(2, 3) M_1(4, 5, 6, 7, 8, 1) + 
 W_2^-(2, 4) M_1(1, 3, 5, 6, 7, 8)
 \notag \\
 \mathfrak{N}^{(8)}[12,45] &= W_2^+(1, 4) M_1(2, 3, 5, 6, 7, 8) + 
 W_2^-(1, 5) M_1(2, 3, 4, 6, 7, 8)  \label{cdel8.B} \\
 &\quad + 
 W_2^+(2, 5) M_1(1, 3, 4, 6, 7, 8) + 
 W_2^-(2, 4) M_1(1, 3, 5, 6, 7, 8)
 \notag \\
 \mathfrak{N}^{(8)}[12,56] &=  W_2^+(1, 6) M_1(2, 3, 4, 5, 7, 8) + 
 W_2^-(1, 5) M_1(2, 3, 4, 6, 7, 8)  \notag \\
 &\quad+ 
 W_2^+(2, 5) M_1(1, 3, 4, 6, 7, 8) + 
 W_2^-(2, 6) M_1(1, 3, 4, 5, 7, 8)
 \notag
\end{align}}
The remaining numerators all have four simultaneous poles and are given by,
{\small \begin{align}
 \mathfrak{N}^{(8)}[12345] &= 
 W_4^+(1, 2, 3, 4)  \Delta(5,7|6,8) +  W_4^-(1, 2, 3 | 5)   \Delta(4,7|6,8) 
 \no \\ & \quad 
 +  W_4^-(1, 2 | 4, 5)  \Delta(3,7|6,8) +  W_4^-(3, 4, 5 | 1)  \Delta(2,7|6,8)
 \no \\ & \quad
 +  W_4^+(2, 3, 4, 5)  \Delta(6,8|7,1)
 \notag \\
 \mathfrak{N}^{(8)}[1234,56] &= 
 W_4^-(1, 2, 3 | 5)   \Delta(4,7|6,8)+  W_4^-(1, 2 | 4, 5)   \Delta(3,7|6,8)
 \no \\ & \quad
 +  W_4^-(3, 4, 5 | 1) \Delta(2,7|6,8)  + W_4^+(2, 3, 4, 5) \Delta(6,8|7,1) 
 \no \\ & \quad
 +  W_4^+(1, 2, 3, 6)  \Delta(4,7|5,8) +  W_4^-(6, 1, 2 | 4)  \Delta(3,7|5,8)   
 \no \\ & \quad
 + W_4^-(3, 4 | 6, 1) \Delta(2,7|5,8) +  W_4^-(2, 3, 4 | 6)  \Delta(5,8|7,1) \notag \\
  \mathfrak{N}^{(8)}[1234,67] &= 
 W_4^+(1, 2, 3, 6)   \Delta(4,7|5,8)+  W_4^-(6, 1, 2 | 4)   \Delta(3,7|5,8)
 \no \\ & \quad
 +  W_4^-(3, 4 | 6, 1)   \Delta(2,7|5,8) +  W_4^-(2, 3, 4 | 6)  \Delta(5,8|7,1) 
 \no \\ & \quad 
 +  W_4^-(1, 2, 3 | 7)  \Delta(4,6|5,8)+  W_4^-(1, 2 | 4, 7)  \Delta(3,6|5,8)  
 \no \\ & \quad 
 +  W_4^-(3, 4, 7 | 1)  \Delta(2,6|5,8)+  W_4^+(2, 3, 4, 7)  \Delta(5,8|6,1) \notag \\
  \mathfrak{N}^{(8)}[1234,78] &=  
 W_4^-(1, 2, 3 | 7)  \Delta(4,6|5,8) +  W_4^-(1, 2 | 4, 7)  \Delta(3,6|5,8) 
 \no \\ & \quad
 +  W_4^-(3, 4, 7 | 1)   \Delta(2,6|5,8) +  W_4^+(2, 3, 4, 7)  \Delta(5,8|6,1) 
  \no \\ & \quad
 +  W_4^+(8, 1, 2, 3)  \Delta(4,6|5,7) +  W_4^-(8, 1, 2 | 4)  \Delta(3,6|5,7)  
 \no \\ & \quad
 +  W_4^-(3, 4 | 8, 1)  \Delta(2,6|5,7) +  W_4^-(2, 3, 4 | 8)  \Delta(5,7|6,1) \notag \\
   \mathfrak{N}^{(8)}[123,45,78] &= 
 W_4^-(1, 2 | 4, 7)  \Delta(3,6|5,8) +  W_4^-(8, 1, 2 | 4)  \Delta(3,6|5,7) 
 \no \\ & \quad
 + W_4^-(1, 2, 5 | 7)  \Delta(3,6|4,8) + W_4^+(1, 2, 5, 8)  \Delta(3,6|4,7) 
 \no \\ & \quad
 +  W_4^-(3, 4, 7 | 1)  \Delta(2,6|5,8) +  W_4^-(8, 1 | 3, 4)  \Delta(2,6|5,7) 
 \no \\ & \quad 
 + W_4^-(1|3|5| 7)   \Delta(2,6|4,8) +  W_4^-(5, 8, 1 | 3)  \Delta(2,6|4,7) 
 \no \\ & \quad 
 + W_4^+(2, 3, 4, 7)  \Delta(1,6|5,8) +  W_4^-(2, 3, 4 | 8)  \Delta(1,6|5,7) 
 \no \\ & \quad
 +  W_4^-(7, 2, 3 | 5)  \Delta(1,6|4,8) +  W_4^-(2, 3 | 5, 8)  \Delta(1,6|4,7)
 \end{align}}
as well as
{\small \begin{align}
   \mathfrak{N}^{(8)}[123,56,78] &= 
W_4^-(1, 2, 5 | 7)    \Delta(3,6|4,8)+  W_4^-(1|3|5| 7)   \Delta(2,6|4,8) 
\no \\ & \quad
+  W_4^-(7, 2, 3 | 5)  \Delta(1,6|4,8)  +  W_4^+(8, 1, 2, 5)  \Delta(3,6|4,7) 
\no \\ & \quad 
+  W_4^-(5, 8, 1 | 3)  \Delta(2,6|4,7) 
+ W_4^-(2,3|5,8) \Delta(1, 6 | 4, 7)
\no \\ & \quad
+  W_4^-(1, 2 | 6, 7)  \Delta(3,5|4,8) +  W_4^-(3, 6, 7 | 1)  \Delta(2,5|4,8) 
\no \\ & \quad 
+  W_4^+(2, 3, 6, 7)  \Delta(1,5|4,8)  +  W_4^-(8, 1, 2 | 6)  \Delta(3,5|4,7) 
\no \\ & \quad
+  W_4^-(8, 1 | 3, 6)  \Delta(2,5|4,7)+  W_4^-(2, 3, 6 | 8)  \Delta(1,5|4,7)
 \notag \\
 \mathfrak{N}^{(8)}[123,45,67] &= 
 W_4^-(6, 1, 2 | 4)  \Delta(3,7|5,8) +  W_4^-(6, 1 | 3, 4)  \Delta(2,7|5,8) 
 \no \\ & \quad
 + W_4^-(2, 3, 4 | 6)  \Delta(1,7|5,8) +  W_4^-(1, 2 | 4, 7)  \Delta(3,6|5,8) 
 \no \\ & \quad
 +  W_4^-(3, 4, 7 | 1)  \Delta(2,6|5,8)+  W_4^+(2, 3, 4, 7)  \Delta(1,6|5,8) 
 \no \\ & \quad 
 + W_4^+(1, 2, 5, 6)  \Delta(3,7|4,8)+  W_4^-(5, 6, 1 | 3) \Delta(2,7|4,8) 
 \no \\ & \quad
 +  W_4^-(2, 3 | 5, 6)  \Delta(1,7|4,8) +  W_4^-(1, 2, 5 | 7)  \Delta(3,6|4,8) 
 \no \\ & \quad
 + W_4^-(1|3|5| 7) \Delta(2,6|4,8) +  W_4^-(7, 2, 3 | 5)  \Delta(1,6|4,8)
\notag \\
      \mathfrak{N}^{(8)}[123,456] &= 
W_4^-(1, 2 | 4, 5)  \Delta(3,7|6,8) +  W_4^-(6, 1, 2 | 4)  \Delta(3,7|5,8) 
\no \\ & \quad
+  W_4^+(1, 2, 5, 6)  \Delta(3,7|4,8) +  W_4^-(3, 4, 5 | 1)  \Delta(2,7|6,8) 
\no \\ & \quad
+  W_4^-(6, 1 | 3, 4)  \Delta(2,7|5,8) +  W_4^-(5, 6, 1 | 3)  \Delta(2,7|4,8) 
\no \\ & \quad
+  W_4^+(2, 3, 4, 5) \Delta(1,7|6,8) +  W_4^-(2, 3, 4 | 6)  \Delta(1,7|5,8) 
\no \\ & \quad
+  W_4^-(2, 3 | 5, 6)  \Delta(1,7|4,8) \notag \\
 \mathfrak{N}^{(8)}[123,567]&= 
 W_4^+(1, 2, 5, 6)  \Delta(3,7|4,8) +  W_4^-(5, 6, 1 | 3)  \Delta(2,7|4,8) 
 \no \\ & \quad
 + W_4^-(2, 3 | 5, 6)  \Delta(1,7|4,8)  +  W_4^-(1, 2, 5 | 7)  \Delta(3,6|4,8) 
 \no \\ & \quad
 +  W_4^-(1|3|5| 7)  \Delta(2,6|4,8) +  W_4^-(7, 2, 3 | 5)  \Delta(1,6|4,8) 
\no \\ & \quad
+  W_4^-(1, 2 | 6, 7)  \Delta(3,5|4,8) +  W_4^-(3, 6, 7 | 1) \Delta(2,5|4,8) 
\no \\ & \quad 
+  W_4^+(2, 3, 6, 7)  \Delta(1,5|4,8) \notag \\
  \mathfrak{N}^{(8)}[12,34,56,78]&= 
W_4^-(1|3|5| 7)  \Delta(2,6|4,8) +  W_4^-(5, 8, 1 | 3)  \Delta(2,6|4,7) 
\no \\ & \quad
+  W_4^-(3, 6, 7 | 1)  \Delta(2,5|4,8)  +  W_4^-(8, 1 | 3, 6)  \Delta(2,5|4,7) 
\no \\ & \quad
+  W_4^-(1, 4, 5 | 7)  \Delta(2,6|3,8) +  W_4^+(1, 4, 5, 8)  \Delta(2,6|3,7) 
\no \\ & \quad
+  W_4^-(6, 7 | 1, 4)  \Delta(2,5|3,8) +  W_4^-(8, 1, 4 | 6)  \Delta(2,5|3,7) 
\no \\ & \quad
+  W_4^-(7, 2, 3 | 5)  \Delta(1,6|4,8) + W_4^-(5, 8 | 2, 3)  \Delta(1,6|4,7) 
\no \\ & \quad 
+  W_4^+(2, 3, 6, 7)  \Delta(1,5|4,8)+  W_4^-(2, 3, 6 | 8)  \Delta(1,5|4,7)  
\no \\ & \quad 
+  W_4^-(4, 5 | 7, 2)  \Delta(1,6|3,8)+  W_4^-(4, 5, 8 | 2) \Delta(1,6|3,7) 
\no \\ & \quad
+  W_4^-(6, 7, 2 | 4)  \Delta(1,5|3,8) +  W_4^-(2 | 4 | 6 | 8)  \Delta(1,5|3,7)
\label{cdel8.E}
\end{align}}

\newpage
 
\section{Symmetrized cyclic products of Szeg\"o kernels}
\label{sec:group}
\setcounter{equation}{0}

The purpose of this appendix is to obtain an $SL(2,\CC)$ group-theoretic decomposition of the symmetrized cyclic product  of Szeg\"o kernels, which we define  as follows, 
\begin{align}
\SZ (1,2, \cdots, n) &= { 1 \over n!} \sum _{\sigma \in \mS_n} S_\delta (\sigma (1) , \sigma (2)) \, S_\delta (\sigma (2) , \sigma (3))
\cdots S_\delta (\sigma (n), \sigma (1)) 
\notag
\\
&=
{ 1 \over (n{-}1)!} \sum _{\sigma \in \mS_{n-1}} S_\delta (1 , \sigma (2)) \, S_\delta (\sigma (2) , \sigma (3))
\cdots S_\delta (\sigma (n), 1) 
\label{defsymmcyc}
\end{align}
where $\mS_n$ is the group of permutations of $n$ elements, and we have inserted the customary $n!$ normalization factor. By construction, $\SZ$ is a symmetric function of its arguments $z_1, \cdots, z_n$.   The expression in the second line follows from the cyclic invariance of the product over $S_\delta (\sigma (j) , \sigma (j{+}1))$ in the first line.
The simplicity of the function $\SZ$ is illustrated by the following lemma.

{\lem
The symmetrized cyclic product  $\SZ(1,\cdots, n)$ vanishes for  $n$  odd and is a holomorphic $(1,0)$ form in each point $z_i$ for even $n\geq 4$. It may be decomposed as follows,
\bea
\SZ(1,\cdots, n)= \LL_\delta ^{a_1 \cdots a_n} \, \varpi_{a_1} (1) \cdots \varpi_{a_n}(n)
\label{defltensor}
\eea
where $\LL_\delta^{a_1 \cdots a_n}$ is a $z_i$-independent rank $n$ symmetric $SL(2,\CC)$ tensor whose components are polynomials in $\mu_m$}.

\sm

The proof proceeds as follows. If $n$ is odd, the sum over permutations of every concatenated product 
includes also the product where the cycle is traversed in the opposite direction. But since $S_\delta$ is odd under interchange of its arguments, and the number of Szeg\"o kernels in the product is odd, we conclude that the whole sum must vanish.  

\sm

For even $n$ in turn, we start from the sum over permutations  $\mS_{n-1}$ of $\{ 2,3, \cdots, n\}$ in the second line of (\ref{defsymmcyc}). To show holomorphicity in the $z_i$ for $n\geq 4$, it suffices to show that $\SZ$ has no poles in $z_1$ at the point $z_2$ since the function $\SZ$ is symmetric in all $z_i$. The pole in $z_1$ at $z_2$ receives contributions from those permutations $\sigma \in \mS_{n-1}$  that have either $\sigma(2)=2$ or $\sigma(n)=2$, 
 \bea
\SZ (1,2, \cdots, n) & = & 
{S_\delta(1,2) \over (n{-}1)!}    \sum _{\sigma \in \mS_{n-2}} S_\delta (2 , \sigma (3)) \cdots S_\delta (\sigma (n), 1)
\label{holoprf} \\ && 
- {S_\delta(1,2) \over (n{-}1)!}   \sum _{\sigma \in \mS_{n-2}} S_\delta (1 , \sigma (2)) 
\cdots S_\delta (\sigma (n{-}1), 2) + {\rm regular} \ {\rm in} \ z_1{-}z_2
\notag
\eea
where $\sigma$ in the first sum permutes the points $z_3, \cdots, z_n$ while $\sigma$ in the second sum maps the points $z_2, \cdots,  z_{n-1}$ to the points $z_3, \cdots, z_n$ and then permutes those points as  in the first sum.
To evaluate the residue at the pole, we set $z_2=z_1$ under the two summation signs, and verify that the sums cancel one another. Being a holomorphic $(1,0)$-form in each $z_i$, it is immediate that $\SZ$ admits the decomposition into the basis of holomorphic $(1,0)$ forms $\varpi_a$ of a single variable. Since each Szeg\"o kernel is $SL(2,\CC)$ invariant, so is $\SZ$ and therefore $\LL_\delta$ is a tensor under $SL(2,\CC)$ whose $z_i$-independence will become clear from Lemma \ref{6:lem.4} below. This completes the proof for even $n\geq 4$. The degenerate case of $n=2$, however, does not admit any distinction or cancellation between 
terms with $\sigma(2)=2$ and $\sigma(n)=2$, and we find a double pole in $\SZ (1,2)
= - S_\delta(1,2)^2$ as $z_1 \rightarrow z_2$.

\subsection{$SL(2,\CC)$ building blocks of symmetrized cyclic products}

Henceforth, we shall assume that $n=2m \geq 4$ is even with $m \in \NN$. In the hyper-elliptic representation, this object takes the following form,
\bea
\SZ (1,2, \cdots, n) & = & \Lambda_\delta(1,2,\cdots, n) \, \prod_{i=1}^n { dx_i \over  s_i} 
\no \\
\Lambda_\delta(1,2,\cdots, n) & = & {1 \over 2^n \, n!} \sum _{\sigma \in \mS_n} 
{N_\delta \big (\sigma (1), \sigma (2), \cdots, \sigma (n) \big ) \over (x_{\sigma(1)} - x_{\sigma(2)})  \cdots (x_{\sigma(n)} - x_{\sigma(1)})}
\label{deflamb.1}
\eea
where we have introduced $N_\delta$ earlier, and repeat it here for convenience, 
\bea
N_\delta (1,2, \cdots, n) & = & \prod _{i=1}^{n} \Big ( s_A(i) s_B(i{+}1) + s_B(i) s_A(i{+}1) \Big )
\eea
The following lemma gives a simplification in terms of the polynomials $Q_\delta$ introduced earlier,
\bea
Q_\delta(i_1, \cdots, i_m | j_1, \cdots, j_m)
& = & s_A(i_1)^2 \cdots s_A(i_m)^2 \, s_B(j_1)^2 \cdots s_B(j_m)^2
\no \\ &&
+ s_B(i_1)^2 \cdots s_B(i_m)^2 \, s_A(j_1)^2 \cdots s_A(j_m)^2
\eea

{\lem The function $\Lambda_\delta$ admits the following expression as a rational function of $x_1, \cdots, x_n$ in terms of the polynomials $Q_\delta$ where $n=2m$ is even, 
\bea
\Lambda_\delta (1,2, \cdots, n) & = &  \sum _{\sigma \in \mS_n} 
{Q_\delta \big (\sigma (1), \sigma (3), \cdots, \sigma (n{-}1) \big | \sigma(2) , \sigma(4), \cdots, \sigma(n) \big )  
\over 2^n \, n! \, (x_{\sigma(1)} - x_{\sigma(2)})  \cdots (x_{\sigma(n)} - x_{\sigma(1)})}
\label{deflamb.2}
\eea}

With the above arrangement of arguments, the polynomial $Q_\delta$ shares the cyclic 
symmetry of the Parke-Taylor factor.
It is also invariant under permutations of the left set of $m$ arguments and the right set of $m$  arguments separately, but these symmetries are not shared by the Parke-Taylor factor. 

\sm

The function $\Lambda_\delta$ may be decomposed into permutation sums of the form, 
\bea
\FF[a_1, \cdots, a_n ](x_1, \cdots x_n)
= \sum _{\sigma \in \mS_n} { x_{\sigma(1)}^{a_1} \cdots x_{\sigma(n)} ^{a_n} \, 
\over (x_{\sigma(1)} - x_{\sigma(2)})  \cdots (x_{\sigma(n)} - x_{\sigma(1)})}
\label{defofF}
\eea
Introducing the symmetric degree-$m$ polynomials $\kappa^{(n)}_m$ in
$n$ variables $x_1,\cdots,x_n$ which are at most linear in each $x_i$,
\bea
\mux _0 & = & 1
\no \\
\mux_1 & = & x_1 + \cdots + x_n
\no \\
\mux_2 & = & x_1 x_2 + \cdots + x_{n-1} x_n
\no \\
& \cdots &
\no \\
\mux_n & = & x_1 \cdots x_n
\eea
we  list the transformation properties of $F$ under $SL(2,\CC)$,
\bea
\BL \, \FF[a_1, \cdots, a_n]  & = & 
\sum_{i=1}^n a_i \FF[a_1, \cdots, a_i{-}1, \cdots a_n] 
\notag \\
\cS \, \FF[a_1, \cdots, a_n]  & = & ( \mux_n)^{-1} \, 
\FF[3{-}a_1, \cdots, 3{-}a_n] 
\label{6.SLactual}
\eea
We shall now establish the following lemma, valid for $0 \leq a_i \leq 3$, which is the only range of the exponents required here since the degrees of the polynomials $s_A^2$ and $s_B^2$ is three. 

{\lem 
\label{6:lem.4}
The function $\FF[a_1, \cdots, a_n] (x_1, \cdots, x_n)$, for exponents $0 \leq a_i \leq 3$, and sum of exponents $N = a_1 + \cdots + a_n$ with $0 \leq N \leq 3n$ and $n\geq 4$ even, has the following properties, 
\begin{enumerate}
\itemsep=-0.07in
\item $\FF$ is a symmetric function of $x_1, \cdots, x_n$ which is homogeneous of degree $N-n$;
\item $\FF$ is holomorphic in $x_1, \cdots, x_n$;
\item $\FF$ vanishes for odd $n$ when the array $[a_1, \cdots, a_n]$ is invariant under $a_i \to a_{n-i}$ reversal;
\item $\FF$ vanishes for $N < n$;
\item $\FF$ vanishes for $N > 2n$;
\item For $n \leq N \leq 2n$ the function $\FF$ equals the symmetric polynomial $\mux_{N-n}$ times an integer that depends on the array $[a_1, \cdots, a_n ]$.
\end{enumerate}}

Item 1 holds by construction. To prove item 2, we reorganize the sum over permutations by first summing over the group of cyclic permutations $\mC_n$ of order $n$, 
\bea
\FF[a_1, \cdots, a_n ](x_1, \cdots x_n)
& = & \sum _{\sigma \in \mS_n/\mC_n} { \hat F [a_1, \cdots, a_n] (x_{\sigma(1)}, \cdots , x_{\sigma (n)})
\over (x_{\sigma(1)} - x_{\sigma(2)})  \cdots (x_{\sigma(n)} - x_{\sigma(1)})}
\no \\
\hat F [a_1, \cdots, a_n] (x_1, \cdots x_n) & = &
\sum _{\rho \in \mC_n} x_{\rho(1)}^{a_1} \cdots x_{\rho(n)} ^{a_n} 
\eea
We then proceed as for the proof of the holomorphicity of $\SZ$ around (\ref{holoprf}). To prove that $F$ has no poles in $x_i$, it suffices to prove that it has no poles in $x_1$ at $x_2$ in view of item 1. To do so we fix $\sigma (1)=1$, in which case the poles at $x_2$ arise from $\sigma(2)=2$ and $\sigma (n)=2$. Thus, the residue of the pole in $x_1$ at $x_2$ is given by,
\bea
{\rm Res}_{x_1=x_2}  \FF[a_1, \cdots, a_n ](x_1, \cdots x_n)
& = & \sum _{\sigma \in \mS_{n-2}} { \hat F [a_1, \cdots, a_n] (x_2, x_2, x_{\sigma(3)} , \cdots ,  x_{\sigma (n)})
\over (x_2 - x_{\sigma(3)})  \cdots (x_{\sigma(n)} - x_2)}
 \\ && 
- \sum _{\sigma \in \mS_{n-2}} { \hat F [a_1, \cdots, a_n] (x_2, x_{\sigma(2)} , \cdots ,  x_{\sigma (n-1)}, x_2)
\over (x_2 - x_{\sigma(2)})  \cdots (x_{\sigma(n-1)} - x_2)}
\no
\eea
The cyclic property of $\hat F$ guarantees that the two sums on the right cancel one another for $n\geq 4$, which proves item 2. If $\hat F$ is also invariant under reversal of the ordering of the exponents, then we use the same argument as we did for the product of the Szeg\"o kernels to conclude that $\FF$ vanishes for odd $n$, which proves item 3.  To prove item 4, we use the fact that a holomorphic rational function $\FF$  of the $x_i$ which is  homogeneous of degree $N-n<0$ must vanish since $F$ admits a Taylor expansion at $x_i=0$. Item 5 then follows from item 4 by using the action (\ref{6.SLactual}) of inversion.  Finally, to prove item 6, we use the fact that $F$ is at most of degree 1 in each variable $x_i$. To see this, it suffices to fix all variables but $x_i$ and let $x_i \to \infty$. The numerator is of degree at most 3 in $x_i$ since $a_i \leq 3$, and the denominator is always of degree 2. Hence, $\FF$ grows at most linearly in $x_i$ as $x_i \to \infty$. Since $F$ is a symmetric polynomial in the $x_i$, at most linear in each $x_i$ and of degree $N-n$, it must be proportional to $\mux_{N-n}$.

\subsection{Procedure of evaluation}

To evaluate the functions $\Lambda_\delta$ in (\ref{deflamb.1}) and (\ref{deflamb.2}), for any given even $n$ it suffices to evaluate either the highest or lowest weight components, $\LL_\delta^{2 \cdots 2}$ or $\LL_\delta ^{1 \cdots 1}$, respectively, which are related to one another by  inversion. Consider the case of highest weight $\LL_\delta^{1 \cdots 1}$ which is the component of degree zero in each one of the $x_i$. Using  Lemma \ref{6:lem.4}, we see that this component must arise entirely from those contributions in $Q_\delta$ that are of total combined degree $n$ in the variables $x_i$, namely from linear combinations of the following permutation sums,
\bea
\FF[a_1, \cdots, a_n](x_1, \cdots , x_n) 
\hskip 1in 
a_1 + \cdots + a_n=n
\eea
But this function evaluates to a combination that is, in fact, independent of the $x_i$. Thus, to evaluate it, we may take advantage of a convenient choice of the $x_i$, such as $x_j = j$ for $j=1,\cdots, n$. This renders evaluation essentially straightforward. Alternatively, to compute $\LL_\delta ^{2 \cdots 2}$ directly, instead of by inversion from $\LL_\delta ^{1 \cdots 1}$, we may retain in $Q_\delta$ those contributions that are of total combined degree $2n$ in the variables $x_i$. The result of summation against the Parke-Taylor factor will then produce terms that are all proportional to the product $x_1 \cdots x_n$. One may again set $x_j=j$ for all $j=1,\cdots, n$ to evaluate $\LL_\delta ^{2 \cdots 2}$ upon dividing by an extra factor of $n!$ in order to account for the evaluation of the factor  $x_1 \cdots x_n$. 

\sm 

We do not have a closed formula for arbitrary values of $n$, but it is possible to evaluate this quantity for low values of $n$, which we consider to be even.

\subsubsection{Evaluating $\Lambda_\delta$ for $n=4$}

For $n=4$, the function $\Lambda_\delta$ takes the form, 
\bea
\Lambda_\delta(1,2,3,4) & = &  { s_A(1)^2 s_B(2)^2 s_A(3)^2 s_B(4)^2  + (A \leftrightarrow B)
\over 2^4 \, 3! \, x_{12} x_{23} x_{34} x_{41}}  + {\rm perm }(2,3,4)
\eea
where the sum is over all the six permutations of $2,3,4$. To compute $\LL_\delta^{2222}$, it suffices to retain in the numerator only the terms that have combined degree 8 in the $x_i$. Upon summation over all permutations,  they will give a term proportional to $\mux_4=x_1x_2x_3x_4$, and provide the component $\LL_\delta ^{2222}$. {\sc maple} gives,
\bea
2^4 \, 3! \, \LL_\delta ^{2222} & = & 6 \phi_2^2 -12 \mu_0 \phi_4 + 4 \mu_1 \phi_3 - 8 \mu_2 \phi_2 + 4 \mu_0 \mu_4- 2 \mu_1 \mu_3 + 2 \mu_2^2
\no \\
2^4 \, 3! \, \LL_\delta ^{1111} & = & 6 \phi_4^2 -12 \mu_6 \phi_2 + 4 \mu_5 \phi_3 - 8 \mu_4 \phi_4 + 4 \mu_2 \mu_6- 2 \mu_3 \mu_5 + 2 \mu_4^2
\eea
It is readily verified that $\BL \LL^{2222}_\delta=0$, and that the two entries are related to one another by inversion, as expected.   The remaining components of $\LL_\delta^{a_1 a_2 a_3 a_4}$ may be obtained from $\LL_\delta ^{1111}$ by successively applying $\BL$. Eliminating the $\phi_m$ and $\mu_m$ variables in favor of $\see^{ab} _\delta$ via (\ref{4.see}) as well as the tensors $\CM_1$ and $\CM_2$ in (\ref{MXcomps}) and (\ref{5.M2}), the expression for all the components of the tensor $\LL_\delta$ becomes,
\bea
\LL_\delta^{a_1a_2a_3a_4} =  \see_\delta ^{(a_1a_2} \see_\delta ^{a_3a_4)} - \half \CM_1^{a_1a_2a_3a_4 b_1b_2} \, \see_\delta ^{c_1c_2} \ep _{b_1c_1} \ep_{b_2c_2}- \half \CM_2^{a_1a_2a_3a_4}
\label{symcyc.4}
\eea

\subsubsection{Comparison with earlier reductions of $C_\delta$ at $n=4$}

The holomorphicity of the symmetrized cyclic product is not manifest from the decomposition of $C_{\delta}(1,2,3,4)$ in (\ref{Cdel4}). We shall now pinpoint the non-trivial identities between $\CM_w, Z(a,b)$ and rational functions of $x_{ij}$ needed to establish the agreement of (\ref{Cdel4}) with the expression for the tensor $\LL_\delta^{a_1a_2a_3a_4}$ in (\ref{symcyc.4}),
\bea
\LL_\delta^{a_1a_2a_3a_4} \varpi_{a_1}(1) \varpi_{a_2}(2) \varpi_{a_3}(3) \varpi_{a_4}(4)
= \frac{1}{3} \big[ C_\delta(1,2,3,4)  +  C_\delta(1,2,4,3)  +  C_\delta(1,3,2,4) \big] \quad \ 
\label{matchsy.01}
\eea
where we have used the reflection symmetry $C_\delta(1,2,3,4)= C_\delta(1,4,3,2)$ to simplify
the symmetrization (\ref{defsymmcyc}) to three terms.

\sm

In order to match the contributions to (\ref{matchsy.01}) linear in $\see_\delta$, 
one needs to demonstrate that,
\begin{align}
&- \frac{1}{2} \bx_{1a_1}\bx_{2a_2}\bx_{3a_3}\bx_{4a_4}
\CM_1^{a_1a_2a_3a_4 b_1b_2} \, \see_\delta ^{c_1c_2} \ep _{b_1c_1} \ep_{b_2c_2}
\notag \\
&= \frac{1}{6} \bigg\{ \frac{  Z(2,3) \bL_\delta(1,4)  +  Z(1,4) \bL_\delta(2,3)
-    Z(1,3) \bL_\delta(2,4)-   Z(2,4) \bL_\delta(1,3)   }{x_{12} x_{34}}+ {\rm cycl}(2,3,4) \bigg\}
\notag \\
&\quad +  \frac{1}{6} \bigg\{ \frac{  Z(1,3) \bL_\delta(2,4) {-}   Z(2,3) \bL_\delta(1,4)   }{x_{31} x_{12}} +  \frac{  Z(1,3) \bL_\delta(2,4) {-}  Z(1,2) \bL_\delta(3,4)    }{x_{13} x_{32}} + {\rm cycl}(1,2,3,4) \bigg\} \notag \\
&= \frac{Z(1,2) \bL_\delta(3,4) x_{34}^2 + Z(3,4) \bL_\delta(1,2) x_{12}^2 }{6 x_{13} x_{32} x_{24} x_{41}} + {\rm cycl}(2,3,4)
\label{matchsy.02}
\end{align}
where the components $\bx_{ia_i}$ of the two-vectors $(1,-x_i)^t$ in (\ref{defbx})
arise from $\varpi_{a_i}(i)=\bx_{ia_i} \frac{ dx_i }{s_i}$. The differences within the numerators
in the second and third line ensure that the residues of the poles in $x_{ij}$ cancel.
Upon isolating the coefficients of $\see_\delta$, this is equivalent to,
\begin{align}
\CM_1^{ab c_1 c_2 c_3 c_4} \bx_{1c_1}\bx_{2c_2}\bx_{3c_3}\bx_{4c_4}
= -  \frac{ \bx_3^{(a} \bx_4^{b)} x_{34}^2 Z(1,2) + \bx_1^{(a} \bx_2^{b)} x_{12}^2 Z(3,4) }{3 x_{13} x_{32} x_{24} x_{41}} + {\rm cycl}(2,3,4)
\label{matchsy.03}
\end{align}
which we have verified via {\sc mathematica}.

\sm 

The contributions to (\ref{matchsy.01}) independent on $\see_\delta$ in turn agree if,
\begin{align}
& \bx_{1a_1}\bx_{2a_2}\bx_{3a_3}\bx_{4a_4}
\CM_2^{a_1a_2a_3a_4} = \frac{1}{6} \bigg\{ \frac{  Z(1,3) Z(2,4)  }{x_{12} x_{23} x_{34} x_{41}}+ 
{\rm cycl}(2,3,4)\bigg\}
\notag \\
&= \frac{1}{6} \bigg\{ \frac{  Z(1,3) Z(2,4) -  Z(1,4) Z(2,3)  }{x_{12} x_{23} x_{34} x_{41}}+ 
 \frac{    Z(1,2) Z(3,4)- Z(1,4) Z(2,3)  }{x_{13} x_{32} x_{24} x_{41}} \bigg\}  
\label{matchsy.04}
\end{align}
which is readily established via {\sc mathematica} as well.

\sm

Finally, the contributions to (\ref{matchsy.01}) bilinear in $\see_\delta$ are readily seen to match, which completes our comparison of the two different expressions for $\SZ (1,2,3,4) $.

\subsubsection{Evaluating $\Lambda_\delta$ for $n=6$}

For $n=6$, the function $\Lambda_\delta$ takes the form,
\bea
\LL_\delta(1,2,3,4,5,6) & = &  { s_A(1)^2 s_B(2)^2 s_A(3)^2 s_B(4)^2 s_A(5)^2 s_B(6)^2  + (A \leftrightarrow B)
\over  2^6 \, 5! \,  x_{12} x_{23} x_{34} x_{45} x_{56} x_{61}}  
\no \\ &&
+ {\rm perm}(2,3,4,5,6)
\label{denomins}
\eea
where the sum is over all the 120 permutations of $2,3,4,5,6$. To compute $\LL_\delta^{222222}$, it suffices to retain in the numerator only the terms that have combined degree 12 in the $x_i$. Upon summation over all permutations, this will give a term proportional to $\mux_6=x_1x_2x_3x_4x_5x_6$, and provide the component $\LL_\delta ^{222222}$. {\sc maple} gives,
\bea
2^6 \, 5! \,  \LL_\delta^{222222}
& = & 120 \mu_0 \phi_2 \phi_4 -120 \mu_0 \phi_3^2 
+ \big ( 120 \mu_0 \mu_2 -80 \mu_1^2 \big )  \phi_4  
+ \big ( 32 \mu_1 \mu_2 + 72 \mu_0 \mu_3 \big )  \phi_3 
\no \\ && 
 + \big ( 72 \mu_4 \mu_0 -48 \mu_1 \mu_3 + 16 \mu_2^2 \big ) \phi_2 
 +96 \mu_0^2 \mu_6 -16 \mu_0 \mu_1 \mu_5 
\no \\ &&
-56 \mu_0 \mu_2 \mu_4 + 8 \mu_1^2 \mu_4 - 24 \mu_0 \mu_3^2 + 24 \mu_1 \mu_2 \mu_3 -16 \mu_2^3
\eea
We have chosen a final form for $\LL_\delta^{222222}$ in which no terms cubic or trilinear in $\phi_2, \phi_3,\phi_4$ occur. This is done by eliminating $\phi_2^3$ in favor of terms of order at most two using the trilinear equation for $\phi_2^3$.  Note that we have included the factors of $\mu_0=1$ to render the relation homogeneous, and allow for its immediate  inversion,
\bea
2^6 \, 5! \,  \LL_\delta^{111111}
& = & 120 \mu_6 \phi_2 \phi_4 -120 \mu_6 \phi_3^2 
+ \big ( 120 \mu_4 \mu_6 - 80 \mu_5^2 \big )  \phi_2  
+ \big ( 32 \mu_4 \mu_5 + 72 \mu_3 \mu_6 \big )  \phi_3
\no \\ && 
  + \big ( 72 \mu_2 \mu_6 -48 \mu_3 \mu_5 + 16 \mu_4^2 \big ) \phi_4 
  +96 \mu_0 \mu_6^2 -16 \mu_1 \mu_5 \mu_6 
\no \\ &&
-56 \mu_2 \mu_4 \mu_6 + 8  \mu_2 \mu_5^2 - 24 \mu_6 \mu_3^2 + 24 \mu_3  \mu_4 \mu_5 - 16 \mu_4^3
\eea
Converting $\phi_m$ into $\see_\delta$, and $\mu_m$ into $\CM_w$ tensors, we obtain, 
\bea
\LL_\delta^{a_1 \cdots a_6 }
& = & 
{ 1 \over 4} (\det \see _\delta) \, \CM_1^{a_1 \cdots a_6 } 
+{45 \over 56} \, \CM_2^{(a_1 \cdots a_4 } \see _\delta ^{a_5a_6) } 
+{3 \over 2}  \, \CM_2^{a_1 \cdots a_6 b_1b_2} \see_\delta ^{c_1c_2}  \ep_{b_1 c_1} \ep_{b_2 c_2}
\no \\ &&
+{27 \over 160} \, \CM_2 \, \CM_1^{a_1 \cdots a_6 }  
- { 15 \over 16} \, \CM_3^{a_1 \cdots a_6 }
\label{symcyc.6}
\eea

\subsubsection{Evaluating $\Lambda_\delta$ for $n=8$}

For $n=8$, the function $\Lambda_\delta$ takes the form, 
\bea
\Lambda_\delta(1,\! 2,\! 3,\! 4,\! 5,\! 6,\! 7, \! 8) 
& = &
  { s_A(1)^2 s_B(2)^2 s_A(3)^2 s_B(4)^2 s_A(5)^2 s_B(6)^2 s_A(7)^2 s_B(8)^2  + (A \leftrightarrow B)
\over  2^8 \, 7! \,  x_{12} x_{23} x_{34} x_{45} x_{56} x_{67} x_{78} x_{81} }  
\no \\ && 
+ {\rm perm }(2,3,4,5,6,7,8)
\eea
where the sum is over all the $7!$ permutations of $2,3,4,5,6,7,8$. To compute $\LL_\delta^{11111111}$ it suffices to retain in the numerator only the terms that have combined degree 8 in the $x_i$. Upon summation over all permutations, this will give a term independent of $x_i$ and provide the component $\LL_\delta ^{11111111}$. {\sc maple} gives, 
\bea
2^8 \, \cdot 7!  \, \LL_\delta ^{11111111}
& = & 
\big (4032 \mu_2 \mu_6  + 336 \mu_4^2  -1008 \mu_3 \mu_5 \big )  \phi_4^2 
+ \big ( 672 \mu_4 \mu_5   - 6048 \mu_3 \mu_6  \big ) \phi_3 \phi_4
\no \\ &&
+ \big (280 \mu_5^2    + 3360 \mu_4 \mu_6   \big ) \phi_3^2 
+ \big (6720 \mu_4 \mu_6  -1960 \mu_5^2  \big ) \phi_2 \phi_4
\no \\ &&
-10080 \mu_5 \mu_6 \phi_2 \phi_3 
+15120 \mu_6^2 \phi_2^2 
\no \\ &&
+ \big (3888 \mu_3^2 \mu_6   -6336 \mu_2 \mu_4 \mu_6   -200 \mu_2 \mu_5^2  - 17280 \mu_6^2  
- 608 \mu_4^3 
\no \\ && \hskip 0.6in
 +1392 \mu_3 \mu_4 \mu_5 + 2400 \mu_1 \mu_5 \mu_6  \big ) \phi_4
 \no \\ &&
+ \big ( 2880 \mu_1 \mu_6^2  +5088 \mu_2 \mu_5 \mu_6 -2304 \mu_3 \mu_4 \mu_6 -312 \mu_3 \mu_5^2 - 480 \mu_4^2 \mu_5 \big ) \phi_3
\no \\ &&
+ \big ( 1176 \mu_4 \mu_5^2 +9072 \mu_3 \mu_5 \mu_6  -18144 \mu_2 \mu_6^2  - 4032 \mu_4^2 \mu_6   \big ) \phi _2 
\no \\ &&
-288 \mu_1 \mu_3 \mu_6^2
-1088 \mu_1 \mu_4 \mu_5 \mu_6 
-360 \mu_1 \mu_5^3 
+5328 \mu_2^2 \mu_6^2 
\no \\ &&
-5232 \mu_2 \mu_3 \mu_5 \mu_6
+ 2368 \mu_2 \mu_4^2 \mu_6 
+ 264 \mu_2 \mu_4 \mu_5^2 
+ 384 \mu_3^2 \mu_4 \mu_6 
\no \\ &&
+ 176 \mu_3^2 \mu_5^2 
- 544 \mu_3 \mu_4^2 \mu_5 
+ 272 \mu_4^4
-960 \mu_4 \mu_6^2
+ 4720 \mu_5^2 \mu_6
\eea
which can be converted to the following tensorial expression:
\begin{align}
\label{symmszg.05} 
 \LL_\delta ^{a_1\ndots a_8} = &
{-}\frac{5}{8} (\det \see_\delta)  \CM_2^{a_1 a_2 \ndots a_8}
+  \frac{3}{16} \CM_1^{b_1 b_2(a_1 a_2 a_3 a_4} \CM_1^{a_5 a_6 a_7 a_8) b_3 b_4}
 \see_\delta^{c_1 c_2}  \see_\delta^{c_3 c_4}
 \varepsilon_{b_1 c_1} \varepsilon_{b_2 c_2}
  \varepsilon_{b_3 c_3} \varepsilon_{b_4 c_4} 
\no \\ & 
+ \frac{3}{2} \CM_2^{b_1 b_2 ( a_1 a_2 a_3 a_4 a_5 a_6}  \see_\delta^{a_7 a_8)}  \see_\delta^{c_1 c_2}
 \varepsilon_{b_1 c_1} \varepsilon_{b_2 c_2}  
 + \frac{3}{7} \CM_2^{( a_1 a_2 a_3 a_4 } \see_\delta^{a_5 a_6} \see_\delta^{a_7 a_8)}
  \notag \\
&  -\frac{43}{56}\CM_3^{(a_1 a_2 \ndots a_6}   \see_\delta^{a_7 a_8)} 
-\frac{3}{56} \CM_2^{b_1 b_2(a_1 a_2 } \CM_1^{a_3 a_4 \ndots   a_8)} \see_\delta^{c_1 c_2} 
 \varepsilon_{b_1 c_1} \varepsilon_{b_2 c_2}
\no \\ &
+  \frac{1}{12} \CM_2 \CM_1^{(a_1 a_2 \ndots a_6}   \see_\delta^{a_7 a_8)} 
+ \frac{3}{56} \CM_2^{(a_1 a_2 a_3 a_4} \CM_1^{a_5 a_6 a_7 a_8)b_1 b_2}   \see_\delta^{c_1 c_2} 
 \varepsilon_{b_1 c_1} \varepsilon_{b_2 c_2} \notag \\
 & 
-\frac{27}{448} \CM_2 \CM_2^{a_1 a_2 \ndots a_8 }
 +\frac{55}{448} \CM_2^{(a_1 a_2a_3 a_4 } \CM_2^{a_5 a_6 a_7 a_8) } 
   -\frac{1}{28} \CM_3^{(a_1 a_2 } \CM_1^{a_3 a_4 \ndots a_8) }  
 \end{align}

\newpage

\subsubsection{Evaluating $\Lambda_\delta$ for $n=10$}

The strategy of the previous sections leads to the following
expression for the ten-point instance of the $ \LL_\delta$ 
tensor in (\ref{defltensor})
\begin{align}
 \LL_\delta^{a_1 a_2 \ndots a_{10}} &=
\frac{19}{144} \CM_2 \CM_1^{(a_1 \ndots a_6} \see_\delta^{a_7 a_8} \see_\delta^{a_9 a_{10})}
+ \frac{5}{16} \CM_1^{(a_1 \ndots a_6} \see_\delta^{a_7 a_8} \CM_2^{a_9 a_{10}) b_1 b_2}
  \see_\delta^{c_1 c_2}   \varepsilon_{b_1 c_1} \varepsilon_{b_2 c_2}
 \notag \\
&\! \! \! \! \! \! \! \! \! \! \! \! \! \! 
 -\frac{25}{42} \CM_3^{(a_1 \ndots a_6} \see_\delta^{a_7 a_8} \see_\delta^{a_9 a_{10})}
+  \frac{25}{112} \CM_2^{(a_1 a_2 a_3 a_4} \see_\delta^{a_5 a_6} \CM_1^{a_7\ndots a_{10}) b_1 b_2}
  \see_\delta^{c_1 c_2}   \varepsilon_{b_1 c_1} \varepsilon_{b_2 c_2} \notag \\
 &\! \! \! \! \! \! \! \! \! \! \! \! \! \!+  \frac{5}{16}  \CM_1^{(a_1 \ndots a_6}\CM_2^{a_7\ndots a_{10})b_1 b_2 b_3 b_4} 
  \see_\delta^{c_1 c_2}   \see_\delta^{c_3 c_4} 
   \varepsilon_{b_1 c_1} \varepsilon_{b_2 c_2}
  \varepsilon_{b_3 c_3} \varepsilon_{b_4 c_4} \notag \\
&\! \! \! \! \! \! \! \! \! \! \! \! \! \!+ \frac{5}{8} \CM_2^{(a_1 \ndots a_8} \CM_1^{a_9 a_{10}) b_1 b_2 b_3 b_4}
  \see_\delta^{c_1 c_2}   \see_\delta^{c_3 c_4} 
   \varepsilon_{b_1 c_1} \varepsilon_{b_2 c_2}
  \varepsilon_{b_3 c_3} \varepsilon_{b_4 c_4}
 +  \frac{15}{448} (\det   \see_\delta)
 \CM_2^{(a_1 \ndots a_4} \CM_1^{a_5\ndots a_{10}) } \notag \\
&\! \! \! \! \! \! \! \! \! \! \! \! \! \!+ \frac{1}{36} \CM_2 \CM_1^{(a_1 \ndots a_6} \CM_1^{a_7\ndots a_{10} )b_1 b_2}
\see_\delta^{c_1 c_2}   \varepsilon_{b_1 c_1} \varepsilon_{b_2 c_2}
+ \frac{5}{32} \CM_2^{(a_1\ndots a_4} \CM_2^{a_5\ndots a_{10})b_1 b_2}
\see_\delta^{c_1 c_2}   \varepsilon_{b_1 c_1} \varepsilon_{b_2 c_2} \notag \\
&\! \! \! \! \! \! \! \! \! \! \! \! \! \! -\frac{15}{224}  \CM_2^{(a_1\ndots a_8} \CM_2^{a_9  a_{10})b_1 b_2}
\see_\delta^{c_1 c_2}   \varepsilon_{b_1 c_1} \varepsilon_{b_2 c_2}
- \frac{395}{1344}  \CM_3^{(a_1 \ndots a_6} \CM_1^{a_7\ndots a_{10} )b_1 b_2}
\see_\delta^{c_1 c_2}   \varepsilon_{b_1 c_1} \varepsilon_{b_2 c_2} \notag \\
&\! \! \! \! \! \! \! \! \! \! \! \! \! \! -\frac{13}{48} \CM_2   \CM_2^{(a_1\ndots a_8}\see_\delta^{a_9 a_{10})}
+ \frac{1355}{2688} \CM_2^{(a_1\ndots a_4} \CM_2^{a_5\ndots a_8} \see_\delta^{a_9 a_{10})}
+  \frac{ 95}{1344} \CM_1^{(a_1 \ndots a_6} \CM_3^{a_7 a_8} \see_\delta^{a_9 a_{10})} \notag \\
&\! \! \! \! \! \! \! \! \! \! \! \! \! \! + \frac{155}{4608} \CM_2  \CM_2^{(a_1\ndots a_4} \CM_1^{a_5\ndots a_{10})}
 -\frac{1625}{5376} \CM_2^{(a_1\ndots a_4} \CM_3^{a_5\ndots a_{10})} \notag \\
&\! \! \! \! \! \! \! \! \! \! \! \! \! \! -\frac{25}{128} \CM_2^{(a_1\ndots a_8} \CM_3^{a_9  a_{10})} 
 -\frac{75}{448} \CM_4^{(a_1\ndots a_4} \CM_1^{a_5\ndots a_{10})}  
 \label{symmszg.06} 
 \end{align}
 where the tensor $\CM_4$ in the last line generalizing $\CM_2, \CM_3$ in
 (\ref{5.M2}), (\ref{5.M3}) is defined by (\ref{defm4ten}).

\newpage

\section{Hyper-elliptic form of the superstring measure}
\label{sec:F}
\setcounter{equation}{0}

In this appendix, we translate the genus-two superstring measure for even spin structures, which was derived in  the language of Riemann $\tet$-functions \cite{DP1, DP4}, into the language of the hyper-elliptic formulation used in this paper. We begin by reviewing key results on Riemann $\tet$-functions for arbitrary spin structures, and use the Thomae formulas to carry out the translation. The Riemann $\tet$-functions with arbitrary characteristics $\kappa$ are defined by, 
\bea
\tet [\kappa] ( \zeta | \Omega) 
= 
\sum _{n  \in \mathbb Z^2  } 
\exp \Big (i \pi (n+\kappa ')^t  \Omega (n+\kappa ') + 2\pi i (n+\kappa ')^t  (\zeta+  \kappa '') \Big ) 
\eea
where $\Omega $ takes values in the Siegel upper half space $\cH_2$, the two-component vector $\zeta \in \CC^2$ is often taken to live in the Jacobian variety of a genus-two surface, and $\kappa = \left [ \kappa', \kappa '' \right ]$ is an array of column matrices $\kappa, \kappa '' \in \{0,\thalf\}^2$.  The parity of the $\tet$-function, and the spin structure $\kappa$,  is defined to be the parity of the integer $4 \kappa ' \cdot \kappa ''$.   For every odd spin structure $\nu$, there exists a Riemann identity between the $\tet$-constants $\tet[\delta](0|\Omega)$ of even spin structures $\delta$, 
\begin{align}
\label{G.Riem}
\sum_{\delta } \langle \nu | \delta \rangle \tet[\delta](0|\Omega) ^4 = 0
\end{align}
where the pairing $\< \kappa |\lambda \> = \overline{ \<  \lambda | \kappa \> }$ between two arbitrary characteristics $\kappa = [ \kappa', \kappa'']$ and $\lambda = [ \lambda ', \lambda '']$ is given by,
\bea
\<\kappa |\lambda \> = \exp \big \{ 4 \pi i (\kappa ' \lambda '' - \lambda ' \kappa '') \big \}
\eea
For arbitrary half-integer characteristics $\kappa, \lambda$, needed here to represent spin structures, the pairing is symmetric and takes the possible value $\<\kappa |\lambda \> = \pm 1$.

\subsection{Modular transformations}
\label{sec:F.1}

Genus-two modular transformations form the group $Sp(4,\ZZ)$ which is defined as follows,
\bea
\mM= \left ( \bma A & B \cr C & D \ema \right ) \in Sp(4,\ZZ) \hskip 0.8in \mM^t \, \mJ \, \mM= \mJ \hskip 0.8in 
\mJ = \left ( \bma 0 & -I \cr I & 0 \ema \right ) 
\label{modtrfs}
\eea
Its action on the period matrix is given by, 
\bea
\Omega \to \tilde \Omega = ( A \Omega +B)(C \Omega +D)^{-1}
\eea
and on an arbitrary spin structure $\kappa$ may be found in \cite{fay},
\bea
\left ( \bma  \kappa '' \cr  \kappa ' \ema \right ) \to
\left ( \bma \tilde \kappa '' \cr \tilde \kappa ' \ema \right )
= \left ( \bma A & -B \cr -C & D \ema \right ) \left ( \bma  \kappa '' \cr  \kappa ' \ema \right )
+ \half \, {\rm diag} \left ( \bma AB^t \cr CD^t \ema \right )
\label{trfkappa}
\eea
The congruence subgroup $\Gamma(2)$, defined by,
\bea
\Gamma (2) = \big \{ \mM \in Sp(4,\ZZ) \hbox{ such that } \mM \equiv I \hbox{ (mod 2)} \big \}
\eea
is a normal subgroup of $Sp(4,\ZZ)$, whose quotient gives the following isomorphisms, 
\bea
Sp(4, \ZZ) / \Gamma (2) \approx Sp(4, \ZZ_2) \approx \mS_6
\eea
where $\ZZ_2$ is the cyclic group $\ZZ_2=\{ 0,1\}$ and $\mS_6$ is the permutation group on 6 elements.  
The group $\Gamma(2)$ leaves each spin structure invariant, while $Sp(4,\ZZ_2)$ acts transitively on the set of spin structures, transforming even into even and odd into odd spin structures. In fact, one may view this action of $\mS_6$ directly on the six odd spin structures, and then deduce the action on even spin structures by expressing each even spin structure as a partition of the six distinct odd spin structures into two subsets of three distinct odd spin structures each.

\sm

The following product of pairings forming a closed 3-cycle of arbitrary spin structures,
\bea
e(\kappa_1, \kappa_2, \kappa_3) = \< \kappa_1 |\kappa_2\> \< \kappa_2 |\kappa_3\>\< \kappa_3 |\kappa_1\>
\eea
is invariant under $Sp(4,\ZZ)$ and thus under $Sp(4,\ZZ_2)$. For an arbitrary triplet of distinct odd spin structures $\nu_1, \nu_2, \nu_3$, the product obeys $e(\nu_1, \nu_2, \nu_3) =-1$. While $Sp(4,\ZZ_2)$ acts transitively on even spin structures, its action on a triple $(\delta_1, \delta_2, \delta_3)$ of distinct even spin structures decomposes into two distinct orbits, referred to as a {\sl syzygous} triple when $e(\delta_1, \delta_2, \delta_3)=+1$  and an {\sl asyzygous} triple when $e(\delta_1, \delta_2, \delta_3)=-1$.

\sm

Finally, the $\tet$-function transforms as follows under $Sp(4, \ZZ)$, 
\bea
\label{tetmod}
\tet [\tilde \delta ] ( 0 | \tilde \Omega) & = &
\epsilon (\delta, \mM) \, \det (C\Omega + D)^{\half} \tet [\delta ](0| \Omega) 
\no \\
\tet [\tilde \delta ] ( 0 | \tilde \Omega)^8 & = & \det (C\Omega + D)^4 \, \tet [\delta ](0| \Omega)^8 
\eea
provided the half-integer characteristics $\delta $  transforms into $\tilde \delta$  according to (\ref{trfkappa}) (see \cite{Igusa}, page 85).  The factor  $\eps(\delta, \mM)$ is independent of $\Omega$ and satisfies $\eps(\delta, \mM)^8=1$. Its explicit expression is complicated and may be found in \cite{Igusa}, but will not be needed here. The last equation shows that $\tet [ \delta ] ( 0 | \Omega)^8$ is a Siegel modular form of weight $4$ under $\Gamma (2)$, while $\tet [ \delta ] ( 0 | \Omega)^4$ transforms as a Siegel modular form, up to a sign factor (also referred to as a multiplier system). In the Riemann identities (\ref{G.Riem}) this sign is compensated for a corresponding sign produced by the transformation of the pairing $\< \nu | \delta \>$ so that the system of Riemann identities transforms into itself under modular transformations.

\subsection{Igusa classification of $Sp(4,\ZZ)$ modular forms}

We define the following combinations of $\tet$-constants involving even spin structures \cite{Igusa},
\bea
\Psi_{4k} = \tfrac{1}{4}  \sum_{\delta} \tet[\delta]^{8k}
\hskip 1in
\Psi_{10} = \prod _\delta \tet [\delta]^2
\label{defpsifour}
\eea
as well as the combination, 
\bea
\Psi _6 = \tfrac{1}{4} \sum_{{\delta_1, \delta_2, \delta_3 \atop e(\delta_1, \delta_2, \delta_3)=1}} 
\sigma(\delta_1, \delta_2, \delta_3) \, \tet [\delta_1]^4 \, \tet[\delta_2]^4 \, \tet[\delta_3]^4 
\label{defpsi6}
\eea
where the sign factor $\sigma(\delta_1, \delta_2, \delta_3)$ are chosen to be consistent with modular transformations. 
The syzygous triplets are given in \cite{Igusa} and explicitly in appendix B of \cite{DHoker:2004qhf} in the basis of spin structures adopted from \cite{DP4}. The functions $\Psi_{4k}(\Omega) , \Psi_{10}(\Omega), \Psi_6(\Omega)$ are holomorphic in $\Omega$ and transform as follows under $Sp(4,\ZZ)$ modular transformations,
\bea
\Psi _{4k} (\tilde \Omega) & = & \det (C \Omega +D)^{4k} \, \Psi _{4k}(\Omega)
\no \\
\Psi _{10} (\tilde \Omega) & = & \det (C \Omega +D)^{10} \, \Psi _{10}(\Omega)
\no \\
\Psi _{6} (\tilde \Omega) & = & \det (C \Omega +D)^{6} \, \Psi _{6}(\Omega)
\eea
so that they are Siegel modular forms of weights $4k, 10$ and $6$, respectively.
It is well-know that $\Psi _8 (\Omega) = \Psi_4(\Omega)^2$ and that $\Psi_{10}(\Omega)$ is a cusp form, namely it vanishes on the separating degeneration. At weight 12, there are 3 linearly independent  modular forms, namely $\Psi_4^3, \Psi_6^2 , \Psi_{12}$. Igusa has shown that the space of $Sp(4,\ZZ)$ modular forms is a polynomial ring generated by $\Psi_4, \Psi_6, \Psi_{10},\Psi_{12}$ and a generator $\Psi_{35}$ whose square $\Psi_{35}^2$ is a polynomial in $\Psi_4, \Psi_6, \Psi_{10},\Psi_{12}$.

\subsection{The Thomae formulas}

The $\tet$-constants for spin structures given by a partition $\delta \equiv A \cup B$ of the branch points may be expressed in terms of the hyper-elliptic representation by \cite{fay,Mumford2}, 
\bea
\tet[\delta](0)^8 = (\det \sigma )^{-4} \, \prod _{i< j \in A} (u_i-u_j)^2 \prod _{k<l \in B} (u_k-u_l)^2 
\label{thetfourth}
\eea
The action of $Sp(4,\ZZ_2) \approx \mS_6$ is given on the $\tet$-constants by (\ref{tetmod}), while on the branch points it is given by permutations. The results are summarized in the following table, where the overall sign has been arbitrarily chosen to be $+$  for the first entry,
{\small \begin{align}
 (\det \sigma )^{2} \tet[\nu_1{+}\nu_2{+}\nu_3](0)^4 &= 
 + (u_1{-}u_2)(u_1{-}u_3)(u_2{-}u_3) \cdot (u_4{-}u_5)(u_4{-}u_6)(u_5{-}u_6) \notag \\
  (\det \sigma )^{2} \tet[\nu_1{+}\nu_2{+}\nu_4](0)^4 &= 
  + (u_1{-}u_2)(u_1{-}u_4)(u_2{-}u_4) \cdot (u_3{-}u_5)(u_3{-}u_6)(u_5{-}u_6) \notag \\
  (\det \sigma )^{2} \tet[\nu_1{+}\nu_2{+}\nu_5](0)^4 &= 
  + (u_1{-}u_2)(u_1{-}u_5)(u_2{-}u_5) \cdot (u_3{-}u_4)(u_3{-}u_6)(u_4{-}u_6) \notag \\
   (\det \sigma )^{2} \tet[\nu_1{+}\nu_2{+}\nu_6](0)^4 &= 
   + (u_1{-}u_2)(u_1{-}u_6)(u_2{-}u_6) \cdot (u_3{-}u_4)(u_3{-}u_5)(u_4{-}u_5) \notag \\
  (\det \sigma )^{2} \tet[\nu_1{+}\nu_3{+}\nu_4](0)^4 &= 
  - (u_1{-}u_3)(u_1{-}u_4)(u_3{-}u_4) \cdot (u_2{-}u_5)(u_2{-}u_6)(u_5{-}u_6) \label{sping2.8} \\
 (\det \sigma )^{2} \tet[\nu_1{+}\nu_3{+}\nu_5](0)^4 &= 
 - (u_1{-}u_3)(u_1{-}u_5)(u_3{-}u_5) \cdot (u_2{-}u_4)(u_2{-}u_6)(u_4{-}u_6) \notag \\
 (\det \sigma )^{2} \tet[\nu_1{+}\nu_3{+}\nu_6](0)^4 &= 
 - (u_1{-}u_3)(u_1{-}u_6)(u_3{-}u_6) \cdot (u_2{-}u_4)(u_2{-}u_5)(u_4{-}u_5) \notag \\
 (\det \sigma )^{2} \tet[\nu_1{+}\nu_4{+}\nu_5](0)^4 &= 
 - (u_1{-}u_4)(u_1{-}u_5)(u_4{-}u_5) \cdot (u_2{-}u_3)(u_2{-}u_6)(u_3{-}u_6) \notag \\
(\det \sigma )^{2} \tet[\nu_1{+}\nu_4{+}\nu_6](0)^4 &= 
- (u_1{-}u_4)(u_1{-}u_6)(u_4{-}u_6) \cdot (u_2{-}u_3)(u_2{-}u_5)(u_3{-}u_5) \notag \\
 (\det \sigma )^{2} \tet[\nu_1{+}\nu_5{+}\nu_6](0)^4 &= 
 - (u_1{-}u_5)(u_1{-}u_6)(u_5{-}u_6) \cdot (u_2{-}u_3)(u_2{-}u_4)(u_3{-}u_4) \notag 
 \label{sqrteq}
\end{align}}
The matrix $\sigma$ in these formulas was defined in (\ref{2.sigma}). We may now use these expressions to translate various modular  forms that are given by sums of products  $\tet$-constants into the hyper-elliptic formulation. For example, the Igusa cusp form $\Psi_{10}$ of \cite{Igusa} may be defined either as the discriminant of the curve, or as the product over all even $\tet$-functions squared,
\bea
\Psi_{10} = \prod _{\delta \, {\rm even}} \tet[\delta](0)^2 = (\det \sigma)^{-10} \prod _{i<j} (u_i-u_j)^2
\eea
As derived from the formulas for $\tet[\delta](0)^4$, the right-most expression would be determined only up to a sign. This sign may be fixed, however, by inspection of the various degenerations, and was determined  to be + in \cite{DP4}. This guarantees that  the right-hand side is positive for real values of the branch points $u_i$, as is the left side since the period matrix is then purely imaginary and the entries of $\sigma$ are real.

\subsection{Translating the superstring measure $\Upsilon_8$}

For an even spin structure $\delta$ with the following decomposition, 
\bea
\delta = \sum _{i \in A} \nu_i
\eea
the superstring measure  $\Upsilon_8[\delta]  = \tet[\delta](0)^4 \, \Xi_6[\delta] $, in the notations of \cite{DP4}, is given as follows in terms of Riemann $\tet$-constants, 
\bea
\Upsilon _8 [\delta]  & = & \sum _{i<j \in A} \< \nu_i |\nu_j \> \prod _{b \not= i,j} \tet[\nu_i+\nu_j+\nu_b](0)^4
\label{defUps}
\eea
The modular transformation law of  $\Upsilon _8[\delta](\Omega)$ is given by,
\bea
\Upsilon _8[ \tilde \delta]( \tilde \Omega) = \det (C \Omega+D )^8 \, \Upsilon _8[\delta](\Omega)
\eea
The quantity $\Upsilon _8[\delta]$ may be expressed uniquely  in terms of the hyper-elliptic representation: since the number of $\tet^4$ factors in each term is even the overall sign choice, that was made earlier in the translation,  drops out.  Denoting the partition of the branch points $u_1, \cdots, u_6$ corresponding to the spin structure $\delta$ by $A=\{ a_1, a_2, a_3 \} $ and $B = \{ b_1, b_2,  b_3\}$, 
\bea
\label{F.Ups}
(\det \sigma)^8 \, \Upsilon_8 [\delta] & = & 
(a_1-a_2)^4(a_2-a_3) (a_3-a_1) \prod_{1\leq i<j \leq 3} (b_i-b_j)^2 \prod _{k=1}^3 (a_1-b_k) (a_2-b_k) 
\no \\ &&
+\hbox{ cycl} (a_1, a_2, a_3)
\eea
Although the expression on the right side might appear to be asymmetrical in $A$ and $B$, it actually is invariant under swapping $A$ and $B$.

\newpage

\section{Higher-order spin structure sums}
\label{highM}
\setcounter{equation}{0}

This appendix generalizes the spin structure sums of section \ref{sec:SUSY} involving the
supersymmetric measure $\Upsilon_8[\delta]$ in (\ref{F.Ups}) to higher orders in $\see_\delta$.

\subsection{Six powers of $\ell_\delta$}
\label{highM.1}

Similar to the presentation of the spin structure sums
over four and five powers of $\see_\delta$ in (\ref{proj.02})
and (\ref{proj.03}), we project six powers of $\see_\delta$
into the irreducible representations ${\bf 1} \oplus {\bf 5} \oplus {\bf 9} \oplus {\bf 13}$
of $SL(2,\mathbb C)$. The four independent spin sums are given by
\begin{align}
\frac{1}{(\det \sigma)^2}  \sum_{\delta \ {\rm even}} \frac{  \Upsilon _8 [\delta] }{ \Psi_{10}}
\, (\det \see_\delta)^3 &=
\frac{189}{256}\,  \CM_2^2 - \frac{117}{64} \,  \CM_4
\label{proj.11} \\
\frac{1}{(\det \sigma)^2 }  \sum_{\delta \ {\rm even}}  \frac{ \Upsilon _8 [\delta] }{\Psi_{10}}
\, (\det \see_\delta)^2 \see^{(a_1 a_2}_\delta \see^{a_3 a_4)}_\delta  &= 
\frac{423}{32}\,  \CM_4^{a_1 a_2 a_3 a_4} + \frac{9}{4}\,  \CM_2 \CM_2^{a_1 a_2 a_3 a_4}
\notag
\end{align}
in terms of $\CM_4$ scalars and tensors defined in (\ref{defm4ten})
as well as
\begin{align}
&\frac{1}{(\det \sigma)^2}  \sum_{\delta \ {\rm even}} \frac{  \Upsilon _8 [\delta] }{ \Psi_{10}}
\, \det \see_\delta \, \see^{(a_1 a_2}_\delta  \see^{a_3 a_4}_\delta
 \see^{a_5 a_6}_\delta   \see^{a_7 a_8)}_\delta   \notag \\
 &\quad  = \frac{ 459}{256}\,  \CM_2 \CM_2^{a_1 a_2 \ndots a_8} 
 + \frac{243}{128} \,  \CM_2^{(a_1 a_2 a_3 a_4} \CM_2^{a_5 a_6 a_7 a_8)} + 
\frac{99}{64} \,  \CM_1^{(a_1 a_2 \ndots a_6} \CM_3^{a_7 a_8)}  \label{proj.12} \\
&\frac{1}{(\det \sigma)^2}  \sum_{\delta \ {\rm even}} \frac{  \Upsilon _8 [\delta] }{ \Psi_{10}}
 \, \see^{(a_1 a_2}_\delta \see^{a_3 a_4}_\delta \see^{a_5 a_6}_\delta 
 \see^{a_7 a_8}_\delta \see^{a_9 a_{10}}_\delta \see^{a_{11} a_{12})}_\delta \notag \\
 &\quad =
\frac{4185}{512} \,  
\CM_2^{(a_1 a_2  \ndots  a_8} \CM_2^{a_9 a_{10} a_{11} a_{12})}
 - \frac{855}{512} \, 
\CM_1^{(a_1 a_2  \ndots a_6} \CM_3^{a_7 a_8\ndots a_{12})}
 +  \frac{441}{512} \,  \CM_2
\CM_1^{(a_1 a_2  \ndots a_6} \CM_1^{a_7 a_8\ndots a_{12})}
\notag
 \end{align}

\subsection{Higher powers of $\ell_\delta$}
\label{highM.2}

The following spin structure sums illustrate the appearance of
the higher-weight scalar $\CM_6$ and the two-tensor $\CM_5^{ab}$ 
defined in (\ref{clean.23}), namely
\begin{align}
\frac{1}{(\det \sigma)^2 }  \sum_{\delta \ {\rm even}}  \frac{ \Upsilon _8 [\delta] }{\Psi_{10}}
\, (\det \see_\delta)^4 &=
\frac{405}{512} \CM_2^3 - \frac{639}{256}\CM_2\CM_4 + \frac{ 4833}{512} \CM_6
 \label{proj.13} \\
\frac{1}{(\det \sigma)^2}  \sum_{\delta \ {\rm even}} \frac{ \Upsilon _8 [\delta]  }{ \Psi_{10}}
\, (\det \see_\delta)^5 &=
\frac{14823}{16384}\, \CM_2^4 - \frac{9315}{1024}\, \CM_4 \CM_2^2  
+ \frac{ 64395}{1024}\, \CM_4^2 + \frac{125145}{4096}\, \CM_2\CM_6
\notag
\end{align}
as well as
\begin{align}
\frac{1}{(\det \sigma)^2 }  \sum_{\delta \ {\rm even}}  \frac{ \Upsilon _8 [\delta] }{\Psi_{10}}
\, (\det \see_\delta)^3\,  \see^{a b}_\delta &=
\frac{243}{64}\, \CM_5^{ab} + \frac{1269}{512}\, \CM_2 \CM_3^{ab}
 \label{proj.14} 
\end{align}

\end{document}